\documentclass[twoside,openright,12pt]{CUEDthesisPSnPDF1}
\usepackage{float}
\usepackage{appendix}
\usepackage{times}
\usepackage{comment}
\usepackage{color}
\usepackage{t1enc}
\usepackage{graphicx}
\usepackage{longtable}
\usepackage{xspace}\usepackage{psfrag}
\usepackage{amsmath, amssymb}
\usepackage{amscd,amsfonts,color,amsthm}
\usepackage{latexsym, graphicx, pstricks,rotating,enumerate}
\usepackage{wrapfig}\usepackage{framed,graphicx,xcolor}
\usepackage{caption}
\usepackage{tikz}
\usepackage{mhchem}
\usetikzlibrary{calc,positioning,fit,backgrounds}
\pgfdeclarelayer{background}
\pgfsetlayers{background,main}

\newcommand{\tr}{\mathrm{Tr}}

\usepackage{graphicx}
\usepackage{watermark}
\begin{document}
\renewcommand\baselinestretch{1.2}
\baselineskip=18pt plus1pt
\setcounter{secnumdepth}{3}
\setcounter{tocdepth}{3}
\frontmatter
\thispagestyle{empty}
\baselineskip=18pt
\begin{center}
{\Large \bf Generation, estimation, and protection of novel quantum states of spin systems} \\
\vspace*{1cm}
{\large{\bf Thesis}} \\
\vspace*{0.5cm}
{For the award of the degree of}\\
\vspace{0.5cm}
{\large{\bf DOCTOR OF  PHILOSOPHY}} \\
\vspace{0.25cm}
\end{center}
\vspace*{3cm}
\begin{tabular}{lp{6cm}l}
{{\it Supervised by:}}  &&
{{\it Submitted by:}} \\
\\
{\bf Prof. Arvind  \&} &&
{\bf Harpreet Singh} \\
{\bf Dr. Kavita Dorai} &&\\

\end{tabular}
\begin{center}
\vspace*{1.5cm}
\hspace*{0cm}
\end{center}
\vspace*{-1cm}
\begin{center}
\includegraphics{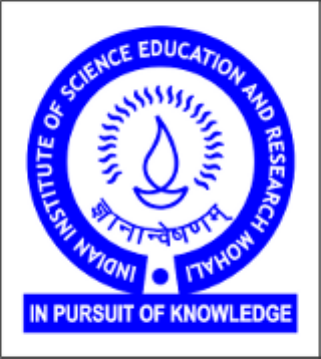}\\
{\bf Indian Institute of Science Education \& Research Mohali\\
Mohali - 140 306\\
India\\ (September 2017)
 }
\end{center}

 \newpage
\thispagestyle{empty}
\centerline{\Large \bf Declaration}
\vspace{0.25cm}
\par 
The work presented in this thesis has been carried
out by me under the guidance of
Prof. Arvind and Dr. Kavita Dorai at the Indian Institute of
Science Education and Research Mohali.

This work has not been submitted in part or in
full for a degree, diploma or a fellowship to
any other University or Institute. Whenever
contributions of others are involved, every effort
has been made to indicate this clearly, with due
acknowledgement of collaborative research
and discussions. This thesis is a bonafide record
of original work done by me and all sources
listed within have been detailed in the
bibliography.
\vspace*{1.5cm}

\hspace*{-0.25in}
\parbox{8in}{
\noindent {{\bf Harpreet Singh}}
}

\vspace*{0.125in}

\hspace*{-0.25in}
\noindent
\parbox{2.5in}{
\noindent Place~:  \\
\noindent Date~:
}

\vspace*{0.5in}

\noindent
In our capacity as supervisors of the
candidate's PhD thesis work, we certify that the
above statements by the candidate are true to the
best of our knowledge.

\vspace*{1.5cm}

\hspace*{-0.25in}
\parbox{8in}{
\noindent {{\bf Dr. Kavita Dorai} \hspace*{5.5cm}  {\bf Dr. Arvind}} \\
\noindent Associate Professor \hspace*{5.2cm} Professor \\
\noindent Department of Physical Sciences \hspace*{3cm} Department of Physical Sciences\\
\noindent IISER Mohali \hspace*{6.2cm}  IISER Mohali
}
\vspace*{0.125in}

\hspace*{-0.25in}
\noindent
\parbox{8in}{
\noindent Place~:  \hspace*{7.4cm} Place~: \\
\noindent Date~:   \hspace*{7.5cm} Date~:
}

\vspace*{0.8in}
\vspace*{1.5cm}

 \newpage
\thispagestyle{empty}
\centerline{\Large \bf Acknowledgements}
\vspace*{12pt}

\baselineskip=15pt

I owe a huge gratitude to my PhD advisors
Dr. Kavita Dorai and Prof. Arvind for teaching the
basics of NMR technique and Quantum Information Processing.
I must admit that without their help and support this thesis 
would not have been possible.

I would like to thank my doctoral committee
member Dr. Abhishek Chaudhuri for his useful advices.
I am grateful to Prof. N Sathyamurthy, the Director of IISER Mohali for 
providing all kind of help needed for my research work.
I am thankful to the faculty of the Department of
Physical Sciences of IISER Mohali for providing their
excellent guidance during my course work.

 I am grateful to the NMR and QCQI group IISER Mohali 
 for discussions and presentations.
 I would like to acknowledge the support from my group
 members: Satnam Singh, Navdeep Gogna,
 Amandeep Singh, Rakesh Sharma, Jyotsana Ojha, Jaskaran Singh,
 Chandan Sharma, Akash Sherawat, Akshay Gaikwad, Rajendera Singh Bhati, 
 Akansha Gautam, Amit Devra, Gaurav Saxena, Dileep Singh and 
 Aditya Mishra, Amrita Kumari and Matsyendra Nath Shukla.
 I am specially grateful to Ritabrata Sengupta, Debmayla Das and
 Shruti Dogra for the useful discussions and providing 
 all kind of support in my research work.

I owe my special thanks to Dr. Paramdip Singh Chandi 
for helping us troubleshooting the errors in the lab machines.
At the same time, I am thankful to Dr B. S. Joshi
(Application, Bruker India) for his technical support and guidance.

I have benefited a lot from discussions with Gopal Verma,
Anshul Choudhary, Vivek Kohar, Kanika Pasrija and Satyam Ravi 
during course work. I am grateful to my hostel 
friends Preetinder Singh, Bhupesh Garg,
Nayyar Aslam, George Thomas, Navnoor Kaur Saran and 
Junaid Khan because of whom I enjoyed my hostel life. 
 
I would like to thank my sisters Hardeep Kaur,
Harveen Kaur and cousin Jasjeet Singh for 
their immense support and encouragement.
I would also like to thank my
brother-in-law  Amandeep Singh for his continuous
encouragement and guidance. I bow my head to
my parents and grandparents for their generous invaluable blessings.

I am obliged to NMR research facility of IISER Mohali for providing
me the platform for carrying out my research.
During this thesis, various experiments were performed
on 600 MHz Bruker NMR spectrometers using QXI, TXI and BBI probeheads.

 I would like to acknowledge the financial support from
 Council of Scientific \& Industrial Research (CSIR)
during my PhD. I would like to thank IISER Mohali for providing fund
for participation in EUROMAR 2015, held in Prague Czech Republic.

Finally, thanks to my wonderful niece Divjot Kaur for keeping the
enthusiasm and positivity alive, which was required for my PhD completion.

\vspace{1cm}

\rightline{\bf \large{Harpreet Singh}}

\newpage
 \thispagestyle{empty}
 \begin{center}
\end{center}
\chapter{Abstract}
This thesis deals with the generation, estimation and 
preservation of novel quantum states of two and three 
qubits on an NMR quantum information processor. Using the 
maximum likelihood ansatz, a method has been developed for 
state estimation such that the reconstructed density matrix 
does not have negative eigenvalues and the errors are within 
the space of valid density operators. Due to interactions 
with the environment, unwanted changes occur in the system, 
leading to decoherence. Controlling decoherence is one of 
the biggest challenges to be overcome to build quantum 
computers. To decouple the quantum system from its environment,
several experimental strategies have been used. These 
strategies are based on our knowledge of
system-environment interaction and states that need to be 
preserved. Considering the first case, where the system state
is known but there is no knowledge about its interaction 
with the environment. To tackle decoherence in this case,
the  super-Zeno scheme is used and its efficacy to preserve quantum states
is demonstrated. The next situation considered is that where only the subspace to 
which the system state belongs is known. To address such a situation, 
the nested Uhrig dynamical decoupling scheme has been used. The later part of the thesis
deals with situations where the state of 
the system as well as its interaction with the environment is known. 
In such situations, since the noise model is known, 
decoupling strategies can be explicitly designed to cancel 
this noise. Using these decoupling strategies, the lifetime
of time-invariant discord of two-qubit Bell-diagonal states 
has been  experimentally extended. The decay of three-qubit entangled 
states namely the GHZ state, the W state and the $\rm W\bar{W}$ state are studied,
and the noise model is constructed for the spin system.
The experimentally observed and theoretical expected entanglement decay rates of these states are compared.
Then, the dynamical decoupling scheme is applied to these states and remarkable protection is observed
in the case of the GHZ state and the $W\bar{W}$ state.

The contents of the thesis have been divided into seven chapters whose brief account is sketched below:
\subsubsection*{Chapter 1}
This chapter provides an introduction to the field 
of NMR quantum computation and quantum information as 
well as the motivation for the present thesis. In 
addition to the basics of NMR and quantum computation, 
recent developments in the field of quantum computation 
and quantum information are discussed.

\subsubsection*{Chapter 2}
This chapter describes the utility of the maximum likelihood (ML) estimation
scheme to estimate quantum states on an NMR quantum information processor. 
Various separable and entangled states of two and three qubits 
are experimentally prepared, and 
the density matrices are reconstructed using both the ML estimation scheme as well as standard quantum
state tomography (QST). Further, an entanglement parameter is defined to
quantify multiqubit entanglement and entanglement is estimated using both the QST and the
ML estimation schemes.

\subsubsection*{Chapter 3}
This chapter experimentally demonstrates the freezing of evolution of
quantum states in one- and two-dimensional subspaces of two
qubits on an NMR quantum information processor. The state
evolution is frozen and leakage of the state from its
subspace to an orthogonal subspace is successfully
prevented using super-Zeno sequences. The super-Zeno scheme
comprises a set of radio frequency (rf) pulses, punctuated by
pre-selected time intervals. The efficacy of
the scheme is demonstrated by preserving different types of states,
including  separable and maximally entangled states in one-
and two-dimensional subspaces of a two-qubit system. The changes in
the experimental density matrices are tracked by carrying
out full state tomography at several time points.  
For the one-dimensional case, the fidelity measure is used and for the two-dimensional case,
the leakage (fraction) into the orthogonal subspace is used
as a qualitative indicator to estimate the resemblance of the
density matrix at a later time to the initially prepared
density matrix. For the case of entangled states, an
entanglement parameter is computed additionally to indicate the presence of
entanglement in the state at different times.  
The experiments demonstrate that the super-Zeno scheme is able to
successfully confine state evolution to the one- or two-dimensional
subspace being protected.
\subsubsection*{Chapter 4}
In this chapter, the efficacy of a three-layer
nested Uhrig dynamical decoupling (NUDD) sequence to
preserve arbitrary quantum states in a two-dimensional
subspace of the four-dimensional two-qubit Hilbert space
is experimentally demonstrated on an NMR quantum 
information processor. The effect of the state preservation
is studied first on four known states,
including two product states and two maximally entangled
Bell states. Next, to evaluate the preservation capacity of
the NUDD scheme, it is applied to eight randomly generated
states in the subspace. Although, the preservation of
different states varies, the scheme on the average performs
very well. The complete tomographs of the states at
different time points are used to compute fidelity. The state
fidelities using NUDD protection are compared with those
obtained without using any protection.
\subsubsection*{Chapter 5}
The discovery of the intriguing phenomenon that certain
kinds of quantum correlations remain impervious to noise up
to a specific point in time and then suddenly decay, has
generated immense recent interest. In this chapter, dynamical
decoupling sequences are exploited to prolong the persistence of
time-invariant quantum discord in a system of two NMR qubits
decohering in independent dephasing environments. 
Noise channels affecting the considered spin system of the molecule are characterized
and each spin of the spin system is mainly affected by 
the independent phase damping channel.
Bell-diagonal quantum states are experimentally prepared on 
a two-qubit NMR processor, and
robust dynamical decoupling schemes are applied for
state preservation. It is demonstrated that these schemes are
able to successfully extend the lifetime of time-invariant quantum discord.

\subsubsection*{Chapter 6}
This chapter demonstrates the experimental 
protection of different classes of tripartite entangled states,
namely the maximally entangled GHZ and W states 
and the ${\rm W \bar{W}}$ state, using
dynamical decoupling. The states are created on a three-qubit NMR quantum information
processor and allowed to evolve in the naturally noisy NMR environment.
The tripartite entanglement is monitored at each time instant during state
evolution, using negativity as an entanglement measure.  It is observed that the
W state is the most robust while the GHZ-type states are the most fragile against the
natural decoherence present in the NMR system. The ${\rm W \bar{W}}$ state
which is in the GHZ-class, yet stores entanglement in a manner akin to the W
state, surprisingly turns out to be more robust than the GHZ state.  
The experimental data are best modeled by considering the
main noise channel to be an uncorrelated phase damping channel acting
independently on each qubit, along with a generalized amplitude damping channel.
Using dynamical decoupling, a significant protection of entanglement for GHZ state is achieved. 
There is a marginal improvement in
the state fidelity for the W state (which is already robust against
natural system decoherence), 
while the ${\rm W \bar{W}}$ state shows a significant improvement 
in fidelity and protection against
decoherence.
\subsubsection*{Chapter 7}
This chapter provides some general remarks on the problems
covered in the thesis. Possible future 
applications of the state protection techniques used
in this thesis and the new avenues of research they open up are described. 
The overall contribution of this thesis in the 
context of the study of decoherence and state preservation
techniques in quantum information processing, is summarized.
\newpage

\thispagestyle{empty}
\centerline{\LARGE \bf List of Publications}
\vspace{1cm}
\vspace*{12pt}
\begin{enumerate}
\addtolength{\itemsep}{12pt}

\item {\bf Harpreet Singh}, Arvind and Kavita Dorai. Experimental protection against evolution of states in a subspace via a super-Zeno
scheme on an NMR quantum information processor.
\href{http://journals.aps.org/pra/abstract/10.1103/PhysRevA.90.052329}{{Phys. Rev. A, 90, 052329 (2014)}}.

\item {\bf Harpreet Singh}, Arvind and Kavita Dorai. Constructing valid density matrices on an NMR quantum information processor via
maximum likelihood estimation.
\href{http://dx.doi.org/10.1016/j.physleta.2016.07.046}{{Phys. Lett. A, 380 3051 (2016)}}.

\item {\bf Harpreet Singh}, Arvind and Kavita Dorai. Experimental protection of arbitrary states in a two-qubit subspace by nested Uhrig dynamical decoupling. 
\href{https://link.aps.org/doi/10.1103/PhysRevA.95.052337}{{Phys. Rev. A, 95, 052337 (2017)}}.

\item {\bf Harpreet Singh}, Arvind and Kavita Dorai. Experimentally freezing quantum discord in a dephasing environment using dynamical decoupling.\href{http://iopscience.iop.org/article/10.1209/0295-5075/118/50001}{{ EPL, 118, 50001 (2017)}}.

\item Rakesh Sharma, Navdeep Gogna,{\bf Harpreet Singh}, Kavita Dorai. Fast profiling of metabolite mixtures using chemometric analysis of a speeded-up 2D heteronuclear correlation NMR experiment. \href{http://pubs.rsc.org/en/content/articlelanding/2017/ra/c7ra04032f#!divAbstract}{{RSC Adv., 7, 29860 (2017)}}.

\item {\bf Harpreet Singh}, Arvind and Kavita Dorai. Evolution of tripartite entangled states in a decohering environment
and their experimental protection using dynamical decoupling. \href{https://journals.aps.org/pra/abstract/10.1103/PhysRevA.97.022302}{{PhysRevA, 97, 022302 (2018)}}.

\item Amit Devra, Prithviraj Prabhu,{\bf Harpreet Singh} and Kavita Dorai. Efficient experimental
design of high-fidelity three-qubit quantum gates via genetic programming.\href{https://link.springer.com/article/10.1007/s11128-018-1835-8}{{Quantum Inf Process, 17, 67 (2018)}}.

\item {\bf Harpreet Singh}, Arvind and Kavita Dorai. Multiple-quantum relaxation in three-spin homonuclear and heteronuclear
systems: An integrated Redfield and Lindblad master-equation approach. {\bf (Manuscript in preparation 2017)}.

\end{enumerate}

\thispagestyle{empty}
\newpage
\thispagestyle{empty}
 \begin{center}
 \end{center}
 \tableofcontents
 \listoffigures
 \newpage
 \thispagestyle{empty} 
\pagebreak
 \listoftables
 \thispagestyle{empty}
 \mainmatter
\chapter{Introduction}\label{intro_chap1}
Quantum computing and quantum information is an area which
has grown tremendously over the past two decades; it
comprises the study and implementation of the information
processing tasks that can be efficiently performed using a
quantum mechanical system.  Quantum computers are able to
accomplish computational tasks which are not possible to
carry out on classical computers.  The encoding of $n$ bits
of classical information requires at least $n$ bits of
classical resources. However, because of the quantum
superposition principle, quantum mechanical systems can in
principle have a better encoding efficiency than classical
systems~\cite{nielsen-book-02,suter-book-04}.  In 1981, R.
Feynman proposed the idea of a `quantum computer' and showed
that a classical computer would experience an exponential
slowdown while simulating a quantum phenomenon, while a
quantum computer would not~\cite{feynmann-ijqi-1982}. In
1985, D. Deutsch, took Feynman's ideas further and defined
two models of quantum computation; he also devised the first
quantum algorithm.  One of Deutsch's ideas is that quantum
computers could take advantage of the computational power
present in many ``parallel universes'' and thus outperform
conventional classical algorithms~\cite{Deutsch97}. In 1994,
P. Shor demonstrated two important problems; the problem of
finding the prime factors of an integer, and the so-called
`discrete logarithm' problem, both of which could be solved
efficiently on a quantum
computer~\cite{nielsen-book-02,shor}. Shor's results clearly
indicate the power of quantum computers. Further in 1996, L.
Grover showed that a search algorithm for an unsorted
database on a quantum computer is quadratically faster then
its classical counterpart~\cite{grover-prl-97}.  The most
popular model of a quantum computer is based on qubits which
are two-level quantum systems, with a qubit being a basic
unit of quantum information. In 2000, D. P. DiVincenzo
proposed a list of requirements for the realization of an
actual quantum computer~\cite{divincenzo}: a scalable
physical system, ability to initialize the system to any
quantum state, a universal set of quantum gates that can be
implemented, qubit-specific measurement and sufficiently
long coherence times (relative to the gate implementation
times).

Till date, no quantum hardware completely fulfills these
criteria.  Several quantum computing experiments have been
performed using optical photons~\cite{barz-jpb-15}, optical
cavity~\cite{grangier-fp-00}, ion traps\cite{hanffer-pr-08},
superconducting qubits~\cite{devoret-sc-13}
nitrogen-vacancy centers~\cite{weber-pnas-10} and nuclear magnetic resonance
(NMR) techniques~\cite{serra-proc-12}. 
In an optical photon quantum computer,
the qubits are represented by the polarization of a photon.
The initial state is prepared by creating single photon
states by attenuating light.  Quantum gates are applied
using beam-splitters, phase shifters and nonlinear Kerr
media.  Measurement is done by detecting single photons
using a photomultiplier tube~\cite{nielsen-book-02}.  In
an optical cavity, qubits are represented by the
polarizations of a photon and initial state preparation is similar
to that of an optical photon quantum computer.  Quantum gates are
applied using beam-splitters, phase shifters and a cavity
QED system, comprised of a Fabry-Perot cavity containing a few
atoms, to which the field is coupled.  In trapped ion
quantum computers, ions are allowed to be cooled down to the extent
that their vibrational state is sufficiently close to having
zero photons and a qubit is realized by the hyperfine state of an
atom and lowest level vibrational modes of the trapped
atoms~\cite{walther-rpp-06}.  Quantum gates here are
constructed by applying laser pulses. Measurement is
done by measuring populations of hyperfine
states~{\cite{monz-prl-11}}. In superconducting quantum
computers, qubits are represented by the phase, charge and
flux qubits.  In the charge qubit, different energy levels
correspond to an integer number of Cooper pairs on a
superconducting island~\cite{averin-prl-92}.  In the
flux qubit, the energy levels correspond to different
integer numbers of magnetic flux quanta trapped in a
superconducting ring.  In the case of a phase qubit, the
energy levels correspond to different quantum charge
oscillation amplitudes across a Josephson junction, where
the charge and the phase are analogous to momentum and
position correspondingly of a quantum harmonic
oscillator~\cite{Barends-prl-13}.  Quantum gates are
implemented using microwave pulses.  The nitrogen-vacancy center 
(NV center) is a point defect in a diamond which offers access to
an isolated quantum system that can be controlled at room temperature.
The $^{14}$N NV$^-$ center has an electronic spin component of
$S = 1$ and a nuclear spin component of $I = 1$, 
which gives spin eigenvalue level count of nine.
Using these energy levels we can realize qubits.
Resonant microwave pulses allow full quantum control of the
state of the center. Measurement can be done using optical
and electrical detection methods. The NV center is of course affected by the
absolute temperature and temperature changes, it has useful properties at room 
temperature, which make it suitable for a range of applications such as 
quantum sensors at room temperature, including quantum 
computing~\cite{Schirhagl-2014,Hui2016,suter-pnmrs-17}.

In May 2016, IBM Corporation has placed a five-qubit quantum computer on 
the IBM Cloud to run algorithms and experiments, and explore tutorials and 
 simulations around what might be possible with quantum computing.
It is a universal five-qubit quantum computer based on superconducting transmon qubits. 
After 18 months, IBM has also brought online a five- and sixteen-qubit system for public access 
through the IBM Q experience and is currently working further for upgradation of qubits.
Recently, Intel has also announced a 49-qubit quantum chip. So quantum computer is 
no more a theoretical concept but now a physical computational machine and in the near
future will be ready to solve real problems~\cite{ibmq}.

This thesis uses NMR as a tool for performing quantum
information processing tasks. 
NMR quantum computing has provided a good testbed for implementing
various quantum information processing protocols. In NMR,
the chemical shifts of different spins are used to address
the spins individually in frequency space 
and 
external radio frequency
pulses are used for quantum control~\cite{oliveira-book-07,levitt-book-2008}.  For
quantum information processing we require pure quantum states.
However, an NMR spin system at room temperature is far
from pure, since the
separation between the spin energy levels is $\hbar \omega$
which much less than $k_{B}T$. Therefore
the initial state of an ensemble of nuclear spins is nearly
random. However, for performing computational tasks we can
initialize the system into a pseudopure
state~\cite{cory-physicad} which mimics a pure state. Using
radio frequency pulses and the couplings between the spins
any unitary operator can be implemented. Further, the
compensations of errors due to pulse imperfections and
offset error can be performed via numerically optimized
pulses using GRAPE and genetic
algorithms~\cite{tosner-jmr-09,glaser-book-07,manu-pra-12,manu-pra-14}
which make the NMR technique an excellent test bed for the
implementation of quantum
algorithms~\cite{vandersypen-review,jones-nature-98,dorai-pra-2000,dorai-pra-2001,stadelhofer-pra-2005,xio-pra-05},
quantum simulations~\cite{li-prx-17,du-prl-10,shankar-pla-14,anjusha-pla-16,peng-prl-09,zhang-prl-11}, the
study of decoherence~\cite{arindam-jmr-05,hegde-pra-14,kawamura-ijqc-06} and
many other quantum information processing applications~\cite{rao-pra-13,dogra-pla-16,katiyar-epl-16,chen-pra-11,criger4620,Jones-review,dorai-review,pan-pra-14,li-prl-15,ramanathan2004}.

In this thesis, we first tackle the problem of negative eigenvalues occurring
during the reconstruction of density matrix from the experimental data. We
experimentally prepared quantum states relevant for 
quantum information processing and reconstruct valid
state density matrices on an NMR quantum information processor of two and three
qubits. In NMR quantum information
processing~\cite{nielsen-book-02,oliveira-book-07}, information is encoded in
the quantum state of an ensemble of nuclei.  
Theoretically, reconstruction of the state
density matrix is possible if we have infinite copies of the spin system.
However, only a finite but large number of copies of the spin system are
available. Furthermore, due to experimental errors such as detection pulse
errors and temperature fluctuations, copies of the spin system are slightly
different~\cite{suter-review}.  
If not properly handled, it can lead to a situation where the standard
state tomography may give rise to an unphysical state.
To tackle this problem, we use the maximum likelihood
method~\cite{banaszek-pra-99,hradil-pra-97,kohout-prl-10,plenio-mle} which
always gives a valid state density matrix close to the experimental data and
resolves this issue of unphysical states~\cite{singh-pla-16}.  In the 
rest of the thesis, we focus on the different strategies to cancel out
system-environment interactions.  First we deal with a situation
where we are aware of the system state but have no knowledge about its
interaction with the environment. It is then required to consider all the
possible interactions by which system in a given state can interact with the
environment. We use the super-Zeno
scheme for state
protection~\cite{dhar-prl-06,mukhtar-pra-10-1,ting-chinese-09,singh-pra-14}, In
this scheme, we construct an inverting pulse 
which has information about the state and
use a train of these inverting pulses punctuated by unequal intervals of time to
protect the system state.  Then we consider a situation where only the subspace
is known to which system state belongs instead of the exact state and its
interaction with the environment, and to resolve this problem we use nested
Uhrig dynamical decoupling (NUDD)
schemes~\cite{mukhtar-pra-10-2,zhen-pra-16,singh-pra-17}.  The NUDD scheme
consists of nesting of protection layers to cancel all the possible
interactions that the state can have with
the environment.  We next move on to 
situations where we have knowledge of the state of the system as well as
its interaction with the environment. We 
study the evolution of the state of the system in the presence of intrinsic NMR
noise and then fit its decay to a noise model to characterize the noise.
In the two- and three-qubit systems 
studied, each qubit of the system is modeled as being 
affected by an independent phase and
amplitude damping noise channel~\cite{childs-pra-03}, with the noise being
dominated by the phase damping channel.  We use the Knill dynamical decoupling
(KDD$_{xy}$ ) scheme and XY16~\cite{souza-prl-11,souza-pra-12} dynamical
decoupling scheme to tackle the dephasing noise. These pulse sequences are
robust against pulse angle errors and offsets errors.  We apply KDD$_{xy}$  and
XY16 sequences on experimentally prepared two-qubit Bell-diagonal states
and see the effect on the lifetime of time-invariant
discord~\cite{mazzola-prl-10,singh-epl-17}.  We also apply these sequences on
experimentally prepared three-qubit GHZ, W and $\rm W\bar{W}$ 
states and observe the
decay of entanglement and its subsequent suppressing using
dynamical decoupling.

\section{Quantum computing and quantum information processing}
Although computational algorithms are conceived
mathematically, a computer which executes these
algorithms has to be a physical device.
The most common model of quantum computation is a generalization of the
classical circuit model known as quantum circuit model. A quantum circuit is an
instruction for carrying out the preparation of an input state, applying 
a set of
quantum gates which cause a unitary evolution and measuring the output state.
The input state is prepared on a quantum register, 
which is the quantum analog of the classical
processor register. A classical register of size of $n$ comprises of $n$ flip
flops which can have $2^n$ possible classical states.  A quantum register of
size $n$ comprises of $n$ two-level quantum systems which are interacting with
each other and due to superposition can have infinite possible states. 

\subsection{Quantum bit}
The basic unit of classical information is a bit.
Classical digital computers process information in a 
discrete form. It operates on data that are 
expressed in binary code i.e. 0 and 1. A bit can have 
two states either 0 or 1 and therefore it can be easily 
physically realized on a two-state device. Quantum computing and quantum information
 are built upon an analogous concept, the qubit i.e. quantum bit~\cite{nielsen-book-02}.
The two possible logical states for a qubit can
be $\vert0\rangle$ and $\vert1\rangle$ states.
However, the most general qubit state is given by:
\begin{equation}
 \vert \psi \rangle = \alpha \vert 0\rangle + \beta \vert 1 \rangle
 \label{psi1_chap1}
\end{equation}
The state of a qubit is a vector 
in a two-dimensional complex vector space;
$\vert0\rangle$ and $\vert1\rangle$ form 
the orthogonal basis for this vector space.
The complex numbers $\alpha$ and $\beta$ are such that 
$\vert\alpha\vert^2+\vert\beta\vert^2=1$.  
We cannot determine the values of $\alpha$ and 
$\beta$ by measurements on a single qubit.

We can rewrite Eq~\ref{psi1_chap1} as 
\begin{equation}
 \vert\psi\rangle= e^{i\gamma}(cos\frac{\theta}{2}\vert0\rangle+e^{i\phi} sin\frac{\theta}{2}\vert1\rangle)
\label{psi2_chap1}
 \end{equation}
\begin{center}
\begin{figure}[h]
\centering
\includegraphics[angle=0,scale=1.0]{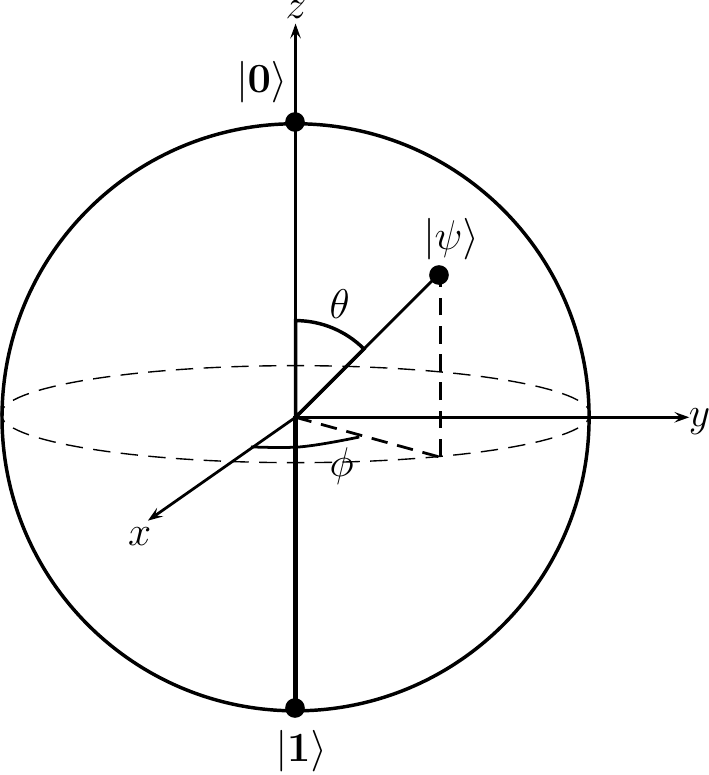}
\caption{Bloch sphere representation of a qubit.}
\label{blochsphere_chap1}
\end{figure}
\end{center}
with new real variables $\theta$, $\phi$, and $\gamma$.
Global phase $e^{i\gamma}$ can be ignored 
because it has no observable effects and we can write 
\begin{equation}
 \vert\psi\rangle= cos\frac{\theta}{2}\vert0\rangle+e^{i\phi} sin\frac{\theta}{2}\vert1\rangle
\label{psi3_chap1}
 \end{equation}
where $\theta$ and $\phi$ define a point on the unit three-dimensional sphere,
known as the
Bloch sphere, as shown in Fig.[\ref{blochsphere_chap1}]. 
Each of the infinite points on the surface of the sphere
corresponds to a state of the qubit.
\subsection{$N$-qubit quantum register}
A quantum register of size $n$ comprises of $n$ qubits 
which are interacting with each other. The most general
state of such a register is a superposition of $2^n$ basis elements which is given by,
\begin{equation}
 \vert\phi\rangle=\sum_j \alpha_j \vert \phi^1_{j}\rangle\otimes \vert \phi^2_{j} \rangle \otimes \dots \otimes \vert \phi^n_{j} \rangle
\end{equation}
where $\vert\phi^i_j\rangle$ refers to $i$th qubit in
$j$th term of the superposition, $\vert\phi^i_j\rangle\in \{\vert0\rangle,\vert1\rangle\}$ 
and $\alpha_j$ are the complex coefficients such that 
$\sum_j \vert \alpha_j \vert^2=1$. If a state $\vert \psi \rangle$
can be expressed as $ \vert \psi_1 \rangle \otimes \vert \psi_2 \rangle \otimes \dots \otimes \vert \psi_n \rangle$ 
where $\vert \psi_i \rangle =\alpha_i 
\vert 0 \rangle +\beta_i \vert 1 \rangle$, 
then the state is called separable and if not then it
is entangled.
Entanglement is an intrinsically quantum mechanical phenomenon 
and it plays a crucial role in various QIP protocols.
Experimental realization of quantum registers is 
one of the biggest challenges in building a quantum computer. 
Up to now only a few qubit quantum registers
have been physically realized. 
For instance in linear optics ten qubits~\cite{wang-prl-16},
in trapped ion fourteen qubits~\cite{monz-prl-11} and 
in NMR twelve qubits have been realized~\cite{Negrevergne-prl-06}.
\subsection{Density matrix representation}
The state of the system can not be reconstructed
if we have a single copy of a qubit. On a measurement of a 
qubit with state $\vert \psi\rangle =\alpha \vert0\rangle +\beta \vert 1 \rangle$
possible outcomes will be 0 or 1 with 
probability $\vert \alpha \vert^2$ and $\vert \beta \vert^2$, 
respectively. 
To compute probabilities $\vert \alpha \vert^2$ and $\vert \beta \vert^2$, 
we need either a large number of measurements on a qubit with repeated state preparation 
as is done in single-photon quantum computing or a simultaneous 
measurement of a large number of copies of the qubit as 
done in NMR quantum computing.
In an ensemble it may be possible that all the spins are in same state $\vert
\psi \rangle$ and this type of ensemble is called pure ensemble. It may be
possible that with $p_1$ probability spins are in $\vert \psi_1 \rangle$, with
$p_2$ probability spins are in $\vert \psi_2 \rangle$ and so on, this type of
ensemble is called a mixed ensemble.  The density matrix formulation is very
useful in describing the state of an ensemble quantum system such as an
ensemble of spins in NMR~\cite{macmohan-book-08}.

For an ensemble with the $p_i$ probability to be in $\vert\psi_i\rangle$ state,
the density operator is given as 
\begin{equation}
 \rho=\sum_i p_i \vert \psi_i \rangle\langle \psi_i \vert
\end{equation}
where $\sum_ip_i=1$. If all the members of an ensemble are in the same state $\vert\psi\rangle$ or for a pure ensemble,
the density operator is given as 
\begin{equation}
 \rho_{pure}=\vert\psi\rangle\langle \psi \vert
\end{equation}
A density operator $\rho$ has to satisfy three important properties: $\rho$ 
is Hermitian, i.e., $\rho=\rho^\dagger$, all the eigenvalues of 
the $\rho$ is positive, and $\tr[\rho]=1$. For a pure state ensemble $\tr[\rho^2]=1$
and for a mixed state ensemble $\tr[\rho^2] < 1$.
The most general state of a single qubit can be written as,
\begin{equation}
 \rho=\frac{I+\vec{r}.\vec{\sigma}}{2}
\end{equation}
where $I$ is a identity matrix, $\vec{r}$ is a three-dimensional 
Bloch vector with
$\vert\vec{r}\vert \le1$, $\vec{\sigma}={\sigma_x} \hat{x}+{\sigma_y} \hat{y}+{\sigma_z} \hat{z}$  and $\sigma_i$s are the Pauli matrices. All the pure states can be represented
as points on the surface of the Bloch sphere and 
mixed states are represented by points inside
the Bloch sphere.
\subsection{Quantum gates}
\label{theory_qg_chap1}
The building block of a digital circuit of a classical computer are 
the logic gates
for e.g. NOT, OR, NOR and NAND. 
The analogous building blocks of a quantum circuit are
quantum gates. 
Quantum gates being unitary, are reversible, as
opposed to the classical logic gates which may be
irreversible and hence dissipative. 
The action of quantum gates can be realized
by a unitary operator ${\rm U}$ (${\rm{UU}}^{\dagger}={\rm I}$).  
It has been shown
that a set of gates that consists of all one-qubit quantum gates $\rm[U(2)]$
and the two-qubit exclusive-OR gate is universal in the sense that all unitary
operations can be expressed as compositions of such
gates~\cite{barenco-pra-95}.  One such set of universal
quantum gates is the Hadamard gate (H), a phase rotation gate
$R(cos^{-1}(\frac{3}{5}))$ and a two-qubit 
controlled-NOT gate. 
Once a basis is chosen,
quantum
gates are represented as matrices.
The following are some of the
important quantum gates:\\

\noindent{\textbf{Hadamard gate}} \\
The Hadamard gate is a single-qubit gate 
and it maps the basis state $\vert0\rangle$ to
$\vert+\rangle=\frac{\vert0\rangle+\vert1\rangle}{\sqrt{2}}$ and $\vert1\rangle$ to
$\vert-\rangle=\frac{\vert0\rangle-\vert1\rangle}{\sqrt{2}}$.
It creates a superposition, which means that a measurement
will have equal probability to become either 1 or 0.
The matrix representation of Hadamard gate is:
\begin{equation}
 {\rm H}=\frac{1}{\sqrt{2}}\left(
\begin{array}{cc}
 1 & 1 \\
 1 & -1 \\
\end{array}
\right)
\end{equation}

\noindent{\textbf{Pauli-X gate (NOT gate)}} \\
The Pauli-X gate maps the basis state $\vert0\rangle$ to $\vert1\rangle$ and
$\vert1\rangle$ to $\vert0\rangle$. The matrix representation of Pauli-X gate is:
\begin{equation}
 {\rm X}=\left(
\begin{array}{cc}
 0 & 1 \\
 1 & 0 \\
\end{array}
\right)
\end{equation}

\noindent{\textbf{Pauli-Y gate}} \\
The Pauli-Y gate maps the basis state $\vert0\rangle$ to $i\vert1\rangle$ and
$\vert1\rangle$ to $-i\vert0\rangle$. The matrix representation of Pauli-Y gate is:
\begin{equation}
 {\rm Y}=\left(
\begin{array}{cc}
 0 & -i \\
 i & 0 \\
\end{array}
\right)
\end{equation}

\noindent{\textbf{Pauli-Z gate}} \\
The Pauli-Z gate leaves the basis state $\vert0\rangle$ unchanged and maps
$\vert1\rangle$ to $-\vert1\rangle$. The matrix representation of Pauli-Z gate is:
\begin{equation}
{\rm Z}=\left(
\begin{array}{cc}
 1 & 0 \\
 0 &-1 \\
\end{array}
\right)
\end{equation}

\noindent{\textbf{Square root of NOT gate ($\sqrt{\rm NOT}$)}} \\
The $\sqrt{\rm NOT}$ gate maps the basis state $\vert0\rangle$ to
$\frac{1}{2}\left((1+i)\vert0\rangle+(1-i)\vert1\rangle \right)$ and $\vert1\rangle$ to
$\frac{1}{2}\left((1-i)\vert0\rangle+(1+i)\vert1\rangle \right)$.
The matrix representation of square root of NOT gate is:
\begin{equation}
 \sqrt{\rm NOT}=\frac{1}{2}\left(
\begin{array}{cc}
 1+i & 1-i \\
 1-i & 1+i \\
\end{array}
\right)
\end{equation}

\noindent{\textbf{Phase shift gate}} \\
The phase shift gate leaves the basis state $\vert0\rangle$ unchanged and maps
$\vert1\rangle$ to $e^{i\phi}\vert1\rangle$. The matrix representation of 
the phase shift gate is:
\begin{equation}
 {\rm R_{\phi}}=\left(
\begin{array}{cc}
 1 & 0 \\
 0 &e^{i\phi} \\
\end{array}
\right)
\end{equation}

\noindent{\textbf{SWAP gate}} \\
The SWAP gate is a two-qubit gate which leaves the basis states $\vert00\rangle$ and $\vert11\rangle$ unchanged.
It maps $\vert01\rangle$ to $\vert10\rangle$ and $\vert10\rangle$ to $\vert01\rangle$.
The matrix representation of the SWAP gate is:
\begin{equation}
 {\rm SWAP}=\left(
\begin{array}{cccc}
 1 & 0 & 0 & 0 \\
 0 & 0 & 1 & 0 \\
 0 & 1 & 0 & 0 \\
 0 & 0 & 0 & 1 \\
\end{array}
\right)
\end{equation}

\noindent{\textbf{Controlled NOT gate}} \\
The controlled NOT (CNOT) gate is two-qubit gate 
which leave the basis state $\vert00\rangle$ and $\vert01\rangle$ unchanged. 
It maps $\vert10\rangle$ to $\vert11\rangle$ and $\vert11\rangle$ to $\vert10\rangle$.
The matrix representation of CNOT gate is:
\begin{equation}
{\rm CNOT}=\left(
\begin{array}{cccc}
 1 & 0 & 0 & 0 \\
 0 & 1 &0 & 0 \\
 0 & 0 & 0 & 1 \\
 0 & 0 & 1 & 0 \\
\end{array}
\right)
\end{equation}
\subsection{Quantum measurement}
The standard measurement schemes in quantum information and
quantum computation use projective measurements which is described
below. Later we will take up the issue of ensemble measurements on
an NMR quantum information processor, which are non-projective in nature.

Consider a quantum system in a pure state specified by
a vector $\vert \psi \rangle$ in an $n$-dimensional Hilbert space. 
Let us suppose that one performs a projective measurement of an 
observable $M$ on it. In the formalism of quantum mechanics,
associated with the observable $M$ is a Hermitian operator $\hat{M}$, where
$\vert m_1 \rangle$, $\vert m_2 \rangle$, \dots, $\vert m_n \rangle$
denote the eigenvectors of the operator $\hat{M}$ with $m_1$, $m_2$,\dots, $m_n$
as the respective eigenvalues. If the eigenvalue spectrum of the observable $M$ is
nondegenerate then
\begin{equation}
 \vert \psi \rangle=c_1\vert m_1 \rangle+ c_2 \vert m_2 \rangle+\dots +c_n\vert m_n \rangle \ \  \ {\text{with}} \sum_i \vert c_i\vert^2=1
\end{equation}
where $c_1$, $c_2$, \dots, $c_n$ are complex numbers. 
Upon a projective measurement of the observable $M$
on such a system, an outcome $m_i$ is obtained with a probability $\vert c_i\vert^2$ 
and the state of the system collapses to the corresponding eigenvector $\vert m_i \rangle$.

A projective measurement is described by a complete set of projectors
$\{\hat{\Pi}_{n} \}$  where $\hat{\Pi}_{n}=\vert m_n\rangle\langle m_n \vert$ with
$ \sum_{n}\hat \Pi_n^{\dagger}\hat \Pi_n=1$. If the state of the quantum system is
$\vert \psi \rangle$ immediately before the measurement
then the probability that $m$ occurs
is given by
\begin{equation}
 p(n)=\langle \psi \vert \hat\Pi_n^{\dagger}\hat\Pi_{n}\vert \psi \rangle, 
\end{equation}
and the state of the system after measurement is 
\begin{equation}
 \frac{\hat\Pi_n\vert\psi\rangle}{\sqrt{\langle \psi \vert \hat\Pi_n^{\dagger}\hat\Pi_{n}\vert \psi \rangle}}.
\end{equation}

For instance, 
consider the measurement of 
a qubit with state $\vert\phi\rangle=\alpha\vert0\rangle+\beta\vert1\rangle$ in the computational basis. The measurement 
is defined by the two measurement operators $\hat\Pi_{0}=\vert0\rangle\langle0\vert$ and 
$\hat\Pi_{1}=\vert1\rangle\langle1\vert$. The measurement operators are Hermitian i.e. 
$\hat\Pi_1^{\dagger}=\hat \Pi_1$ and $\hat \Pi_2^{\dagger}=\hat\Pi_2$. 
The probability of obtaining the measurement outcome 0 is 
\begin{equation}
 p(0)= \langle \phi \vert \hat\Pi_0^{\dagger}\hat\Pi_{0}\vert \phi \rangle=\langle\phi\vert0\rangle\langle0\vert0\rangle\langle0\vert\phi\rangle=|\alpha|^2 
\end{equation}
and the qubit state will collapse to $\vert0\rangle$. 
Similarly, the probability of obtaining the measurement
outcome 1 is p(1)=$|\beta|^2$ and the qubit state will 
collapse to  $\vert1\rangle$.

\section{Nuclear Magnetic Resonance}
\label{nmrtheory_chap1}
Nuclear magnetic resonance (NMR) 
describes a phenomenon wherein, an ensemble of nuclear spins
precessing in a static magnetic field, absorb and emit radiation
in the radiofrequency range in resonance with their
Larmor frequencies~\cite{oliveira-book-07}.
In the quantum  mechanical formalism,
the spin magnetization is a vector operator represented by $\hbar {I}$
where ${I}$ is a dimensionless operator 
representing the total angular momentum of the nuclear spin. 
Atomic nuclei with non-zero spin also possess 
a magnetic dipole moment $\mu$ which is given as
\begin{equation}
 \mu=\gamma_n \hbar {I},
\end{equation}
where $\gamma_n$ is called the gyromagnetic ratio of the nucleus, which is a fundamental property of the nucleus.
\begin{center}
\begin{figure}[h]
\centering
\includegraphics[angle=0,scale=1.0]{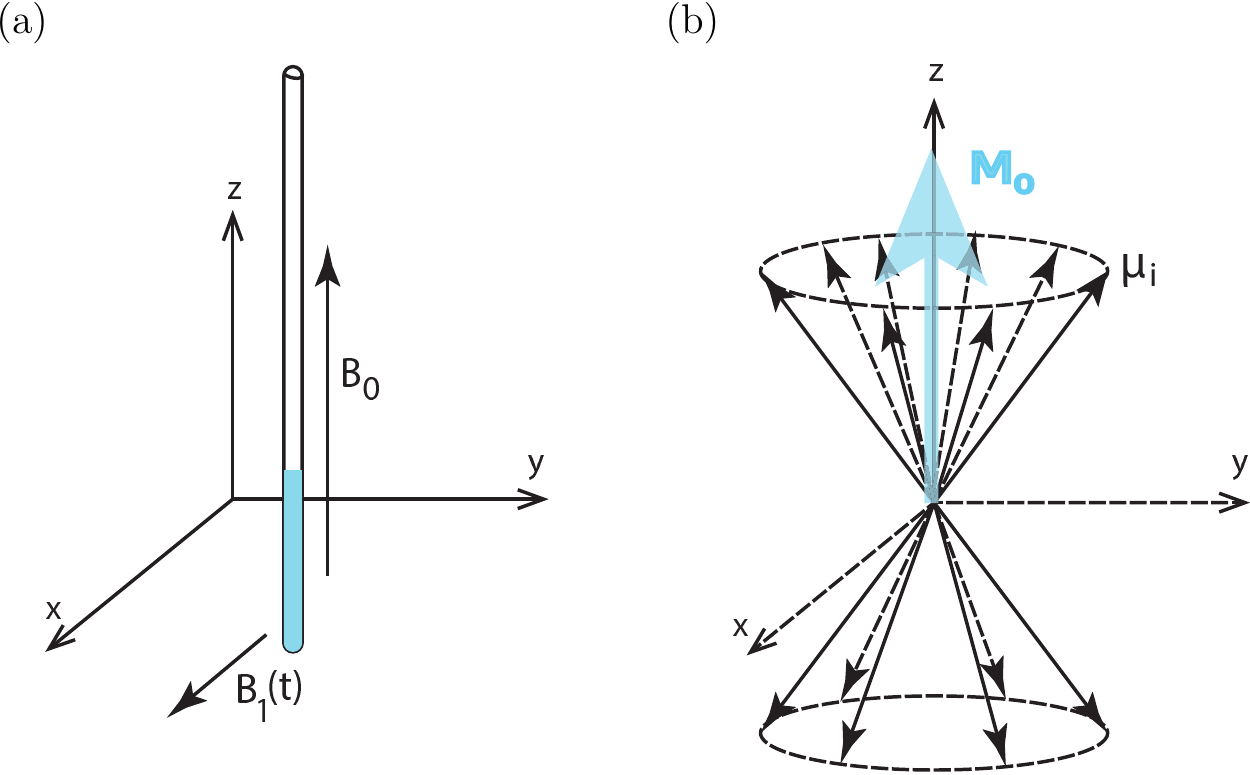}
\caption{(a) NMR tube with sample 
oriented in a strong static magnetic field $B_0$ 
along the $z$-axis and time-dependent magnetic field $B_1(t)$ along
the $x$-axis. (b) The number of spins precessing around
the direction parallel to the field are more than the
number antiparallel to the field direction, 
which creates a bulk magnetization $M_0$.}
\label{nmr_setup_chap1}
\end{figure}
\end{center}

A nuclear spin with ${I} \neq 0$ when placed in a magnetic field of strength
${B_0}$ applied along the $z$-axis 
precesses as shown in Fig.~\ref{nmr_setup_chap1}(a).  The
Hamiltonian of interaction between the spin and the magnetic field is given by,
\begin{equation}
 {H}=-{\mu}.{B_0}\hat{z}=-\gamma_n\hbar{ B_0}{I_z}=-\hbar \omega_{n}{I_z}
\end{equation}
The spins precess about the $z$-axis with 
a characteristic frequency called Larmor
frequency $\omega_n=-\gamma_n{B_0}$ (in rad s$^{-1}$) as shown in Fig.~\ref{nmr_setup_chap1}(b).
The magnetic field ${B_0}$ is applied along the $z$-direction 
and all the quantum operators
act in the subspace spanned by 
the magnetic quantum number 
$\vert m \rangle$ where $m=-{I},-{I}+1,\dots,{I}-1,{I}$.
Under the action of the Hamiltonian ${H}$, 
the expectation values of the angular momentum
operators in the plane perpendicular to
the $z$-direction i.e. $\langle{I_x}\rangle$ and $\langle{I_y}\rangle$ 
show oscillatory behavior with time, with a frequency $\omega_n$, 
whereas $\langle{I_z} \rangle$ is stationary.
The eigenvalues of the Hamiltonian ${H}$ are given by:
\begin{equation}
 E_m=-m\hbar \omega_n
\end{equation}
For a nucleus with spin $I$, there are $(2I+1)$ energy levels equally spaced by the amount $\hbar \omega_n$.

For an ensemble of identical nuclei which are not
perfectly isolated from the environment and surrounded by the lattice
is at a temperature $T$. The interactions of nuclei with the lattice   
lead to a thermal equilibrium state and the population of
each energy level in this state is given by the Boltzmann distribution. 
For a two-level system $I=\frac{1}{2}$, with the population $n_{-}$
and $n_{+}$ of the $m=-\frac{1}{2}$ and $m=\frac{1}{2}$ levels, respectively
\begin{equation}
 \frac{n_{-}}{n_{+}}=e^{-\hbar \omega_n/k_{B}T}
\end{equation}
where $k_{B}$ is the Boltzmann constant and T is
the absolute temperature of the ensemble.
The Boltzmann factor $e^{-\hbar \omega_n/k_{B}T}$ 
for protons ($^{1}{ H}$) in a magnetic field of 14.1 
Tesla at room temperature is very close to unity. 
The fractional difference of populations is about 1 part in $10^5$. 
This slight difference in the populations of $m=-\frac{1}{2}$ 
and $m=\frac{1}{2}$ levels cause the net magnetization along 
the $z$-direction. For $n$ spin-1/2 nuclei the thermal equilibrium 
magnetization is given by:
\begin{equation}
 {M_0}=\frac{\mu_0 \gamma_n^2 \hbar^2 {B_0}}{4k_{B}T}
\end{equation}
Since 
the Larmor frequency depends on the gyromagnetic ratio $\gamma_n$, each 
nucleus has its own characteristic Larmor frequency. Nuclear spins
in a molecule are surrounded by the electronic environment, which 
leads to 
shielding of the magnetic field, the so
called ``chemical shift'', 
with the effective magnetic field being given by
\begin{equation}
 {B}_{{\rm eff}}= {B_0}(1-\sigma_0)
\end{equation}
where $\sigma_0$ is the isotropic chemical shift tensor.

There are several terms in the nuclear spin Hamiltonian which encompass
different spin-spin interactions such as the scalar coupling term 
${{H_J}}$, the dipolar
coupling term ${H_{DD}}$, and the quadrupolar coupling 
term ${H_Q}$.  The scalar coupling
interaction ${{H_J}}$ arise from the hyperfine interactions between the nuclei
and local electrons. A pair of nuclei exhibit dipole-dipole interaction
${H_{DD}}$ by inducing local magnetic fields at the site of each other through
space. In an isotropic liquid at room temperature, molecules tumble very
fast, thus averaging the intramolecular dipolar coupling to zero. The
quadrupolar coupling ${H_Q}$ is exhibited by nuclei with spin $> 1/2$ which
possess an asymmetric charge distribution~\cite{keeler-book}.

\noindent{\bf Radio frequency field interaction and the resonance phenomenon}:-
The Larmor frequencies of the nuclear spins in a static magnetic field of a few
Tesla are of the order of MHz. The transition between the different spin states
can be induced by a radio frequency (rf) 
oscillating magnetic field~\cite{suter-book-04}.
\begin{equation}
\vec{B}_{rf}=2B_1cos(\omega_{rf}t+\phi)\hat{x},
\end{equation}
where $\omega_{rf}$ is the frequency of the magnetic field 
and $\phi$ is the phase. 

\begin{equation}
  H_{rf}=-\mu.\vec{B}_{rf}=-\gamma_n\hbar{I_x}\left(2B_1cos(\omega_{rf}t+\phi)\right)
\end{equation}
We can rewrite $\vec{B}_{rf}$ as a superposition of 
two fields rotating in opposite 
directions.
\begin{equation}
\label{brot}
 \vec{B}_{rf}=B_1(cos(\omega_{rf}t+\phi)\hat{x}+sin(\omega_{rf}t+\phi)\hat{y})+
 B_1(cos(\omega_{rf}t+\phi)\hat{x}-sin(\omega_{rf}t+\phi)\hat{y}),
\end{equation}
For the simplicity, we assume $\phi=0$ and analyze Eq. \ref{brot} in a coordinate system
that rotates around the static magnetic field at the frequency 
$\omega_{{rf}}$. In this rotating frame 
\begin{equation}
 \vec{B}_{rf}^{rot}=B_1 \hat{x}+ B_1(cos(2\omega_{rf}t)\hat{x}-sin(2\omega_{rf}t)\hat{y})
\end{equation}
We can observe that 
one of the two components is now static and the other is rotating at twice
the rf field frequency  (which can be neglected)~\cite{bloch-pr-1940}.
We 
can transform $ H_{rf}$ into rotating frame using the unitary operator
\begin{equation}
  U(t)=e^{i\omega_n t{I_{z}}/\hbar},
\end{equation}
\begin{center}
 $ H_{rf}^{rot}=U^{-1}H_{rf}U+ 
i\hbar \dot{U}^{-1}U=-\hbar(\omega_n-\omega_{rf}) 
{I_z}-\hbar\omega_1{I_x}$                                                       
\end{center}
where $\omega_1=\gamma_n B_1$. If the phase $\phi\ne0$ then 
\begin{equation}
 H_{rf}^{rot}=-\hbar(\omega_n-\omega_{rf}) {I_z}-\hbar\omega_1\{{I_x} cos\phi+ I_y sin\phi \}.                                                      
\end{equation}
The evolution of the quantum ensemble under the 
effective field in the rotating frame is described by
\begin{equation}
\rho_{rot}(t)=e^{-iH_{rf}^{rot} \ t}\rho_{rot}(0)e^{iH_{rf}^{rot} \ t},
\end{equation}
where $\rho_{rot}(0)$ is density matrix of state at time $t$.
\section{NMR quantum computing}
In 1997, D. G. Cory and I. L. Chuang independently proposed a
NMR quantum computer that can be programmed much like a quantum
computer~\cite{cory-pnas-97,gershenfeld-sc-97}. Their computational model uses
an ensemble quantum computer wherein the results of a
measurement are the expectation values of the observables.
This computational model can be realized by NMR
spectroscopy on macroscopic ensembles of nuclear spins.  
Several quantum algorithms have been implemented on an NMR quantum computer such as 
the Grover search algorithm~\cite{jones-nature-98}, realization of Shor
algorithm~\cite{vandersypen-nature-01}, implementation of the Deutsch-Jozsa
algorithm using noncommuting  selective pulses~\cite{dorai-pra-2000} and many
more till date. A qubit in an NMR quantum computer is realized by a spin-1/2
nucleus. 
The NMR spectrometer consists of a superconducting magnet which applies a high
magnetic field in the $z$-direction and rf coils for exciting the spins and
receiving the  NMR signal from the relaxing spin ensemble. When the sample is
placed in the magnetic field, the spins interact with the magnetic field, and
energy levels split depending upon the size of the spin system.  At room
temperature, these energy levels are populated according to the Boltzmann
distribution and thus the system is in a mixed state at thermal equilibrium.
This poses a difficult challenge for quantum computing, which requires pure
states as initial quantum states. This difficulty is circumvented in NMR
quantum computing by creating a ``pseudopure'' state as an initial state, which
mimics a pure state.  Using the rf pulses and interaction between the spins,
quantum gates are implemented and as a result of the computation the  NMR
signal was recorded which is an average magnetization the in $x$ and $y$
directions.  This signal is directly proportional to the expectation values of
some elements of the basis set of the qubits.  With the application of rf
pulses rotating individual spins, the expectation of all the elements in the
basis set can be calculated. From these expectation values, we can reconstruct
the density matrix. Further, recent developments in NMR in the area of
control of spin dynamics via rf pulses makes it possible to 
implement quantum gates for NMR quantum computing with high fidelities. A nuclear 
spin is well separated from its environment due to which 
it exhibits long coherence times. Even with all these merits, one major
limitation of liquid state NMR quantum computers is scalability.
In the following sections, state initialization,
implementation of quantum gates and measurement in NMR quantum computing are discussed.

\subsection{NMR qubits}
Consider an ensemble of $N$ spin-1/2 nuclei tumbling in a liquid and
placed in a magnetic field $B_0$. The Hamiltonian $H$ of this system is
given as
\begin{equation}
 H=-\omega_{0}I_z
\end{equation}
where $I_z=\sigma_z/2$. The eigenstate and eigenvalues of $H$ are
$\{\vert0\rangle,\vert1\rangle\}$ and $\{\omega_0/2,-\omega_0/2\}$ respectively.
The energy difference between the two 
levels is given by $\Delta E=\hbar\omega_0$
Hence such a two-level system acts as a single NMR qubit. 
For a system of $n$ interacting spins-1/2 in a magnetic field the
Hamiltonian is given by:
\begin{equation}
 H_0=\sum_{i=1}^n\omega_iI_z^i+2\pi\sum_{i<j}^nJ_{ij}I^i.I^j
\end{equation}
where $J_{ij}$ is the scalar coupling between the spins and $\omega_i$ is the
Larmor frequency. If $\vert \omega_i-\omega_j\vert >> 2\pi\vert j_{ij} \vert$ then the
NMR qubits are weakly coupled and the Hamiltonian for such a system is
\begin{equation}
 H_0=\sum_{i=1}^n\omega_iI_z^i+2\pi\sum_{i<j}^nJ_{ij}I_z^i.I_z^j
\end{equation}
\subsection{Initialization}
\label{chap1_Initialization}
Any QIP task begins by initializing the system into a pure state. In NMR QIP,
an $N$-qubit ensemble of spins at room temperature  has a
population distribution of energy levels given by  
 the Boltzmann distribution~\cite{nielsen-book-02}.
 All the energy levels  are almost equally populated and
the initial state is mixed. Under the high temperature
 approximation the initial state of the system is given by:
 \begin{equation}
  \rm \rho_{eq}\approx\frac{1}{2^N}(I+\epsilon \Delta \rho_{eq})
 \end{equation}
where $\rm I$ is 
an identity matrix of $\rm 2^N \times 2^N$, $\epsilon(\approx 10^{-5})$
is a purity factor and $\Delta \rho_{eq}$ is a deviation density matrix.
The problem of pure states in NMR can be overcome by
preparing a pseudopure state
which 
is isomorphic to a pure state~\cite{cory-pnas-97}.
An ensemble of a pure state is given by $\rho_{pure}=\vert\psi\rangle\langle\psi\vert$
and the corresponding pseudopure state is given by
\begin{equation}
 \rm \rho_{pps}=\frac{1-\epsilon}{2^N} I+\epsilon \vert\psi\rangle\langle\psi\vert
\end{equation}
\begin{center}
A pseudopure state in NMR can be prepared by several methods
such as spatial averaging, temporal averaging and logical labelling; 
all based on the idea of preparing  $\rm 2^N-1$ energy levels 
with equal population
and with one energy level being more  populated
than the other energy levels as shown 
for two qubits in Fig.{\ref{energy_level_chap1}}.
\begin{figure}[h]
\centering
\includegraphics[angle=0,scale=1]{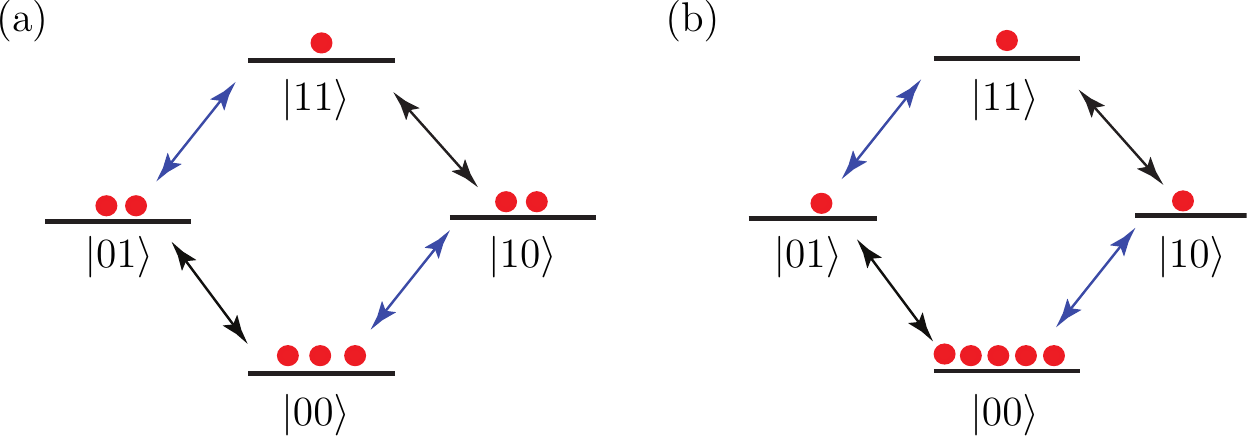}
\caption{Populations of energy levels of a
two-qubit system of a (a) thermal equilibrium state and
(b) a pseudopure state.}
\label{energy_level_chap1}
\end{figure}
\end{center}

To implement most quntum algorithms we need to create 
an entangled state. The power of quantum computers largely 
depends on entanglement. There has been a longstanding debate about 
the existence of entanglement in spin ensembles at high temperature 
as encountered in NMR experiments. There are two ways to look at the
situation. Entangled states in such ensembles are obtained via
unitary transformations on pseudopure states. If we consider
the entire spin ensemble, given that the number of spins that
are involved in the pseudopure state is very small compared to
the total number of spins, it has been shown that the overall
ensemble is not entangled~\cite{braunstein-prl-99,yu-pra-05}. However, one can take a
different point of view and only consider the subensemble of
spins that have been prepared in the pseudopure state, and
as far as these spins are concerned entanglement genuinely
exists~\cite{soares-pinto-proc-12,serra-proc-12}. The states 
that we have created in this thesis for our experiments are entangled in this sense, and hence 
may not be considered as entangled if one works with the entire 
ensemble. Therefore, one has to be aware and cautious about this aspect while dealing with these
states. These states are sometimes referred to as 
being pseudoentangled.

\noindent {\bf Temporal averaging technique} 
is based on the fact that quantum operations are
linear and the observables measured in NMR are traceless. 
Experimentally, the temporal
averaging scheme relies on adding the 
computational results of multiple experiments, where each 
experiment starts off with a different state preparation 
pulse sequence which permutes
the populations~\cite{oliveira-book-07}. For a two-spin system 
this technique begins with the
density matrix
\begin{center}
 $\rho_1=\left(
\begin{array}{cccc}
 p_1 & 0 & 0 & 0 \\
 0 & p_2 & 0 & 0 \\
 0 & 0 & p_3 & 0 \\
 0 & 0 & 0 & p_4 \\
\end{array}
\right)$
\end{center}
where $p_1$, $p_2$, $p_3$ and $p_4$ are populations of the normalized 
density operator $\rho_1$, with $\sum_{i=1}^4p_i=1$. $U_1$ and $U_2$
are operators constructed from controlled-NOT gates to obtain 
a state with the permuted populations:
\begin{center}
 $\rho_2=U_1.\rho_1.U_1^{\dagger}=\left(
\begin{array}{cccc}
 p_1 & 0 & 0 & 0 \\
 0 & p_3 & 0 & 0 \\
 0 & 0 & p_4 & 0 \\
 0 & 0 & 0 & p_2 \\
\end{array}
\right)$
\end{center}
and
\begin{center}
 $\rho_3=U_2.\rho_1.U_2^{\dagger}=\left(
\begin{array}{cccc}
 p_1 & 0 & 0 & 0 \\
 0 & p_4 & 0 & 0 \\
 0 & 0 & p_2 & 0 \\
 0 & 0 & 0 & p_3 \\
\end{array}
\right)$
\end{center}
Since the readout is linear with respect to the initial state,
all three permuted density matrices 
are added  to 
realize the pseudopure state 
$\rho = \rho_1 + \rho _2+ \rho_2$.  
\\
\begin{center}
 $\rho=\left(
\begin{array}{cccc}
 3p_1 & 0 & 0 & 0 \\
 0 & p_2+p_3+p_4 & 0 & 0 \\
 0 & 0 &p_2+p_3+p_4 & 0 \\
 0 & 0 & 0 & p_2+p_3+p_4 \\
\end{array}
\right)$
\end{center}
Rewriting $\rho$
\begin{center}
 $\rho=\frac{p_2+p_3+p_4}{3}I+\frac{1}{3}\left(
\begin{array}{cccc}
 4p_1-1 & 0 & 0 & 0 \\
 0 & 0 & 0 & 0 \\
 0 & 0 &0 & 0 \\
 0 & 0 & 0 & 0 \\
\end{array}
\right)$
\end{center}
\begin{equation}
 \rho=\frac{1}{3} \left\{(1-p_1)I+(4p_1-1)\vert00\rangle\langle00\vert \right\}
\end{equation}
where $\rho$ is the effective pure state corresponding to $\vert00\rangle$.

\noindent {\bf Spatial averaging technique} uses rf pulses
and pulsed field gradients (PFG) to prepare pseudopure states.
The PFG kills the magnetization in the plane
perpendicular to its applied direction by randomizing the spin 
magnetization in that plane and spin magnetization is retained 
only in the direction along which the PFGs are applied. For a two-qubit
homonuclear system (homonuclear meaning spins belonging
to the same species) the pseudopure state $\rho_{00}$ can 
be prepared from an initial thermal state using the following steps:

\begin{center}
 \ce{$\rm I_z^1$ +$\rm I_z^2$ ->[$\rm (\frac{\pi}{3})^2_x$] $\rm I_z^1$ + $\rm \frac{1}{2}I_z^2$ + $\rm \frac{\sqrt{3}}{2}I_y^2$} \\
 \ce{->[$\rm G_z$] $\rm I_z^1$ + $\rm \frac{1}{2}I_z^2$} \\
 \hspace{1.5cm}\ce{->[$\rm (\frac{\pi}{4})^1_x$] $\rm \frac{1}{\sqrt{2}}I_z^1$-$\frac{1}{\sqrt{2}}I_y^1$ + $\rm \frac{1}{{2}}I_z^2$} \\
 \hspace{1.8cm}\ce{->[$\rm \frac{1}{2J_{12}}$] $\rm \frac{1}{\sqrt{2}}I_z^1$ + $\rm \frac{1}{\sqrt{2}}2I_x^1I_z^2$ + $\rm \frac{1}{{2}}I_z^2$} \\
 \hspace{3.1cm}\ce{->[$\rm (\frac{\pi}{4})^1_{-y}$] $\rm \frac{1}{\sqrt{2}}I_z^1$-$\frac{1}{\sqrt{2}}I_x^1$ + $\rm \frac{1}{\sqrt{2}}2I_x^1I_z^2$ + $\rm \frac{1}{{2}}2I_z^2I_z^2$ + $\rm \frac{1}{{2}}I_z^2$} \\
 \ce{->[$\rm G_z$] $\rm \frac{1}{2}( I_z^1$ + $\rm I_z^2 + 2I_z^1I_z^2)$} \\
\end{center}
where $I_i^1=\frac{1}{2}\sigma_i\otimes I$,
$I_i^2=\frac{1}{2}I\otimes\sigma_i$, $\sigma_i$ with $i=x,y,z$ are Pauli
matrices, ${J_{12}}$ is the scalar coupling constant between two spins and
$\rm G_z$ is a PFG along the $z$-axis which kills all the magnetization in
the $xy$-plane.

\noindent{\bf Logical labeling technique} uses one qubit of $n$-qubits
to
label the state while 
the other $n-1$ qubits are placed in a pseudopure
configuration~\cite{gershenfeld-sc-97}. To illustrate the logical labeling
technique, let us consider a homonuclear three-qubit system at thermal
equilibrium with its deviation density matrix 
\begin{center}
$\Delta\rho_{eq}={}\left(
\begin{array}{cccccccc}
 3 & 0 & 0 & 0 & 0 & 0 & 0 & 0 \\
 0 & 1 & 0 & 0 & 0 & 0 & 0 & 0 \\
 0 & 0 & 1 & 0 & 0 & 0 & 0 & 0 \\
 0 & 0 & 0 & -1 & 0 & 0 & 0 & 0 \\
 0 & 0 & 0 & 0 & 1 & 0 & 0 & 0 \\
 0 & 0 & 0 & 0 & 0 & -1 & 0 & 0 \\
 0 & 0 & 0 & 0 & 0 & 0 & -1 & 0 \\
 0 & 0 & 0 & 0 & 0 & 0 & 0 & -3 \\
\end{array}
\right)$
\end{center}
The relative population of the eigenstates are: 
\begin{center}
$\begin{array}{ccccccccc}
 {\text{ State}} & \vert000\rangle & \vert001\rangle & \vert010\rangle& \vert011\rangle & \vert100\rangle&\vert101\rangle & \vert110\rangle & \vert111\rangle\\
 {\text{ Relative population}} & 6 & 4 & 4 & 2 & 4 & 2 & 2 & 0 \\
\end{array}$
\end{center}
Assuming the first qubit as a label, the 
first four eigenstates can be perceived as a two-qubit system 
with the label qubit in the state
$\vert 0 \rangle$ and 
the other four eigenstates 
can be considered as a two-qubit system with the label qubit 
in the state $\vert 1 \rangle$. First a
$\rm CNOT_{21}$ gate is applied (the second qubit 
being a control qubit and the first qubit being the target qubit) 
and then a
$\rm CNOT_{31}$ gate is
applied (the third qubit being the control 
qubit and the first qubit as the target qubit). The action of these
two gates results in a final relative state population: 
\begin{center}
$\begin{array}{ccccccccc}
 {\text{ State}} & \vert000\rangle & \vert001\rangle & \vert010\rangle& \vert011\rangle & \vert100\rangle&\vert101\rangle & \vert110\rangle & \vert111\rangle\\
 {\text{ Relative population}} & 6 & 2 & 2 & 2 & 4 & 4 & 4 & 0 \\
\end{array}$
\end{center}
The deviation part of the pseudopure state density matrix 
of the two qubits corresponding to label 0 is
$\Delta\rho^0=4\vert00\rangle\langle00\vert-I$ and 
corresponding to label 1 is $\Delta\rho^1=I-4\vert11\rangle\langle11\vert$.
In this thesis we will use the spatial averaging 
technique throughout for pseudopure state preparation.
\subsection{Quantum gate implementation in NMR}
\label{NMR_qg_chap1}
Section~\ref{theory_qg_chap1} dealt with the mathematical description
of quantum gates. This section will explain
the physical implementation of unitary gates on an NMR quantum computer.
In traditional NMR techniques, spin states are manipulated 
by using rf pulses or by free evolution under the internal
nuclear spin interactions. It was shown in Section~\ref{nmrtheory_chap1}
that a spin which satisfies the resonance condition 
can be rotated about the $\hat{\phi}$ axis by
applying an rf pulse along the $\hat{\phi}$ axis with high precession.
Due to this, any quantum gate can be implemented in NMR with 
high fidelity using rf excitation pulses and interaction between the spins.
The action of an on-resonance rf pulse with arbitrary phase $\hat{\phi}$ and
duration $t_p$ is given by
\begin{equation}
 (\theta)_{\phi}^I=exp(-i\omega_1 t_p I_{\phi})=exp(-i\theta I_{\phi})
\end{equation}
where $I_{\phi}=I_xcos(\phi)+I_ysin(\phi)$ and $\theta=\omega_1t_p$.
The rf excitation pulse rotates a spin on-resonance with 
an angle $\theta$ along the $\hat\phi$ axis.
A single-qubit gate can hence be implemented using this set of rotations.
Some examples of NMR implementations of single-qubit gates are:
\begin{itemize}
 \item Hadamard gate (H) can be implemented by a 
 spin-selective pulse $\pi$ pulse  along
the $x$-axis and a $\frac{\pi}{2}$ pulse along the $y$-axis. 
 
 \begin{center}
\begin{figure}[h]
\centering
\includegraphics[angle=0,scale=0.75]{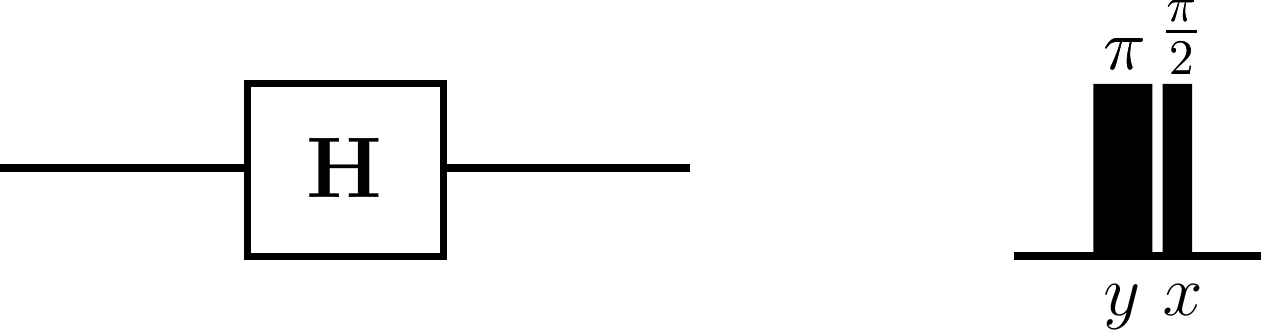}
\end{figure}
\end{center}
 \item Pauli-X gate (NOT gate) can be implemented by
 a single spin-selective $\pi$ pulse along the $x$-axis. 
 \begin{center}
\begin{figure}[h]
\centering
\includegraphics[angle=0,scale=0.75]{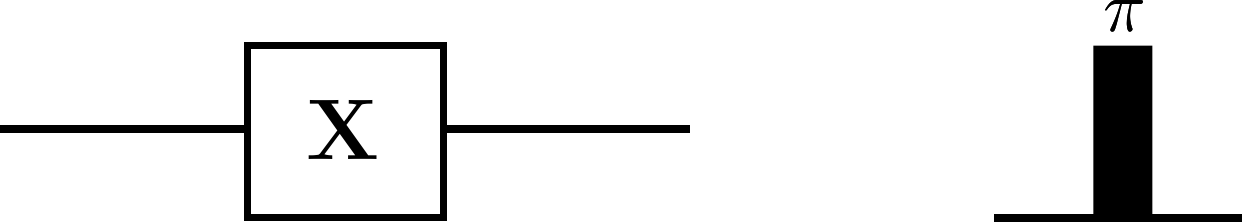}
\end{figure}
\end{center}
 \item Phase shift gate ($\rm R_{\phi}$) can be implemented 
 by a single spin-selective $\phi$ pulse along the $z$-axis.
 However in NMR we can only apply rf pulses along an axis in
the $xy$ plane, so a $\phi$ rotation along the $z$-axis  
is typically 
decomposed as a pulse
cascade $(\frac{\pi}{2})_x(\phi)_y(\frac{\pi}{2})_{-x}$. 
\begin{center}
\begin{figure}[h]
\centering
\includegraphics[angle=0,scale=0.75]{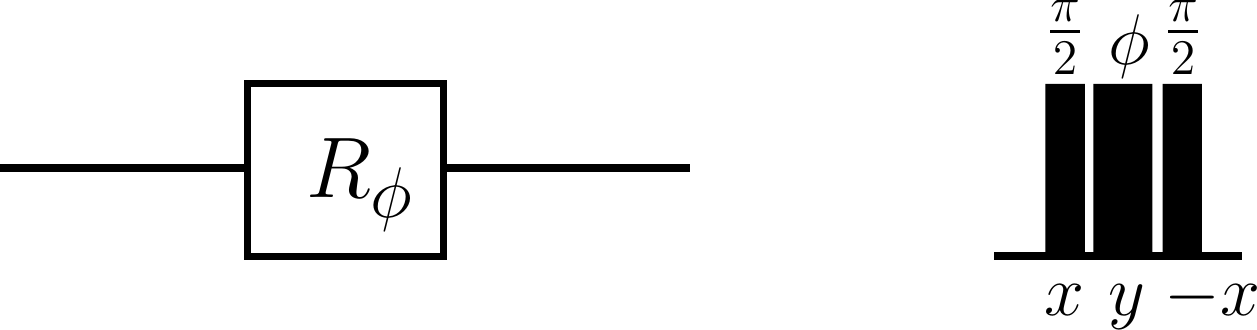}
\end{figure}
\end{center}
\end{itemize}
For the implementation of multi-qubit gates, we use the 
spin-spin interaction term in the Hamiltonian
along with single-qubit gates. An NMR pulse sequence for
the two-qubit CNOT gate 
is $(\frac{\pi}{2})_{-y}^2(\frac{\pi}{2})_{-z}^{1,2}\frac{1}{4J_{12}}{(\pi)^{1,2}_y}\frac{1}{4J_{12}}$
${(\pi)^{1}_y}{(\frac{\pi}{2})_y}$; where $\frac{1}{J_{12}}$ 
denotes an evolution
period under the coupling Hamiltonian. 
\begin{center}
\begin{figure}[h]
\centering
\includegraphics[angle=0,scale=0.8]{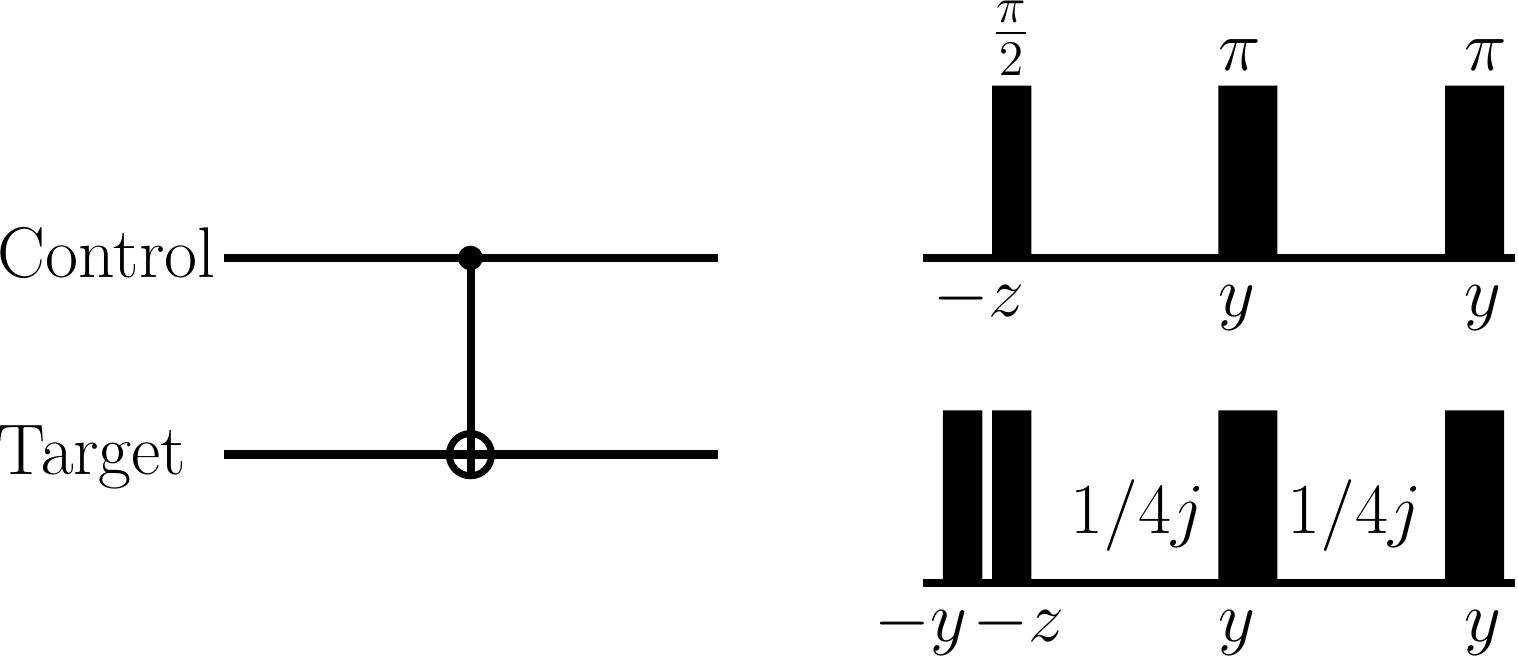}
\label{cnot_chap1}
\end{figure}
\end{center}
\subsection{Numerical techniques for quantum gate optimization}
NMR quantum gates can be realized by the application of rf pulses and
interactions between the spins.  For heteronuclear coupled spins (where the
spins under consideration belong to different nuclear species), due to the
large Larmor frequency difference between the spins and the availability of
multi-channel rf coils in the spectrometer hardware, it is easy to
experimentally implement individual spin-selective rotations.  However, for
homonuclear spins (where the spins under consideration belong to the same
nuclear species), it is difficult to selectively manipulate an individual spin,
due to much smaller differences in the chemical shifts of the spins.  The
traditional way of exciting individual spins in NMR is by using a shaped pulse
which is usually of long duration and results in a low experimental gate
fidelity.  To tackle this problem, one  possible solution is optimization of
quantum gates using numerical techniques.  The most commonly used numerical
optimization techniques are strongly modulated pulses, genetic algorithms and
the gradient ascent pulse engineering (GRAPE) algorithm.  This thesis mainly
uses the GRAPE algorithm for gate optimization so it is discussed in detail,
while the other techniques are discussed briefly.

\noindent{\bf Strongly modulated pulses (SMPs)} is a procedure 
for finding high-power pulses that strongly modulate the dynamics
of the system to precisely craft a desired unitary 
operation~\cite{fortunato-jcp-02}. 
It uses the knowledge of the internal Hamiltonian and the form of the 
external Hamiltonian to generate the parameter values 
to determine the desired gate.
SMPs make use of the Nelder-Mead Simplex algorithm~\cite{Nelder} 
to minimize the quality factor by
searching through the mathematical parameter space. It generates a control sequence
as a cascade of rf pulses with fixed power, 
transmitter frequency, initial phase and
pulse duration.

\noindent{\bf Genetic algorithms (GAs)}: These are stochastic 
search algorithms based on the concept of natural selection, 
a process which drives the biological evolution~\cite{holland-book}. 
GAs modify
the population of the individual solution at each step using the biological 
inspired operations such as selection, mutation, crossover etc. to 
evolve towards 
an optimal solution. At each step, 
the algorithm calculates the fitness of every individual 
solution and the algorithm runs until the desired fitness is achieved.
In quantum information processing, GAs have been used 
to optimize quantum algorithms \cite{suter-ia-08,hardy-2010,bang-korean},
for quantum entanglement~\cite{navarro-pra-06} and for optical dynamical 
decoupling~\cite{lidar-pra}. GAs have also been 
used to optimize the pulse 
sequences for unitary transformations on  an NMR quantum information 
processor~\cite{manu-pra-12,manu-pra-14}.

\noindent{\bf Gradient ascent pulse engineering}: To construct the desired 
unitary quantum gate $U_{tgt}$ using the GRAPE algorithm~\cite{khaneja-jmr-05},
we assume a closed system, 
with the propagator $U$ evolving under the Hamiltonian $H$ according to
\begin{equation}
 \frac{d}{dt}U=-iHU.
\end{equation}
Solving this equation leads to
\begin{equation}
 U_{opt}=\prod_{j=1}^{N} U_j 
\end{equation}
and 
\begin{equation}
 U_j=exp \left\{-i\Delta t \left (H_0+\sum_{k=1}^m u_k(j)H_k \right ) \right \}
\end{equation}
where $H_0$ is the system Hamiltonian, $H_k$ is the rf
control Hamiltonian and the control amplitudes $u_k$ are constant, i.e., 
during the $j$th step the amplitude $u_k(j)$ 
of the $kth$ control Hamiltonian is given
by $u_k(j)$. If $T$ is the total pulse duration of the unitary gate
then for simplicity the total time $T$ is discretized in 
$N$ equal steps and 
$\Delta t= T/N$. So, the problem is to find the optimal amplitudes $u_k(j)$
of the rf fields. 
The actual propagator $U_{opt}$ is identical to the desired operator
$U_{D}$ when $||U_{opt}-U_{trg}||^2=0$ and in an optimization we will search for its minimum.
Expanding further 
\begin{eqnarray}
 ||U-U_{D}||^2&=&Tr\{(U_{opt}-U_{tgt})^{\dagger}(U_{opt}-U_{tgt})^{\dagger}\} \nonumber \\  
 &=& 2 Tr\{I\}-2Re\{Tr\{U_{tgt}^{\dagger}U_{opt}\}\},
\end{eqnarray}
Hence our task is equivalent to maximization of 
\begin{equation}
 \Phi= Re\{Tr\{U_{tgt}^{\dagger}U_{opt}\}\}. 
\end{equation}
Further it is not necessary to exactly 
reproduce $U_{tgt}$. It serves equally well to reproduce
the target operator up to a global phase 
factor $e^{i\phi}U_{tgt}$. Thus the task is equivalent to
the maximization of
\begin{equation}
 \Phi_0= |Tr\{U_{tgt}^{\dagger}U_{opt}\}|^2 
\end{equation}
This performance function increases if we choose
\begin{center}
\ce{$u_{k}(j)$ -> $u_{k}(j)$ + $\epsilon$ $\frac{\delta\Phi_0}{\delta u_k(j)}$}
\end{center}
where $\frac{\delta\Phi_0}{\delta u_k(j)}=2\Delta t 
{\rm Im}\{Tr\{U^{\dagger}_{tgt}U_N \dots U_{j+1}H_kU_{j}\dots U_{1}\}Tr\{U_{opt}^{\dagger}U_{tgt}\}\}$ 
and $\epsilon$ is a small step size.

The basic GRAPE algorithm consist of the following steps:
\begin{enumerate}
 \item Guess initial controls $u_k(j)$.
 \item Evaluate $\Phi_0$.
 \item Evaluate $\frac{\delta\Phi_0}{\delta u_k(j)}$  
and update the $m \times N$ control amplitudes $u_k(j)$.
 \item  With these as the new controls, iterate to step 2.
\end{enumerate}
 The algorithm is terminated if the change in the performance 
 index $\Phi_0$ is smaller than a chosen threshold value.

\begin{center}
\begin{figure}[h]
\centering
\includegraphics[angle=0,scale=1.6]{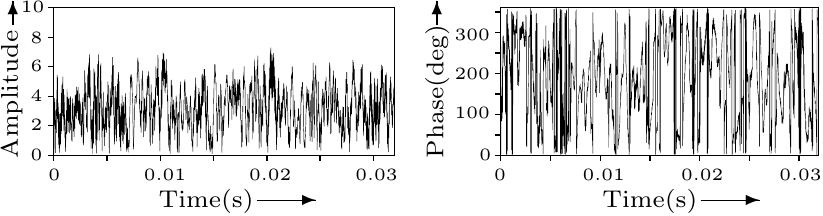}
\caption{ Plot of rf pulse amplitude and phase with time, optimized with GRAPE 
for CSWAP gate.}
\label{grape_fig}
\end{figure}
\end{center}

In Fig.~\ref{grape_fig}, plot of rf pulse amplitude and phase with time for 
GRAPE optimized CSWAP gate 
is shown with fidelity 0.9995. We used the three $\rm ^{19}F$ 
spins of the trifluoroiodoethylene ($\rm C_2F_3I$) molecule as NMR sample.
With GRAPE we can tackle errors due to rf inhomogeneity, off-set
and flip angle by optimization. 
\subsection{Measurement in NMR}
A conventional detection of the NMR signal is a so-called
ensemble weak measurement, as the 
weak interaction of spins with radio-frequency coil does not
change significantly the quantum states of the spins 
in  the  process  of  
measuring  the  total  spin magnetization. 
A direct projective measurement is not possible on NMR 
quantum computer. However, some experiments have been 
done to simulate projective
measurements in NMR~\cite{Khitrin2011,lee-apl-06}.
\begin{center}
\begin{figure}[h]
\centering
\includegraphics[angle=0,scale=0.55]{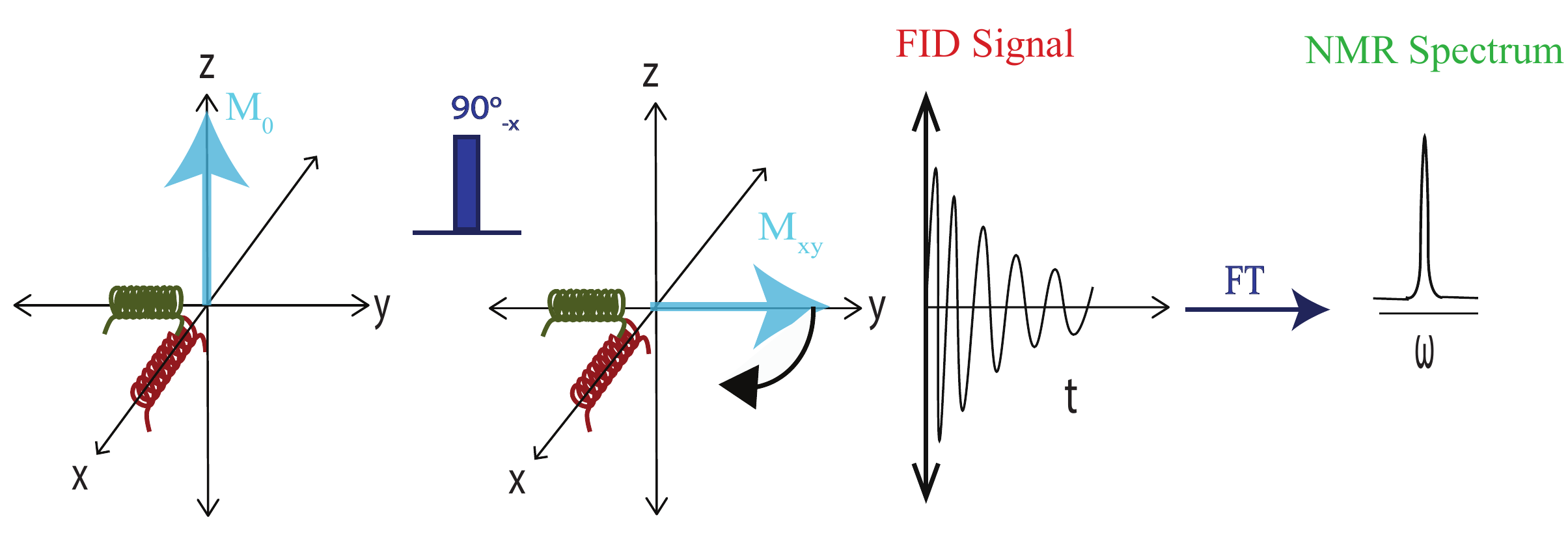}
\caption{Rotation of Bulk magnetization using resonance. 
$B_1(t)$ applied in -x-direction for duration such that total rotation is $90^{\circ}$}
\label{nmr_signal_chap1}
\end{figure}
\end{center}
As described in Section~\ref{nmr_setup_chap1}, 
when the nuclear spins
are placed in the magnetic field $B_0$ along the $z$-axis, the average
of the the magnetic moments $\mu$ of the nuclei at thermal equilibrium 
produces a bulk magnetization. With the application of rf pulses, 
this bulk magnetization is rotated from the $z$-axis to the $xy$ plane,
where this rotated bulk magnetization precesses  
about the $z$-axis with a Larmor frequency $\omega_0$.
The precessing bulk magnetization causes
a change in magnetic flux in the rf coil which in turn 
produces a signal voltage as
shown in Fig.~\ref{nmr_signal_chap1}. 
The recorded signal is proportional to the time rate 
of change of the magnetic flux linking an inductor 
that is a part of a tuned circuit. Due to relaxation processes
with time the magnetization in the $xy$ plane decays and the resultant signal
also decays with time (called free 
induction decay (FID)) as shown in Fig~{\ref{nmr_signal_chap1}}.
If the quality factor of the coil is not too high, the recorded signal may be
regarded as a time record of the instantaneous 
bulk magnetization that is transverse (in the $xy$-plane ) 
to the applied static field (which is in the $z$-axis).
This rf signal is mixed down with a phase-sensitive detector,
and the signal has both real ($x$) and imaginary ($y$) components.
The time-domain signal of the transverse magnetization is given as
\begin{equation}
\label{nmr_signal_eq_chap1}
 S(t)\propto Tr\left\{\rho(t)\sum_{k}(\sigma_{kx}+i\sigma_{ky})\right\}
\end{equation}
where  $\sum_{k}(\sigma_{kx}-i\sigma_{ky}$ is the detection operator,
$\sigma_{kx}$ and $\sigma_{ky}$ are Pauli spin operators 
proportional to the $x$ 
and $y$ components of the magnetization due to $k^{th}$ spin and $\rho(t)$
is a reduced density operator which represents the average state of a single 
molecule~\cite{leskowitz-pra-04}. The Fourier transform of Eq.{\ref{nmr_signal_eq_chap1}}
gives the signal in the frequency domain which represent
spectral lines at well-defined frequencies.
These spectral lines are characteristic of the spin system used.

The state density matrix $\rho$, at any instant $t$, can be reconstructed by systematically
measuring the NMR signal. This process of state reconstruction is called quantum state
tomography (QST)~\cite{long-joptb,leskowitz-pra-04}. Any general normalized state state
density matrix can be written as 
\begin{center}
 $\rho=\left(
\begin{array}{cc}
 a_1 & a_2+ia_3 \\
 a_2-ia_3 & 1-a_1 \\
\end{array}
\right)$
\end{center}
The NMR signal is proportional to $\rho_{12}=a_2+ia_3$
and can be measured by measuring the intensity of the peaks
from the real and imaginary parts of the spectra.
The real intensity is proportional to $a_2$ and the imaginary intensity
is proportional to $a_3$. For measuring $a_1$, a $\pi/2$ pulse 
along the $y$ axis is 
applied and the real intensity of the peak of spectra 
is proportional to $a_1$~\cite{long-joptb}.
The reconstruction of density matrix is discussed 
in detail in Chapter~\ref{chapter_mle}.
\section{Evolution of quantum systems}
Quantum systems which do not interact 
with the outside world are called closed systems.
In reality however, no physical system is an entirely closed system,
except perhaps the universe as a whole. Real systems suffer
from unwanted interactions with their environment. 
These adverse interactions show up as noise in the quantum system. 
So, it is important to understand and control such a noise process
in order to build realistic quantum information processors.
The tools traditionally used by physicists for the description of
open quantum systems are master equations,
Langevin equations and stochastic differential equations.
Another potent tool, which simultaneously addresses a 
broad range of physical scenarios is the mathematical
formalism of quantum operations. With this formalism 
not only nearly closed systems which are weakly coupled
to their environments but also the systems which are
strongly coupled to their environments can be modeled.  
Quantum operations formalism is well adapted to
describe discrete state change, that is, transformations
between an initial state $\rho$ to final state $\rho{'}$,
without explicit reference to the passage of time~\cite{nielsen-book-02}.  

\begin{center}
\begin{figure}[h]
\centering
\includegraphics[angle=0,scale=0.75]{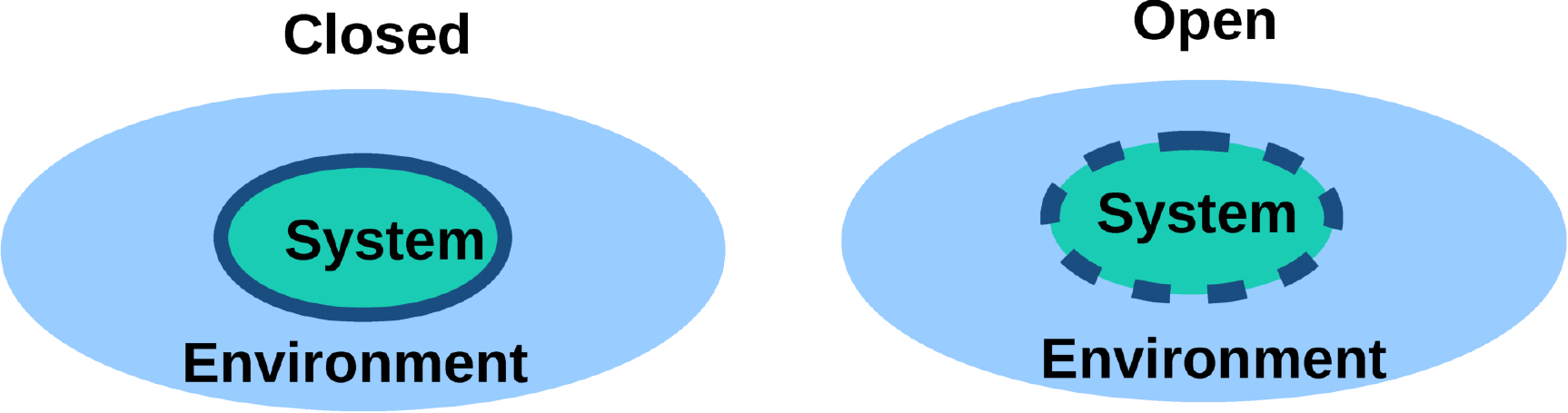}
\caption{(a) Representation of a closed system; the circle around
the system depicts no 
interaction between the system and environment.
(b) Representation of an open system; the dashed circle 
around the system shows that the system and environment are interacting.}
\label{system_type}
\end{figure}
\end{center}
\subsection{Closed quantum systems}
The dynamics of a closed quantum system in a pure state is governed by 
the Schr$\rm \ddot{o}$dinger 
equation
\begin{equation}
 {i \hbar}\dfrac{\partial}{\partial t}{\vert\psi(x,t) \rangle } = {H}_{sys}\vert \psi(x,t)\rangle 
 \label{schrodinger}
\end{equation}
where $\psi(x,t)$ is the wave function, ${H}_{sys}$ the Hamiltonian,
and $\hbar$ is Planck's constant. 
In NMR closed systems, 
unitary evolution is governed by the Liouville-von Neumann
equation
\begin{equation}
 \dot{\rho}_{}(t)=-\frac{i}{\hbar}[\ H_{sys},\rho_{}(t)]\
 \label{new}
\end{equation}
The solution to Eq.~(\ref{schrodinger}) and Eq.~(\ref{new}) is given by
\begin{eqnarray}
 \vert \psi(t)\rangle &=& U(t)\vert\psi(0)\rangle \\
 \rho(t)&=&U(t)\rho(0)U(t)^{\dagger}
\end{eqnarray}
where $\vert \psi(0) \rangle$ and $\rho(0)$ is a state of the system at time t=0 and
\begin{equation}
 U(t)=T exp \left[-\frac{i}{\hbar} \int_0^t H_{sys} dt \right]
\end{equation}
is a unitary operator. The dynamics of a closed quantum system 
can be described by a unitary transformation.

\begin{figure}[h]
\hspace{4cm} \includegraphics[angle=0,scale=1.0]{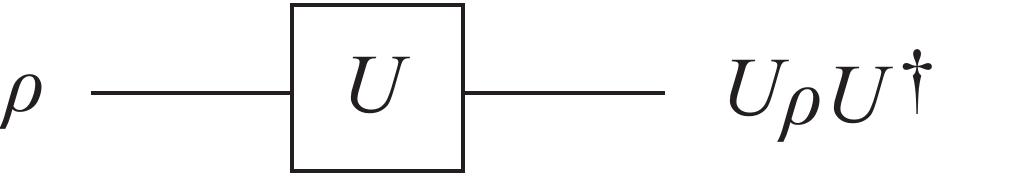}
\caption{Model of closed quantum systems.}
\label{unitary}
\end{figure}

A model of a closed system is presented in Fig.~\ref{unitary} where 
the unitary transformation is represented as a box into which the input state
$\rho$ enters and from which the 
output state $\rho^{'}=U.\rho.U^{\dagger}$ exits. 
\subsection{Open systems}
The standard approach for deriving the equations of 
motion for a system interacting with its environment 
is to expand the scope of the system to include the environment. 
The combined quantum system is then closed, and its evolution
is governed by the Von Neumann equation.
\begin{equation}
 \dot{\rho}_{tot}(t)=-\frac{i}{\hbar}[\ H_{tot},\rho_{tot}(t)]\
 \label{new1}
\end{equation}
Here, we assume that the initial state of the total system can
be written as a separable state $\rho_{tot}=\rho_{sys}\otimes\rho_{env}$
and $H_{tot}=H_{sys}+H_{env}+H_{int}$ is the total Hamiltonian,
which includes the original system $H_{sys}$, 
the environment $H_{env}$, and interaction between
the system and its environment Hamiltonian $H_{int}$.
The solution to Eq.(\ref{new1}) is given by
\begin{equation}
 {\rho}_{tot}(t)= U(t)\rho_{tot}(0)U(t)^{\dagger}
\end{equation}
Since we are interested in the dynamics of the principal system $\rho_{sys}$, we can extract
the information about the system by taking the partial trace 
over the environment
\begin{equation}
 \rho_{sys}=Tr_{env}[U(t)(\rho_{sys}\otimes\rho_{evn})U(t)^{\dagger}]
\end{equation}
The most general trace-preserving and completely positive form 
of this evolution Eq.(\ref{new1}) is the Lindblad master equation for the reduced
density matrix $ \rho_{sys}=Tr_{env}[\ \rho_{tot}]\ $. The Lindblad equation is 
the most general form for a Markovian master equation, and it is very important for the
treatment of irreversible and non-unitary processes, from dissipation and decoherence to the quantum measurement
process.
\begin{equation}
\label{lbeq_chap1}
 \dot{\rho_{sys}}(t)=-\frac{i}{\hbar}[\ H_{sys},\rho_{tot}(t)]\ + \sum_{i,\alpha}
 ( L_{i,\alpha}\rho L_{i,\alpha}^{\dagger}+\frac{1}{2}\{L_{i,\alpha}^{\dagger}L_{i,\alpha}, \rho \} )
\end{equation}
where the Lindblad operator
$L_{i,\alpha}=\sqrt{k_{i,\alpha}} \sigma_{\alpha}^{(i)}$
acts on the $i$th qubit and describes decoherence,
and $\sigma_{\alpha}^{(i)}$ denotes the Pauli
matrix of the $ith$ qubit with $\alpha$= x, y, z. The constant $k_{i,\alpha}$ is
approximately equal to the inverse of decoherence time.

If we consider only the first term on 
the right hand side of Eq.(\ref{lbeq_chap1}),
we obtain the Liouville-von Neumann equation. This term is the Liouvillian 
and describes
the unitary evolution of the density operator. The second term on the
right hand side of the Eq.(\ref{lbeq_chap1}) is the Lindbladian and
it emerges when we take the partial trace (a non-unitary operation) 
over the degrees of freedom of the environment.
The Lindbladian describes the non-unitary evolution of the density
operator and the Lindblad operators can be
understood to represent the 
system contribution to the system-environment interaction. 
\begin{center}
\begin{figure}
\centering
\includegraphics[angle=0,scale=1.0]{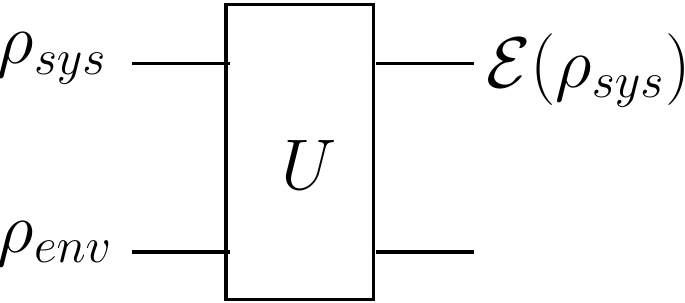}
\caption{Model of open quantum system consisting of two parts: 
the principal system and its environment.}
\label{nonunitary_chap1}
\end{figure}
\end{center}
In Fig.~\ref{nonunitary_chap1}, we have a system in state $\rho_{sys}$
and environment in state $\rho_{env}$ which together form a closed system
and sent into a box which represents the unitary operator on total system
with the final state exiting the box being ${\cal E} (\rho)$. It is to
be noted that
${\cal E} (\rho)$ may not be related by unitary transformation to
the  initial system state $\rho_{sys}$. The reduced state
of the system alone can be obtained by taking a partial trace 
over the environment
\begin{equation}
 {\cal E}(\rho_{sys})=Tr_{evn}[U(\rho_{sys}\otimes\rho_{evn})U^{\dagger}]
 \label{rhosys_chap1}
\end{equation}
Now, let us consider $\{\vert e_k \rangle\}$ to be 
an orthonormal basis for the state space of the environment
and the initial state of the environment 
can be written as $\rho_{evn}=\vert e_0\rangle\langle e_0 \vert$.
Using Eq.(\ref{rhosys_chap1}), we get
\begin{eqnarray}
{\cal E}(\rho)&=&\sum_k\langle e_k\vert U[\rho\otimes\vert e_0\rangle\langle e_0\vert]U^{\dagger}\vert e_k\rangle \nonumber \\
 &=& \sum_k E_k\rho E_k^{\dagger},
 \label{operatorsum_chap1}
\end{eqnarray}
where $E_k=\langle e_k\vert U \vert e_0\rangle$ is 
the Kraus operator. Eq.(\ref{operatorsum_chap1}) is known as the 
operator-sum representation of $\cal{E}$. The operators $\{E_k\}$
are known as operation elements for 
the quantum operation $\cal E$, which satisfy
\begin{equation}
\sum_kE_k^{\dagger}E_k= I
\end{equation}
\subsection{Quantum noise channels}
Quantum channels are convex-linear and completely positive trace 
preserving maps ${\cal E}$
which transform the initial state $\rho$ of a quantum system 
to another state ${\cal E}(\rho)$.
\begin{eqnarray}
 {\cal E}: \rho &\rightarrow& {\cal E}(\rho) \nonumber \\
 {\cal E}(\rho)&=&\sum_k E_k \rho E_k^{\dagger}
\end{eqnarray}
where $E_k$'s are the Kraus operators. 
Quantum channels can be used 
to describe the transformation occurring 
in the state of a system
due to the system-environment interaction. 
The interaction of a single qubit
with its environment can be described by three
quantum noise channels which are: phase damping, 
generalized amplitude damping and depolarizing channel.
\subsubsection{Generalized amplitude damping channel}
The generalized amplitude damping channel describes the dissipative
interactions between the system and its environment which cause 
interconversions of populations
from ground state to excited states and vice versa at finite
temperature~\cite{nielsen-book-02}. For a single qubit, the Kraus operators for
this channel are given by:
\begin{eqnarray}
 E_1&=& \sqrt{p}\left(
\begin{array}{cc}
 1 &  0 \\
  0 & \sqrt{1-a}\\
\end{array} \right)  \\
 E_2&=& \sqrt{p}\left(
\begin{array}{cc}
 0 &  \sqrt{a} \\
  0 & 0 \\
\end{array} \right)  \\
 E_3&=& \sqrt{1-p}\left(
\begin{array}{cc}
 \sqrt{1-a} &  0 \\
  0 & 1\\
\end{array} \right)  \\
 E_4&=& \sqrt{1-p}\left(
\begin{array}{cc}
 0 &  0 \\
  \sqrt{a} & 0\\
\end{array} \right) 
\end{eqnarray}
where $E_1$ and $E_2$ operators are responsible for the process by which 
the population from its excited state will 
decay to its ground state, $E_3$ and $E_4$
operators are responsible for 
the reverse process in which 
populations convert from the ground state to 
the excited state,
$p$ is the probability of finding population in the 
ground state at thermal equilibrium and
$a=1-e^{-\gamma t}$ where `$\gamma$' is the decay constant. The 
action of these Kraus operators on the density matrix
$\rho$ is given by:
\begin{eqnarray} 
 {\cal E}(\rho)&= &E_1\rho E_1^{\dagger}+E_2\rho E_2^{\dagger},\nonumber \\
 \rho &\rightarrow& {\cal E}(\rho), \nonumber \\
 \left(
\begin{array}{cc}
 \rho_{11} &  \rho_{12} \\
  \rho_{21} & \rho_{22}\\
\end{array}
\right) &\rightarrow& \left(
\begin{array}{cc}
 \rho_{11}^{'} & \rho_{12}^{'}  \\
 \rho_{21}^{'} & \rho_{22}^{'}  \\
\end{array} \right), \nonumber \\
\end{eqnarray}
where 
\begin{eqnarray}
 \rho_{11}^{'}&=&e^{-\gamma t}+p(1-e^{-\gamma t})\rho_{11}+p(1-e^{-\gamma t})\rho_{22}, \nonumber \\
 \rho_{22}^{'}&=&(1-p)(1-e^{-\gamma t})\rho_{11}+p e^{-\gamma t}\rho_{22}, \nonumber \\
  \rho_{12}^{'}&=& e^{-\gamma t/2}\rho_{12}, \nonumber \\
    \rho_{21}^{'}&=& e^{-\gamma t/2}\rho_{21}. \nonumber \\
\end{eqnarray}

The Lindblad operator corresponding to the 
generalized amplitude damping channel is given as
\begin{equation}
{\cal L}_{GD}=\sqrt{\frac{\gamma}{2}}\left(\begin{array}{cc}
 0 & \sqrt{p} \\
 \sqrt{(1-p)} & 0 \\
\end{array}
\right)
\end{equation}
In NMR, $\gamma$=1/ T$_1$ where T$_1$ is longitudinal relaxation time 
(explained in Section~{\ref{t1_chap1}}). The calculation 
becomes simple by assuming
a high temperature approximation where $p=1/2$. 
\subsubsection{Phase damping channel}
Phase damping (PD) channel is a non-dissipative channel,
which mainly describes the loss of coherence without loss of energy.
In this channel, the 
relative phase between $\vert0\rangle$ and $\vert 1 \rangle$
remains unchanged with some 
probability $p$ or is inverted $(\phi \rightarrow \phi+\pi)$ with
probability $1-p$. If the system is 
in state $\vert0\rangle$ or $\vert1\rangle$, it will be unaffected
by this channel. However, if it is 
in $\vert \psi\rangle=\alpha\vert0\rangle+\beta\vert1\rangle$ it gets
entangled with the environment which destroys all the coherences 
but the probability of finding the qubit
in state $\vert0\rangle$ or $\vert1\rangle$ does not change. 
The Kraus operators are given by:
\begin{eqnarray}
 {\rm E_1}&=&\sqrt{p}\left(
\begin{array}{cc}
 1 & 0 \\
 0 &1 \\
\end{array}
\right) \\
{\rm E_2}&=&\sqrt{1-p}\left(
\begin{array}{cc}
 1 & 0 \\
 0 &-1 \\
\end{array}
\right)
\end{eqnarray}
where $p=1-exp(-\lambda t)$ and $\lambda$ is the decay rate.
The action of these operators transform the initial 
state $\rho$ to the final state
${\cal E}(\rho)$
\begin{eqnarray} 
 {\cal E}(\rho)&= &E_1\rho E_1^{\dagger}+E_2\rho E_2^{\dagger}, \\
 \rho &\rightarrow& {\cal E}(\rho), \\
 \left(
\begin{array}{cc}
 \rho_{11} &  \rho_{12} \\
  \rho_{21} & \rho_{22}\\
\end{array}
\right) &\rightarrow& \left(
\begin{array}{cc}
 \rho_{11} & exp(-\lambda t)\rho_{12} \\
  exp(-\lambda t) \rho_{21} & \rho_{22}\\
\end{array}
\right),
\end{eqnarray}
Under the action of the phase damping channel, the off-diagonal 
elements decay and diagonal elements remain unaffected.

The Lindblad operator corresponding to the phase damping channel is given as
\begin{equation}
{\cal L}_{ph}=\sqrt{\frac{\lambda}{2}}\left(\begin{array}{cc}
 1 & 0 \\
 0 &-1 \\
\end{array}
\right)
\end{equation}
In NMR, $\lambda$=1/T$_2$ where T$_2$ is 
the transverse relaxation time (explained in Section~{\ref{t2_chap1}}). 
\subsubsection{Depolarizing channel}
Under the action of the depolarizing channel, 
the qubit remains intact
with probability $1-p$ while with probability
$p$ an identity type of noise occurs. The Kraus operator 
for the depolarizing channel is given by:
\begin{eqnarray}
 {\rm E_1}&=&\sqrt{1-p}\left(
\begin{array}{cc}
 1 & 0 \\
 0 &1 \\
\end{array}
\right), \nonumber \\
{\rm E_2}&=&\sqrt{\frac{p}{3}}\left(
\begin{array}{cc}
 0 & 1 \\
 1 & 0 \\
\end{array}
\right),\nonumber \\
{\rm E_3}&=&\sqrt{\frac{p}{3}}\left(
\begin{array}{cc}
 0 & -i \\
 i &0 \\
\end{array}
\right), \nonumber \\
{\rm E_4}&=&\sqrt{\frac{p}{3}}\left(
\begin{array}{cc}
 1 & 0 \\
 0 & -1 \\
\end{array}
\right), \nonumber
\end{eqnarray}
where $p=1-exp(-dt)$ and $d$ is 
the decay rate. The action of these Kraus operators change 
the initial state $\rho$ to the final state ${\cal E}(\rho)$,
\begin{eqnarray}
 \rho &\rightarrow& {\cal E}(\rho), \nonumber \\
  {\cal E}(\rho)&= &E_1\rho E_1^{\dagger}+E_2\rho E_2^{\dagger}+E_3\rho E_3^{\dagger}+E_4\rho E_4^{\dagger},\nonumber \\
 \left(
\begin{array}{cc}
 \rho_{11} &  \rho_{12} \\
  \rho_{21} & \rho_{22}\\
\end{array}
\right) &\rightarrow& \left(
\begin{array}{cc}
 (\frac{2p}{3}+(1-\frac{4p}{3}))\rho_{11} & (1-\frac{4p}{3})\rho_{12}  \\
 (1-\frac{4p}{3})\rho_{21} & (\frac{2p}{3}+(1-\frac{4p}{3}))\rho_{22}  \\
\end{array} \right), \\
\label{dipolar}
\end{eqnarray}
We can further simplify Eq.(\ref{dipolar}) to
\begin{equation}
 {\cal E}(\rho)= \frac{\lambda}{2}I+(1-\lambda)\rho
\end{equation}
where $\lambda=\frac{4p}{3}$ and $I$ is identity matrix.

The Lindblad operator corresponding to the depolarizing channel is
\begin{equation}
{\cal L}_{D}=\sqrt{\frac{\delta}{3}} (\sigma_x+\sigma_y+\sigma_z)
\end{equation}
\subsection{Nuclear spin relaxation}
The bulk spin magnetization which is along the $z$-axis
at thermal equilibrium, can be rotated to some other 
direction by the application of rf pulses.
Over time the magnetization returns to the $z$-axis 
due to relaxation processes, which are 
explained by the famous Bloch equations, describing
T$_1$ and T$_2$ relaxation processes.
\subsection{Longitudinal relaxation}
\label{t1_chap1}
Longitudinal relaxation is the process by which
the longitudinal component of spin magnetization 
returns to its equilibrium value, after 
a perturbation.
In this process, energy is exchanged between the system of
nuclear spins and its environment, which is called the lattice.
This process is also known as spin-lattice relaxation. 
The phenomenological equation describing this process is of the form:
 \begin{equation}
  \frac{d {\text{M}}_z}{dt}=\frac{\text{M}_0-\text{M}_z}{\text{T}_1}
 \end{equation}
 where T$_1$ is known as the longitudinal or the spin-lattice relaxation
 time and M$_0$ is the thermal equilibrium magnetization. The solution 
of the above equation
 is 
 \begin{equation}
  \text{M}_z=\text{M}_0(1-e^{-t/\text{T}_1}), \nonumber
 \end{equation}
when the M$_0$ is tilted to the $xy$ plane, then M$_z(0)=0$. 

For measuring T$_1$, the inversion recovery
experiment is commonly used, where the
spin magnetization is first inverted such that M$_z(0)$=-M$_{0}$:
\begin{equation}
  \text{M}_z=\text{M}_0(1-2e^{-t/\text{T}_1}).
 \end{equation}
\subsection{Transverse relaxation}
\label{t2_chap1}
Transverse relaxation is the process that leads to the disappearance
of  the coherences namely the $xy$-magnetization. 
The phenomenological equation describing the decay of the  transverse 
magnetization in the rotating frame can be written as:
\begin{equation}
   \frac{d\text{M}_{x,y}}{dt}=-\frac{\text{M}_{x,y}}{\text{T}_2}
\end{equation}
where T$_2$ is called the transverse relaxation time.
The solution of this equation is
\begin{equation}
 \text{M}_{x,y}=\text{M}_0e^{-t/\text{T}_2}
 \end{equation}
 where $\text{M}_0$ is the initial value of the transverse magnetization after
 the application of a $90^{\circ}$ rf pulse.
\subsection{Bloch-Wangness-Redfield relaxation theory}
This relaxation model uses a quantum mechanical approach to describe the system
parameters while the surrounding environment is described classically.  The
main limitation of this approximation is that at equilibrium the energy levels
are predicted to be equally populated. The theory is formally valid only in the
high-temperature limit. For finite temperatures, corrections are required to
ensure that the correct equilibrium populations are reached.  However these
corrections are significant only in the case of very low
temperatures~\cite{ernst-book-87,palmer-95,keeler-book,kumar-pnmrs-00}.

The von Neumann-Liouville equation, which describes
the time evolution of the magnetic resonance phenomenon
using spin density matrix $\rho(t)$ is given by
\begin{equation}
 \frac{d\rho(t)}{dt}=-i[H_0+H_1(t),\rho(t)]
 \label{nleq_chap1}
\end{equation}
where $H_0$ is the time-independent part of the
Hamiltonian which contains the spin Hamiltonian
and $H_1(t)$ describes the time-dependent perturbations.

It is convenient to remove the explicit dependence on $H_0$ by
rewriting the density operator $\rho(t)$ in a new reference frame,
called the interaction frame:
\begin{equation}
 \rho^*=exp(iH_0(t))\rho(t)exp(-iH_0(t))
\end{equation}
It is possible to rewrite Eq.({\ref{nleq_chap1}}) in the interaction
frame:
\begin{equation}
 \frac{d\rho^*(t)}{dt}=-i[H^*_1(t),\rho^*(t)]
 \label{nleqint_chap1}
\end{equation}
To solve Eq.(\ref{nleqint_chap1}) the following assumptions are required:
\begin{enumerate}
 \item The ensemble average of $H_1^*(t)$ is zero. 
 \item $\rho^*(t)$ and $H_1^*(t)$ are not correlated, with 
this assumption it is 
 possible to take the ensemble average of the fluctuations of the Hamiltonian
 and quantum states independently.
 \item $\tau_c<<t<<2/R$, where $\tau_c$ is the correction time relevant for $H_1^*(t)$
 and $R$ is the relevant relaxation rate constant.
 \item For the system to relax towards the thermal equilibrium, $\rho^*(t)$ has
 to be replaced by  $\rho^*(t)-\rho_0$, where $\rho_0$ is the density operator
 at equilibrium.
\end{enumerate}
Using these assumption, the R.H.S in Eq.(\ref{nleqint_chap1}) can be replaced
by an integral:
\begin{equation}
 \frac{d\rho^*(t)}{dt}=-\int_0^{\infty}\overline{[H^*_1(t),[H^*_1(t-\tau),\rho^*(t)-\rho_0]]}d\tau
 \label{nleq3_chap1}
\end{equation}
where the overbar represents the ensemble average.  
The third assumption allows the integral to run to infinity
and with the assumption that the 
fluctuations of the Hamiltonian are not correlated 
with the density matrix, we can calculate the
ensemble average over the stochastic 
Hamiltonian independently from $\rho^*(t)$.

For transforming Eq.(\ref{nleq3_chap1}) back in the lab frame, the
stochastic Hamiltonian $H_1^*(t)$ has to be decomposed as the sum of the random
functions of the spatial variable $F_k^q(t)$ and tensor spin operators $A^q_k$:
\begin{equation}
 H_1(t)=\sum_{q=-k}^{k}(-1)^q F_k^{-q}(t)A_k^q
 \label{eq4_chap1}
\end{equation}
The tensor spin operators are chosen to be spherical
tensor operators because of their
transformations properties under rotations.
For the Hamiltonians of interest in NMR
spectroscopy, the rank of the tensor $k$ is one or two. 
These operators can be further
decomposed as a sum of basis operators:
\begin{equation}
 A_k^q=\sum_p A^q_{kp}
\end{equation}
where the components $A^q_{kp}$ satisfy $[H_0, A_{kp}^q]=\omega_p^q A_{kp}^q$.
The transformation of $A_k^q$ in the interaction frame:
\begin{equation}
 A_k^{q*}=exp(iH_0t)A_k^q exp(-iH_0t)=\sum_p A_{kp}^qexp(i\omega_p^qt)
 \label{eq5_chap1}
\end{equation}
Using Eq.(\ref{eq4_chap1}) and Eq.(\ref{eq5_chap1}) we can rewrite Eq.(\ref{nleq3_chap1})
\begin{eqnarray}
 \frac{d\rho^*(t)}{dt}&=&-\sum_{q,q^{\prime}}\sum_{p,p^{\prime}}(-1)^{q+q^{\prime}}exp\{i(\omega_p^q+\omega^{q^{\prime}}_{p^{\prime}})t\}[A^{q^{\prime}}_{kp^{\prime}},[A^{q}_{kp},\rho^*(t)-\rho_0]]\nonumber \\
 &&\int_0^{\infty}\overline{F_k^{-q}(t)F_k^{-q}(t-\tau)}d\tau
 \label{eq6_chap1}
\end{eqnarray}
If $q \neq -q$, the two random processes $F_k^{-q^{\prime}} (t)$ and $F_k^{-q^{\prime}}(t)$ 
are assumed to be
statistically independent, due to which the ensemble average vanishes, 
unless $q^{\prime}=-q$. 
\begin{eqnarray}
 \frac{d\rho^*(t)}{dt}&=&-\sum_{q=-k'}^k\sum_{p,p^{\prime}}exp\{i(\omega_p^q-\omega^{q^{\prime}}_{p^{\prime}})t\}[A^{q}_{kp^{\prime}},[A^{q}_{kp},\rho^*(t)-\rho_0]]\nonumber \\
 &&\int_0^{\infty}\overline{F_k^{-q}(t)F_k^{-q}(t-\tau)exp(i\omega_p^q\tau)}d\tau
 \label{eq7_chap1}
\end{eqnarray}
Further it is to be noted 
that terms in which $\vert\omega_p^q-\omega_p^q\vert>>0$
oscillate much faster than the typical time scales of 
the relaxation phenomena 
will not affect the evolution. In the absence of degenerate eigenfrequencies, 
terms in Eq.(\ref{eq7_chap1}) do not vanish when $p=p^{\prime}$. Hence 
\begin{eqnarray}
 \frac{d\rho^*(t)}{dt}&=&-\sum_{q=-k'}^k\sum_{p}[A^{q}_{kp},[A^{q}_{kp},\rho^*(t)-\rho_0]]\nonumber \\
 &&\int_0^{\infty}\overline{F_k^{-q}(t)F_k^{-q}(t-\tau)exp(i\omega_p^q\tau)}d\tau
 \label{eq8_chap1}
\end{eqnarray}
The terms $\overline{F_k^{-q}(t)F_k^{-q}(t-\tau)}$ are \emph{correlation functions}.
The real part of the integral in Eq.( \ref{eq8_chap1}) 
is the power spectral density
function 
\begin{eqnarray}
 j^q(\omega)&=&2 {\rm Re} \left\{ \int_0^{\infty}\overline{F_k^{-q}(t)F_k^{-q}(t-\tau)exp(i\omega\tau)}   \right\} \nonumber 
\end{eqnarray}
The imaginary part of the 
integral in Eq.( \ref{eq8_chap1}) is the power spectral density
function 
\begin{eqnarray}
 k^q(\omega)&=&{\rm Im} \left\{ \int_0^{\infty}\overline{F_k^{-q}(t)F_k^{-q}(t-\tau)exp(i\omega\tau)}   \right\} \nonumber
\end{eqnarray}
In the high-temperature limit, the equilibrium 
density matrix is proportional to $H_0$. Thus, using Eq. (\ref{eq5_chap1}), the double commutator $[[A^{-q}_{kp},A^{q}_{kp}],\rho_0]]=0$
\begin{equation}
 \frac{d\rho^*(t)}{dt}=-\frac{1}{2} \sum_{q=-k}^k\sum_{p}[A^{-q}_{kp},[A^{q}_{kp},\rho^*(t)-\rho_0]]j^q(\omega^q_p)+i\sum_{q=0}^k\sum_{p}[[A^{-q}_{kp},A^{q}_{kp}],\rho^*(t)]k^q(\omega^q_p)
\end{equation}

By transforming the above equation in lab frame

\begin{equation}
 \frac{d\rho(t)}{dt}=-i[H_0,\rho(t)]-i[\Delta,\rho(t)]-\hat{\Gamma}(\rho(t)-\rho_0)
 \label{eq10_chap1}
\end{equation}
where the \emph{relaxation superoperator} is 
\begin{equation}
 \hat{\Gamma}=-\frac{1}{2} \sum_{q=-k}^k\sum_{p}[A^{-q}_{kp},[A^{q}_{kp},]]j^q(\omega^q_p)
\end{equation}
$\Delta$ is the dynamic frequency shift operator that accounts for second-order
frequency shifts of the resonance lines
\begin{equation}
 \Delta =-\sum_{q=0}^k\sum_{p}[A^{-q}_{kp},A^{q}_{kp}]k^q(\omega^q_p)
\end{equation}
This term can be incorporated into the Hamiltonian to obtain the final result, known as \emph{master equation}:

\begin{equation}
 \frac{d\rho(t)}{dt}=-i[H_0,\rho(t)]-\hat{\Gamma}(\rho(t)-\rho_0)
 \label{eq12_chap1}
\end{equation}
In the calculation of relaxation rates it is often
convenient to expand Eq.(\ref{eq12_chap1}) in terms
of the basis operators used to expand the density operator
\begin{equation}
 \frac{d b_r(t)}{dt}=\sum_s\{-i\Omega_{rs}b_s(t)-{\Gamma_{rs}}[b_{s}(t)-b_{s0}]\}
\end{equation}
where $\Omega_{rs}$ are characteristic frequencies defined as 
\begin{equation}
 \Omega_{rs}=\frac{\langle \mathbf{B}_r|[H_0,\mathbf{B}_s]\rangle}{\langle \mathbf{B}_r|\mathbf{B}_s\rangle}
\end{equation}
$\Gamma_{rs}$ are the rate constant for relaxation between
the operator $\mathbf{B}_s$ and $\mathbf{B}_r$
\begin{eqnarray}
 \Gamma_{rs}&=&\frac{\langle \mathbf{B}_r|\hat\Gamma\mathbf{B}_s\rangle}{\langle \mathbf{B}_r|\mathbf{B}_s\rangle} \nonumber \\
&=&-\frac{1}{2} \sum_{q=-k}^k\sum_{p}\left\{\frac{\langle \mathbf{B}_r|[A^{-q}_{kp},[A^{q}_{kp},\mathbf{B}_s]]\rangle}{\langle \mathbf{B}_r|\mathbf{B}_s\rangle}\right\}j^q(\omega^q_p)
 \end{eqnarray}
and
\begin{equation}
 b_{r}(t)=\frac{\langle \mathbf{B}_r|\rho(t)\rangle}{\langle \mathbf{B}_r|\mathbf{B}_s\rangle}
\end{equation}
The diagonal elements $\Gamma_{rr}$ are \emph{auto}-relaxation, while off-diagonal elements  $\Gamma_{rs}$,
are \emph{cross}-relaxation rates.Because it is assumed that only terms satisfying $q=-q$
give non-zero contributions to Eq.(\ref{eq6_chap1}), cross-relaxation can occur only between
operators with the same coherence order. In addition, because of
the secular approximation in Eq.(\ref{eq8_chap1}),
cross-relaxation between off-diagonal terms is forbidden in the
absence of degenerate transitions.
These two features give rise to a characteristic
block shape in the relaxation superoperator, known as Redfield kite. 
\section{Decoherence suppression}
The coherent superposition of states in combination
with the quantization of observables, represents
the one of the most fundamental features that mark 
the departure of quantum mechanics from the classical realm~\cite{streltsov-rmp-17}. 
Quantum coherence in multi-qubit systems embodies the essence of entanglement and is an
essential ingredient quantum information processing.A coherent quantum state in
contact with environment loses coherence i.e. decoherence and 
entangle states are much more fragile to it. In NMR, $T_2$ gives an estimation  
of the decoherence time of the system and in liquid state NMR it is the of order of 
seconds. There is a longstanding debate on quantumness of NMR spin-systems states, 
the presence of nonzero discord in some of the states of NMR spin-systems indicates the intrinsic
quantumness of nuclear spin systems even at high temperatures~\cite{soares-pinto-proc-12,katiyar-pra-12}.

Preserving quantum coherence is an important 
task in quantum information and different techniques have been developed
to suppress decoherence. These techniques are broadly categorized as
quantum error correction~\cite{preskill385}, decoherence free subspaces \cite{duan-prl-97, geoffrey-book-03} and dynamical 
decoupling (DD) methods~\cite{viola-review,viola-prl-99-2}. In particular,
the DD technique is an important
technique  which suppresses the decoherence by eliminating the
system-bath coupling. 
The idea comes from spin-echo pulses in NMR where 
static but nonuniform couplings can 
be compensated for perfectly by a single $\pi$ pulse 
in the middle of the time interval~\cite{hahn-pr-50}.
The idea of the spin echo was expanded to suppress dynamic
interactions with the environment by using periodic 
$\pi$ pulses or by periodic Carr-Purcell cycles.
The Carr-Purcell sequence was further modified to compensate errors due to 
$\pi$ pulses and Carr-Purcell-Meiboom-Gill sequence (CPMG) was 
devised~\cite{meiboom-rsi-58}.  A more sophisticated technique namely 
the Uhrig dynamical decoupling sequence was devised and 
it was shown that
instead of applying $\pi$ pulses at equal intervals of time
if $\pi$ pulses are applied at unequal intervals of
time then the sequence shows better preservation~\cite{uhrig-prl-07}. 
One of the advantages of the DD technique is that no extra qubits
are required unlike other techniques. Most DD preserving
sequences are constructed to take care of dephasing type noise.
In NMR language, T$_2$ type relaxation is considered and noise due to T$_1$
relaxation is ignored.
\begin{center}
\begin{figure}[h]
\centering
\includegraphics[angle=0,scale=0.7]{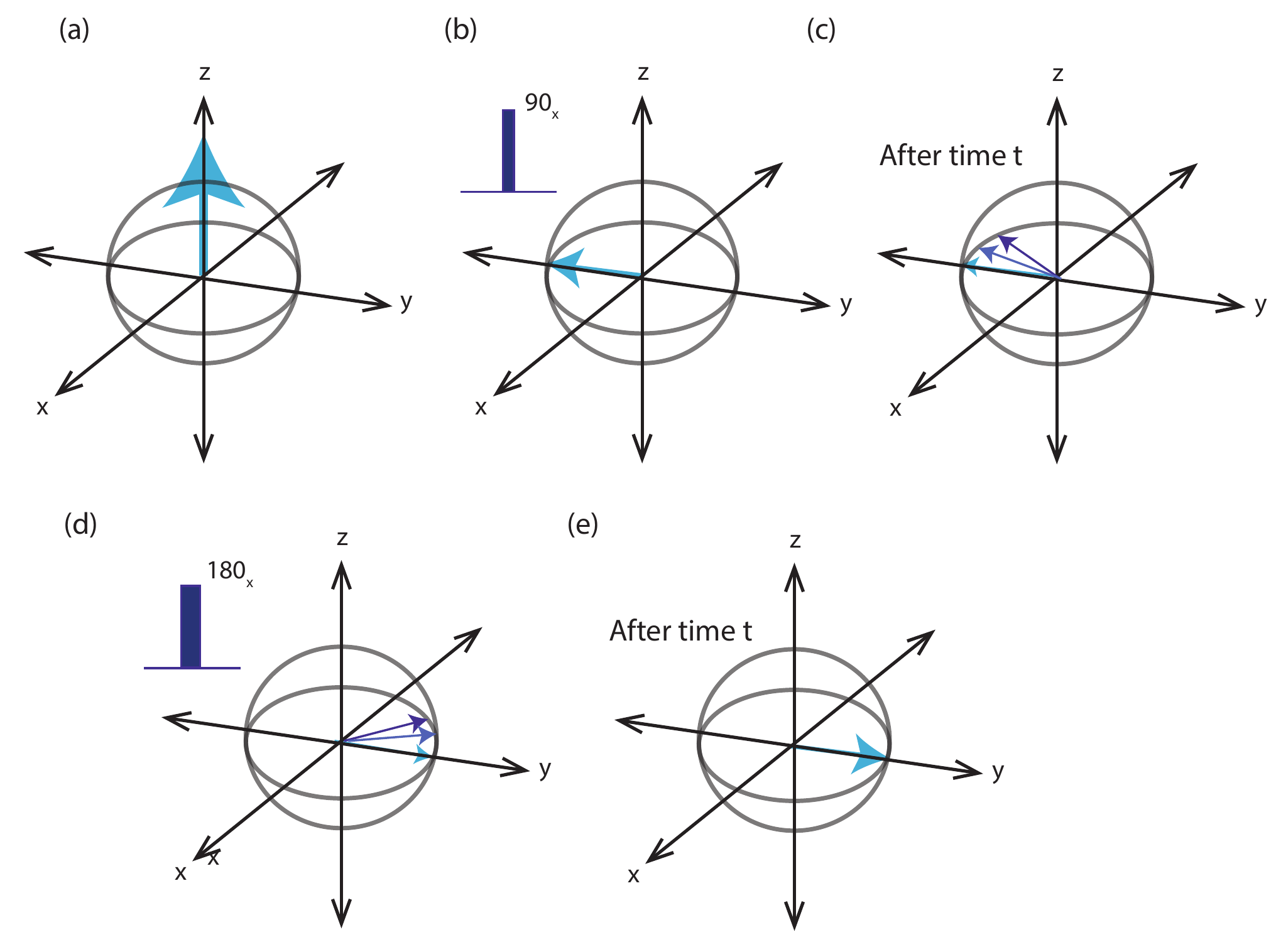}
\caption{Evolution of the bulk magnetization under Hahn echo sequence. (a) Initially
thermal equilibrium bulk magnetization in $z$ direction is represented 
by an arrow.
(b) Bulk magnetization rotated to $-y$ axis using $90^{\circ}$ pulse, 
(c) A delay
of time $t$ is given and arrows represent different spins, (d) $180^{\circ}$
is applied to spins due to which a spin precessing fast will fall behind a 
spin precessing slowly, and (e) after time $t$, all the spins 
are in the same direction. 
}
\label{spinecho_chap1}
\end{figure}
\end{center}
\subsection{Hahn echo}
This technique was constructed 
by E. L. Hahn for suppressing time-independent noise
in a system of isolated spins~\cite{hahn-pr-50}.  In an NMR setup, the static
magnetic field $B_0$ along the $z$ axis has a spatial inhomogeneity due to
which different spins in the ensemble experience different magnetic field and
hence precess with different Larmor frequencies (which cause spin dephasing).
To tackle this problem Hahn devised the spin-echo sequence as shown in
Fig.~\ref{spinecho_chap1}. Initially at thermal equilibrium the bulk
magnetization is in the $z$ direction. With the application of an rf pulse,
a $(\pi/2)_x$ rotation is applied to the bulk magnetization. After time 
$t$ a spin
experiencing a greater 
magnetic field will be ahead of a spin which experiences a smaller
magnetic field as shown in Fig.~\ref{spinecho_chap1}(c). Then, a 
$\pi$ pulse is
applied due to which 
the slow moving spins come close to and the fast moving spins move
away from the $y$ axis.  After time $t$ 
all the spins precess in the same direction as shown in
Fig.~\ref{spinecho_chap1}(e).
\subsection{CPMG DD sequence}
In the Carr-Purcell(CP) sequence a series of rf
pulses are applied: the first pulse flips the magnetization
through $\frac{\pi}{2}$ angle with a $\frac{\pi}{2}$ pulse along the $x$-axis, and the
following train of equidistant pulses flip the magnetization through $\pi$ with the $\pi$ pulse
along the $x$-axis~\cite{carr-pra-54}.
In the actual application of the Carr and Purcell method
for the measurement of long relaxation times, it was found
that the amplitude adjustment of the $\pi$ pulses was critical.
This is because a small deviation from the exact
$\pi$ value gives a cumulative error in the results.
In the Carr-Purcell-Meiboom-Gill(CPMG) sequence:
the first pulse flips the magnetization
through $\frac{\pi}{2}$ angle with a $\frac{\pi}{2}$ 
pulse along a $y$-axis, and a series of $\pi$ pulses along the $x$ axis
are applied at times $t= \frac{(2n+ 1)\tau}{2}$,
$(n= 0, 1, 2\dots )$. The CPMG sequence was able to suppress the time-dependent noise.
\begin{center}
\begin{figure}[h]
\centering
\includegraphics[angle=0,scale=1.7]{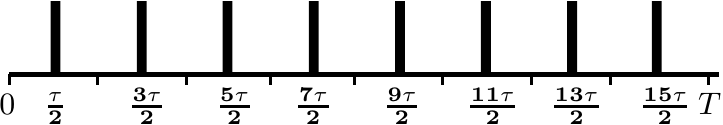}
\caption{Pulse sequence of CPMG DD sequence for a cycle of duration $T$
and in one cycle eight pulses are applied; filled bars
represent $\pi$ rotation pulses and $\tau$ is the duration
between two consecutive $\pi$ pulses. 
}
\label{cpmg_chap1}
\end{figure}
\end{center}
In the CPMG sequence a train of equidistant $\pi$ pulses 
are applied on a qubit. In Fig.~{\ref{cpmg_chap1}}
a train of eight pulses are applied in one cycle of duration $T$.
The more the number of pulses in a cycle, the better 
is the decoherence suppression.
\subsection{Uhrig DD sequence}
The Uhrig DD (UDD) sequence is an optimal DD scheme and was first  constructed
by Uhrig for a pure dephasing spin-boson model~\cite{uhrig-prl-07},
which uses $N$ $\pi$ pulses applied at time intervals $T_j$
\begin{equation}
\label{UhrigTj}
 T_j=Tsin^2\frac{j\pi}{2(N+1)} \ \ \ \text{for} \ \ \ j=1,2, \dots N,
\end{equation}
to eliminate the dephasing up to order $O(T^{N+1})$; hence the
UDD technique suppresses low-frequency noise.
The CPMG sequence is a UDD sequence of order $N=2$. If for 
an interval of duration $T$,
two $\pi$ pulses are applied the CPMG sequence, it will eliminate
dephasing up to order $O(T^{3})$. Further, the proof of the 
universality of the UDD in suppressing the pure dephasing or the
longitudinal relaxation of a qubit coupled to a
generic bath has been given~\cite{yang-prl-08}.
\begin{center}
\begin{figure}[h]
\centering
\includegraphics[angle=0,scale=1.7]{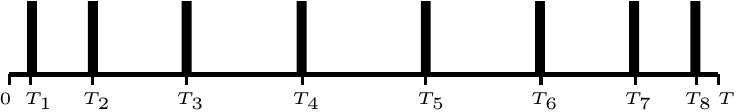}
\caption{Pulse sequence for UDD sequence for a cycle time of $T$ and the number of 
$\pi$ pulses in one cycle of sequence is eight. Filled bars
represent $\pi$ rotation pulses.
}
\label{uddpulses_chap1}
\end{figure}
\end{center}

Yang and Liu considered ideal UDD pulse sequences for a Hamiltonian
of the form
\begin{align}
\hat{H}=\hat{C}+\hat{\sigma}_z\otimes\hat{Z},
\end{align}
where $\hat{\sigma_z}$ is the qubit Pauli matrix along the
$z$-direction, and $\hat{C}$ and $\hat{Z}$ are bath operators. This
Hamiltonian describes a pure dephasing model as it contains no
qubit flip processes and therefore leads to no longitudinal
relaxation but only transverse dephasing.  
Defining two unitary operator $U^{(N)}_{\pm}$ as
follows:
\begin{eqnarray}
\label{oldun}
 U^{(N)}_{\pm}(T)& =& e^{-i[C\pm (-1)^N Z](T-T_N)} \nonumber \\
&&\times\   e^{-i[C\pm (-1)^{(N-1)} Z](T_N-T_{N-1})} \cdots  \nonumber \\
           &&\times\ e^{-i[C\mp Z](T_2-T_1)} e^{-i[C\pm Z]T_1}
\end{eqnarray}
Yang and Liu proved that for $T_j$ satisfying Eq. (\ref{UhrigTj}), we must have
\begin{equation}
\left(U^{(N)}_{-}\right)^{\dagger} U^{(N)}_{+} = 1+O(T^{N+1}),
\label{Tn1}
\end{equation}
i.e., the product of $\left(U^{(N)}_{-}\right)^{\dagger}$ and
$U^{(N)}_{+}$ differs from unity only by the order of $O(T^{N+1})$
for sufficiently small $T$~\cite{yang-prl-08}. 
In Fig.{\ref{uddpulses_chap1}}, eight $\pi$ pulses are applied 
with UDD timing preserving qubit coherence up to order $O(T^{9})$,
whereas the CPMG sequence in Fig.{\ref{cpmg_chap1}} 
is able to preserve coherences up to order $O(T^{3})$.
\subsection{Super-Zeno Scheme}
The super-Zeno scheme is an algorithm for suppressing
the transitions of a quantum mechanical system, initially
prepared in a subspace $\cal{P}$ of the full Hilbert space
of the system, to outside this subspace by subjecting it
to a sequence of unequally spaced short-duration pulses~\cite{dhar-prl-06}.
These durations were calculated numerically the
leakage probability from subspace $\cal{P}$ to its orthogonal subspace
was minimized, 
and surprisingly the durations matched with Eq.(\ref{UhrigTj}).
This scheme is experimentally implemented in 
Chapter{\ref{chapter_superzeno}}.
This scheme efficiently cancels all the
noise affecting
the state, and the preservation is up to the order $O(T^{N+1})$, where 
$N$ denotes the
number of inverting pulses $J$ which are applied. The construction of this
inverting pulse $J$ depends the on subspace $\cal{P}$.
\subsection{Nested Uhrig dynamical decoupling}
The nested Uhrig dynamical decoupling (NUDD) scheme is
an extension of the super-Zeno scheme for the case where
instead of a state, only the subspace to which the state
belongs is known~\cite{mukhtar-pra-10-2}. For protecting an unknown state
in a known subspace, nesting of UDD protecting sequences are done in a
a smart way such that these nesting of layers cancels all the possible
interactions which affect states belonging to the $\cal{P}$ subspace.
The inverting pulses of each layer are constructed
on the basis of subspace to be protected, and these inverting pulses are
applied at the UDD time points given by Eq.(\ref{UhrigTj}).
The NUDD scheme is very sensitive to the nesting of layers, and
for the time interval $T$ with $N$ pulses in each layer,
protection is achieved of $O(T^{N+1})$. 
The NUDD scheme is discussed in
detail in Chapter~\ref{chapter_nudd}.
\section{Organization of the thesis}
This thesis deals with the estimation of experimentally prepared quantum states
and protection of these states using different decoupling strategies. The
thesis is organized as follows: Chapter~2 demonstrates the utility of the
Maximum Likelihood (ML) estimation scheme to estimate quantum states on an NMR
quantum information processor. We experimentally prepare separable and
entangled states of two and three qubits, and reconstruct the density matrices
using both the ML estimation scheme as well as standard quantum state
tomography (QST). Further, we define an entanglement parameter to quantify
multiqubit entanglement and estimate entanglement using both the QST and the ML
estimation schemes. Chapter 3 demonstrates the efficacy of the super-Zeno
scheme for the preservation of a state by freezing state evolution
(one-dimensional subspace protection) and subspace preservation by preventing
leakage of population to an orthogonal subspace (two-dimensional subspace
protection). Both kinds of protection schemes are experimentally demonstrated
on separable as well as on maximally entangled two-qubit states.  Chapter 4
demonstrates the efficacy of NUDD scheme for the protection of arbitrary states
of the known subspace. Chapter 5 
focuses on extending the lifetime of time-invariant discord using
dynamical decoupling schemes.  
Chapter 6 is based on modeling the noise and
experimental protection of a three-qubit system using dynamical decoupling. 
Chapter 7 summarizes the work done in this thesis and discusses 
the future prospects.
\newpage

\chapter{State tomography on an NMR quantum information processor via maximum likelihood estimation}\label{chapter_mle}
\section{Introduction}
Classically, a state is assigned to a physical system by determining
the phase space point corresponding to the configuration of the system 
and measuring the relevant system parameters 
in a non-invasive manner. For quantum
systems, non-invasive measurements are not possible and
therefore an ensemble of identically
prepared systems are required for state estimation. 
The quantum state cannot be known from a single measurement and the 
`no-cloning' theorem renders impossible the possibility of making
several copies of the state and using them to make different
measurements on the same state. Quantum state estimation is
hence intrinsically a statistical process~\cite{massar-prl-95,derka-prl-98}.
In quantum information and experimental quantum computing,
the complete estimation of a quantum state from a set of
measurements on a finite ensemble is a hot topic of research.
Several schemes have been proposed and implemented for 
quantum state estimation~\cite{rehacek-book,plenio-naturecomm,rehacek-pra-15}. 
Due to the finite size of ensembles, a physical situation may 
even have two different candidate states and there is always some ambiguity
associated with the estimated state. Such uncertainties and ambiguities, 
if not treated properly can lead to a self-contradictory estimation, where
the estimated state is not even a valid quantum state.
The density operator
provides a convenient way to describe a quantum
system whose state is not a pure state. It is a positive, Hermitian 
operator with $Tr[\rho]=1$ and $Tr[\rho^2]\leq 1$. For a pure 
ensemble $Tr[\rho^2]=1$ and for a mixed ensemble $Tr[\rho^2] < 1$.
In NMR, the quantum state tomography (QST) of $n$ 
coupled spins 
is carried out by
measuring free induction decays (FIDs)
after applying a set of 
preparatory 
pulses
and applying a 
Fourier transformation to obtain spectral 
lines~\cite{chuang-proc-98,lee-pla-02,long-joptb,leskowitz-pra-04}. 
Fitting these spectral lines, yields complex amplitudes at well defined
frequencies which are characteristic of the spin system. These complex
amplitudes are measured experimentally for each pulse in 
the preparatory pulse set.
Using this experimental data a state density matrix is reconstructed.
However, the reconstructed density matrix with the standard QST procedure 
leads to an unphysical density matrix i.e., some eigenvalues could turn out to be negative. 
This unphysical density matrix makes no sense, as even if there are experimental
errors they should be within the space of allowed density operators.
An \emph{ad hoc} way to circumvent this problem is to
add a multiple of identity to this density matrix
so that the eigenvalues are positive. A scheme that redresses this
issue of reconstructed density matrices that are unphysical, is the
maximum-likelihood (ML) estimation scheme, which obtains a positive definite
estimate for the density matrix by optimizing a likelihood functional that
links experimental data to the estimated density
matrix along with the constraint that the density
matrix should be positive at every point of optimization
~\cite{kohout-prl-10,plenio-mle,james-pra-01}.

The ML estimation scheme begins with a guess
quantum state and improves the estimate based on the measurements made;
the more the number of measurements, the better
is the state estimate.

This issue of 
unphysical density matrices was first pointed out 
in quantum optics where state reconstruction based
on the inversion of measured data 
could not guarantee the positive definiteness of
the reconstructed density matrix. 
Hence an algorithm for quantum-state 
estimation based on the maximum-likelihood (ML) estimation was proposed~\cite{hradil-pra-97,tan-joptb-99}. 
A general method proposed for quantum state estimation based on the
ML approach can be applied to multi-mode radiation fields as well as to spin systems~\cite{banaszek-pra-99}.
An ensemble of spin-$\frac{1}{2}$ particles was observed repeatedly using Stern-Gerlach devices
with varying orientations and the state of an ensemble 
was reconstructed via ML estimation~\cite{hradil-pra-00} .
A simple iterative algorithm for ML 
estimation of the quantum state was derived~\cite{rehacek-pra-01}.
A tomographic protocol for a two-qubit system was
recently constructed based on the measurement of 16 generalized
Pauli operators which is maximally 
robust against errors~\cite{miranowicz-pra-14}. A 
refined iterative ML algorithm was also proposed
to reconstruct a quantum state and 
applied to the tomography
of optical states and 
entangled spin states of trapped ions~\cite{rehacek-pra-07}.
Other quantum state estimation algorithms include Bayesian mean
estimation~\cite{huszar-pra-12}, 
least-squares inversion~\cite{opatrny-pra-97}, numerical strategies
for state estimation~\cite{opatrny-pra-97} and 
linear regression estimation~\cite{kaznady-pra-09}.
If the size of the ensemble is infinite, the estimation procedure
will yield the unique true quantum state of the system. However,
such an ensemble is never achievable in any laboratory setting,
as one can only perform measurements on a finite ensemble.
As a result, the estimated state will
be different from the true state and depends on the details of
the estimation procedure. Quantum state reconstruction on a finite 
number of copies of a quantum system with informationally
incomplete measurements, as a rule, does not yield a unique result.
A reconstruction scheme was derived where both the likelihood and
the von Neumann entropy functionals were maximized (MLME) in order to
systematically select the most-likely estimator with the largest entropy,
that is, the least-bias estimator, was 
consistent with a given set of measurement data.
This MLME estimation protocol was applied to time-multiplexed 
detection tomography and light-beam tomography~\cite{teo-prl-11,teo-pra-12}.

In this chapter, the utility of the ML estimation scheme 
has been demonstrated to perform quantum state estimation on
an NMR quantum information processor. Separable and entangled
states of two and three qubits are experimentally prepared,
and the density matrices are reconstructed using both the standard 
QST and the ML estimation schemes. For the quantification 
of entanglement in multiqubit systems an entanglement parameter
is defined and it is shown that the standard QST method overestimates
the residual state entanglement at a given time, while
the ML estimation method gives a correct estimate of the amount 
of entanglement present in the state. 

\subsection{NMR quantum state tomography}
The basic aim of quantum state tomography (QST) is to completely
reconstruct an unknown state via a set of
measurements on an ensemble of identically
prepared states. Any density matrix $\rho$ of $n$
qubits in a $2^n$-dimensional Hilbert space  can
be uniquely determined using $4^n-1$ independent
measurements and the state of the system as described
by its density operator $\rho$ can be reconstructed by performing
a set of projective measurements on multiple
copies of the
state~\cite{chen-pra-13,vandersypen-review}.
Determining all the elements of $\rho$ would involve
making repeated measurements of the same state in different
measurement bases, until all the elements of $\rho$ are
determined ~\cite{smithey-prl-93,thew-pra-02,james-pra-01}.

In NMR we cannot perform projective measurements
and instead measure the expectation values of
certain fixed operators over the entire ensemble.
Therefore, we rotate the state via different
unitary transformations before performing the
measurement to collect information about different
elements of the density
matrix ~\cite{vandersypen-review}.
The standard tomographic
protocol for NMR  uses the
Pauli basis to expand an $n$ qubit $\rho$,
\begin{equation}
\rho = \sum_{i=0}^{3} \sum_{j=0}^{3} ...
\sum_{k=0}^{3} c_{ij...k} \sigma_i \otimes \sigma_j
\otimes ....\sigma_k
\label{rho-compact}
\end{equation} 
where $c_{00...0}=1/2^{n}$ and $\sigma_0$ denotes the
$2\times 2 $ identity matrix while $\sigma_1$,
$\sigma_2$ and $\sigma_3$ are standard Pauli
matrices. The measurements of expectation values
allowed in an NMR experiment combined with unitary
rotations leads to the determination of the
coefficients $c_{ij..k}$.

In an NMR experiment, we measure the signal
induced in the detection coils while the nuclear
spins precess freely in a strong applied magnetic field.
This signal is called the free induction decay
(FID) and is proportional to the time rate of change of
magnetic flux. This time-domain signal can be expressed in
terms of the expectation values of the transverse
magnetization~\cite{leskowitz-pra-04}:
\begin{equation}
S(t)\propto \mbox{Tr}\left
\{\rho(t)\sum_{k}(\sigma_{kx}-i\sigma_{ky})\right \}
\label{nmrsignal1}
\end{equation}
where $\sigma_{kx}$ and $\sigma_{ky}$ are the Pauli spin operators proportional
to the $x$ and $y$ components of the magnetization of the $k^{th}$
spin and $\rho(t)$ is the instantaneous density
operator at time $t$ during 
the FID.
The recorded signal represents an average over a large
number of identical molecules of the sample.

The transformation of an initial density operator $\rho_{0}$
under applied pulses $U_{P}$ and under free-evolution 
Hamiltonian $H$ for time $t$ is given by:
\begin{equation}
 \rho(t)=e^{-iHt}U_{P}\rho_{0}U_{P}^{\dagger}e^{iHt}.
\end{equation}
Then Eq.(\ref{nmrsignal1}) reduces to
\begin{equation}
S_p(t)\propto \mbox{Tr}\left
\{e^{-iHt}U_{P}\rho_{0}U_{P}^{\dagger}e^{iHt}\sum_{k}(\sigma_{kx}-i\sigma_{ky})\right \},
\label{nmrsignal2}
\end{equation}
using the linearity of the trace and its invariance with respect to cyclic
permutation of the operators, we can write 
\begin{equation}
S_p(t)\propto \mbox{Tr}\left
\{\rho_{0} U_{P}^{\dagger}e^{iHt}\sum_{k}(\sigma_{kx}-i\sigma_{ky})e^{-iHt}U_{P}\right \}
\label{nmrsignal3}
\end{equation}

In NMR spectroscopy, the initial density operator $\rho_{0}$ is known
and it represents the thermal equilibrium state then from the 
signal we can determine the Hamiltonian. However in state tomography the reverse 
is true i.e. the Hamiltonian is known and the state $\rho_{0}$ is unknown.
The Hamiltonian of $n$ weakly
coupled spins-$\frac{1}{2}$ is given by,
\begin{equation}
H = -\sum_{k=1}^{n} \omega_k I_{kz} + 2 \pi\sum_{k=2
}^{n} \sum_{j =1}^{k-1} J_{jk} I_{jz} I_{kz}
\end{equation}
where $\omega_k/2\pi$ is the Larmor frequency of the $k$th spin,
and $J_{jk}$ is the spin-spin coupling constant between $j$th and $k$th
spins. 

\noindent For a single spin system, the Hamiltonian $H_1$ consists of a 
Zeeman term only, which can be written as

\begin{equation}
 H_1=\frac{1}{2}\omega \sigma_z
\end{equation}

where $\omega$ is the Larmor frequency of the nuclear spin
in the external magnetic field. Then, the NMR signal can be written as

\begin{equation}
S^1_p(t)\propto \mbox{Tr}\left
\{\rho_{0} U_{P}^{\dagger}e^{i\frac{1}{2}\omega \sigma_zt}\sum_{k}(\sigma_{kx}-i\sigma_{ky})e^{-i\frac{1}{2}\omega \sigma_z t}U_{P}\right \}
\label{nmrsignal4}
\end{equation}

Without applying any pulse i.e. $U_p=I$, the NMR signal is

\begin{equation}
S^1_I(t)\propto \left
\{ \mbox{Tr}[\rho_{0}\sigma_x]-i  \mbox{Tr}[\rho_{0}\sigma_y]\right \} e^{i\omega t}.
\label{nmrsignal5}
\end{equation}

after applying $90^o$ pulse along $x$ i.e. $U_p=90^o_x$, the NMR signal 
is

\begin{equation}
S^1_X(t)\propto \left
\{ \mbox{Tr}[\rho_{0}\sigma_x]-i  \mbox{Tr}[\rho_{0}\sigma_z]\right \} e^{i\omega t}.
\label{nmrsignal6}
\end{equation}

after applying $90^o$ pulse along $y$ i.e. $U_p=90^o_y$, the NMR signal can be written as

\begin{equation}
S^1_Y(t)\propto \left
\{ \mbox{Tr}[\rho_{0}\sigma_z]-i  \mbox{Tr}[\rho_{0}\sigma_y]\right \} e^{i\omega t}.
\label{nmrsignal7}
\end{equation}

Applying a Fourier transformation on Eq.(\ref{nmrsignal5}, \ref{nmrsignal6} and \ref{nmrsignal7})
then the ensemble average of operators $\sigma_x$, $\sigma_y$ and $\sigma_z$ can be obtained

\begin{eqnarray}
 \langle \sigma_x\rangle &=& c~{\rm avg}({\rm Re}[S_I(\omega)],{\rm Re}[S_X(\omega)]), \\
  \langle \sigma_y\rangle &=& c~{\rm avg}(-{\rm Im}[S_I(\omega)],-{\rm Im} S_Y(\omega)]), \\
 \langle \sigma_z\rangle &=&c~{\rm avg}(-{\rm Im}[S_X(\omega)],{\rm Re}[S_Y(\omega)]). 
 \end{eqnarray}
 
 Where ${\rm avg}(a,b)$ means an average of $a$ and $b$. The factor $c$
 depends on the experimental details such as the receiver gain and the 
number of spins. After determining the factor $c$, the density operator of a 
 single spin can be estimated as
 
 {\begin{equation}
       \rho=\frac{1}{2}I+\langle \sigma_x\rangle \sigma_x+\langle \sigma_y\rangle \sigma_y+\langle \sigma_z\rangle \sigma_z 
      \end{equation}}

\noindent For a two-spin system, the Hamiltonian can be written as
 \begin{equation}
  H_2=\frac{1}{2}\omega_1 \sigma_{1z}+\frac{1}{2} \omega_2 \sigma_{2z}+\frac{\pi}{2}J_{12}\sigma_{1z}\sigma_{2z}
 \end{equation}
where $J_{12}$ is the scalar coupling constant.
After inserting $H_2$ in Eq.(\ref{nmrsignal3}) and solving few steps
the NMR signal due to spin 1 can be written as

\begin{equation}
 S_{P,1}(t)\propto \frac{1}{2}(e^{i(\omega_{1}-\pi J_{12})t},e^{i(\omega_{1}+\pi J_{12})t})
 \times \left(
\begin{array}{cc}
 1 & 1 \\
 1 & -1 \\
\end{array}
\right)
\left(
\begin{array}{c}
 \tr \{\rho_{0}\tilde{\sigma}_{1-}\}\\
 \tr \{\rho_{0}\tilde{\sigma}_{1-}\tilde{\sigma}_{2z}\}\\
\end{array}
\right)
\label{sig_in_t_peak1}
\end{equation}

 Where $\tilde{\sigma}_{1-}= U_{P}^{\dagger}(\sigma_{1x}-i\sigma_{1y})U_{P}$.
Fourier transformation of $S_{P,1}(t)$ leads to,

\begin{equation}
 \left(\begin{array}{c}
  \bar{S}_{P,1}(\omega_1-\pi J_{12})\\
  \bar{S}_{P,1}(\omega_1+\pi J_{12})\\
 \end{array}\right) \propto \frac{1}{2}
 \left(
\begin{array}{cc}
 1 & 1 \\
 1 & -1 \\
\end{array}
\right)
\left(
\begin{array}{c}
 \tr \{\rho_{0}\tilde{\sigma}_{1-}\}\\
 \tr \{\rho_{0}\tilde{\sigma}_{1-}\tilde{\sigma}_{2z}\}\\
\end{array}
\right)
\label{sig_in_w_peak1}
\end{equation}

Similarly for spin 2,

\begin{equation}
 \left(\begin{array}{c}
  \bar{S}_{P,2}(\omega_2-\pi J_{12})\\
  \bar{S}_{P,2}(\omega_2+\pi J_{12})\\
 \end{array}\right) \propto \frac{1}{2}
 \left(
\begin{array}{cc}
 1 & 1 \\
 1 & -1 \\
\end{array}
\right)
\left(
\begin{array}{c}
 \tr \{\rho_{0}\tilde{\omega}_{2-}\}\\
 \tr \{\rho_{0}2 \tilde{\omega}_{1z}\tilde{\omega}_{2-}\}\\
\end{array}
\right)
\label{sig_in_w_peak2}
\end{equation}

The density matrix of a two-spin system
can be expanded in terms of Pauli basis operators $I_1 \otimes I_2 \dots
\sigma_{1z} \otimes \sigma_{2z}$, as follows:
\begin{eqnarray}
\rho
&=& \frac{1}{4}(I_1\otimes
I_2+\langle\sigma_{1x}\otimes I_2\rangle
\sigma_{1x}\otimes I_2+
\dots \nonumber \\
&&+\langle\sigma_{1z}\otimes \sigma_{2y}\rangle
\sigma_{1z}\otimes \sigma_{2y}
+ \langle\sigma_{1z}\otimes \sigma_{2z}\rangle
\sigma_{1z}\otimes \sigma_{2z})
\label{2qrho}
\end{eqnarray}
and for the estimation of $\rho$ we need to calculate the 
expectation values $\langle\sigma_{1x}\otimes I_2\rangle$,$\dots$,
$\langle\sigma_{1z}\otimes \sigma_{2z}\rangle$. From the NMR spectra
we can calculate the peaks intensities i.e. 
$\bar{S}_{P,1}$ and $\bar{S}_{P,2}$ and rewriting the
Eq.(\ref{sig_in_w_peak1}), and Eq.(\ref{sig_in_w_peak2})

\begin{equation}
 \left(
\begin{array}{c}
 \tr \{\rho_{0}\tilde{I}_{1-}\}\\
 \tr \{\rho_{0}2\tilde{I}_{1-}\tilde{I}_{2z}\}\\
\end{array}
\right)\propto
 \left(
\begin{array}{cc}
 1 & 1 \\
 1 & -1 \\
\end{array}
\right)
\left(\begin{array}{c}
  \bar{S}_{P,1}(\omega_1-\pi J_{12})\\
  \bar{S}_{P,1}(\omega_1+\pi J_{12})\\
 \end{array}\right) 
\label{exp_in_w_peak1}
\end{equation}

\begin{equation}
 \left(
\begin{array}{c}
 \tr \{\rho_{0}\tilde{I}_{2-}\}\\
 \tr \{\rho_{0}2 \tilde{I}_{1z}\tilde{I}_{2-}\}\\
\end{array}\right)\propto
 \left(
\begin{array}{cc}
 1 & 1 \\
 1 & -1 \\
\end{array}
\right)
\left(\begin{array}{c}
  \bar{S}_{P,2}(\omega_2-\pi J_{12})\\
  \bar{S}_{P,2}(\omega_2+\pi J_{12})\\
 \end{array}\right) 
\label{exp_in_w_peak2}
\end{equation}
we can calculate
the expectation values. Taking  $U_{P}=\mathbf{I}\mathbf{I}$ and inserting
the experimentally measured peak 
intensities of spin 1 in Eq.(\ref{exp_in_w_peak1}),
we get the 
expectation values 
$\langle \sigma_{1x} \otimes I_2 \rangle $, $\langle \sigma_{1y} \otimes I_2 \rangle$,
$\langle \sigma_{1x}\otimes \sigma_{2z} \rangle$, and $\langle \sigma_{1y}\otimes \sigma_{2z} \rangle$. On inserting the peak intensities of spin 2 in
Eq.(\ref{exp_in_w_peak2}) we get the expectation values $\langle I_1 \otimes  \sigma_{2x} \rangle  $, $\langle I_1 \otimes  \sigma_{2y} \rangle  $,
$\langle \sigma_{1z}\otimes \sigma_{2x} \rangle$, and $\langle \sigma_{1z}\otimes \sigma_{2y} \rangle$. Similarly we can measure other expectation values by changing
the preparatory pulse. With the set 
$U_{P}=\{\mathbf{I}\mathbf{I},\mathbf{I}X,
\mathbf{I}Y,XX\}$, we can measure all 
the expectation values required to reconstruct
the density matrix, where $\mathbf{I}\mathbf{I}$ corresponds to ``no
operation'' on both spins,
$\mathbf{I}X(Y)$ corresponds to a ``no operation''
on the first spin and a
$90^{\circ}$ rf pulse of phase $X(Y)$ on the 
second spin, and $XX$ corresponds to a
$90^{\circ}$ rf pulse of phase $X$ on both
spins.

As an
example for a two-qubit system, we created
the quantum state
$\frac{1}{\sqrt{2}}(\vert 00\rangle
+\vert 01\rangle)$ 
and reconstructed it using standard QST. 
We experimentally generated
twenty-five density matrices for this state, and
computed the mean and the variance.
The 
reconstructed density matrix 
$\rho_{_{QST}}$ turned out to be 

\begin{equation}
{\rho_{_{QST}}}= \left(\begin{array}{cccc}
0.484 &   0.508+i0.028 &  -0.025-i0.029 &  -0.025+i0.019 \\
 0.508-i0.027 & 0.516 & 0.025+i0.003 & 0.009+i0.030 \\
-0.025+i0.029 & 0.025-i0.003 & -0.039+i0.000 & -0.025-i0.011 \\
-0.025-i0.019 & 0.009-i0.030 & -0.025+i0.011 & 0.039+i0.000
\end{array}
\right)
\label{qstmatrix}
\end{equation}

The above density matrix $\rho_{_{QST}}$,
reconstructed using the standard QST protocol, is
normalized and Hermitian  and its eigenvalues are 
$\{1.011 \pm 0.008, 0.052 \pm 0.025, 0.016 \pm
0.008,-0.079 \pm 0.018\}$.  
The errors in the reconstructed density matrix using the
QST method show up in the third decimal place.
As is evident from the
eigenvalues, the reconstructed density matrix is
not positive, and furthermore,
$\mbox{Tr}(\rho^2_{_{QST}}) = 1.031 \pm 0.009$. 
It is clear from the above data that the
negativity of the eigen value is statistically
significant and is due to the way we have carried
out state estimation.
Density matrices that represent physical
quantum states must have the property of positive
definiteness which, in conjunction with the
properties of normalization and Hermiticity,
implies that all the eigenvalues must lie in
the interval [0,1] and  sum to 1 i.e. $0 \leq \mbox{Tr}
(\rho^2 ) \leq 1$. Clearly, the above density matrix
which is reconstructed by the standard QST protocol violates these
conditions. Due to its negative eigenvalues it has
as a strange feature that  $\mbox{Tr} (\rho^2) >
\mbox{Tr} (\rho)$.
The obvious reasons for this problem are
experimental inaccuracies, which implies that the
actual magnetization values recorded in an NMR
experiment differ from those that can be obtained
from the Eq.~(\ref{2qrho}). However, in a correct
estimation scheme the experimental inaccuracies
should lead to an error in the state estimation by
giving a state which is close to the actual state
with some confidence level and should not give a
non-state!  An {\em ad hoc} way to circumvent this
problem is to add a multiple of identity to this
density matrix so that the eigenvalues are
positive. However, this kind of addition is
completely {\em ad hoc}, and leads to non-optimal
estimates and one should be able to do better. We
turn to this issue in the next section via the
maximum likelihood estimation method.
\subsection{Maximum likelihood estimation}
\label{mle}
The example in the previous section
illustrates that density matrices which are tomographed using
standard QST may not correspond to a physical
quantum state.
To address this
problem, the
maximum likelihood (ML) estimation method was
developed to ensure that the reconstructed
density matrix is always positive and
normalized~\cite{james-pra-01}.  The ML estimation method
estimates the entire quantum state, by finding the
parameters that are most likely to match the
experimentally generated data and maximizing a
specific target function; {\em a priori} knowledge
of the density matrix can also be incorporated
into the method. The main advantage of this method
is that at every stage of the estimation process
the density matrix is positive and normalized and
therefore represents a valid physical situation.
The construction of a valid density operator through
maximum likelihood estimation consists of 
the following steps:
\begin{enumerate}
 \item The density operator is first obtained 
from a lower triangular matrix $T$ such that 
 $\rho=T^{\dagger}T$, here $T$ is a function of real variables 
 $\{t_1,t_2,t_3,\dots,t_{4^n} \}$ and $n$ is the number of qubits. With this 
 $\rho(t_1,t_2,t_3,\dots,t_{4^n})$ will be always Hermitian and positive. 
 
 \item A ``likelihood function'' is then constructed which quantifies how close
 the density operator $\rho(t_1,t_2,t_3,\dots,t_{4^n})$ is with respect to
 the experimental data. This likelihood function is a function of $t_i$
 and experimental data $n_i$ and can be written as $\mathcal{L}(t_1,t_2,\dots,t_{4^n}; n_1,n_2,\dots,n_{4^n})$.
 
 \item Using standard numerical optimization techniques, 
the optimum set of variables
 $\{ t_1^{(opt)},t_2^{(opt)},t_3^{(opt)},\dots,t_{4^n}^{(opt)}\}$ 
is obtained, 
for which the function $\mathcal{L}_(t_1,t_2,\dots,t_{4^n};n_1,n_2$ 
 
 $,\dots,n_{16})$
 has the maximum value. The best estimated density operator is $\rho(t_1^{(opt)},t_2^{(opt)},$

 $ t_3^{(opt)},\dots,t_{4^n}^{(opt)})$.

\end{enumerate}

For a system of two qubits, the density matrix can
be written in a compact form following
Eq.~(\ref{rho-compact}): 
\begin{equation} \rho=
\sum_{j=0}^{3} \sum_{k=0}^{3} n_{jk} \sigma_{j}
\otimes \sigma_{k} \label{rho2} \end{equation}
where $n_{jk}$ are real coefficients determining
the state.

A physical  density matrix $\rho$  has to be
Hermitian, positive  and must have trace equal to
unity. Such a density matrix can  be written
in terms of a lower triangular matrix $T$~\cite{james-pra-01}
\begin{equation}
\rho(t_{1},t_{2},\dots,t_{16}) = T^{\dag} T\quad \mbox{Tr} (T^{\dag} T)=1
\label{tmatrix}
\end{equation}

For a two-qubit system the lower triangular matrix
$T$ from which we obtain $\rho$
has  15
independent real parameters (one parameter from
the 16 is eliminated due to the trace
condition), and can be written as
\begin{equation}
T= \left( \begin{array}{cccc}
t_{1}&0&0&0 \\
t_{5}+it_{6}&t_{2}&0&0 \\
t_{11}+it_{12}&t_{7}+it_{8}&t_{3}&0 \\
t_{15}+it_{16}&t_{13}+it_{14}&t_{9}+it_{10}&t_{4}
\end{array}
\right)
\end{equation}

Given a valid density matrix as described
in~\cite{james-pra-01}, it is possible to 
invert Eq.~(\ref{tmatrix}) to obtain the matrix $T$ 
\begin{equation}
T= \left( \begin{array}{cccc}
\sqrt{\frac{\Delta}{\mathcal{M}_{11}^{(1)}}}&0&0&0 \\
{\frac{\mathcal{M}_{12}^{(1)}}{\sqrt{\mathcal{M}_{11}^{(1)}\mathcal{M}_{11,22}^{(2)}}}}&\sqrt{\frac{\mathcal{M}_{11}^{(1)}}{\mathcal{M}_{11,22}^{(2)}}}&0&0
\\
{\frac{\mathcal{M}_{12,23}^{(2)}}{\sqrt{\rho_{44}}\sqrt{\mathcal{M}_{11,23}^{(2)}}}}&\frac{\mathcal{M}_{11,22}^{(2)}}{\sqrt{\rho_{44}}\sqrt{\mathcal{M}_{11,22}^{(2)}}}&\sqrt{\frac{\mathcal{M}_{11,22}^{(2)}}{\rho_{44}}}&0
\\
\frac{\rho_{41}}{\sqrt{\rho_{44}}}&\frac{\rho_{42}}{\sqrt{\rho_{44}}}&\frac{\rho_{43}}{\sqrt{\rho_{44}}}&\sqrt{\rho_{44}}
\end{array}
\right)
\label{rho2t}
\end{equation}
where $\Delta=\mbox{Det}(\rho)$, $\mathcal{M}_{ij}^{(1)}$ is
the first minor of $\rho$ (the determinant of
the $3\times3$ matrix formed by deleting the $i$th and $j$th
columns
of the $\rho$ matrix), $\mathcal{M}_{ij,kl}^{(2)}$ is the second
minor of $\rho$ (the determinant of the $2\times2$
matrix formed by deleting the $i$th and $k$th rows and $j$th and
$l$th columns of the $\rho$ matrix with $i \neq j$ and $k
\neq l$). 
From the experimental data we obtain a set of
expectation values $\bar{n}_{jk}=\langle\sigma_{j}
\otimes \sigma_{k}\rangle= \mbox{Tr}((\sigma_{j}
\otimes \sigma_{k})\rho)$.

The noise in a complex NMR signal acquired from a single receiving
coil using quadrature detection is uncorrelated (white) and Gaussian~\cite{garge-jmr-03}. 
So, it is assumed that the experimental
noise has a Gaussian probability distribution and
the probability of obtaining a set of 
measurement results for the set of expectation values
$\{n_{jk}\}$ is \begin{equation}
P(n_{11},\cdots n_{33})= A \prod_{j=0,k=0}^{3,3}
exp\left[-\frac{(n_{jk}-\bar
n_{jk})^2} {2\sigma_{jk}^2}\right]
\end{equation} where $A$ is a normalization
constant and $\sigma_{jk}$ is the standard
deviation of the measured variable $n_{jk}$
(approximately
given by $\sqrt{\bar n_{jk}}$).   
The next step in the ML estimation method is to maximize the
likelihood that the physical density matrix $\rho$ will give
rise to the experimental data $\{n_{jk}\}$.
Since $\rm ln(x)$ is an increasing function, the maxima of the
likelihood and the log of the likelihood coincide.
Rather than finding the
maximum value of the probability $P$, the optimization
problem gets simplified by finding the maximum of its
logarithm. Here we neglect the dependence of the 
normalization constant on $t_1, t_2, \dots, t_{16}$ , 
which only weakly affects the solution
for the most likely state. So, we need to maximize
\begin{center}
$-\sum_{j=0,k=0}^{3,3}\frac{\left(n_{jk}(t_1,\cdots
t_{16}) - \bar
n_{jk}(t_1,\cdots t_{16})\right)^2}{2\bar\sigma_{jk}^2}$.
\end{center}

Mathematically, if $x_0$ is a maxima of function $f$ 
then $\max (f(x_0))= \min(-f(x_0))$, 
thus the optimization problem is reduced to
finding the minimum of  a ``likelihood function''
\begin{equation}
 \mathcal{L}(t_1,\cdots t_{16})=
\sum_{j=0,k=0}^{3,3}\frac{\left(n_{jk}(t_1,\cdots
t_{16}) - \bar
n_{jk}(t_1,\cdots t_{16})\right)^2}{2\bar\sigma_{jk}^2}
\label{functional}
\end{equation}
Strictly speaking, for a system of NMR coupled qubits, the
functional defined in Eq.~(\ref{functional}) is not a
``likelihood'' since the NMR experiment measures 
expectation values but can be considered to be a
Gaussian approximation of
likelihood~\cite{james-pra-01}.

For a system of two qubits, the optimum set of variables
$\{t_{1}^{opt},t_{2}^{opt},\dots,t_{16}^{opt}\}$ which
minimizes this likelihood function can be determined using
numerical optimization techniques.
We used the MATLAB routine ``lsqnonlin''~\cite{matlab} to
find the minimum of the likelihood function. To execute this
routine, one requires the initial estimation of the value of
$t_{1},t_{2},\dots,t_{16}$. Since a sixteen 
parameter optimization can be tricky, it is important to
use a good initial guess for parameters. A
reasonable way is to first estimate the state
using the standard method, and obtain the values
of $t'_{i}$s using the
Eq.~(\ref{rho2t}). Since the state may not be
a physically allowed state, the parameters obtained
in this manner are not necessarily real. Thus for
our initial guess we drop the imaginary part and
use the real parts of each of the $t'_{i}$s as the
initial estimate to go into the optimization
routine.  
We used the same 
experimentally generated $\frac{1}{\sqrt{2}}(\vert
00\rangle +\vert 01\rangle)$ state (as a mean of
twenty-five experimental density matrices as described in the
example given in the earlier subsection), 
and re-computed the density matrix now
using the ML estimation
method, and obtained:
\begin{equation}
{\rho_{_{ML}}}= \left(\begin{array}{cccc}
0.488 & 0.487+i0.001 & 0.002-i0.012 & -0.002+i0.012 \\
0.487-i0.001 & 0.487 & 0.002-i0.012 & -0.002 +i0.012 \\
0.002+i0.012 & 0.002+i0.012 & 0.013 & -0.013 +i0.000 \\
-0.002-i0.012 & -0.002-i0.012 & -0.013-i0.000 & 0.013
\end{array}
\right)
\label{mlematrix}
\end{equation}

The eigen values of this matrix are 
$\{0.975 \pm 0.001, 0.025 \pm 0.001, 
0.001 \pm 0.000,0.000$ $ \pm 0.000 \}$
and are all positive and
furthermore $\mbox{Tr}(\rho^2_{\mbox{ML}})=
0.950 \pm 0.004$.  
The errors in the reconstructed density matrix using the
ML estimation method show up in the third decimal place.
While the density matrix reconstructed
using QST was unphysical, the ML reconstruction
led to a valid density matrix.
\section{Comparison of quantum state estimation via ML estimation and standard QST schemes}
\label{expt}
\begin{figure}[ht]
\begin{center}
\includegraphics[angle=0,scale=0.84]{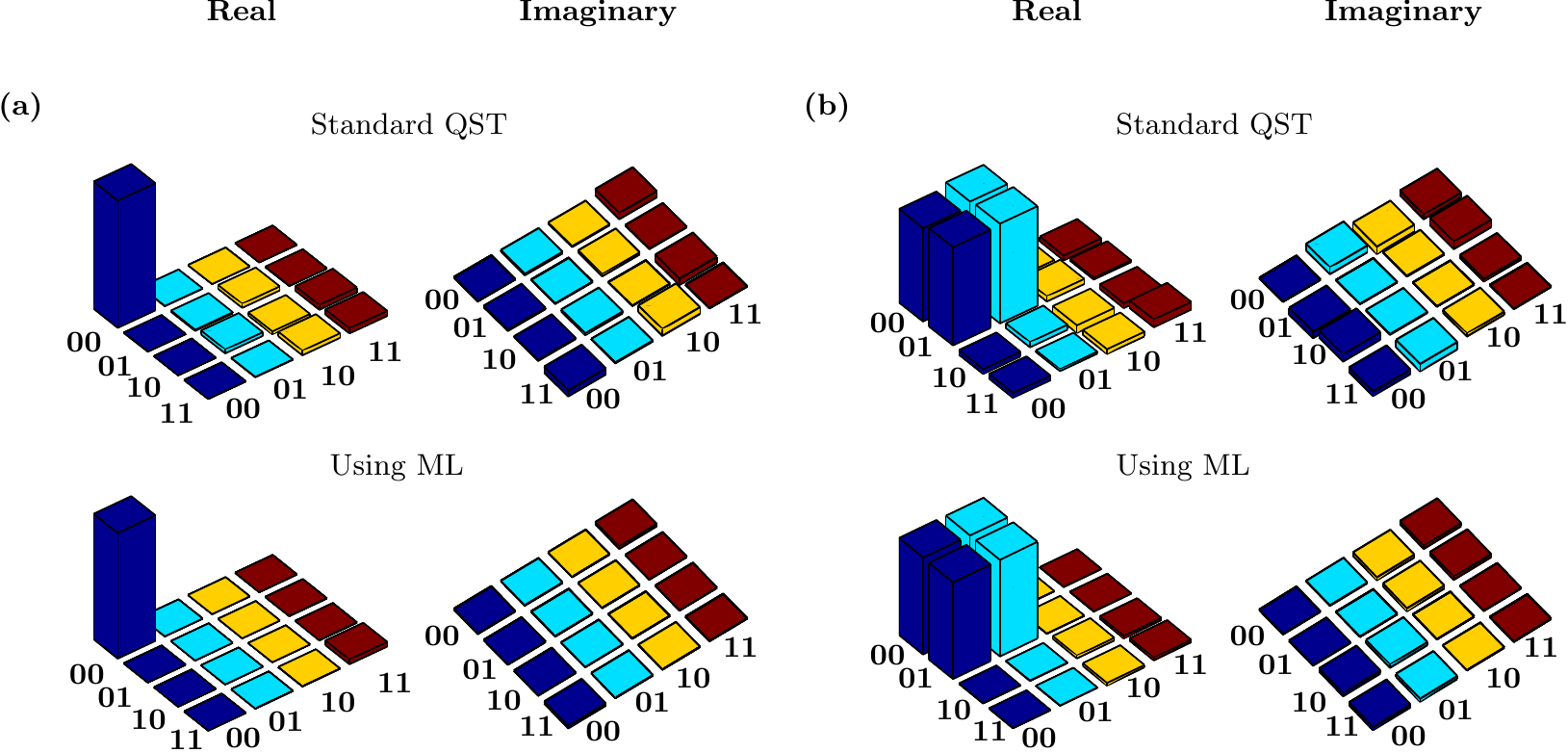}
\caption{Real (left) and imaginary (right) parts of
the experimental tomographs of the (a) $\vert 00 \rangle$
state, with a
computed
fidelity of 0.9937 using standard QST and 
a computed fidelity of 0.9992 using ML method for state estimation.
(b)
$\frac{1}{\sqrt{2}}(\vert 00 \rangle+\vert 01
\rangle)$ state, with a
computed
fidelity of 0.9928 using standard QST and 
a computed fidelity of 0.9991 using ML  method for
state estimation.
The rows and columns are labeled in the
computational basis ordered from $\vert 00 \rangle$ to
$\vert 11 \rangle$.}
\label{2tomo}
\end{center}
\end{figure}
\begin{figure}[ht]
\centering
\includegraphics[angle=0,scale=1.1]{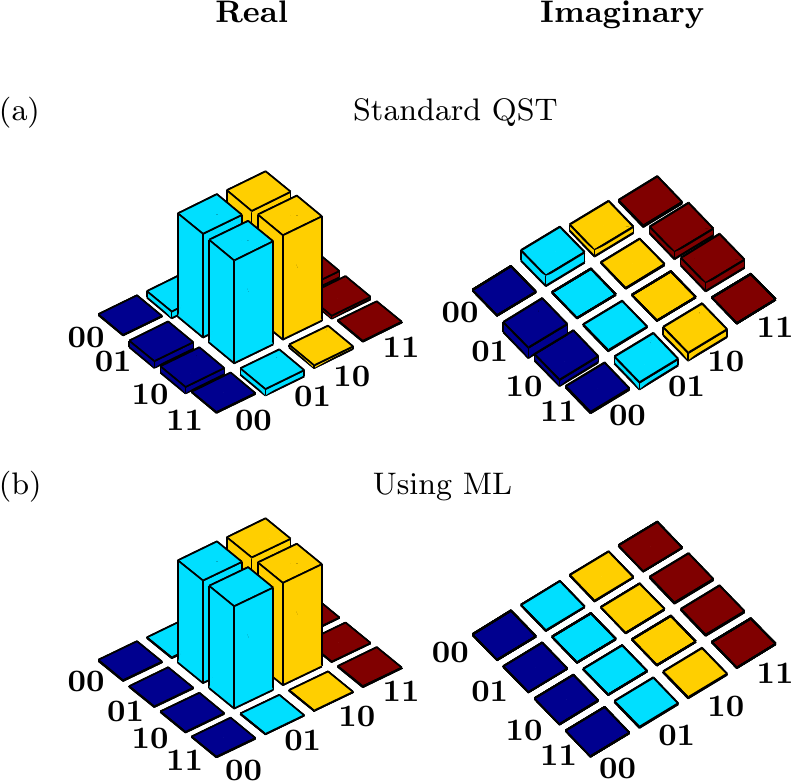}
\caption{ 
Real (left) and imaginary (right) parts of
the experimental tomographs of the
entangled state $\frac{1}{\sqrt{2}}(\vert 01
\rangle+\vert 10 \rangle)$ reconstructed
(a) using standard QST and (b) using ML estimation.
The fidelities computed using standard QST
and using ML method for state estimation are
0.9933 
and 0.9999
respectively. 
The rows and
columns are
labeled in the computational basis ordered from $\vert 00
\rangle$ to
$\vert 11 \rangle$.
}
\label{entangtomo}
\end{figure}
We performed state estimation of several different quantum
states of two and three qubits, constructed on an NMR
quantum information processor, using the ML estimation method.  The
results were compared every time with the results obtained
by reconstruction using the standard QST protocol. 

\subsubsection{Fidelity measure and state estimation}
The fidelity measures commonly used in NMR quantum computing
are:
\begin{enumerate}

 \item The fidelity measure $F_1$ is computed by measuring the
overlap between theoretically expected and
experimentally measured states~\cite{weinstein-prl-01}:
\begin{equation}
F_1 =
\frac{Tr(\rho_{\rm theory}\rho_{\rm expt})}
{\sqrt(Tr(\rho_{\rm theory}^{2}))
\sqrt(Tr(\rho_{\rm expt}^{2}))}
\end{equation}

\item The fidelity measure $F_2$ is computed by
measuring the
projection between the
theoretically expected and experimentally
measured states using the Uhlmann-Jozsa
fidelity measure~\cite{uhlmann-fidelity,jozsa-fidelity}:

\begin{equation}
F_2 =
\left(Tr \left( \sqrt{
\sqrt{\rho_{\rm theory}}
\rho_{\rm expt} \sqrt{\rho_{\rm theory}}
}
\right)\right)^2
\label{mle_fidelity_2}
\end{equation}

\end{enumerate}

where $\rho_{\rm theory}$ and $\rho_{\rm expt}$ denote the
theoretically expected and experimentally reconstructed
density matrices, respectively.

\subsection{Comparison of separable states estimation}
On a system of two qubits, we began by tomographing a pure
state $|00\rangle $, as well as a superposition state
$\frac{1}{\sqrt{2}}(\vert 00\rangle +\vert 01\rangle)$
(which can be written as a tensor product of the first qubit
in the $\vert 0 \rangle$ state  and the second qubit in a
coherent superposition of the $\vert 0 \rangle$ and $\vert 1
\rangle$ states). 
The mean of ten to twenty-five experimentally determined data
matrices were considered, and the reconstructed density matrices using
the ML estimation method and using the standard QST method are shown
as bar tomographs in Figure~\ref{2tomo}, with the states
labeled in the computational basis in the order $\vert 00
\rangle$ to $\vert 11 \rangle$.  Using standard QST, the
reconstructed $\vert 00 \rangle$ state had negative 
eigenvalues: 
$\{0.994 \pm 0.000,0.073 \pm 0.006,-0.001 
\pm 0.002,-0.066 \pm 0.005 \}$, 
and state fidelity 
was computed with measure $F_1$ to be 0.9937 and 
with measure $F_2$ to be 0.9940. 
Reconstructing the state using
ML estimation, we obtained all positive eigenvalues: 
$\{0.965 \pm 0.001, 0.035 \pm 0.001,
0.000 \pm 0.000, 0.000 \pm 0.000 \}$,
while state fidelity was computed with measure $F_1$ to be
0.9992 and with measure $F_2$ to be 0.9652.  For the superposition state
$\frac{1}{\sqrt{2}}(\vert 00\rangle +\vert 01\rangle)$,
state reconstruction using standard QST led to some negative
eigenvalues: 
$\{1.011 \pm 0.002, 0.052 \pm 0.005, 0.016 \pm 0.002,
-0.079 \pm 0.004 \}$
with the fidelity measures $F_1$ and $F_2$  are 0.9928 and 1.0110 respectively. 
Using ML estimation on the other hand, led to all positive
eigenvalues: 
$\{0.975 \pm 0.001, 0.025 \pm 0.001, 
0.001 \pm 0.000, 0.000 \pm 0.000 \}$
with a state
fidelity  with measures $F_1$ and $F_2$ are  0.9991 and 0.9745.
While state fidelities with $F_1$ measure are nearly the same 
(or slightly better when calculated after ML
reconstruction of the density matrix), we find that by using
the ML estimation method for state estimation, we always  obtain a
$\rho$ which is 
physically valid.
\subsection{Comparison of entangled states estimation}
\label{chapmle_entang}
It has been previously  noted~\cite{james-pra-01} that
the standard QST protocol frequently leads to unphysical
density matrices for entangled multiqubit states. Since
entanglement has been posited to lie at the heart of
quantum computational speedup, their construction and
estimation is of prime importance.  We used the ML estimation method
to reconstruct two-qubit and three-qubit entangled states
and evaluated the efficacy of this scheme to construct
valid density matrices.
\begin{figure}[ht]
\centering
\includegraphics[angle=0,scale=1.1]{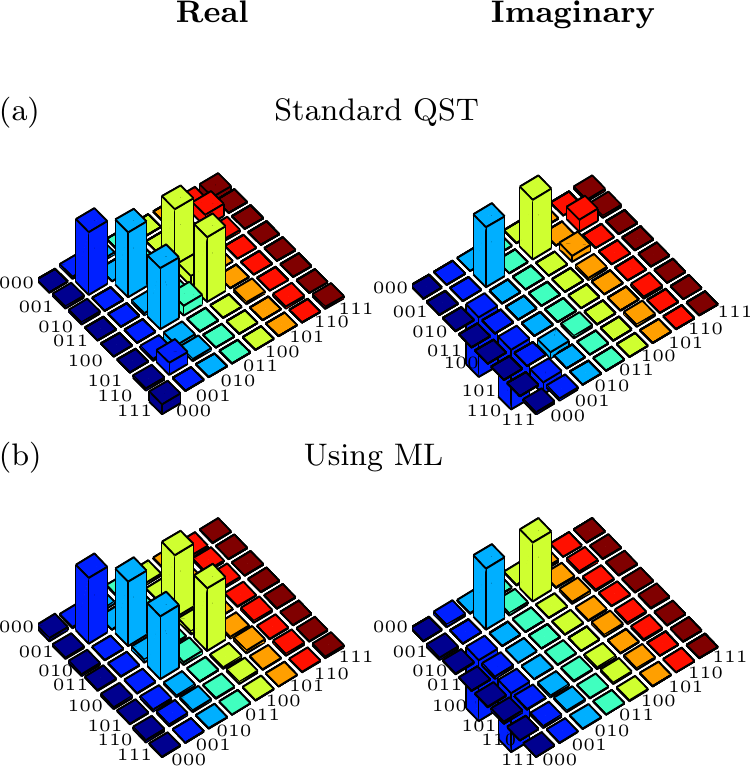}
\caption{ 
Real (left) and imaginary (right) parts of
the experimental tomographs of the three-qubit
maximally entangled state
$\vert W \rangle=\frac{1}{\sqrt{3}}(i\vert
001\rangle+\vert010\rangle+\vert100\rangle)$, 
reconstructed (a) using standard QST
with a
computed
fidelity of 0.9833 and
(b)
using ML estimation with a
computed
fidelity of 0.9968.
The rows and
columns are labeled in the computational basis ordered from
$\vert 000 \rangle$ to $\vert 111 \rangle$.
}
\label{3tomo}
\end{figure}

The state estimation of a two-qubit entangled Bell state
$\frac{1}{\sqrt{2}}(\vert 01\rangle+\vert 10\rangle)$  is
shown in Figure~\ref{entangtomo}, using both QST and ML methods for
density matrix reconstruction.  Using the QST protocol for
tomography, we obtain the eigenvalues: $\{0.996 \pm 0.002, 0.018 \pm 0.001, 0.005 \pm 0.001,
-0.019 \pm 0.002 \}$  with the first
eigenvalue being negative, and with a computed fidelity with measure $F_1$
and $F_2$ are 0.9933 and 0.9964 respectively.  Using
ML method for state estimation leads to all positive eigenvalues:
$\{ 0.993 \pm 0.002, 0.006 \pm 0.001, 0.001 \pm 0.000, 0.001 \pm 0.000 \}$ 
with a computed state fidelity with measures $F_1$ and $F_2$ are
0.9999 and 0.9926 respectively.  
\begin{figure}{H}
\centering
\includegraphics[angle=0,scale=1.5]{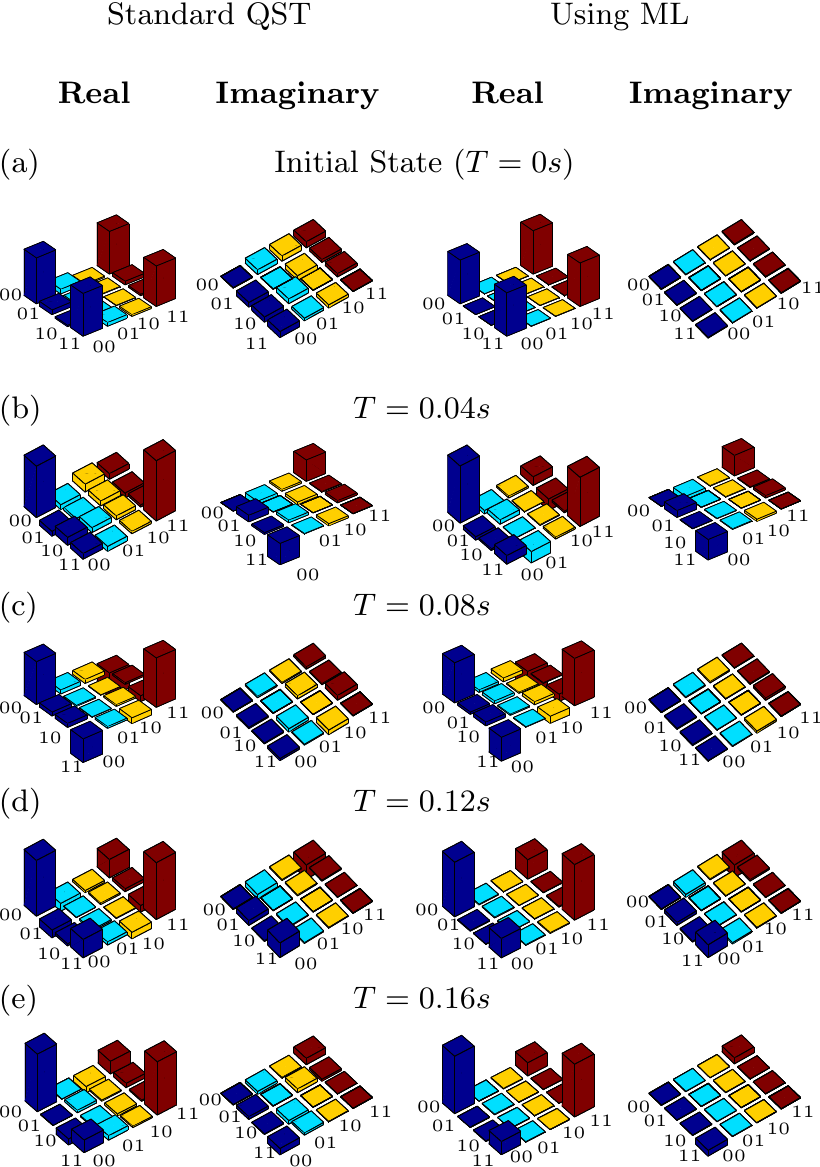}
\caption{ 
Real (left) and imaginary (right) parts of the experimental
tomographs of the (a) 
$\frac{1}{\sqrt{2}}(\vert 00\rangle +\vert 11\rangle)$ state.
Tomographs (b)-(e) depict the state at $T = 0.04,
0.08, 0.12, 0.16 s$, with the tomographs on the left and the
right representing the state estimated using standard
QST and using
ML estimation, respectively.  
The rows and
columns are labeled in the computational basis ordered from
$\vert 00 \rangle$ to $\vert 11 \rangle$.
}
\label{fig5}
\end{figure}

Recently, schemes to construct maximally entangled
three-qubit states from a generic state have been
implemented on an NMR quantum information 
processor~\cite{dogra-pra-15,das-pra-15}.
We used these schemes to construct the maximally entangled
$W$ state on a system of three qubits 
$\vert W \rangle=\frac{1}{\sqrt{3}}(i\vert
001\rangle+\vert010\rangle+\vert100\rangle)$, and
thereafter performed state estimation using both
the standard QST and the ML methods. The experimentally
reconstructed tomographs are depicted in Figure~\ref{3tomo},
with the states being labeled in the computational basis
ordered from $\vert 000 \rangle$ to $\vert 111 \rangle$.
After QST tomography on this three-qubit
state, we obtained the eigenvalues:
\{0.939, 0.104, 0.078, 0.054, -0.002, -0.042, -0.061, 
-0.071\}, and a calculated state 
fidelity $F_1$ is 0.9759  and $F_2$ is 0.9399.   
After performing state estimation using the ML method, the
eigenvalues turned out to be all positive:
\{0.919, 0.036, 0.027, 0.008, 0.006, 0.002, 0.002,
0.000\}, with a calculated state fidelity with $F_1$ is 0.9968
and with $F_2$ is 0.9191.

A topic of much research focus here is the accurate measurement of
the decay of multiqubit entanglement with time. 
To study this, we performed state  estimation
of the entangled two-qubit 
state $\frac{1}{\sqrt{2}}(\vert 00\rangle+\vert 11\rangle)$
using both QST and ML protocols.
The bar tomographs of the reconstructed density matrices 
at different times (T=0, 0.04, 0.08, 0.12, 0.16 sec) 
are shown in Figure~\ref{fig5}.

The amount of entanglement that
remains in the state
after a certain time can be quantified by
an entanglement parameter denoted by $\eta$~\cite{singh-pra-14}.
Since we are dealing with mixed bipartite
states of two-qubit, all entangled states
will be negative under partial
transpose (NPT).  
For such NPT states, a reasonable measure of
entanglement is the minimum eigenvalue of 
the partially transposed density operator.
For a given experimentally tomographed 
density operator $\rho$,
we obtain
$\rho^{PT}$ by taking a partial transpose
with respect to one of the qubits.
The entanglement parameter $\eta$ 
for the state $\rho$ in terms of the
smallest eigenvalue $E^{\rho}_{\rm Min}$  
of $\rho^{PT}$ is defined as~\cite{singh-pra-14}.
\begin{equation}
\eta = \left\{\begin{array}{ll}
-E^{\rho}_{\rm Min} &{\rm if~} E^{\rho}_{\rm Min} <0
\\
&\\
\phantom{-}0   &{\rm if~} E^{\rho}_{\rm Min} > 0
\end{array}\right.
\end{equation}
For the calculation of the entanglement parameter $\eta$,
the purity factor $\epsilon \approx 1$, 
since, we are considering the subensemble of spins that have been
prepared in the pseudopure state, and in which the spins
are genuinely entangled as described in Chap.~\ref{intro_chap1},
Sec.~\ref{chap1_Initialization}.

\begin{figure}[ht]
\centering
\includegraphics[angle=0,scale=1.0]{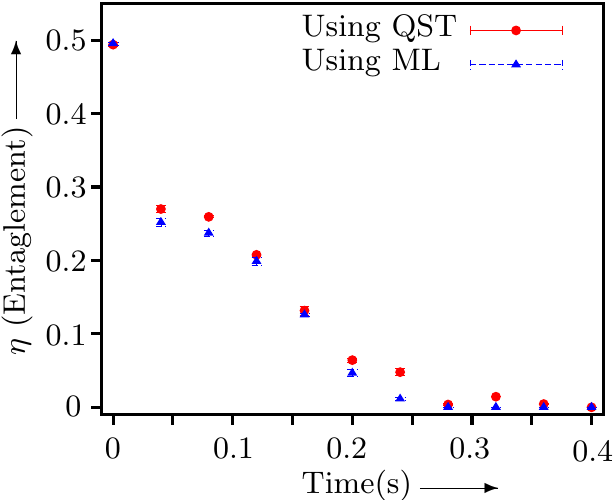}
\caption{Plot of the entanglement parameter $\eta$ with time,
using standard QST and ML protocols for state reconstruction,
computed for the
$\frac{1}{\sqrt{2}}(\vert 0 0 \rangle + \vert 1 1 \rangle)$
state.}
\label{entangfig}
\end{figure}

A plot of the entanglement parameter $\eta$ with time
is depicted in Figure~\ref{entangfig}, for the 
two-qubit maximally entangled Bell state
$\frac{1}{\sqrt{2}}(\vert 00\rangle +\vert 11\rangle)$, 
estimated using both standard QST and the ML method.
As can be seen from Figure~\ref{entangfig}, the QST
method led to negative eigenvalues in the reconstructed
(unphysical) density matrix and hence an overestimation of
the entanglement parameter quantifying the residual
entanglement in the state. The ML estimation method on the other hand,
by virtue of its leading to a physical density matrix
reconstruction every time, gives us a true measure of
residual entanglement, and hence can be used to
quantitatively study the decoherence of multiqubit
entanglement.
The difference between the two methods is statistically
significant, as is evident from the graph, where the error bars 
are shown on each data point. The error is in fact very small,
compared to the anomaly in the estimation of the entanglement. 
\section{Conclusions}
\label{concl}
We used the maximum likelihood estimation method for state estimation on an NMR
quantum information processor, to circumvent the problem of unphysical density
matrices that occur due to statistical errors while using the standard QST
protocol.  It has been previously shown that state reconstruction using QST, of
entangled states and other fragile quantum states are particularly susceptible
to errors, and can lead to unphysical density matrices for such states.  We
showed that the experimental density matrices reconstructed for entangled
states of two and three qubits using the ML estimation method are always
positive, definite and normalized. The state fidelities computed using fidelity
measure $F_1$ are comparable for standard QST and using ML estimation. However
if we use the fidelity measure $F_2$, the obtained state fidelity with ML
estimation is less than the standard QST.  The advantage of 
the ML estimation method
is that it always leads to a valid density matrix and hence is a better
estimator of the state of the quantum system. In the rest of this
thesis we will use the ML
estimation method for quantum state tomography and $F_2$ 
as the measure for state
fidelity.
\newpage

\chapter{Experimental protection  
of quantum states via a super-Zeno 
scheme}\label{chapter_superzeno}
\section{Introduction}
\label{intro_sz}
Unwanted changes occur in quantum systems
due to their interaction with the environment which 
leads to state degradation.
This process, where quantum coherence is lost, is known
as decoherence. 
Decoherence is a major obstacle in implementing
quantum computing and quantum information processing schemes.
In order to tackle this problem 
a number of techniques have been developed.
The dynamical decoupling (DD) approach can be used to
effectively decouple the system  
from its environment and thus decoherence can be
suppressed~\cite{viola-pra-98,viola-prl-99}. 
The idea that frequent
measurements which project a quantum system back
to its initial state can be used to not let
the state evolve  
is known as the quantum Zeno
effect~\cite{sudarshan-jomp-77,sudarshan,facchi-jmp-10,facchi-jphyconf-09}. If
the measurements project the system back into a
finite-dimensional subspace that includes the initial state,
the state evolution remains confined within this subspace
and the subspace can be protected against leakage of
population using a quantum Zeno
strategy~\cite{facchi-prl-02,busch-jphyconf-10}. Zeno-like schemes
have been used for error prevention~\cite{erez-pra-04}, and
to enhance the entanglement of a state and bring it to a
Bell state, even after entanglement sudden
death~\cite{sabrina-prl-08,oliveira-pra-08}. It has been
shown that under certain assumptions, the Zeno effect can be
realized with weak measurements and can protect an unknown
encoded state against environmental
effects~\cite{silva-prl-12}.  
All these strategies are based on our knowledge of 
the system-environment
interaction and the state that needs to be preserved.

In a situation where the system state is known but
there is no knowledge about its interaction with the environment,
decoherence can be tackled by an interesting quantum Zeno-type strategy for
state preservation.  This is
achieved by using a sequence of non-periodic short duration
pulses and is called the super-Zeno scheme~\cite{dhar-prl-06}. The
super-Zeno scheme does not assume any Hamiltonian symmetry, does not involve
projective quantum measurements and achieves a significant reduction of the
leakage probability as compared to standard Zeno-based preservation schemes.
The super-Zeno scheme for state protection is similar to universal dynamical
decoupling schemes for multi-qubit states~\cite{mukhtar-pra-10-1}.  
The only
difference in the super-Zeno scheme is that 
non-periodic short duration between the
unitary kicks are optimized numerically by minimizing 
the leakage probability.
Similar
schemes involving dynamical decoupling have been devised to suppress qubit pure
dephasing and relaxation~\cite{uhrig-njp-08,yang-prl-08}.  Another scheme to
preserve entanglement in a two-qubit spin-coupled system has been constructed,
which unlike the super-Zeno scheme, is based on a sequence of operations
performed periodically on the system  in a given time
interval~\cite{hou-annal-12}.

There are several experimental implementations of the
quantum Zeno phenomenon, including suppressing
unitary evolution driven by external fields between
the two states of a trapped ion~\cite{itano-pra-90}, in
atomic systems~\cite{bernu-prl-08}
and suppressing failure events in a linear optics quantum
computing scheme~\cite{franson-pra-04}. Decoherence control
in a superconducting qubit system has been proposed using
the quantum Zeno effect~\cite{tong-pra-14}.  
Unlike the super-Zeno and dynamical decoupling schemes that
are based on unitary pulses, the quantum Zeno effect
achieves suppression of state evolution using
projective measurements.
The quantum
Zeno effect was first demonstrated in NMR by a set of
symmetric $\pi$ pulses~\cite{xiao-pla-06}, wherein  
pulsed magnetic field gradients and controlled-NOT
gates were used to mimic
projective measurements.
The
entanglement preservation of a Bell state in a two-spin
system in the presence of anisotropy was demonstrated using
a preservation procedure involving free evolution and
unitary operations~\cite{manu-pra-14}.  An NMR scheme to
preserve a separable state was constructed using the
super-Zeno scheme and the state preservation was found to be
more efficient as compared to the standard Zeno
scheme~\cite{ting-chinese-09}.  
The quantum Zeno effect was used to stabilize superpositions
of states of NMR qubits against dephasing, using an
ancilla to perform the measurement~\cite{kondo-qph-14}.
Entanglement preservation
based on a dynamic quantum Zeno effect was demonstrated
using NMR wherein frequent measurements were implemented
through entangling the target and measuring
qubits~\cite{zheng-pra-13}.

In this chapter we demonstrate the use of the super-Zeno
scheme, while the implementation of 
dynamical decoupling schemes will be taken up 
in later chapters of the thesis.
Two applications of the super-Zeno
scheme are described in this
chapter: (i) Preservation of a state by freezing state
evolution (one-dimensional subspace protection) and (ii)
Subspace preservation by preventing leakage of population to
an orthogonal subspace (two-dimensional subspace
protection).  Both kinds of protection schemes are
experimentally demonstrated on separable as well as on
maximally entangled two-qubit state.  
One-dimensional subspace protection is demonstrated
on the separable $\vert 1 1 \rangle$ state and on
the maximally entangled 
$\frac{1}{\sqrt{2}}(\vert 0 1 \rangle - \vert 1 0 \rangle)$
(singlet) state.
Two-dimensional subspace preservation
is demonstrated by choosing the $\{\vert 0 1 \rangle, \vert
1 0 \rangle\}$ subspace in the four-dimensional Hilbert
space of two qubits, and implementing the super-Zeno
subspace preservation protocol on three different states,
namely $\vert 0 1 \rangle$, $\vert 1 0 \rangle$ and
$\frac{1}{\sqrt{2}}(\vert 0 1 \rangle - \vert 1 0 \rangle)$
(singlet) states.  Complete state tomography via maximally likelihood estimation as
described in Chapter~\ref{chapter_mle}
is utilized to compute experimental density matrices at several time points. 
State
fidelities at these time points were computed to evaluate
how closely the states resemble the initially prepared
states, with and without super-Zeno protection.  The success
of the super-Zeno scheme in protecting states in the
two-dimensional subspace spanned by $\{\vert 0 1 \rangle,
\vert 1 0 \rangle \}$ is evaluated by computing a leakage
parameter, which computes leakage to the orthogonal subspace
spanned by $\{\vert 0 0 \rangle, \vert 1 1 \rangle \}$.  For
entangled states, an additional entanglement parameter is
constructed to quantify the residual entanglement in the
state over time.
State fidelities, the leakage parameter and the entanglement
parameter are plotted as a function of time, to quantify the
performance of the super-Zeno scheme.
\section{The super-Zeno scheme}
\label{theory_sz}
The super-Zeno algorithm to preserve quantum states has
been developed along lines similar to bang-bang control
schemes, and limits the quantum system's evolution to
a desired subspace using a series of 
unitary kicks~\cite{dhar-prl-06}. 
A finite-dimensional Hilbert space $\cal{H}$ can be written
as a direct sum of two orthogonal subspaces $\cal{P}$ and
$\cal{Q}$. The 
super-Zeno scheme involves a unitary kick {\bf J},
which can be constructed as 
\begin{equation}
{\bf J = Q-P} 
\label{Jeqn_sz}
\end{equation}
where $\bf P, Q$ are the projection operators
onto the subspaces $\cal{P},\cal{Q}$ respectively.
The action of this specially crafted pulse 
${\bf J}$ on a state
$\vert \psi \rangle \in \cal{H}$
is as follows: 
\begin{eqnarray}
{\bf J} \vert \psi \rangle &=& - \vert \psi \rangle, 
\quad \vert \psi \rangle \in {\cal P} \nonumber \\
{\bf J} \vert \psi \rangle &=&  \vert \psi \rangle, 
\quad \vert \psi \rangle \in {\cal Q} 
\end{eqnarray}
where  ${\cal P}$ is the subspace being preserved~\cite{dhar-prl-06}.

The basic aim of this scheme is to design a sequence 
of appropriately spaced inverting pulses, such that if the system 
is initially in a state $\vert \psi \rangle \in {\cal P}$, then the
leakage of the system state over time to ${\cal Q}$ after this pulse sequence 
is minimum.
The inverting pulse {\bf J} produces destructive interference
of quantum amplitudes and reduces the transition rate 
from the $\cal{P}$ subspace to $\cal{Q}$. Let the system
be prepared in a general state $\vert \psi(0) \rangle=\vert p \rangle \in \cal{P}$, 
the system Hamiltonian {$\mathbf H$} is bounded and the unitary operator 
corresponding to the evolution for a time interval $t$ is given by
\begin{equation}
 \mathbf{ U(t)=  e^{{-\iota H t}/{\hbar}}}.
\end{equation}
For simplicity we use natural units and  $\hbar=1$.
Then, the state $\vert\psi(0) \rangle$ after time $t$ is
\begin{equation}
 \mathbf{\vert \psi(t) \rangle = e^{-\iota  \mathbf{H} t}\vert \psi(0) \rangle }
\end{equation}
The amplitude of the 
system to be 
in a state  $\mathbf{\vert q \rangle} \in \cal{Q}$ 
after a time interval $t$ is given by 
\begin{equation}
 \mathbf{\langle q \vert e^{-\iota \mathbf{H} t}\vert \psi(0) \rangle} 
\end{equation}

However, if we evolve the system for the half interval t/2,
subject the system to an inverting pulse {\textbf{J}}, and
then further evolve by a time t/2, the above amplitude turns 
out to be 
\begin{equation}
 \mathbf{\langle q \vert e^{-\iota \mathbf{H}t/2} \mathbf{J} e^{-\iota \mathbf{H} t/2} \vert \psi(0)\rangle} 
\end{equation}

For a small time interval t and the unitary operator 
$\mathbf{U(t)}$ satisfying $\mathbf{U(t)=\textbf{I}+  \mathcal{O}(t)}$ 

\begin{equation}
\label{sz_cal}
 \mathbf{\langle q \vert \left( I+\frac{-\iota \mathbf{H}t}{2} \right) \mathbf{J} \left(I+\frac{-\iota \mathbf{H}t}{2} \right) \vert \psi(0)\rangle} 
\end{equation}

\begin{equation}
 \mathbf{ =\langle q \vert \mathbf{J}\vert p\rangle + \langle q \vert \mathbf{J}\frac{-\iota \mathbf{H}t}{2}  +\frac{-\iota \mathbf{H}t}{2}  \mathbf{J} \vert p\rangle + \mathcal{O}(t^2)} \nonumber
\end{equation}

\begin{equation}
 \mathbf{ = \langle q \vert \mathbf{J}\frac{-\iota \mathbf{H}t}{2} \vert p\rangle  +  \langle q \vert \frac{-\iota \mathbf{H}t}{2}  \mathbf{J} \vert p\rangle + \mathcal{O}(t^2)} \nonumber
\end{equation}

\begin{equation}
  \mathbf{= \langle q \vert\frac{-\iota \mathbf{H}t}{2} \vert p\rangle  -  \langle q \vert \frac{-\iota \mathbf{H}t}{2}  \vert p\rangle + \mathcal{O}(t^2)} \nonumber
\end{equation}
Due to the action of $\mathbf{J}$, 
destructive interference takes place between
the two amplitudes of $\mathcal{O}(t)$ in Eq.(\ref{sz_cal}) and
we are left with terms of $\mathcal{O}(t^2)$ term and higher, which are
negligible for a small time interval $t$. In general, 
if the matrix elements of $\mathbf{U(t)}$ i.e $\mathbf{U_{qp}}$  and 
$\mathbf{U_{pq}}$  are of $\mathcal{O}(t^{(r)})$ then the transition amplitude 
\begin{equation}
 \mathbf{\langle q\vert U(t).J.U(t)\vert p \rangle= -\sum_{p' \in \cal{P}} U_{qp'}U_{p'p}+\sum_{q' \in \cal{Q}} U_{qq'}U_{q'p}-U_{qp}U_{pp}-U_{qq}U_{qp}} 
 \nonumber
 \end{equation}
\begin{equation}
 \mathbf{=\mathcal{O}(t^{r+1})-U_{qp}U_{pp}-U_{qq}U_{qp}=\mathcal{O}(t^{r+1})}
 \label{Ur_sz}
 \end{equation}
where a precise cancellation of the amplitude
of $\mathbf{\mathcal{O}(t^r)}$ term occurs~\cite{dhar-prl-06}.

It is important to note that $\mathbf{V(t)=U(t).J.U(t)}$ tends to $\mathbf{J}$ 
 as $t \rightarrow 0 $ but $\mathbf{V(t)^2}$ tends to 
 $\mathbf{I}$ and if $\mathbf{V(t)=J+\mathcal{O}(t)}$ and $\mathbf{V(t)_{qp}=\mathcal{O}(t^{r})}$ 
\begin{eqnarray}
\mathbf{\langle q\vert V(t)^2 \vert p \rangle} &=& \mathbf{-\sum_{p'' \in \cal{P}} V(t)_{qp'}V(t)_{p'p}+\sum_{q' \in \cal{Q}} V(t)_{qq'}V(t)_{q'p}} \nonumber \\
&&\mathbf{-V(t)_{qp}V(t)_{pp}-V(t)_{qq}V(t)_{qp}} \nonumber\\
&=&\mathbf{\mathcal{O}(t^{r+1})+V(t)_{qp}(V(t)_{qq}+V(t)_{pp})} \nonumber 
\label{Vt_sz}
\end{eqnarray}

\begin{equation}
\begin{aligned}
\mathbf{\langle q\vert V(t)^2 \vert p \rangle= \mathcal{O}(t^{r+1})}
\end{aligned}
\end{equation}

Hence, if for $\mathbf{U_0(t)}$ the transition amplitude to an 
orthogonal state is proportional to $t$, then for the operator
$\mathbf{U_1(t)=U_0(t).J.U_0(t)}$ from Eq.~(\ref{Ur_sz}), 
the transition amplitude is $\mathcal{O}(t^2)$. For 
$\mathbf{U_1^2}$ using Eq.~(\ref{Vt_sz}), the transition
amplitude is $\mathcal{O}(t^3)$~\cite{dhar-prl-06}. Using this result,
it is straight forward to construct a pulse sequence
where the transition amplitude is $\mathcal{O}(t^r)$ for any  positive 
integer $r$ by
recursion. Defining operators $\mathbf{U}_{m}$ by the recursion relations
\begin{eqnarray}
{\bf U}_{m+1}(t) &=& {\bf U}_{m}(t/2) ~{\mathbf J} ~{\bf U}_{m}(t/2), {\rm 
~for~} 
m 
{\rm~ even}
\nonumber \\
& =& {\bf U}_{m}(t/2) ~{\bf U}_{m}(t/2),{\rm ~for} ~m  {\rm ~odd}
\end{eqnarray}
with ${\bf U}_{0} = {\bf U}_{0}(t)$. Then, by induction,
it follows that  the transition amplitude 
$\langle q|{\bf U}_m|p\rangle$ is of order ${\mathcal O}(t^{m+1})$.

 $\mathbf{U}_{m}(t)$ can be written explicitly as a product of 
$\mathbf{U}_{0}(t/2^{m})%
$'s and $\mathbf{J}$'s. For example
\begin{eqnarray}
\mathbf{U}_{1}(t)&= &\mathbf{U}_{0}(t/2) {\mathbf J}
\mathbf{U}_{0}(t/2), \nonumber\\
\mathbf{U}_{2}(t)  & =& [\mathbf{U}_{0}(t/4) {\mathbf J}
\mathbf{U}_{0}(t/4)]^{2},\nonumber \\
\mathbf{U}_{3}(t)&=& 
[\mathbf{U}_{0}(t/8) {\mathbf J}
\mathbf{U}
_{0}(t/8)]^{2} {\mathbf J} [\mathbf{U}_{0}(t/8) {\mathbf J} 
\mathbf{U}_{0}(t/8)]^{2}. 
\end{eqnarray}

If $N_m$ are the  number of pulses 
used in the sequence ${\mathbf U}_m$, then it is easily verified  that
\begin{eqnarray}
N_{m}=(2^{m+1}-2)/3, {\rm ~for }~m  {\rm ~even},\nonumber\\
     =(2^{m+1}-1)/3, {\rm ~for}~ m {\rm ~odd} .\label{twentynine_sz}%
\end{eqnarray}

The leakage probability $L_{m}$ from subspace $%
\mathcal{P}$ to subspace $\mathcal{Q}$ for operator $\mathbf{U}_{m}$ 
applied for a total  time $T$ is
\begin{equation}
L_{m} = \sum_{q \in Q} |\langle q|\mathbf{U}_{m}|p\rangle|^{2} \leq
2^{-m(m+1)} [ET]^{2m+2}  \label{eq10_sz}
\end{equation}
where $E = |\mathbf{H}|$ is the norm of the Hamiltonian as defined
by 
\begin{equation}
|\mathbf{H}| = {%
\genfrac{}{}{0pt}{}{_{}\mathrm{ sup}} {\psi}%
} \ ||\mathbf{H}|\psi\rangle||~/~||~|\psi\rangle||%
\end{equation}

where the $\mathrm{sup}$ notation is defined as 
(for any function $f$ and constraint set ${\cal X}$)

\begin{equation}
 \genfrac{}{}{0pt}{}{_{}\mathrm{ sup}} {x\in{\cal X}}%
 \ f(x)\ge f(v) \ \  \forall \ \ v \ \in \ {\cal X}
\end{equation}

The total number of inverting pulses 
$N_{m}$ required to keep the leakage probability less 
than $\epsilon$ up to time $T$ grows as 
$\frac{2}{3}ET$ $ 2^{{log_2(E^2T^2/\epsilon)}^{1/2}}$~\cite{dhar-prl-06}. It is 
possible to decrease the required number of pulses significantly 
by allowing the intervals between the inverting pulses to be varied
continuously independent of each other.

For a given $N$, further improvements are possible and one can express $\langle q|{W}_N(t)|p \rangle$ as a Taylor series
in powers of $t$. The coefficients of different powers of $t$ are sums of
matrix-elements of the type $\langle q| {\bf H}^{n_1} {\mathbf J} {\bf
H}^{n_2}{\mathbf J} ..|p \rangle$, with coefficients that are polynomials of $\{x_j\}$. Thus, the total super-Zeno sequence for $N$ pulses is given by
\begin{equation}
W_N(t) = U(x_{N+1}t) {\bf J} \dots {\bf J} U(x_2 t) 
{\bf J} U(x_1 t)
\label{superzenoeqn_sz}
\end{equation}
where $U$ denotes unitary evolution under
the system Hamiltonian and $x_i t$ is the time interval between the
$i$th and $(i+1)$th pulse.
The sequence $\{ x_i t\}$ of time intervals between pulses
is optimized such that if the system
starts out in the subspace $\cal P$, 
after measurement the probability of finding the
system in the orthogonal subspace $\cal Q$ is minimized. 
In this work we
used four inverting pulses interspersed with 
five unequal time intervals in each repetition
of the preserving super-Zeno sequence. The
optimized sequence 
is given by
$\{ x_i \} = \{ \beta, 1/4, 1/2-2\beta, 1/4,
\beta \}$ with $\beta = (3 - \sqrt{5})/8,
i=1 \dots 5$ and $t$ is a fixed time interval
(we used the $x_i$ as worked out in Ref.~\cite{dhar-prl-06}).

The explicit form of the unitary kick ${\bf J}$ depends
on the subspace that needs to be preserved, and in the
following section, we implement 
several illustrative examples for both separable
and entangled states embedded in one- and two-dimensional
subspaces of two qubits.
\section{Experimental implementations of 
super-Zeno scheme}
\label{expt_sz}
\begin{figure}[h]
\centering
\includegraphics[angle=0,scale=1.1]{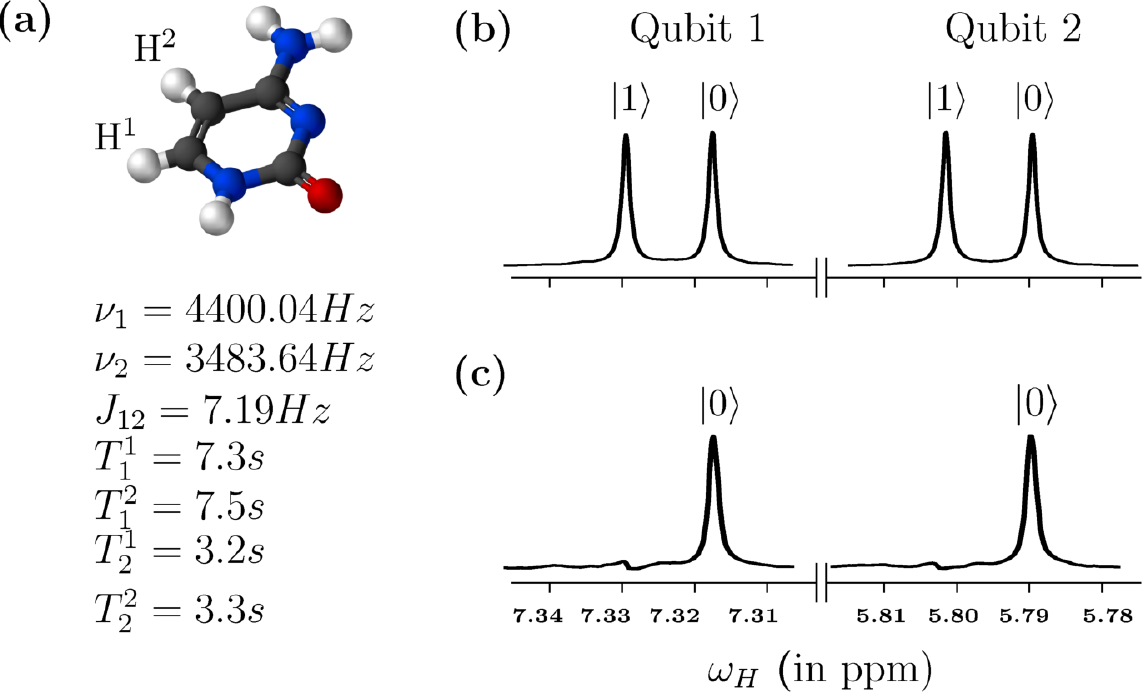}
\caption{  (a) Molecular
structure of cytosine with the two qubits labeled
as $H^{1}$ and $H^{2}$
and tabulated system parameters 
with chemical shifts $\nu_i$ and scalar coupling $J_{12}$ (in
Hz) and relaxation times $T_{1}$ and $T_{2}$ (in seconds)
(b) 
NMR spectrum obtained after a $\pi/2$
readout pulse on the thermal equilibrium state.
The resonance lines of each qubit are labeled by
the corresponding logical states of the other
qubit and
(c) NMR spectrum of the pseudopure $\vert 0 0 \rangle$
state.
}
\label{molecule_sz}

\end{figure}
\subsection{NMR system details}
The two protons of the molecule cytosine 
encode the two qubits.  The two-qubit molecular
structure, system parameters and 
NMR spectra of the pseudopure and thermal
initial states are shown in Figs.~\ref{molecule_sz}(a)-(c).  
The Hamiltonian of a two-qubit system in the
rotating frame is given by
\begin{equation}
H =  -(\omega_1-\omega_{rf}) I_{1z} - (\omega_2-\omega_{rf}) I_{2z} + 2\pi J_{12} I_{1z}I_{2z}
\end{equation}
where $\omega_i=2\pi\nu_i$ are the chemical shift of the
spins in rad s$^{-1}$, $\omega_{{\rm rf}}$ reference chemical shift
of rotating frame, and $J_{12}$ is the spin-spin coupling
constant. An average longitudinal T$_1$ relaxation time
of 7.4 s and an average transverse T$_2$ relaxation time of
3.25 s was experimentally measured for both the
qubits.
The experiments were performed at an ambient temperature of
298 K on a Bruker Avance III 600 MHz NMR spectrometer equipped with a QXI
probe.  
The two-qubit system
was initialized into the pseudopure state $\vert 00 \rangle$
using the spatial averaging technique~\cite{cory-physicad},
with the
density operator given by
\begin{equation}
\rho_{00} = \frac{1-\epsilon}{4} I
+ \epsilon \vert 00 \rangle \langle 00 \vert
\label{ppure_sz}
\end{equation}
with a thermal polarization $\epsilon \approx
10^{-5}$ and $I$ being a $4 \times 4$
identity operator.  The experimentally created
pseudopure state $\vert 00 \rangle$ was
tomographed with a fidelity of $0.99$.
The pulse propagators for selective excitation were
constructed using the GRAPE algorithm~\cite{tosner-jmr-09} to design
the amplitude and phase modulated rf profiles.  
Selective excitation was typically achieved with pulses of
duration 1 ms.
Numerically
generated GRAPE pulse profiles were optimized to be robust
against rf inhomogeneity and had an average fidelity of 
$ \ge 0.99$.
All experimental density matrices were reconstructed using a
 quantum state tomography via maximum likelihood protocol 
(Chapter~\ref{chapter_mle}).
 The fidelity of an experimental density matrix was computed using Eq.~(\ref{mle_fidelity_2}).

\subsection{Super-Zeno scheme for state preservation}
\label{statesection_sz}
When the subspace $\cal P$ is a one-dimensional subspace,
and hence consists of a single state, the super-Zeno
scheme becomes a state preservation scheme.

\subsubsection{Preservation of product states:}
We begin by 
implementing the super-Zeno scheme on the product state $\vert
11 \rangle$ of two qubits, where the Hilbert space can be
decomposed as a direct sum of the
subspaces ${\cal P}=\{\vert 11 \rangle\}$
and ${\cal Q}=\{ \vert 00 \rangle, \vert 01 \rangle, \vert
10 \rangle)\}$. The super-Zeno pulse $\bf J$ to
protect the state $\vert 11 \rangle \in {\cal P}$  
is given by Eqn.~(\ref{Jeqn_sz}):
\begin{equation}
{\bf J}= I - 2 \vert 11 \rangle \langle 11 \vert
\end{equation}
with the corresponding matrix form 
\begin{equation}
{\bf J}= \left(\begin{array}{cccc}
1&0&0&0 \\
0&1&0&0 \\
0&0&1&0 \\
0&0&0&-1
\end{array}
\right)
\end{equation}
\begin{figure}[hbtp]
\centering
\includegraphics[angle=0,scale=1.5]{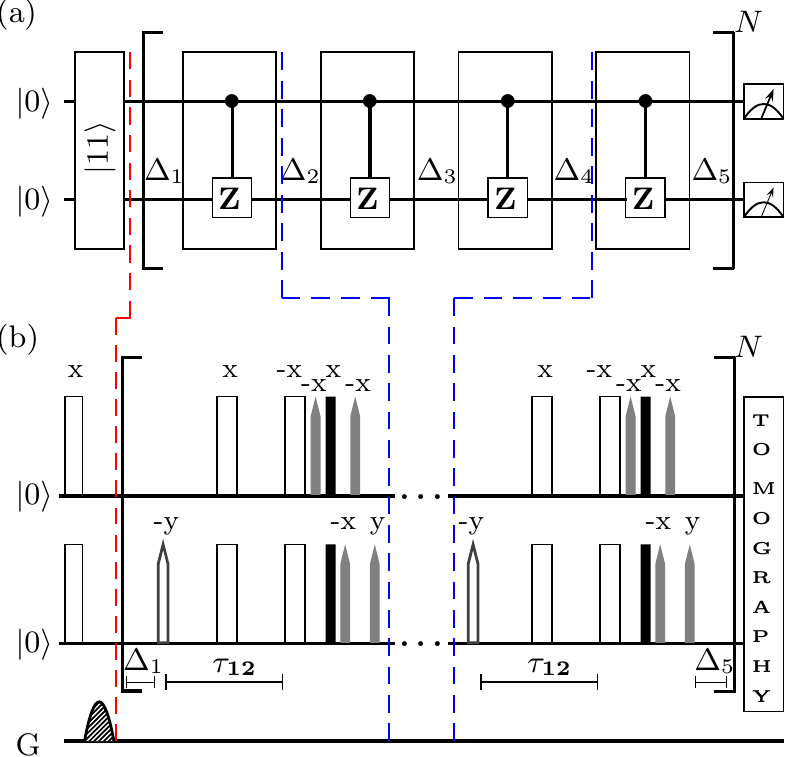}
\caption{  
(a) Quantum circuit for preservation of the state $\vert 1 1
\rangle$ using the super-Zeno scheme.  
$\Delta_i = x_i t, (i=1...5)$
denote time intervals punctuating the unitary
operation blocks.  Each unitary operation block
contains a controlled-phase gate ($Z$), with the first (top) qubit 
as the control and the second (bottom) qubit as the
target. 
The entire scheme is repeated $N$ times
before measurement (for our experiments
$N=30$).  (b) Block-wise depiction of the
corresponding NMR pulse sequence. A $z$-gradient is applied
just before the super-Zeno pulses, to clean up 
undesired residual magnetization.  The unfilled 
and black rectangles represent
hard $180^{0}$ and $90^{0}$
pulses respectively, while
the unfilled  
and gray-shaded conical shapes represent
$180^{0}$ and $90^{0}$ pulses
(numerically optimized using GRAPE) respectively;
$\tau_{12}$ is
the evolution period under the $J_{12}$ coupling. Pulses are
labeled with their respective phases and unless explicitly
labeled, the phase of the pulses on the second (bottom)
qubit are the same as those on the first (top) qubit.
}
\label{11ckt_sz}
\end{figure}
\begin{figure}[htbp] 
\centering
\includegraphics[angle=0,scale=1.5]{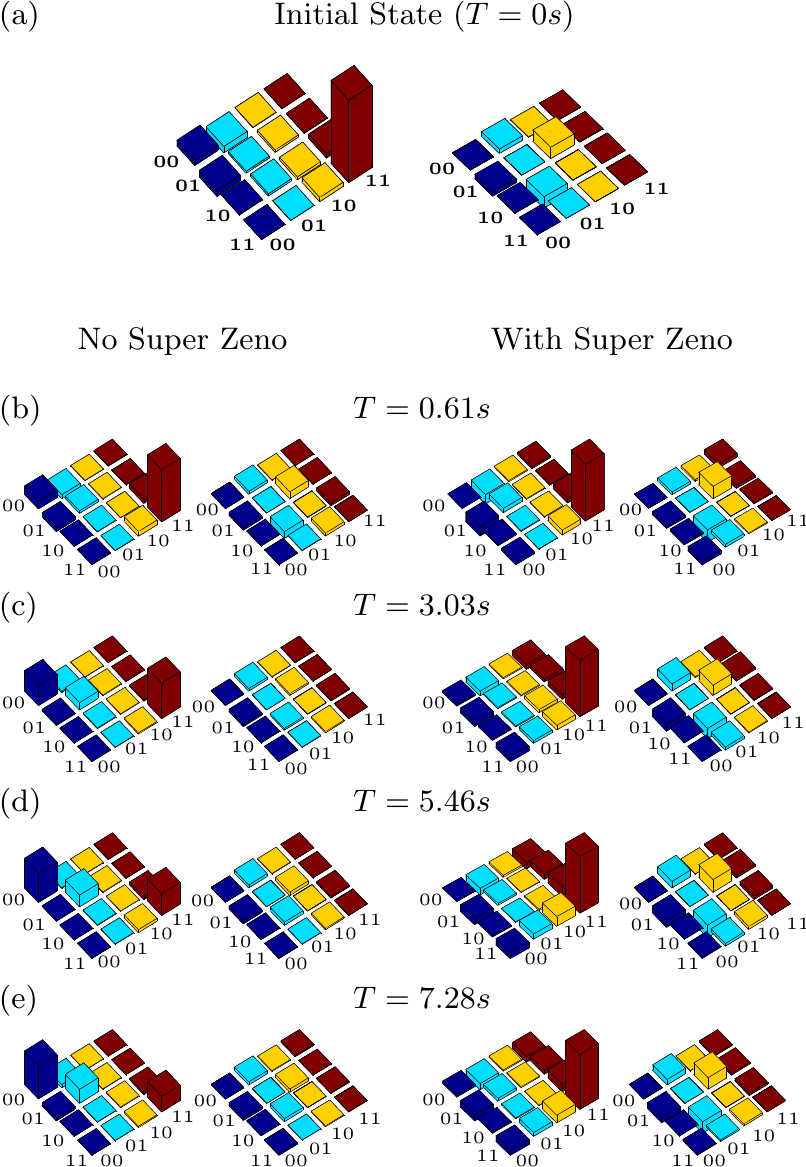} 
\caption{  
Real (left) and imaginary (right) parts of the experimental
tomographs of the (a) $ \vert 11\rangle$ state, with a computed
fidelity of 0.99.  (b)-(e) depict the state at $T = 0.61,
3.03, 5.46, 7.28$ s, with the tomographs on the left and the
right representing the state without and after applying the
super-Zeno preserving scheme, respectively.  The rows and
columns are labeled in the computational basis ordered from
$\vert 00 \rangle$ to $\vert 11 \rangle$.
}
\label{11tomo_sz}
\end{figure}
The super-Zeno circuit to preserve the $\vert 11 \rangle$
state, and the corresponding NMR pulse sequence is
given in Fig.~\ref{11ckt_sz}. The controlled-phase gate
($Z$) in Fig.~\ref{11ckt_sz}(a) which replicates the unitary kick ${\bf J}$ for
preservation of the $\vert 11 \rangle$ state is implemented
using a set of three sequential gates: two Hadamard gates
on the second qubit sandwiching a controlled-NOT gate
({\rm CNOT}$_{12}$), with the first qubit as the control
and the second qubit as the target. 
The $\Delta_i$ time interval in Fig.~\ref{11ckt_sz}(a)
is given by $\Delta_i = x_i t$, with $x_i$ as
defined in Eq.~(\ref{superzenoeqn_sz}).
The five $\Delta_i$ time intervals were worked
to be $0.095$ ms, $0.25$ ms, $0.3$ ms, $0.25$ ms, 
and $0.095$ ms respectively, for $t=1$ ms.
One run of the super-Zeno circuit (with four inverting ${\bf J}$s
and five $\Delta_i$ time evolution periods) takes
approximately $300$ ms and 
the entire preserving sequence $W_N(t)$ 
in Eq.~(\ref{superzenoeqn_sz}) was applied
$30$ times.
The final state of the system was reconstructed using
state tomography and the real and imaginary
parts of the tomographed experimental density matrices without
any preservation and after applying the
super-Zeno scheme,
are shown in
Fig.~\ref{11tomo_sz}. 
The initial $\vert 11 \rangle$ state (at time $T=0$ s)
was created (using the spatial 
averaging scheme) with a fidelity of 0.99.
The tomographs (on the right in
Fig.~\ref{11tomo_sz}) clearly show that state evolution 
has been frozen with the super-Zeno scheme.

\subsubsection{Preservation of entangled states}

\begin{figure}[htbp]
\centering
\includegraphics[angle=0,scale=1.5]{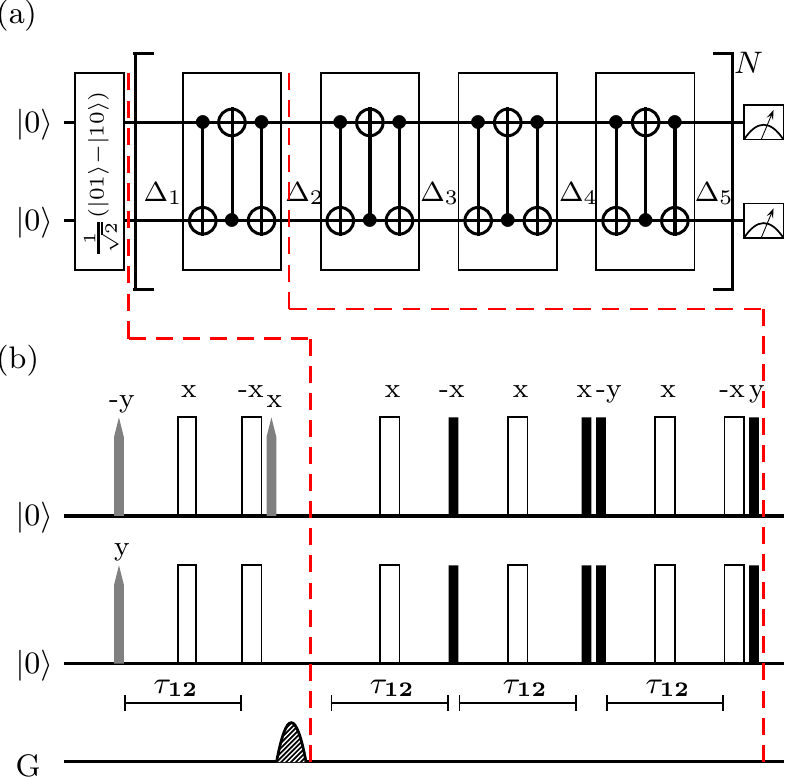}
\caption{  
(a) Quantum circuit for preservation of the
singlet state using the super-Zeno scheme.  $\Delta_i,
(i=1...5)$ denote time intervals punctuating
the unitary operation blocks.  
The entire scheme is repeated $N$ times before
measurement (for our
experiments $N=10$).  (b) NMR pulse sequence corresponding to one
unitary block of the circuit in (a). 
A $z$-gradient is applied just
before the super-Zeno pulses, to clean up undesired
residual magnetization.  The unfilled rectangles represent hard
$180^{0}$ pulses, the black filled rectangles representing hard $90^{0}$ pulses,
while the shaded shapes represent numerically optimized (using GRAPE) pulses
and the gray-shaded shapes representing $90^{0}$ pulses respectively; 
$\tau_{12}$ is
the evolution period under the $J_{12}$ coupling. Pulses are
labeled with their respective phases and unless explicitly
labeled, the phase of the pulses on the second (bottom)
qubit are the same as those on the first (top) qubit.
}
\label{singletckt_sz}
\end{figure}
\begin{figure}[htbp]
\centering
\includegraphics[angle=0,scale=1.5]{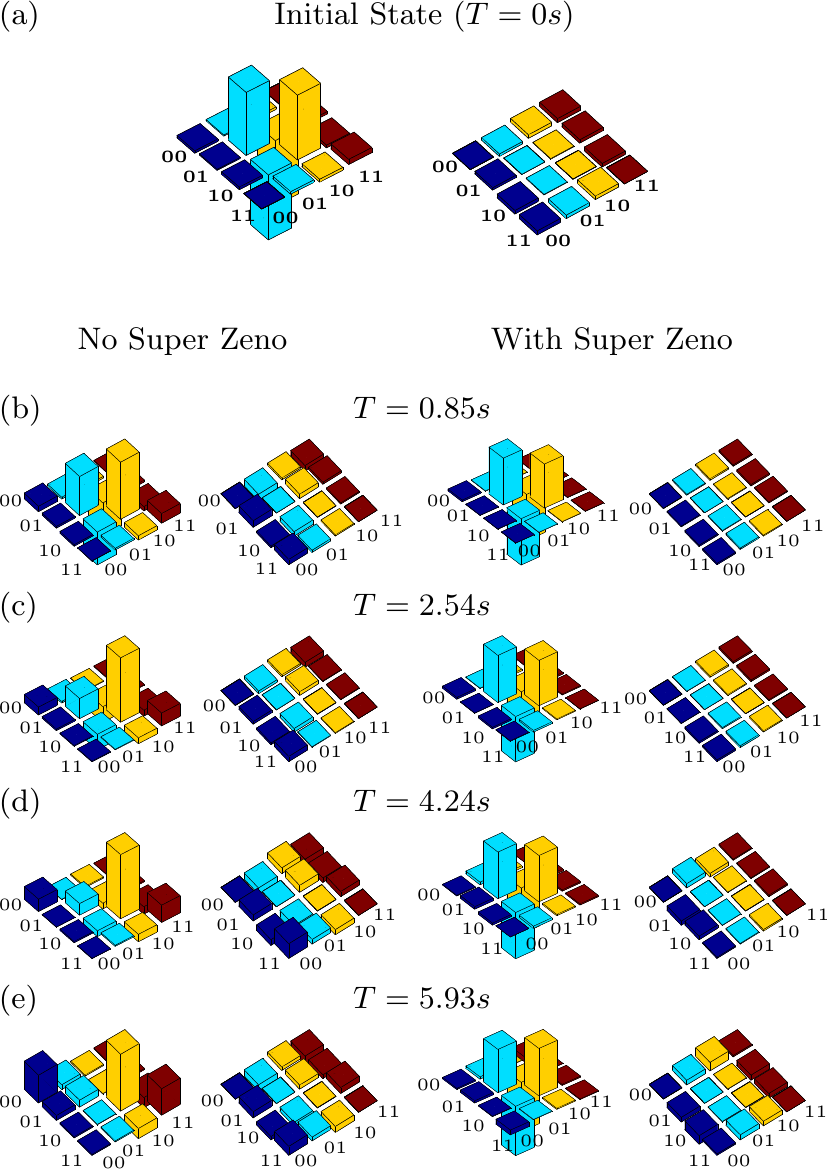}
\caption{  
Real (left) and imaginary (right) parts of the experimental
tomographs of the (a) $\frac{1}{\sqrt{2}}(\vert 01\rangle
-\vert 1 0 \rangle)$ (singlet) state, with a computed
fidelity of 0.99.  (b)-(e) depict the state at $T = 0.85,
2.54, 4.24, 5.93$ s, with the tomographs on the left and the
right representing the state without and after applying the
super-Zeno preserving scheme, respectively.  The rows and
columns are labeled in the computational basis ordered from
$\vert 00 \rangle$ to $\vert 11 \rangle$.
}
\label{singlettomo_sz}
\end{figure}

We next apply the super-Zeno scheme to
preserve an entangled state in our system of two qubits.
We chose the singlet state
$\frac{1}{\sqrt{2}}(\vert 0 1 \rangle - \vert 1 0
\rangle)$ as the entangled state to be
preserved.  It is well known that
entanglement is an important but fragile resource
for quantum information processing
and constructing schemes to protect entangled
states from evolving into other states, is of considerable
interest in quantum information processing
~\cite{nielsen-book-02}. 

We again write the Hilbert space as a 
direct sum of two subspaces: the subspace being
protected and the subspace orthogonal to it.
In this case, the one-dimensional subspace 
${\cal P}$ being protected is  
\begin{equation}
{\cal P} = \left \{
\frac{1}{\sqrt{2}}\left(\vert 01 \rangle- \vert 10\rangle
\right) \right\}
\end{equation}
and the orthogonal subspace ${\cal Q}$ into which
one would like to prevent leakage is 
\begin{equation}
{\cal Q} = \left \{
\frac{1}{\sqrt{2}}(\vert 01 \rangle +\vert 10 \rangle),
\vert 00 \rangle, \vert 11 \rangle
\right \}
\end{equation}

The super-Zeno pulse to protect the singlet state 
as constructed using Eq.~(\ref{Jeqn_sz}) is:
\begin{equation}
{\bf J}= I -  \left(
\vert 01\rangle \langle01 \vert+ \vert 10
\rangle\langle10 \vert- \vert 01\rangle\langle10
\vert- \vert 10\rangle\langle 01\vert \right)
\end{equation}
with the corresponding matrix form:
\begin{equation}
{\bf J}= \left(\begin{array}{cccc}
1&0&0&0 \\
0&0&1&0 \\
0&1&0&0 \\
0&0&0&1
\end{array}
\right)
\end{equation}

The quantum circuit and the NMR
pulse sequence for preservation of the 
singlet state using the super-Zeno scheme are given in
Fig.~\ref{singletckt_sz}. Each {\bf J} inverting pulse
in the unitary block in the circuit is decomposed
as a sequential operation of three non-commuting controlled-NOT
gates:~{\rm CNOT$_{12}$-CNOT$_{21}$-CNOT$_{12}$},
where {\rm CNOT$_{ij}$} denotes a controlled-NOT with $i$
as the control and $j$ as the target qubit. 
The five $\Delta_i$ time intervals were worked
to be $0.95$ ms, $2.5$ ms, $3$ ms, $2.5$ ms, 
and $0.95$ ms respectively, for $t = 10$ ms.
One run of the super-Zeno circuit (with four inverting ${\bf J}$s
and five $\Delta_i$ time evolution periods) takes
approximately $847$ ms
and the entire super-Zeno preserving sequence 
$W_N(t)$ 
in Eq.~(\ref{superzenoeqn_sz}), is applied
$10$ times.

The singlet state was prepared from an initial
pseudopure state $\vert 00 \rangle$ by a sequence
of three gates: a 
non-selective NOT gate (hard $\pi_x$ pulse) on both qubits, 
a Hadamard gate and a {\rm CNOT}$_{12}$ gate. The 
singlet state thus prepared
was computed to have a fidelity of 0.99.
The effect of chemical shift evolution during the delays was
compensated for with refocusing pulses.
The final singlet state has been reconstructed using
state tomography, and the real and imaginary
parts of the tomographed experimental density matrices without
any preservation and after applying the
super-Zeno scheme,
are shown in
Fig.~\ref{singlettomo_sz}. As can be seen from the
experimental tomographs in Fig.~\ref{singlettomo_sz}, the evolution of the
singlet state is almost completely frozen by the
super-Zeno sequence upto nearly 6 s, while without
any preservation the state has leaked into the
orthogonal subspace within 2 s. 

\begin{figure}[htbp!] 
\centering
\includegraphics[angle=0,scale=1.55]{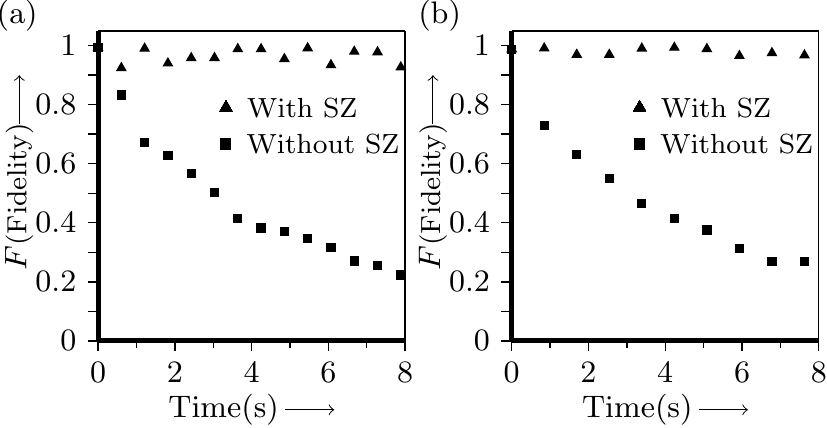} 
\caption{
Plot of fidelity versus time of (a) the $\vert 11\rangle$
state and (b) the $\frac{1}{\sqrt{2}}(\vert 0 1 \rangle -
\vert 1 0 \rangle)$ (singlet) state, without any preserving
scheme and after the super-Zeno preserving sequence. The
fidelity of the state with the super-Zeno preservation
remains close to 1.
}
\label{fidelityfig_sz}
\end{figure}

\subsubsection{Estimation of state fidelity}
\label{chapsz_fidelity_esti}
\begin{figure}[htbp]
\centering
\includegraphics[angle=0,scale=1.3]{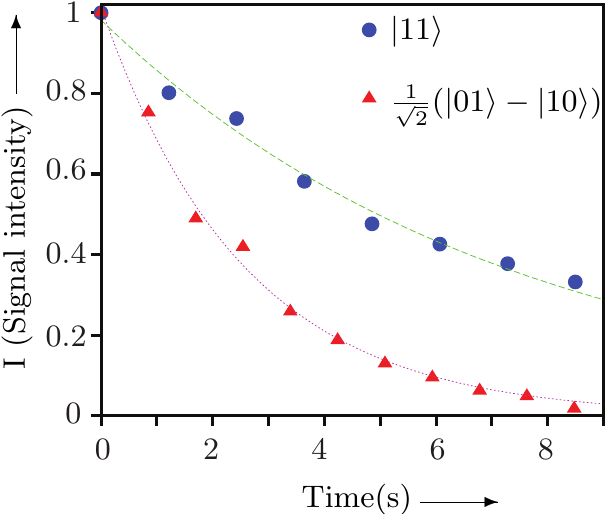} 
\caption{
Plot of signal intensity versus time of  the $\vert 11\rangle$
state and the $\frac{1}{\sqrt{2}}(\vert 0 1 \rangle -
\vert 1 0 \rangle)$ (singlet) state, after the super-Zeno
preserving sequence.
}
\label{Intensityfig_sz}
\end{figure}

The plots of state fidelity 
versus time are shown in
Fig.~\ref{fidelityfig_sz} for the state $\vert 11 \rangle$ and
the singlet state, with and without the super-Zeno
preserving sequence. The signal attenuation has been ignored and 
the deviation density matrix is renormalized
at every point and the state fidelity is estimated 
using the definition in Eq.~(\ref{mle_fidelity_2}).
Renormalization is performed since our
focus here is on the quantum state
of the spins contributing to the signal 
and not on the number \textit{per se} of participating
spins~\cite{brazil-norm}. The norm of deviation density matrix
is proportional to the intensity of the tomographed spectra and 
is proportional to the number of spins contributing to the signal intensity.
Due to super-Zeno preserving pulses, an attenuation in 
signal intensity takes place and we renormalize the intensity of the
of the tomographed spectra in order to compensate for this attenuation.
The plot of signal intensity versus time is shown in Fig.~\ref{Intensityfig_sz} for
the state $\vert 11 \rangle$ and
the singlet state, with the super-Zeno preserving sequence.
The exponential intensity decay
constant is $0.136\pm 0.006$ s$^{-1}$
for the $\vert11\rangle$ state and $0.393\pm 0.013$ s$^{-1}$ for the singlet state.

The plots in Fig.~\ref{fidelityfig_sz} and the tomographs
in Fig.~\ref{11tomo_sz} and Fig.~\ref{singlettomo_sz} show that
with super-Zeno protection, the state remains confined
to the $\vert 11 \rangle$ (singlet) part of the density
matrix, while without the protection scheme, the state
leaks into the orthogonal subspace.
As seen from both plots in Fig.~\ref{fidelityfig_sz}, the
state evolution of specific states can be arrested
for quite a long time using the super-Zeno preservation
scheme, while leakage probability of the state to other
states in the orthogonal subspace spanned by ${\cal Q}$ is
minimized. A similar renormalization procedure is adopted in the 
subsequent sections where we plot the leak fraction and
entanglement parameters (Fig.~\ref{leakagefig_sz}
and Fig.~\ref{entangfig_sz}).

\subsection{Super-Zeno for subspace preservation}
\label{subspacesection_sz}
\begin{figure}[htbp]
\centering
\includegraphics[angle=0,scale=1.5]{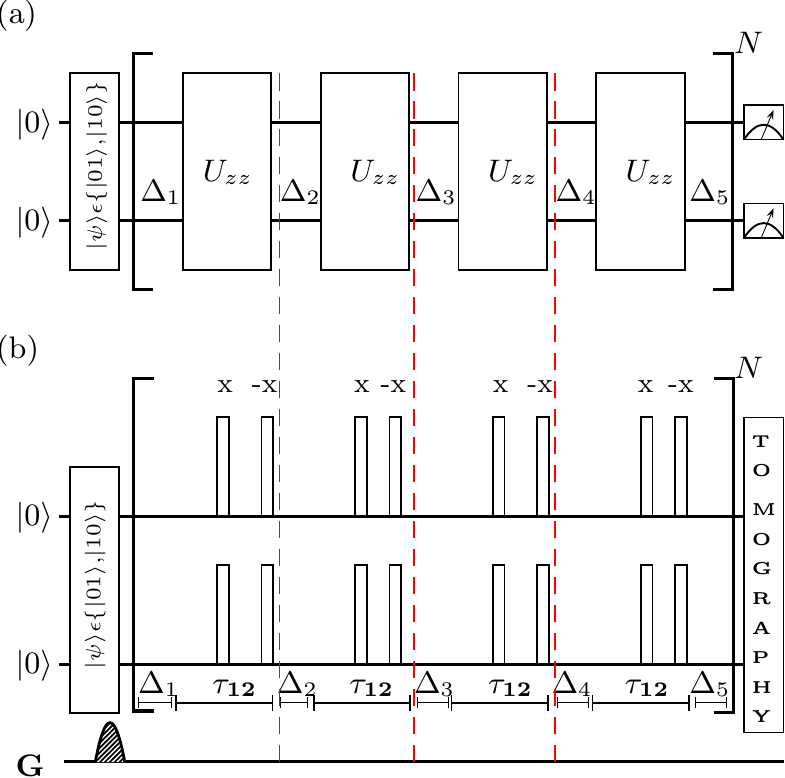}
\caption{  
(a) Quantum circuit for preservation of the $\{01, 10\}$
subspace using the super-Zeno scheme.  $\Delta_i, (i=1...5)$
denote time intervals punctuating the unitary
operation blocks.  
The entire scheme is repeated $N$ times
before measurement (for our experiments
$N=30$).  (b) NMR pulse sequence corresponding to
the circuit in (a).  A $z$-gradient is applied just before
the super-Zeno pulses, to clean up undesired
residual magnetization.  The unfilled rectangles represent
hard $180^{0}$ pulses;
$\tau_{12}$ is the evolution period under the $J_{12}$
coupling. Pulses are labeled with their respective phases.
}
\label{subspaceckt_sz}
\end{figure}
While in the previous subsection, the super-Zeno scheme was shown to be
effective in arresting the evolution of 
a one-dimensional subspace (as applied to
the cases of a product and an entangled state),
the scheme is in fact more general.
For example, if we choose a two-dimensional subspace in the
state space of two qubits and protect it by the
super-Zeno scheme, then any state in this subspace
is expected to remain within this subspace and not
leak into the orthogonal subspace. While the state
can meander within this subspace, its evolution out of the
subspace is frozen.

We now turn to implementing 
the super-Zeno scheme for subspace preservation, by
constructing the {\bf J} operator to preserve a
general state embedded in a two-dimensional subspace.
We choose the subspace  
spanned by ${\cal P}=\{\vert 01\rangle, \vert 10\rangle\}$
as the subspace to be preserved, with its orthogonal
subspace now being 
${\cal Q}=\{\vert 00 \rangle, \vert 11 \rangle\}$.
It is worth noting that within the subspace being
protected, we have product as well as entangled states.

The super-Zeno pulse {\bf J} to protect a general state 
$\vert \psi\rangle \in P$ can be constructed as:
\begin{equation}
{\bf J}= I-2 \left( 
\vert01\rangle\langle01\vert+\vert10\rangle\langle10\vert 
\right)
\end{equation}

with the corresponding matrix form 
\begin{equation}
{\bf J}= \left( \begin{array}{cccc}
1&0&0&0 \\
0&-1&0&0 \\
0&0&-1&0 \\
0&0&0&1
\end{array}
\right)
\end{equation}
\begin{figure}[htbp]
\centering
\includegraphics[angle=0,scale=1.5]{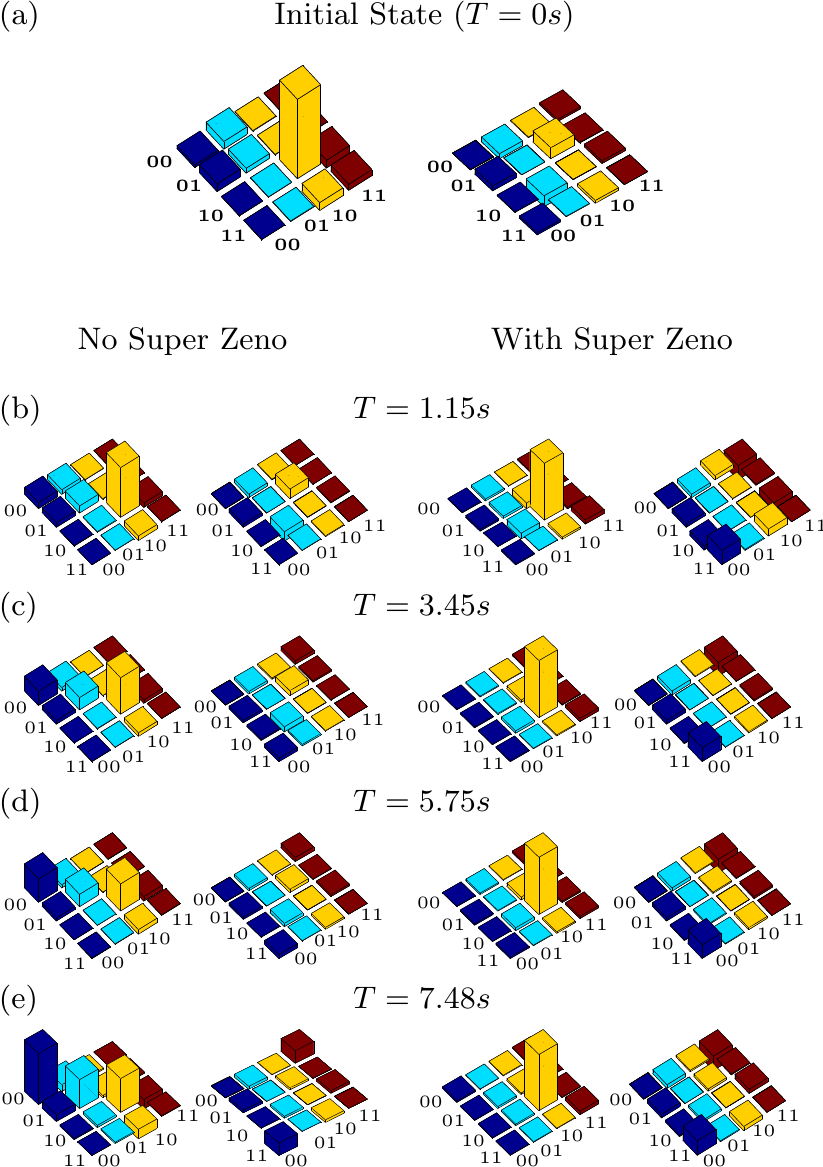}
\caption{  
Real (left) and imaginary (right) parts of
the experimental tomographs of the (a) $\vert 1 0 \rangle$ 
state in the two-dimensional subspace $\{ 01, 10 \}$,
with a computed fidelity of 0.98.
(b)-(e) depict the state at $T = 1.15,
3.45, 5.75, 7.48$ s, with the tomographs on the left and the
right representing the state without and after applying the
super-Zeno preserving scheme, respectively.  The rows and
columns are labeled in the computational basis ordered from
$\vert 00 \rangle$ to $\vert 11 \rangle$.
}
\label{sz_10tomo_sz}
\end{figure}
\begin{figure}[htbp]
\centering
\includegraphics[angle=0,scale=1.5]{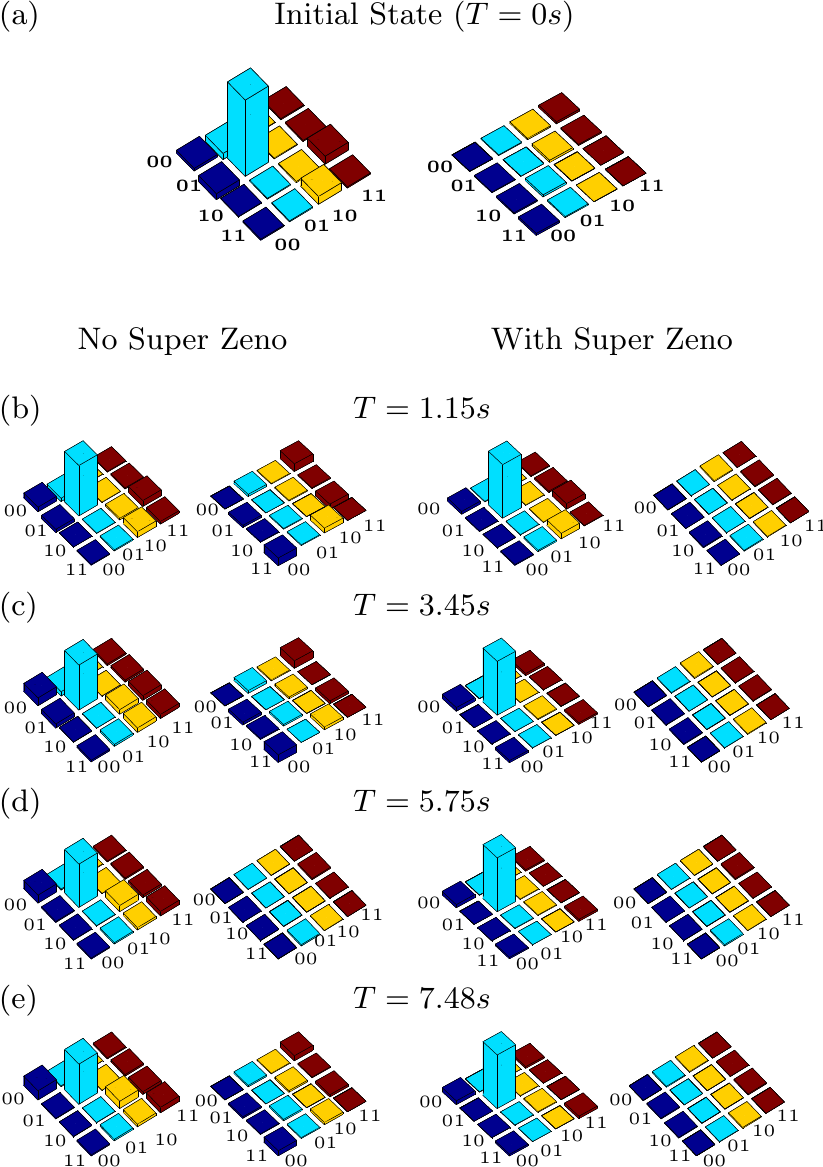}
\caption{  
Real (left) and imaginary (right) parts of
the experimental tomographs of the (a) $\vert 0 1 \rangle$ 
state in the two-dimensional subspace $\{ 01, 10 \}$,
with a computed fidelity of 0.99.
(b)-(e) depict the state at $T = 1.15,
3.45, 5.75, 7.48$ s, with the tomographs on the left and the
right representing the state without and after applying the
super-Zeno preserving scheme, respectively.  The rows and
columns are labeled in the computational basis ordered from
$\vert 00 \rangle$ to $\vert 11 \rangle$.
}
\label{sz_01tomo_sz}
\end{figure}

The quantum circuit and corresponding NMR pulse sequence to
preserve a general state in the $\{\vert 01 
\rangle, \vert 10 \rangle \}$ subspace is
given in Fig.~\ref{subspaceckt_sz}. The unitary kick (denoted
as $U_{zz}$ in the unitary operation block in
Fig.~\ref{subspaceckt_sz}(a)) is  
implemented by  tailoring the gate time to the $J$-coupling
evolution interval of the system Hamiltonian, sandwiched by
non-selective $\pi$ pulses (NOT gates), to refocus undesired
chemical shift evolution during the action of the
gate. 
The five $\Delta_i$ intervals were worked
to be $0.95$ ms, $2.5 $ ms, $3 $ ms, $2.5$ ms 
and $0.95$ ms respectively, for $t = 10$ ms.
One run of the super-Zeno circuit (with four inverting ${\bf J}$s
and five $\Delta_i$ time evolution periods) takes
approximately $288$ ms
and the entire super-Zeno preserving sequence 
$W_N(t)$ in
Eqn.~(\ref{superzenoeqn_sz}), is applied
$30$ times.
\begin{figure}[htbp]
\centering
\includegraphics[angle=0,scale=1.5]{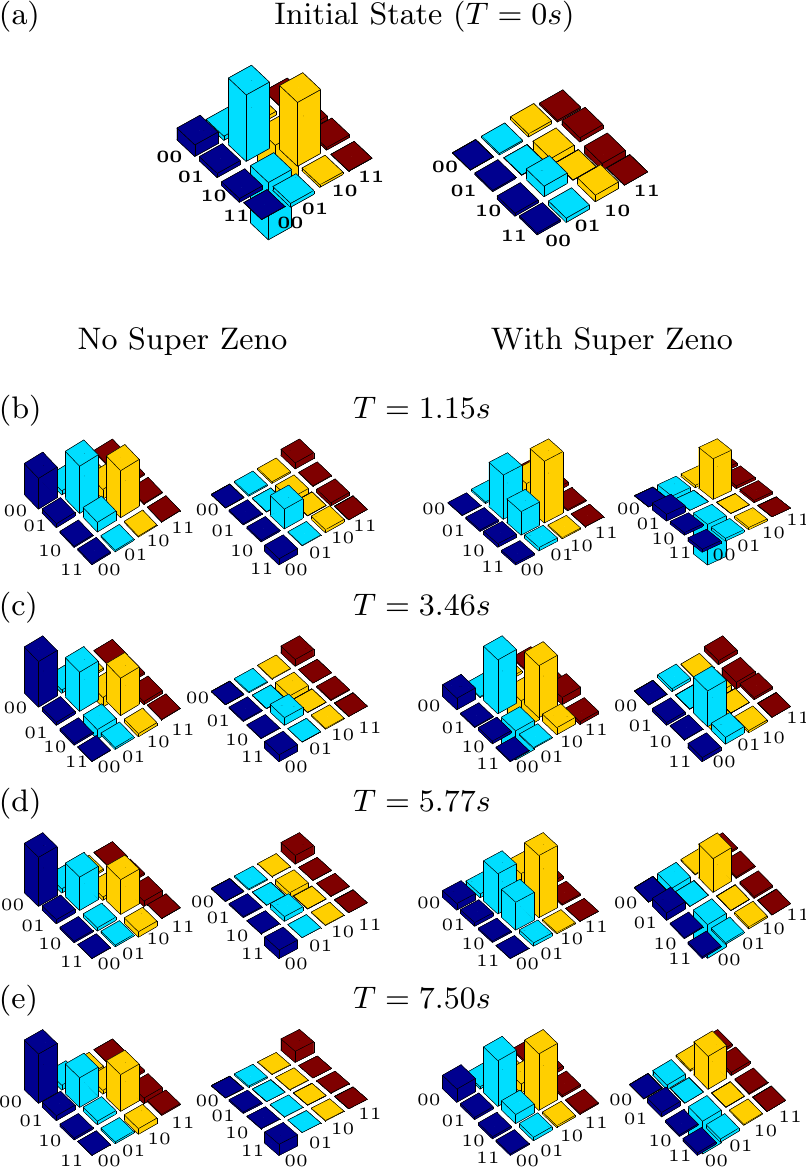}
\caption{   Real (left) and imaginary (right) parts of
the experimental tomographs of the (a) 
$\frac{1}{\sqrt{2}}(\vert 0 1 \rangle - \vert 1 0 \rangle)$
(singlet) 
state in the two-dimensional subspace $\{ 01, 10 \}$,
with a computed fidelity of 0.98.
(b)-(e) depict the state at $T = 1.15,
3.46, 5.77, 7.50$ s, with the tomographs on the left and the
right representing the state without and after applying the
super-Zeno preserving scheme, respectively.  The rows and
columns are labeled in the computational basis ordered from
$\vert 00 \rangle$ to $\vert 11 \rangle$.
}
\label{singletsubspacetomo_sz}
\end{figure}


\subsubsection{Preservation of  product states in
the subspace}
We implemented the subspace-preserving scheme on
two different (separable) states $\vert 0 1 \rangle$ and
$\vert 1 0 \rangle$ in the subspace ${\cal P}$.
The efficacy of the preserving unitary is
verified by tomographing the experimental density
matrices at different time points and computing the
state fidelity. Both the $\vert 0 1 \rangle$ and
$\vert 1 0 \rangle$ states remain within the
subspace ${\cal P}$ and
do not leak out to the orthogonal subspace ${\cal Q} =
\{ \vert 00 \rangle , \vert 11 \rangle \}$. 

The final $\vert 1 0 \rangle$ state has been reconstructed using
state tomography, and the real and imaginary
parts of the experimental density matrices without
any preservation and after applying the
super-Zeno scheme,
tomographed at different time points, are shown in
Fig.~\ref{sz_10tomo_sz}. As can be seen from the
experimental tomographs, the evolution of the
$\vert 1 0 \rangle$ state out of the
subspace is almost completely frozen by the
super-Zeno sequence upto nearly 7.5 s, while without
any preservation the state has leaked into the
orthogonal subspace within 3.5 s. 
The tomographs showed in Fig.~\ref{sz_01tomo_sz} 
for the $\vert 0 1 \rangle$ state, show a
similar level of preservation.


\subsubsection{Preservation of an entangled state in 
the subspace}
We now prepare an entangled state (the singlet state)
embedded in the two-dimensional ${\cal P}=\{\vert 01
\rangle, \vert 10 \rangle \}$
subspace, and used the subspace-preserving scheme described
in Fig.~\ref{subspaceckt_sz} to protect ${\cal P}$.
The singlet state was reconstructed using
state tomography, and the real and imaginary
parts of the tomographed experimental density matrices without
any preservation and after applying the
super-Zeno scheme,
are shown in
Fig.~\ref{singletsubspacetomo_sz}. 
As can be seen from the experimental
tomographs, the state evolution remains
within the ${\cal P}$ subspace but the state
itself does not remain maximally entangled. 

\begin{figure}[htbp]
\begin{center}
\includegraphics[angle=0,scale=1.55]{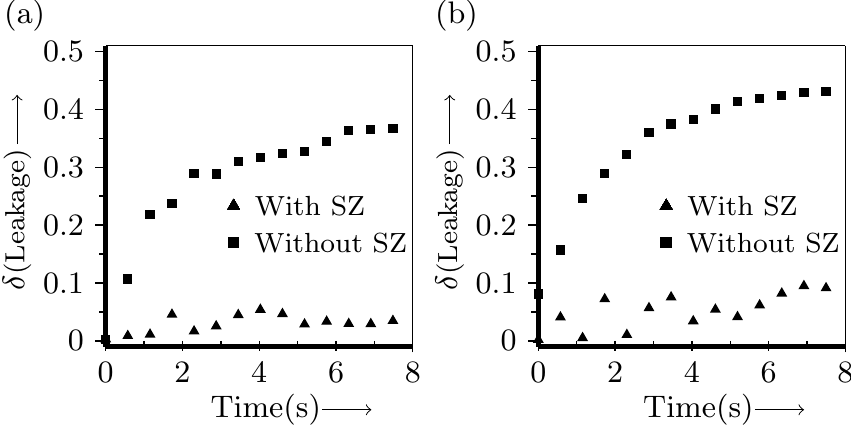}
\caption{Plot of leakage fraction 
from the $\{\vert 01\rangle, \vert 10\rangle \}$ subspace to its orthogonal
subspace $\{\vert 00\rangle, \vert 11\rangle \}$ of (a) the $\vert 1 0 \rangle$
state and (b) the
$\frac{1}{\sqrt{2}}(\vert 0 1 \rangle - \vert 1 0 \rangle)$
(singlet) state, without any preservation and after
applying the super-Zeno sequence. The leakage to the
orthogonal subspace is minimal (remains close to zero)
after applying the super-Zeno scheme.
\label{leakagefig_sz}
}
\end{center}
\end{figure} 
\subsubsection{Estimating leakage outside subspace}
An ensemble of spins initially prepared in a state belonging to the subspace 
${\cal P}=\{\vert 01 \rangle, \vert 10 \rangle \}$, 
can evolve to the orthogonal subspace
${\cal Q}=\{\vert 00 \rangle, \vert 11 \rangle \}$ 
due to unwanted
interactions with its environment.
This evolution of the state
to the orthogonal subspace is called leakage. 
We applied super-Zeno scheme to protect   
leakage of a state from subspace $\cal{P}$ to subspace $\cal{Q}$.
The subspace-preserving capability of the circuit
given in Fig.~\ref{subspaceckt_sz} was quantified by
computing a leakage parameter that defines the amount of
leakage of the state to the orthogonal
${\cal Q}=\{\vert 00 \rangle, \vert 11 \rangle \}$ subspace. 

For a given density operator $\rho$ the ``leak
(fraction)'' $\delta$, into the subspace ${\cal Q}$ is defined 
as
\begin{equation}
\delta = 
\langle 00 \vert \rho \vert 00 \rangle + 
\langle 11 \vert \rho \vert 11 \rangle  
\end{equation}
The leak (fraction) $\delta$ represents the number 
of spins of the ensemble that have migrated
to the subspace $\cal{Q}$ divided by 
the total number of spins in the ensemble;
$\delta$ is equal to one when
all the states remain in the subspace
$\cal{P}$ and is equal to zero if all the 
states have leaked to the subspace $\cal{Q}$.
The leak (fraction) $\delta$ versus time is plotted in Figs.~\ref{leakagefig_sz}(a) and (b), for
the $\vert 1 0\rangle$ and the singlet state respectively, with
and without applying the super-Zeno subspace-preserving
sequence. The leakage parameter remains close to zero
for both kinds of states, proving the success 
and the generality of the
super-Zeno scheme.
\subsection{Preservation of entanglement}
The amount of entanglement that remains in the state
after a certain time is quantified by
an entanglement parameter denoted by $\eta$ as described in 
Chap.~\ref{chapter_mle}, Sec.~\ref{chapmle_entang}.
We will use this entanglement parameter $\eta$ to
quantify the amount of entanglement at different
times. The maximally entangled singlet state was created
and its evolution studied in two different
scenarios. In the first scenario described
in Sec.~\ref{statesection_sz}, the 
singlet state was protected against evolution by the
application of the super-Zeno scheme. In the second
scenario described
in Sec.~\ref{subspacesection_sz}, a two-dimensional subspace containing
the singlet state was protected using the super-Zeno
scheme. For the former case, one expects that the state will
remain a singlet state, while in the latter case, it
can evolve within the protected two-dimensional subspace.
Since in the second case, the protected subspace contains
entangled as well as separable states, one does not expect
preservation of entanglement to the same extent as expected
in the first case, where the one-dimensional subspace
defined by the singlet state itself is protected.
The experimental tomographs at different times and
fidelity for the case of state protection and 
the leakage fraction for the case of subspace protection  
have been discussed in detail in the previous subsections.
\begin{figure}[htbp]
\centering
\includegraphics[angle=0,scale=1.55]{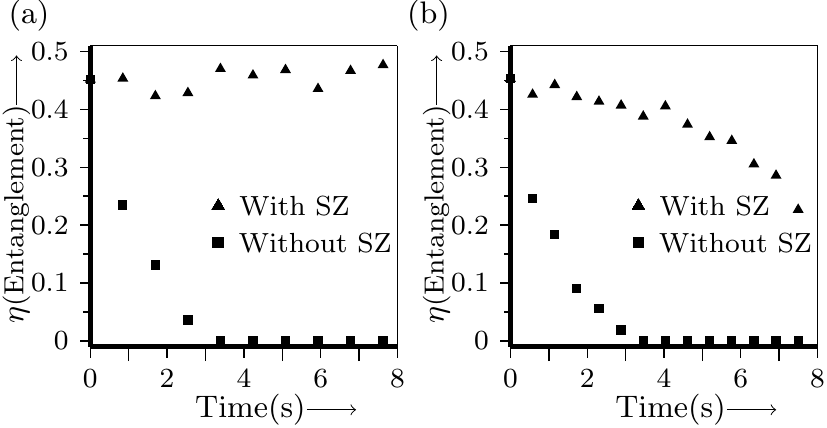}
\caption{Plot of entanglement parameter $\eta$ with time,
with and without applying the super-Zeno sequence,
computed for (a) the
$\frac{1}{\sqrt{2}}(\vert 0 1 \rangle - \vert 1 0 \rangle)$
(singlet) state, and (b) the same singlet state when
embedded in the subspace $\{\vert 01 \rangle, 
\vert 10 \rangle \}$ being
preserved.}
\label{entangfig_sz}
\end{figure} 
Here we focus our attention on the entanglement present in
the state at different times.
The entanglement parameter $\eta$ for the
evolved singlet state 
is plotted as
a function of time and is
shown in Figs.~\ref{entangfig_sz}(a) and (b),
after 
applying the state-preserving and the subspace-preserving
super-Zeno sequence respectively. 
In both cases,
the state becomes disentangled very quickly (after
approximately 2 $s$) if no super-Zeno preservation
is performed. After applying the state-preserving
super-Zeno sequence (Fig.~\ref{entangfig_sz}(a)), the
amount of entanglement in the state remains 
close to maximum for a long
time (upto 8 $s$). After applying the
subspace-preserving super-Zeno sequence
(Fig.~\ref{entangfig_sz}(b)), the state shows some
residual entanglement over long times but it is
clear that the state is no longer maximally
entangled. This implies that 
the subspace-preserving sequence does not
completely preserve the entanglement of the singlet state,
as expected.
However, while the singlet state becomes mixed over time,
its evolution
remains confined to states within the 
two-dimensional subspace (${\cal P} = \{\vert 
01 \rangle, \vert 10 \rangle \}$) 
being preserved as is shown in Fig.~\ref{leakagefig_sz},
where we calculate the leak (fraction). 
\section{Conclusions}
\label{concl_sz}
In this chapter it was experimentally demonstrated that the
super-Zeno scheme can efficiently preserve states in one
and two-dimensional subspaces, by preventing leakage to
a subspace orthogonal to the subspace being preserved.
The super-Zeno sequence was implemented on product as well as on
entangled states, embedded in one- and two-dimensional
subspaces of a two-qubit NMR quantum information processor.
The advantage of the super-Zeno protection lies in its ability to 
preserve the state such that while the number of spins in that particular
state reduces with time, the state remains the same.
Without the super-Zeno protection, the number of spins in the state
reduces with time and the state itself migrates towards
a thermal state, reducing the fidelity.
This work adds to the arsenal of real-life attempts to
protect against evolution of states in quantum computers and
points the way to the possibility of developing hybrid
 strategies (combining the super-Zeno scheme with other
schemes such as dynamical decoupling sequences) to tackle
 preservation of fragile computational resources such as
entangled states.
\newpage
\chapter{Experimental protection 
of unknown states using 
nested Uhrig dynamical decoupling sequences}\label{chapter_nudd}
\section{Introduction}
\label{intro_nudd}
Different decoupling strategies can be used 
to decouple a quantum system 
from its environment, thus controlling the effect
of undesired changes. In Chapter \ref{chapter_superzeno},
we protected a known state of the system and also prevented leakage 
from a known subspace of the system via the super-Zeno scheme. While
protecting against leakage from the subspace, the state was not
leaking out to the orthogonal subspace but 
was 
evolving within the subspace.
Hence, the obvious question comes to mind: can one freeze
this evolution within the subspace? The answer is yes; 
an unknown state in a known subspace can be frozen
using nested Uhrig dynamical decoupling sequences. 
DD schemes rely on the repeated 
application of control pulses and delays to remove
the unwanted contributions due to system-environment
interaction~\cite{viola-review}. For a quantum system
coupled to a bath, the DD sequence decouples the system and
bath by adding a suitable decoupling interaction, periodic
with cycle time $T_c$ to the overall system-bath 
Hamiltonian~\cite{viola-prl-99-2}.  After $N$ applications of
the cycle for a time $N T_c$, the system is governed by a
stroboscopic evolution under an effective average
Hamiltonian, in which system-bath interaction
 terms are no longer present.

More sophisticated DD schemes are of the
Uhrig dynamical decoupling (UDD) type, 
wherein the pulse timing in the DD
sequence is tailored to produce higher-order cancellations in the Magnus
expansion of the effective average Hamiltonian,
thereby achieving system-bath decoupling to a 
higher order and hence stronger noise
protection~\cite{uhrig-njp-08,hodgson-pra-10,schroeder-pra-11,yang-fpc-11,liu-nc-13}.
A UDD sequence can suppress 
decoherence up to  $O(T^{N+1})$ with only $N$ pulses. In a UDD sequence, the $k^{th}$ control pulse is applied at the time 
\begin{equation}
 T_k=T {\rm sin}^2\left(\frac{k\pi}{2N+2}\right); k=1,2,\dots,N.
\end{equation}

UDD schemes are applicable when the control pulses can be considered as ideal
(i.e. instantaneous) and when the environment noise has a sharp frequency
cutoff~\cite{dhar-prl-06,yang-prl-08,uhrig-prl-09,khodjasteh-pra-11}.  These
initial UDD schemes dealt with protecting a single qubit against different
types of noise, and were later expanded to a whole host of optimized sequences
involving nonlocal control operators, to protect multi-qubit systems against
decoherence~\cite{mukhtar-pra-10-1,pan-jpb-11,cong-ijqi-11,alvarez-pra-12,west-njp-12,ahmed-pra-13}.

While UDD schemes can well protect states against 
single- and two-axis noise (i.e. pure dephasing and/or pure bit-flip),
they are not able to protect against general three-axis
decoherence~\cite{kuo-jmp-12}.
Nested UDD (NUDD) schemes were hence proposed to
protect multiqubit systems in generic quantum baths to
arbitrary decoupling orders,
by nesting several UDD layers. 
The structure of the Hamiltonian was exploited
and an appropriate 
set of mutually orthogonal operation set (MOOS) was designed. 
On the basis of commutation and anticommutation property of operators in an MOOS
with Hamiltonian, the NUDD scheme was constructed such that after nesting of operator layers
only those terms remained 
in the generating algebra of Hamiltonian which did not affect the state. 
It was  shown that the NUDD scheme can preserve a set of
unitary Hermitian system operators (and hence all operators
in the Lie algebra generated from this set of operators)
that mutually either commute or 
anticommute~\cite{wang-pra-11,jing2015}.
Furthermore, it was proved that the NUDD scheme is universal i.e. it can
preserve the coherence of $m$ coupled qubits by suppressing
decoherence upto order $N$, independent of the nature of the
system-environment coupling~\cite{jiang-pra-11}.

The efficiency of NUDD schemes in 
protecting unknown randomly generated two-qubit 
states was shown to be a powerful approach
for protecting quantum states against
decoherence~\cite{mukhtar-pra-10-2}. Numerical
simulations on a five-spin system
were carried out in this context. Two of the five spins were 
identified as a two-qubit system and the other 
three spins were regarded as the bath. To show the efficacy of the NUDD scheme,
it was applied on ten arbitrary states 
belonging to a known subspace. For the correct 
nesting of UDD layers a remarkably high fidelity was achieved in 
locking the initial 
unknown superposition state~\cite{mukhtar-pra-10-2}.

In this chapter, the efficacy of protection of the NUDD scheme
is first evaluated by applying it on four specific states of the subspace ${\cal
P}=\{\vert 01 \rangle, \vert 10 \rangle \}$ i.e.  two
separable states: $|01\rangle$ and $|10\rangle$, and two
maximally entangled singlet and triplet Bell states:
$\frac{1}{\sqrt{2}}(|01\rangle-|10\rangle)$ and
$\frac{1}{\sqrt{2}}(|01\rangle+|01\rangle)$ in a
four-dimensional two-qubit Hilbert space.  Next, to evaluate
the effectiveness of the NUDD scheme on the entire subspace,
randomly states are generated  in the subspace ${\cal P}$
(considered as a superposition of the known basis states
$\vert 01 \rangle, \vert 10 \rangle$) and protected them using
NUDD scheme. Eight states in the two-qubit
subspace randomly generated and protected using a three-layer NUDD sequence.
Full state tomography is used to compute the experimental
density matrices.  Each state is allowed to decohere, and
the state fidelity is computed at each time point without
protection and after NUDD protection.  The results are
presented as a histogram and showed that while NUDD is always
able to provide some protection, the degree of protection
varies from state to state. 

\section{The NUDD scheme}
\label{theory_nudd}
Consider a two-qubit quantum system with its state
space spanned by the states $\{ \vert 00 \rangle,$ \\
$\vert 01\rangle, \vert 10 \rangle, \vert 11 \rangle \}$, the
eigenstates of the Pauli operator $\sigma_{z}^{1} \otimes 
\sigma_{z}^{2}$.
Our interest is in protecting states in the subspace
$\cal{P}$ spanned by states $\{\vert 01 \rangle,\vert 10
\rangle\}$, against decoherence.
The density matrix 
corresponding to an arbitrary pure state $\vert \psi \rangle
=\alpha \vert 01 \rangle + \beta \vert 10 \rangle$
belonging to the subspace $\cal{P}$
is given by
\begin{equation}
\rho(t)= \left(\begin{array}{cccc}
0&0&0&0 \\
0& \vert \alpha \vert^2&\alpha \beta^{*}&0 \\
0& \beta \alpha^{*}& \vert \beta \vert^2&0 \\
0&0&0&0
\end{array}
\right)
\end{equation}
with the coefficients $\alpha$ and $\beta$ satisfying 
$ \vert \alpha \vert^2+ \vert \beta \vert^2=1$ at 
time $t=0$.
We describe here the theoretical construction
of a three-layer NUDD scheme to protect arbitrary states
in the two-qubit subspace $\cal{P}$~\cite{mukhtar-pra-10-1,mukhtar-pra-10-2}.

The general total Hamiltonian of a two-qubit system 
interacting with an arbitrary bath can be written as
\begin{equation}
H_{\rm total} =H_{S}+ H_{B} + H_{jB} + H_{12}  
\label{totham1_nudd}
\end{equation}
where $H_S$ is the system Hamiltonian, $H_{B}$ is
the bath Hamiltonian, $H_{jB}$ is qubit-bath
interaction Hamiltonian and $H_{12}$ is the
qubit-qubit interaction Hamiltonian (which can be
bath-dependent). 
Our interest here is in bath-dependent terms and
their control, which can be expressed using a
special basis set for the two-qubit system as
follows~\cite{mukhtar-pra-10-1,mukhtar-pra-10-2}:
\begin{eqnarray}
H &=& H_{B} + H_{jB} + H_{12}  \nonumber \\
  &=& \sum_{j=1}^{16} W_j Y_j 
\label{newbasis_nudd}
\end{eqnarray}
where the coefficients $W_j$ contain arbitrary bath
operators.  
$Y$ are the special basis computed from the
perspective of preserving the subspace spanned
by the states $\{ \vert 01 \rangle, \vert 10 \rangle \}$
in the two-qubit
space~\cite{mukhtar-pra-10-1,mukhtar-pra-10-2}:
\begin{eqnarray}
&&Y_{1}=I,
\quad 
\quad 
\quad 
\quad 
\,\,\,\,\,
Y_{2}=|01\rangle\langle01|+|10\rangle\langle10|, 
\nonumber \\
&&Y_{3}=
|00\rangle\langle11|,
\quad 
\quad
Y_{4}=
|00\rangle\langle00|-|11\rangle\langle11|,
\nonumber \\
&&Y_{5}=|11\rangle\langle00|,
\quad
\quad
Y_{6}=|01\rangle\langle01|-|10\rangle\langle10|, 
\nonumber \\
&&Y_{7}=|10\rangle\langle00|, 
\quad
\quad
Y_{8}=|00\rangle\langle10|,
\nonumber \\
&&Y_{9}=|10\rangle\langle11|,
\quad
\quad
Y_{10}=|11\rangle\langle10|,
\nonumber \\
&&Y_{11}=|01\rangle\langle00|,
\quad \,\,\,\,
Y_{12}=|00\rangle\langle01|,
\nonumber \\
&&Y_{13}=|01\rangle\langle11|,
\quad \,\,\,\,
Y_{14}=|11\rangle\langle01|, 
\nonumber \\
&&Y_{15}=|01\rangle\langle10|+|10\rangle\langle
01|,  \nonumber \\
&&Y_{16} =-i(|10\rangle\langle01|-|01\rangle\langle10|).
\end{eqnarray}

To protect a general two-qubit
state $\vert \psi \rangle \in \cal{P}$
against decoherence using NUDD, we are required to
protect diagonal populations $|\alpha|^2$, $|\beta|^2$
and off-diagonal coherences $\alpha\beta^{*}$. Hence the
locking scheme requires the
nesting of three layers of UDD sequences.

\noindent{$\bullet$ \bf Innermost UDD layer:} 
The diagonal populations
$\text{Tr}[\rho(t) \vert 01\rangle\langle
01 \vert]\approx \vert \alpha \vert^2 $ are locked by
this UDD layer 
with the control operator 
\begin{equation} 
X_{0}= I - 2 \vert 01\rangle\langle01 \vert.
\end{equation}

We can write the total Hamiltonian,
\begin{eqnarray}
H&=&H_0+H_1, \nonumber\\
H_{0}&=&\sum_{i=1}^{10}W_{i}Y_{i}, \nonumber \\
H_1&=&\sum_{i=11}^{16}W_{i}Y_{i},
\end{eqnarray}
such that the $X_0$ commute with $H_0$, i.e. $[X_0,H_0]=0$ and 
anti-commute with $H_1$, i.e. $\{X_0,H_1\}_{+}=0$.
An inverting pulse with control Hamiltonian,
\begin{eqnarray}
\label{control1_nudd}
H_{c}=\sum_{j=1}^{N}\pi \delta(t-T_j)\frac{X_0}{2},
\label{UDDcontrol_nudd}
\end{eqnarray}
applied with the UDD timing $T_{j}$, defined as : 
\begin{eqnarray} T_{j}  =T \sin^{2}(\frac{j\pi}{2N+2}),\
j=1, 2\cdots ,N.
\label{UDDtime_nudd}
\end{eqnarray}
The unitary evolution operator of the system for 
time period $t=0$ to $t=T$ is given by ($\hbar=1$) : 
\begin{eqnarray}
{U}_{N}(T) & = & X_{0}^{N} e^{-i[H_{0}+H_1](T-T_N)} (-iX_0) \nonumber \\
& &\ \times\ e^{-i[H_{0}+H_1](T_N-T_{N-1})} (-iX_0) \nonumber \\
& &\ \cdots \nonumber \\
 & &\ \times\  e^{-i[H_{0}+H_1](T_3-T_2)} (-iX_0) \nonumber \\
& &\ \times\ e^{-i[H_{0}+H_1](T_2-T_1)} (-iX_0)\nonumber \\
& &\ \times\ \ e^{-i[H_{0}+H_1]T_1}. \label{Ut_nudd}
  \end{eqnarray}
Using the UDD universality proof~\cite{mukhtar-pra-10-1} and 
the fact that 
 $H_0$ and $H_1$ commute and anticommute respectively with
  $X_0$, it can be shown that 
 \begin{eqnarray}
 U_{N}(T)=U_{N}^{\text{even}}+O(T^{N+1}),
\end{eqnarray}
where
\begin{eqnarray}
U_N^{\text{even}}=\exp(-iH_0 T) \sum_{k=0}^{+\infty} (-i)^{2k}\Delta
_{2k}, \label{Deltaform_nudd}
\end{eqnarray}
with $\Delta_{2k}$ containing only even powers of $H_1^I(t)$, defined
by
\begin{center}
$H_1^I(t)\equiv \exp(iH_0 t) H_1\exp(-iH_0 t)$.
\end{center}
$\Delta_{2k}$ can be expanded as a linear superposition of all
possible basis operators that commute with $X_0$. i.e.
\begin{equation}
 \Delta_{2k}=\sum_{i=1}^{10} A_{i} Y_i
\end{equation}
where $A_i$ are the expansion coefficients containing bath operators.
The $N$th order, $U_{N}(T)$ can be expressed as a combination of
$Y_1$, $Y_2$, $\cdots$, $Y_{10}$ only. 
Using the closure of this set of operators, i.e.,
\begin{eqnarray}
\left(\sum_{i=1}^{10}A_{i}Y_{i}\right) \left(
\sum_{k=1}^{10}B_{k}Y_{k}\right)&=&\sum_{l=1}^{10}C_{l}Y_{l}
\end{eqnarray}
we further obtain
\begin{eqnarray}
U_{N}(T)= \exp(-iH_{\text{eff}}^{\text{UDD-1}} T)+ O(T^{N+1}),
\end{eqnarray}
where
\begin{eqnarray}
H_{\text{eff}}^{\text{UDD-1}} =  \sum_{i=1}^{10} D_{1,i} Y_i,
\end{eqnarray}
where $D_{1,i}$ refer to the expansion coefficients of
this first UDD layer. Terms containing basis operators
$Y_{11} \cdots Y_{16}$ are efficiently decoupled.

\noindent{$\bullet$ \bf Second UDD layer:}
The diagonal populations
$\text{Tr}[\rho(t) \vert 10\rangle\langle10
\vert]\approx
\vert \beta \vert^2$ are locked  by this second UDD layer
with the control operator
\begin{equation}
X_{1}= I -2 \vert 10\rangle\langle10 \vert
\end{equation}
We can further decompose 
the effective Hamiltonian after 
the first layer $H_{\text{eff}}^{\text{UDD-1}}$ into 
\begin{eqnarray}
H_{\text{eff}}^{\text{UDD-1}} &= & H_{\text{eff,0}}^{\text{UDD-1}}+
H_{\text{eff,1}}^{\text{UDD-1}},
\end{eqnarray}
such that $[H_{\text{eff,0}}^{\text{UDD-1}},X_1]=0$ and $\{H_{\text{eff,1}}^{\text{UDD-1}},X_1\}_+=0$.
where \begin{eqnarray}
 H_{\text{eff,0}}^{\text{UDD-1}} &\equiv & \sum_{i=1}^{6}D_{1,i} Y_i;  \nonumber \\
 H_{\text{eff,1}}^{\text{UDD-1}} &\equiv &  \sum_{i=7}^{10}D_{1,i} Y_i.
 \end{eqnarray}
It is straightforward to see that
the operators $Y_i$, $i=1-6$ form a closed algebra. Hence when a second layer
of UDD sequence of $X_1$ is applied to the $N$th order,
the dynamics of $H_{\text{eff}}^{\text{UDD-1}}$ reduce to
\begin{eqnarray}
H_{\text{eff}}^{\text{UDD-2}}=\sum_{i=1}^{6}D_{2,i} Y_i,
\end{eqnarray}
where $D_{2,i}$ refer to the expansion coefficients of
this second UDD layer. Terms containing basis operators
$Y_{7} \cdots Y_{10}$ are efficiently decoupled.

\noindent{$\bullet$ \bf Outermost UDD layer:}
The off-diagonal coherences
$\text{Tr}[\rho(t) \vert 01\rangle\langle 
10 \vert] \approx \alpha \beta^{*}$ are locked by this
final UDD layer 
with the control operator
\begin{equation}
X_{\phi}= I - [\vert 01\rangle+ \vert 10\rangle][\langle01\vert+\langle
10 \vert].
\end{equation}
Again we can write the effective Hamiltonian after the second 
layer  as
\begin{equation}
 H_{{\rm eff}}^{{\rm UDD-2}}= H_{{\rm eff,0}}^{{\rm UDD-2}}+H_{{\rm eff,1}}^{{\rm UDD-2}},
\end{equation}
such that $[H_{\text{eff,0}}^{\text{UDD-2}},X_\phi]=0$ and $\{H_{\text{eff,1}}^{\text{UDD-2}},X_\phi\}_+=0$. 
In the self-closed set of the operators that form 
$H_{{\rm eff}}^{{\rm UDD-2}}$, the only term 
$D_{2,6}Y_6=D_{2,6}[|01\rangle\langle01|-|10\rangle\langle 10|]$
can affect $\vert \psi(0)\rangle$, which is effectively decoupled 
by this third layer. The final reduced effective Hamiltonian after the
three-layer NUDD contains
five operators:
$H_{{\rm eff}}^{{\rm UDD-3}}=\sum_{i=1}^{5} D_{3,i}Y_i$,
where $D_{3,i}$ are the coefficients due to three UDD layers.

The innermost UDD control $X_0$ pulses are applied at the 
time intervals
$T_{j,k,l}$, the middle layer UDD control $X_1$ pulses are
applied at the time intervals $T_{j,k}$ and the outermost
UDD control $X_{\phi}$ pulses are applied at the time
intervals $T_j$ ($j,k,l=1,2,...N$) given by:
\begin{eqnarray}
T_{j,k,l} &=& T_{j,k} + (T_{j,k+l} - T_{j,k})
\sin^2{\left(\frac{l \pi}{2N+2}\right)} \nonumber \\
T_{j,k} &=& T_{j} + (T_{j+1} - T_{j})
\sin^2{\left(\frac{k \pi}{2N+2}\right)} \nonumber \\
T_{j} &=& T \sin^2{\left(\frac{j \pi}{2N+2}\right)} 
\end{eqnarray}
The total time interval in the $N^{th}$ order sequence is
$(N+1)^3$ with the total number of pulses in one run being
given by $N((N+1)^2+N+2)$ for even
$N$~\cite{mukhtar-pra-10-2}. 

\noindent{\bf Summary of the NUDD scheme:}

The recipe to design UDD protection for a two-qubit state
(say $\vert \chi \rangle$) is given in the following
steps:~(i) First a control operator $X_c$ is constructed
using $X_c = I-2 \vert \chi \rangle\langle \chi \vert$ such
that $X_c^2=I$, with the commuting relation $[X_c,H_0]=0$
and the anticommuting relation $\{X_c,H_1\}=0$;  (ii)
The control UDD Hamiltonian  is then applied so that
system evolution is now under a UDD-reduced effective Hamiltonian
thus achieving state protection upto order $N$; (iii)
Depending on the explicit commuting or anticommuting
relations of $X_c$ with $H_0$ and $H_1$, the UDD sequence
efficiently removes a few operators $Y_{i}$ from the initial
generating algebra of $H$ and hence suppresses all couplings
between the state $\vert \chi \rangle$ and all other states.

\begin{center}
\framebox{$Y_i$, $i=1,2, \cdots, 16$} \\
\vspace{0.2cm}
$\Downarrow${$X_0$, \text{UDD-1}}\\
\vspace{0.2cm}
\framebox{$Y_i$, $i=1,2, \cdots, 10$} \\
\vspace{0.2cm}
$\Downarrow${$X_1$,\text{UDD-2}}\\
\vspace{0.2cm}
\framebox{$Y_i$, $i=1,2,\cdots, 6$}\\
\vspace{0.2cm}
$\Downarrow${$X_\phi$,\text{UDD-3}}\\
\vspace{0.2cm}
\framebox{$Y_i$, $i=1,2, \cdots, 5$}.
\end{center}%

\begin{figure}[htbp]
\centering
\includegraphics[angle=0,scale=1.3]{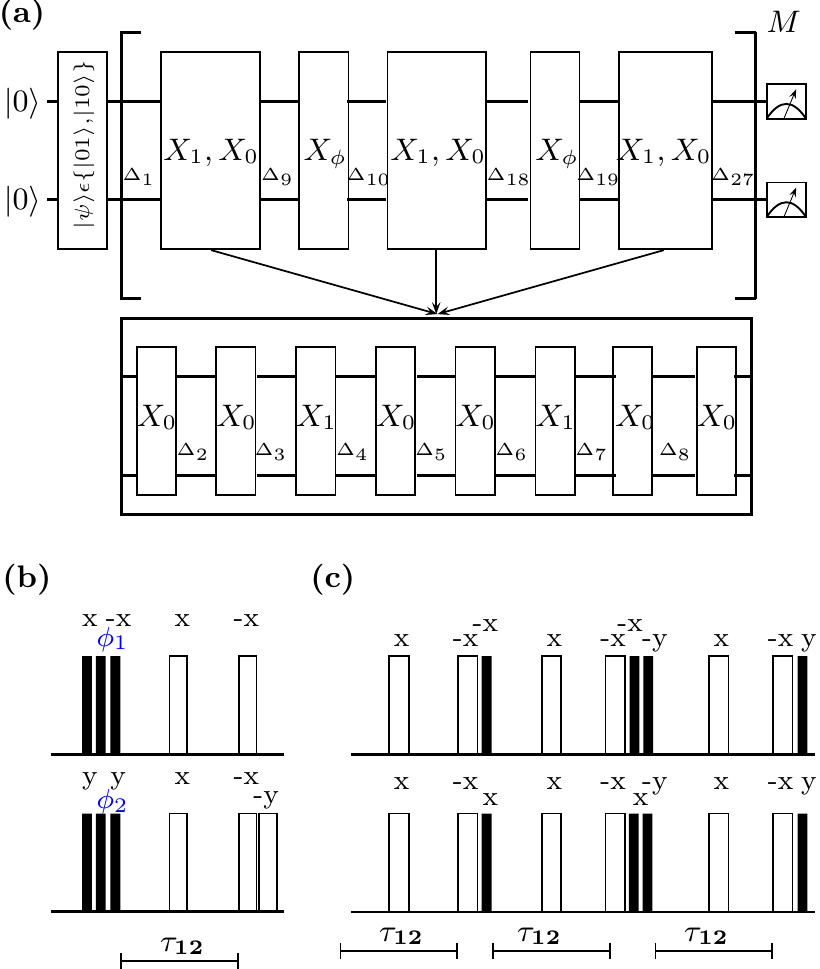} 
\caption{(a) Circuit diagram for
the  three-layer NUDD sequence. 
The innermost UDD layer
consists of $X_0$ control pulses, the middle layer
comprises $X_1$ control pulses and the outermost layer
consists of $X_{\phi}$ pulses. The entire NUDD sequence
is repeated $M$ times;
$\Delta_i$ are time intervals.
(b) NMR pulse sequence to implement the control pulses 
for $X_{0}$ and $X_{1}$ UDD sequences. The values of the
rf pulse phases  
$\phi_1$ and $\phi_2$ are set to $x$ and
$y$ for the $X_0$ and to $-x$ and $-y$ for the
$X_1$ UDD sequence, respectively.
(c) NMR pulse sequence to implement the
control pulses for the $X_{\phi}$ UDD sequence.
The filled rectangles
denote $\pi/2$ pulses while the unfilled rectangles
denote $\pi$ pulses, respectively.
The time period $\tau_{12}$ is set to the value $(2
J_{12})^{-1}$, where $J_{12}$ denotes the strength of the
scalar coupling between the two qubits.
}
\label{nudd_ckt_nudd}
\end{figure}

\begin{figure}[htbp]
\centering
\includegraphics[angle=0,scale=1.15]{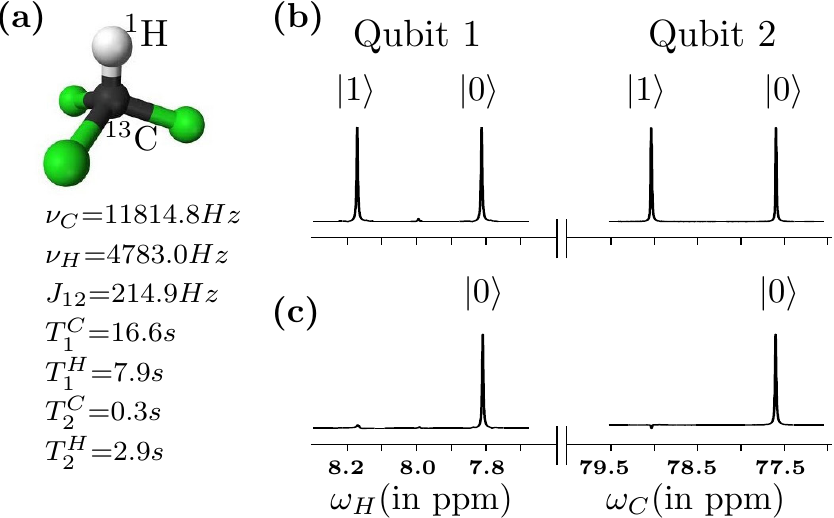}
\caption{(a) 
Structure of isotopically enriched 
chloroform-${}^{13}$C molecule, with the
${}^{1}$H spin labeling the first qubit 
and the ${}^{13}$C spin labeling the second qubit.
The system parameters are tabulated alongside
with chemical shifts $\nu_i$ and scalar coupling $J_{12}$ (in
Hz) and NMR spin-lattice
and spin-spin relaxation times $T_{1}$ and $T_{2}$ (in seconds).
(b) 
NMR spectrum obtained after a $\pi/2$
readout pulse on the thermal equilibrium state
and
(c) NMR spectrum of the pseudopure $\vert 0 0 \rangle$
state.
The resonance lines of each qubit in the spectra are labeled by
the corresponding logical states of the other qubit. 
}
\label{molecule_nudd}
\end{figure}

\section{Experimental protection of two qubits using NUDD}
\subsection{Experimental implementation of the NUDD scheme}
\label{expt1_nudd}
We now turn to the NUDD implementation for $N=2$ on
a two-qubit NMR system.
The entire NUDD sequence can be written in terms of 
UDD control operators 
$X_0, X_1, X_{\phi}$ (defined in the previous section) and
time evolution $U(\delta_i t)$ under the general Hamiltonian
for time interval fractions $\delta_i$:
\begin{eqnarray}
X_{c}(t)
&=& U(\delta_1t)X_{0}U(\delta_2t)X_{0}U(\delta_3t)X_{1}
U(\delta_4t)X_{0}U(\delta_5t) 
X_{0}
\nonumber \\
&&
U(\delta_6t)X_{1}U(\delta_7t)X_{0}
U(\delta_8t)X_{0}U(\delta_9t) X_{\phi}U(\delta_{10}t) 
X_{0} 
\nonumber \\  
&&
U(\delta_{11}t)
X_{0}U(\delta_{12}t)X_{1}
U(\delta_{13}t)X_{0}U(\delta_{14}t)
X_{0}
U(\delta_{15}t)
\nonumber \\
&&
X_{1}
U(\delta_{16}t)X_{0}
U(\delta_{17}t)X_{0}U(\delta_{18}t)
X_{\phi}
U(\delta_{19}t)
\nonumber \\ 
&&
X_{0}
U(\delta_{20}t)
X_{0}
U(\delta_{21}t)X_{1}U(\delta_{22}t)X_{0}
U(\delta_{23}t)
\nonumber \\ 
&&
X_{0}
U(\delta_{24}t)X_{1}
U(\delta_{25}t)
X_{0}U(\delta_{26}t)X_{0}U(\delta_{27}t) 
\label{timing-eqn_nudd}
\end{eqnarray}
In our implementation, the number of
$X_0, X_1$ and $X_{\phi}$ control pulses used 
in one run of the three-layer NUDD sequence are
18, 6 and 2, respectively.

Using the UDD timing intervals defined above and
applying the condition $\sum \delta_i=1$, 
their values are computed to be
\begin{eqnarray}
\{\delta_i \}
&=&\{\beta, 2\beta, \beta, 2\beta, 4\beta,
2\beta,\beta,2\beta, \beta, 2\beta, 4\beta,
2\beta, 4\beta,8\beta,  \nonumber \\ &&4\beta,
2\beta, 4\beta, 2\beta,\beta, 2\beta, \beta,
2\beta,4\beta, 2\beta,\beta,2\beta,\beta \}
\label{delta_nudd}
\end{eqnarray}
where the intervals between the
$X_0, X_1$ and $X_{\phi}$ control pulses 
turn out to be a multiple of $\beta=0.015625 $ .

The NUDD scheme for state protection and the
corresponding NMR pulse sequence is given in
Fig.~\ref{nudd_ckt_nudd}.  The unitary gates $X_{0}$,
$X_{1}$, and $X_{\phi}$ drawn in
Fig.~\ref{nudd_ckt_nudd}(a) correspond to the UDD
control operators already defined in the previous
section.  The $\Delta_i$ time interval in the
circuit given in Fig.~\ref{nudd_ckt_nudd}(a) is defined
by $\Delta_i = \delta_i t$, using the $\delta_i$
given in Eqn.~(\ref{delta_nudd}). The pulses on the top
line in Figs.~\ref{nudd_ckt_nudd}(b) and (c) are
applied on the first qubit (${}^{1}$H spin in
Fig.~\ref{molecule_nudd}), while those at the bottom
are applied on the second qubit (${}^{13}$C spin
in Fig.~\ref{molecule_nudd}), respectively.  All the
pulses are spin-selective pulses, with the
$90^{\circ}$ pulse length being $7.6 \mu$s and
$15.6 \mu$s for the proton and carbon rf channels,
respectively.  When applying pulses simultaneously
on both the carbon and proton spins, care was
taken to ensure that the pulses are centered
properly and the delay between two pulses was
measured from the center of the pulse duration
time.  We note here that the NUDD schemes are
experimentally demanding to implement as they
contain long repetitive cycles of rf pulses
applied simultaneously on both qubits and the
timings of the UDD control sequences were matched
carefully with the duty cycle of the rf probe
being used.

We chose the chloroform-${}^{13}$C molecule as the two-qubit
system to implement the NUDD sequence (Fig.~\ref{molecule_nudd} 
for details of system parameters and
average NMR relaxation times of both the qubits).
The two-qubit system Hamiltonian in the rotating
frame
(which includes the Hamiltonians $H_S$ and $H_{12}$
of Eqn.~\ref{totham1_nudd})
is given by 
\begin{equation}
\label{hamiltonian_nudd}
H_{\rm rot}  = 2\pi[(\nu_{H}-\nu^{rf}_H )I_{z}^{H} +
(\nu_C-\nu^{rf}_C )I_z^C +
J_{12} I_{z}^H I_z^C ] 
\end{equation}
where $\nu_{H}$ ($\nu_{C}$ ) is the chemical shift
of the $^{1}H(^{13}C)$ spin, $ \nu^{rf}_i $ is the
rotating frame frequency ($\nu^{rf}_i = \nu_{i}$
for on-resonance), $I_z^H (I_z^C )$ is the $z$
component of the spin angular momentum operator
for the $^{1}H(^{13}C)$ spin, and $J_{12}$ is the
spin-spin coupling constant.

\begin{figure}[htbp]
\centering
\includegraphics[angle=0,scale=1.5]{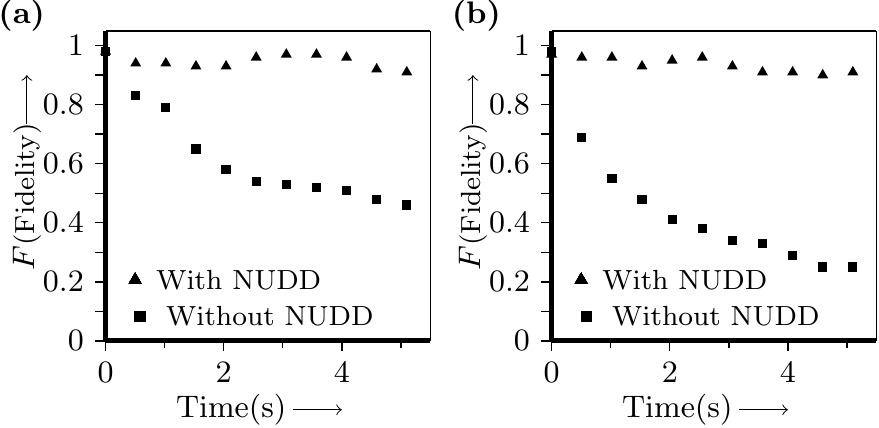}
\caption{Plot of fidelity versus time for (a) the $\vert 01\rangle$
state and (b) the $\vert 10 \rangle)$ state, without any 
protection and after applying NUDD protection. The
fidelity of both the states remains close to 1 for
upto long times, after NUDD
protection.
}
\label{nudd_fidelity0110}
\end{figure}

The two qubits
were initialized into the pseudopure state $\vert 00 \rangle$
using the spatial averaging technique~\cite{sharf-pra-00},
with the corresponding
density operator given by
\begin{equation}
\rho_{00} = \frac{1-\epsilon}{4} I
+ \epsilon \vert 00 \rangle \langle 00 \vert
\label{ppure_nudd}
\end{equation}
with a thermal polarization $\epsilon \approx
10^{-5}$ and $I$ being a $4 \times 4$
identity operator.  
All experimental density matrices were reconstructed using 
quantum state tomography via a maximum likelihood 
protocol (Chapter~\ref{chapter_mle}).
The experimentally created
pseudopure state $\vert 00 \rangle$ was
tomographed with a fidelity of $0.99$. The fidelity of an experimental
density matrix was computed using Eq.~(\ref{mle_fidelity_2}).
\begin{figure}[htbp]
\centering
\includegraphics[angle=0,scale=1.5]{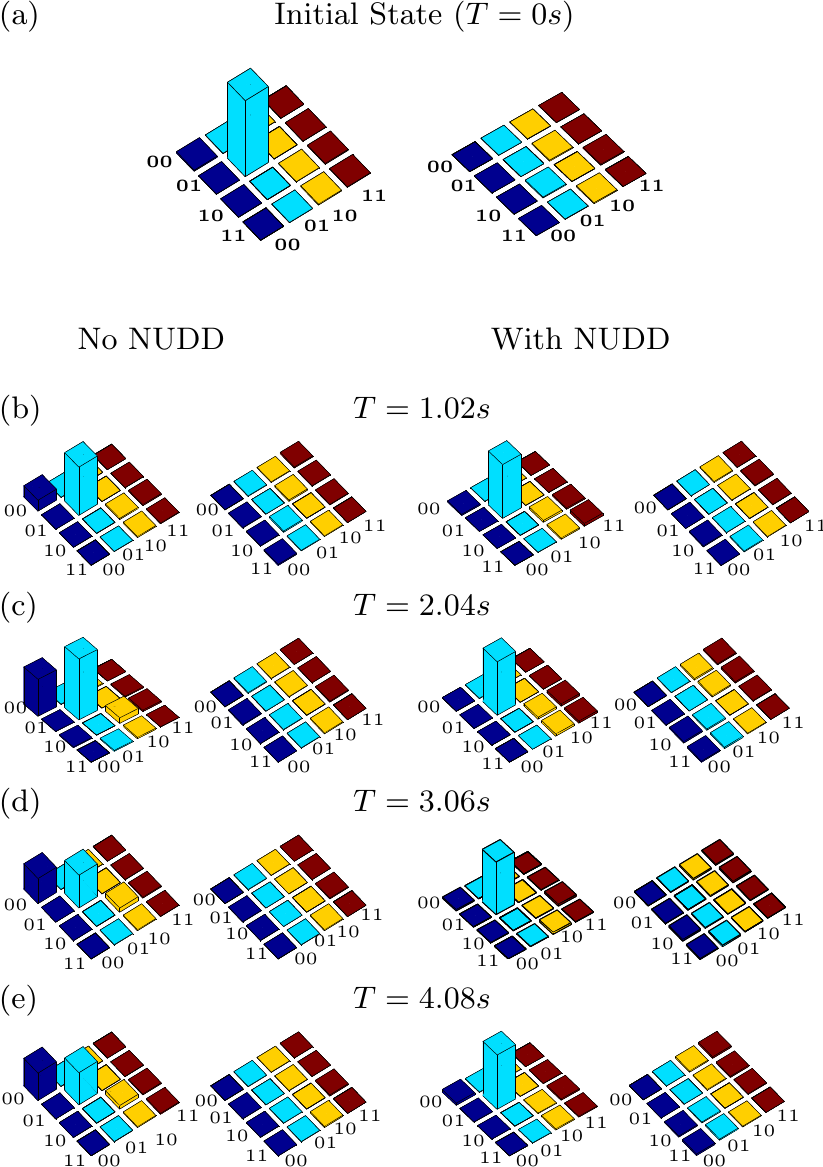}
\caption{
Real (left) and imaginary (right) parts of the experimental
tomographs of the (a)
$|01\rangle$ state, with a
computed fidelity of 0.98.  (b)-(e) depict the state at $T =
1.02, 2.04, 3.06, 4.08$ s, with the tomographs on the left
and the right representing the state without any
protection and after
applying NUDD protection, respectively.  The rows
and columns are labeled in the computational basis ordered
from $\vert 00 \rangle$ to $\vert 11 \rangle$.
}
\label{nudd_tomo01}
\end{figure}
\begin{figure}[htbp]
\centering
\includegraphics[angle=0,scale=1.5]{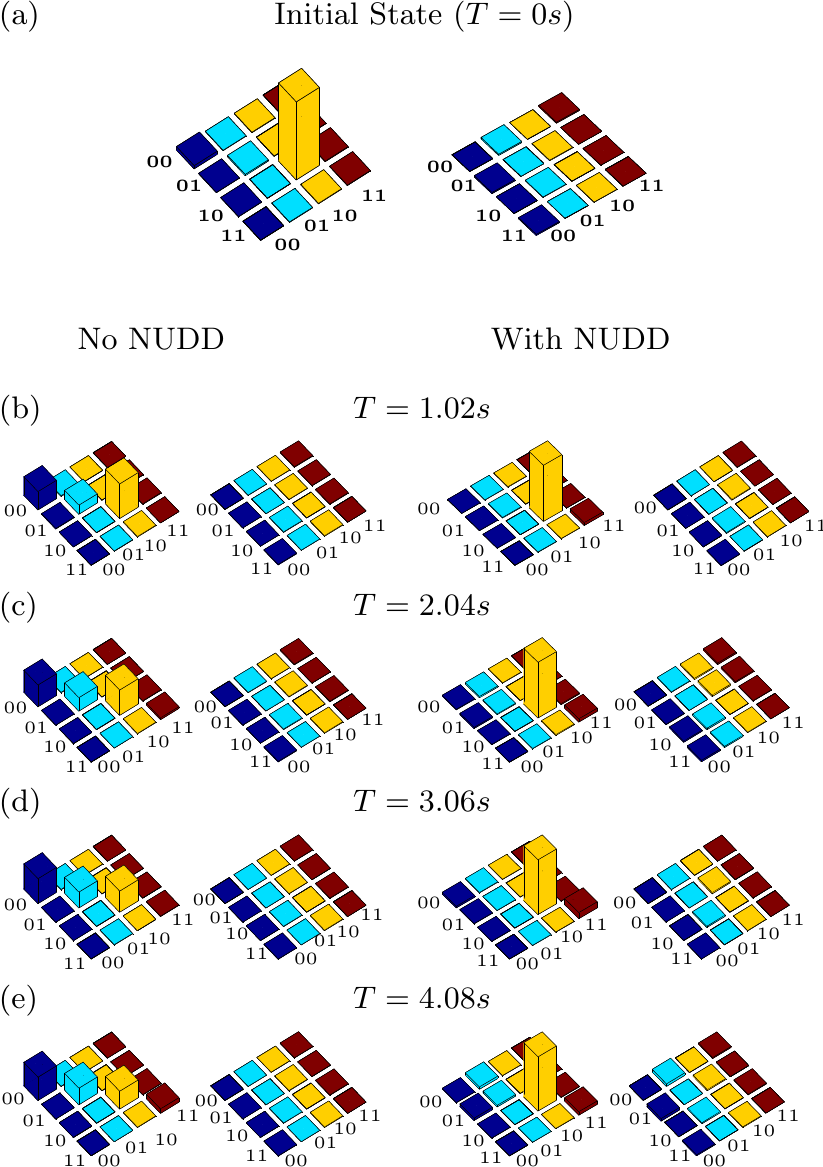}
\caption{
Real (left) and imaginary (right) parts of the experimental
tomographs of the (a)
$|10\rangle$ state, with a
computed fidelity of 0.97.  (b)-(e) depict the state at  $T =
1.02, 2.04, 3.06, 4.08$ s, with the tomographs on the left
and the right representing the state without any
protection and after
applying NUDD protection, respectively.  The rows
and columns are labeled in the computational basis ordered
from $\vert 00 \rangle$ to $\vert 11 \rangle$.
}
\label{nudd_tomo10}
\end{figure}

\begin{figure}[htbp]
\centering
\includegraphics[angle=0,scale=1.5]{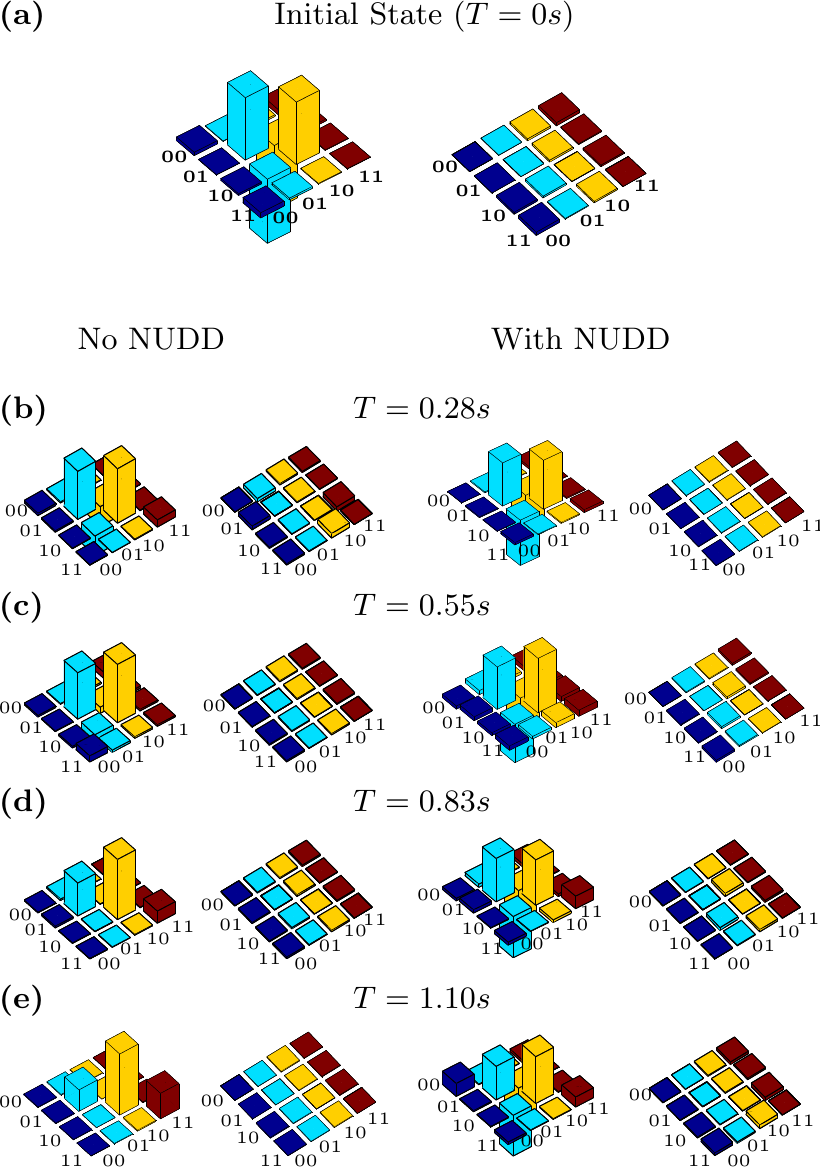}
\caption{
Real (left) and imaginary (right) parts of the experimental
tomographs of the (a)
$\frac{1}{\sqrt{2}}(|01\rangle-|10\rangle)$ state, with a
computed fidelity of 0.99.  (b)-(e) depict the state at $T =
0.28, 0.55, 0.83, 1.10$ s, with the tomographs on the left
and the right representing the state without any
protection and after
applying NUDD protection, respectively.  The rows
and columns are labeled in the computational basis ordered
from $\vert 00 \rangle$ to $\vert 11 \rangle$.
}
\label{nudd_tomosinglet}
\end{figure}
\begin{figure}[htbp]
\centering
\includegraphics[angle=0,scale=1.5]{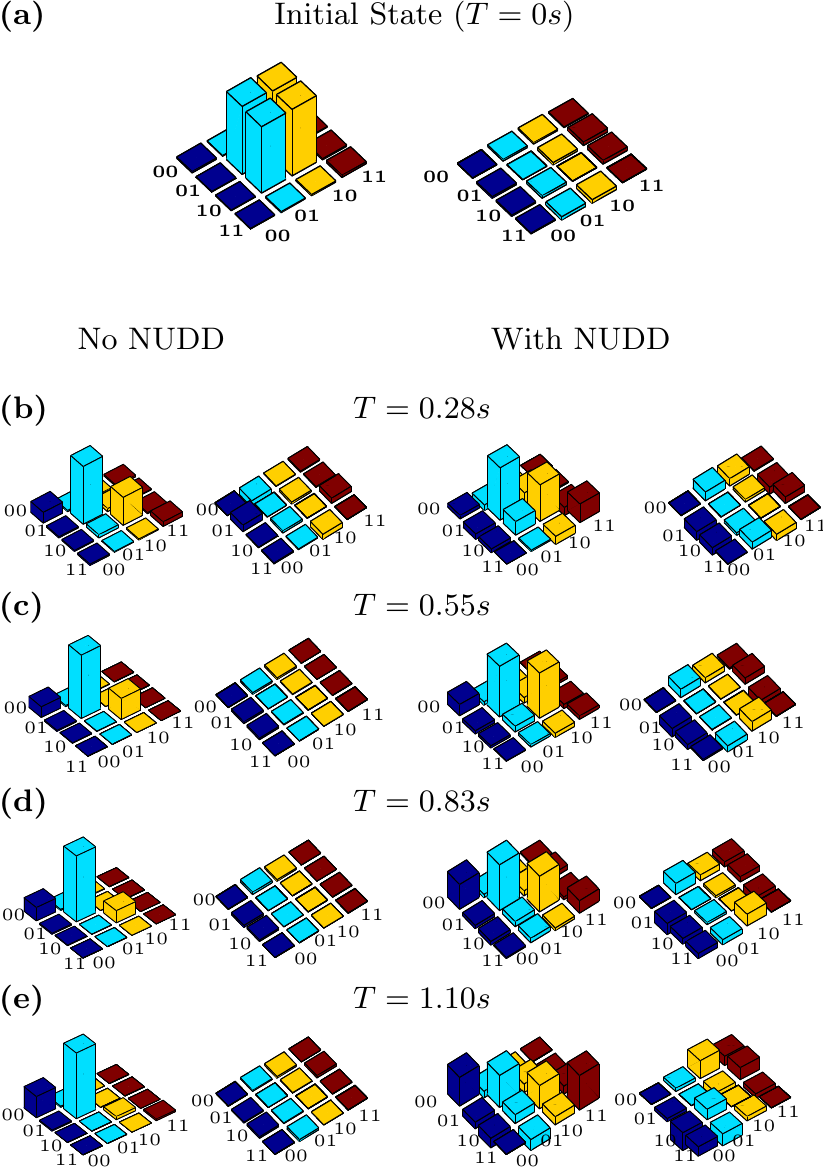}
\caption{
Real (left) and imaginary (right) parts of the experimental
tomographs of the (a)
$\frac{1}{\sqrt{2}}(|01\rangle+|10\rangle)$ state, with a
computed fidelity of 0.99.  (b)-(e) depict the state at $T =
0.28, 0.55, 0.83, 1.10$ s, with the tomographs on the left
and the right representing the state without any
protection and after
applying NUDD protection, respectively.  The rows
and columns are labeled in the computational basis ordered
from $\vert 00 \rangle$ to $\vert 11 \rangle$.
}
\label{nudd_tomotriplet}
\end{figure}

\begin{figure}[htbp]
\centering
\includegraphics[angle=0,scale=1.5]{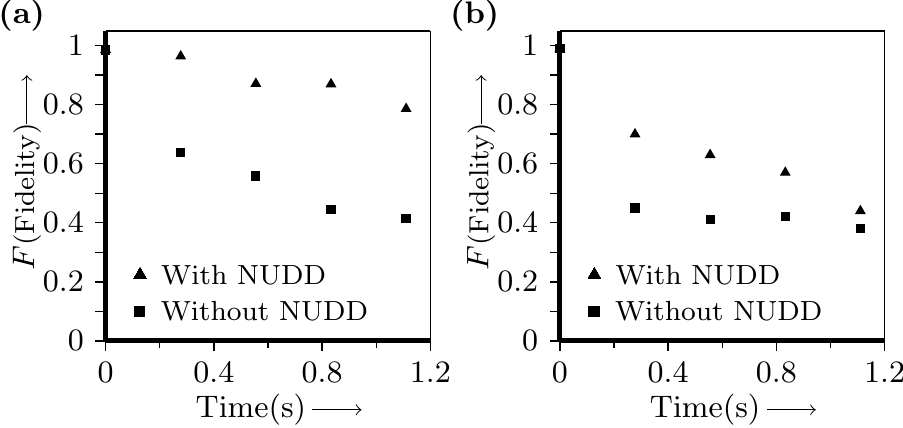}
\caption{Plot of fidelity versus time for (a) the Bell
singlet state and (b) the Bell triplet state, without any 
protection and after applying NUDD protection. 
}
\label{nudd_bell-fidelity}
\end{figure}
\subsection{NUDD protection of known states in the
subspace}
\label{known_nudd}
We begin evaluating the efficiency of the NUDD scheme by
first applying it to protect four known states in the
two-dimensional subspace ${\cal P}$, namely two separable
and two maximally entangled (Bell) states.

\noindent{\bf Protecting two-qubit separable
states:} We experimentally created the two-qubit
separable states $\vert 01 \rangle$ and $\vert 10
\rangle$ from the initial state $\vert00 \rangle$
by applying a $\pi_x$ on the second qubit and on
the first qubit, respectively.  The states were
prepared with a fidelity of 0.98 and 0.97,
respectively.  One run of the NUDD sequence took
0.12756 s 
which included the time taken
to implement the control operators,
and $t=0.05$ s 
(as per
Eqn.~(\ref{timing-eqn_nudd})).  
The entire
NUDD sequence was applied 40 times.  The state
fidelity was computed at different time instants,
without any protection and after applying NUDD
protection.  The state fidelity remains close to
0.9 for long times (upto 5 s) when NUDD is
applied, whereas for no protection the $\vert 01
\rangle$ state loses its fidelity (fidelity
approaches 0.5) after 3 s and the $\vert 10
\rangle$ state loses its fidelity after 2 s. A
plot of state fidelities versus time is displayed
in Fig.~\ref{nudd_fidelity0110}, demonstrating the
remarkable efficacy of the NUDD sequence in
protecting separable two-qubit states against
decoherence. The signal attenuation has been ignored and the density matrix
has been renormalized at every time point in Fig.~\ref{nudd_fidelity0110},
as described in Chap.~\ref{chapter_superzeno},Sec.~\ref{chapsz_fidelity_esti}.

\begin{figure}[h]
\centering
\includegraphics[angle=0,scale=1.5]{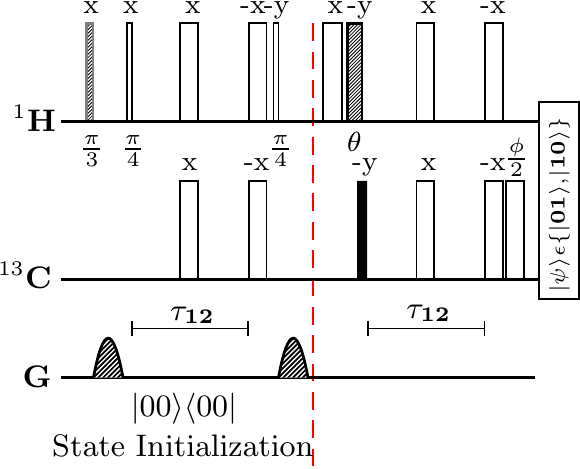}
\caption{ NMR pulse sequence for the
preparation of arbitrary states.  The sequence of pulses before
the vertical dashed line achieve state initialization
into the $\vert 00 \rangle$ state.  The values of flip
angles $\theta$ and $\phi$ of the rf pulses are 
the same as the $\theta$ and $\phi$ values describing
a general state in the two-dimensional subspace
${\cal P} = \{ \vert 0 1 \rangle, \vert 1 0 \rangle \}$.
Filled and unfilled rectangles represent
$\frac{\pi}{2}$ and $\pi$ pulses respectively, while all
other rf pulses are labeled with their respective flip
angles and phases; the interval $\tau_{12}$ is set to $(2
J_{12})^{-1}$ where $J_{12}$ is the scalar
coupling.}
\label{randomckt_nudd}
\end{figure}

\noindent{\bf Protecting two-qubit Bell states:}
We next implemented NUDD protection on the 
maximally entangled singlet state
$\frac{1}{\sqrt{2}}(\vert 01\rangle- \vert 10\rangle)$.
We experimentally constructed
the singlet state from the initial
$\vert00 \rangle$ state via the pulse sequence given 
in Fig.~\ref{randomckt_nudd} with values of $\theta=-\frac{\pi}{2}$
and $\phi=0$.  
The fidelity of the experimentally constructed singlet
state was computed to be 0.99.

One run of the
NUDD sequence took 0.27756 s
and $t$ was kept at $t=0.2$ s.
The
entire NUDD sequence was applied 4 times on the state. 
The singlet state fidelity at different time points was computed 
without any protection and after applying NUDD protection,
and the state tomographs are displayed in
Fig.~\ref{nudd_tomosinglet}. The signal attenuation has been
ignored and the density matrix has been renormalized at every
time point in Fig.~\ref{nudd_tomosinglet}. The fidelity of the singlet state
remained close to 0.8 for 1 s when NUDD protection
was applied, whereas 
when no protection is applied the state decoheres
(fidelity approaches 0.5) after 0.55 s.
We also implemented NUDD protection on the 
maximally entangled triplet state
$\frac{1}{\sqrt{2}}(\vert 01\rangle+\vert 10\rangle)$.
We experimentally constructed
the triplet state from the initial
$\vert 00 \rangle$ state via the pulse sequence given 
in Fig.~\ref{randomckt_nudd} with values of $\theta=\frac{\pi}{2}$
and $\phi=0$.  
The fidelity of the experimentally constructed triplet
state was computed to be 0.99.
The total NUDD time was kept at 
$t=0.2$ s and one run of the
NUDD sequence took 0.27756 s.  The
entire NUDD sequence was repeated 4 times on the state. 
The state fidelity at different time points was computed 
without any protection and after applying NUDD protection
and the state tomographs are displayed in Fig.~\ref{nudd_tomotriplet}.
The signal attenuation has been
ignored and the density matrix has been renormalized at every
time point in Fig.~\ref{nudd_tomotriplet}.

The fidelity of the triplet state
remained close to 0.71 for 0.28 s when NUDD protection
was applied, whereas 
when no protection is applied the state decoheres
quite rapidly (fidelity approaches 0.5) after 0.28 s.
A plot of state fidelities of both Bell
states versus time
is displayed in Fig.~\ref{nudd_bell-fidelity}, In that
the signal attenuation has been
ignored and the density matrix has been renormalized at every
time point.
While the NUDD scheme was able to protect the singlet
state quite well (the time for which the state
remains protected is double as compared to its
natural decay time), it is not able to extend the
lifetime of the triplet state to any appreciable extent.
What is worth noting here is the fact that the
state fidelity remains 
considerably higher 
under NUDD protection compared to no protection, 
implying that there is 
a reduction in the
``leakage'' to other states.

\subsection{NUDD protection of unknown states in the subspace}
\label{unknown_nudd}
\begin{figure}[htbp]
\centering
\includegraphics[angle=0,scale=1]{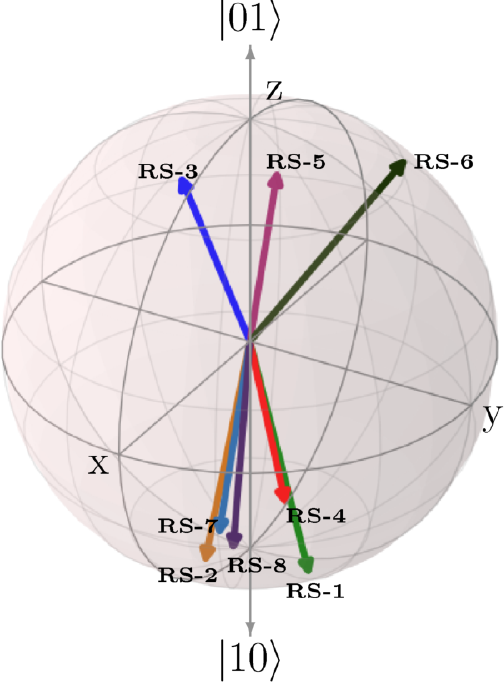}
\caption{Geometrical representation of eight randomly 
generated states on a Bloch sphere belonging to the two-qubit
subspace ${\cal P} =\{\vert 01\rangle,\vert 10\rangle\}$.
Each vector makes angles $\theta,\phi$ with the $z$ and
$x$ axes, respectively.
The state labels RS-$i$ ($i=1..8$) are explained in the text.
}
\label{blochfig_nudd}
\end{figure}
We wanted to carry out an unbiased assessment of the
efficacy of the NUDD scheme for state protection. To this
end, we randomly generated several states in the two-dimensional
subspace ${\cal P}$, and applied  the NUDD sequence on each
state.
\begin{table*}[h]
\caption{\label{table1_nudd}
Results of applying NUDD protection on eight randomly
generated states in the two-dimensional
subspace.
Each random state (RS) is tagged with a number for
convenience, and its corresponding
($\theta$,$\phi$) angles are given in the column alongside.
The fourth column displays the time at which the state
fidelity approaches $\approx 0.8$ without NUDD
protection
and the last column displays the time for which
state fidelity approaches $\approx 0.8$ after
applying NUDD protection.}
\centering

\begin{tabular}{ccccc}
\hline
{\bf State}& {\bf Label} &$(\theta,\phi)$({\bf deg})&
{\bf Time (s)} & {\bf Time (s)} \\
&&&
{($F>0.8$)} & ($F>0.8$) \\
&&&
{\bf \small{Without NUDD}} & {\bf \small{With NUDD}} \\
\hline
$ 0.29|01\rangle + (0.94 + \iota 0.18) |10\rangle$&
RS-1&(147,57)&0.1s&1.0s\\
$ 0.15 |01\rangle - (0.76 + \iota 0.63) |10\rangle$&
RS-2&(163,349)&0.3s&1.1s\\
$ 0.98 |01\rangle + (0.11 - \iota0.17) |10\rangle$&
RS-3&(23,345)&0.1s&1.1s\\
$ 0.14 |01\rangle +  (0.36 -\iota 0.92) |10\rangle$&
RS-4&(164,175)&0.3s&1.1s\\
$ 0.99 |01\rangle +  (0.10 + \iota0.11) |10\rangle$&
RS-5&(18,51)&0.3s&1.1s\\
$ 0.91 |01\rangle + (0.22 +\iota 0.36) |10\rangle$&
RS-6&(50,152)&0.1s&0.9s\\
$ 0.07 |01\rangle + (-0.77 +\iota 0.64) |10\rangle$&
RS-7&(172,285)&0.1s&1.1s\\
$ 0.06 |01\rangle + (0.99 -\iota 0.16) |10\rangle$&
RS-8&(174,346)&0.3s&1.1s\\
\hline
\end{tabular}

\end{table*}
\begin{figure}[htp]
\centering
\includegraphics[angle=0,scale=1.2]{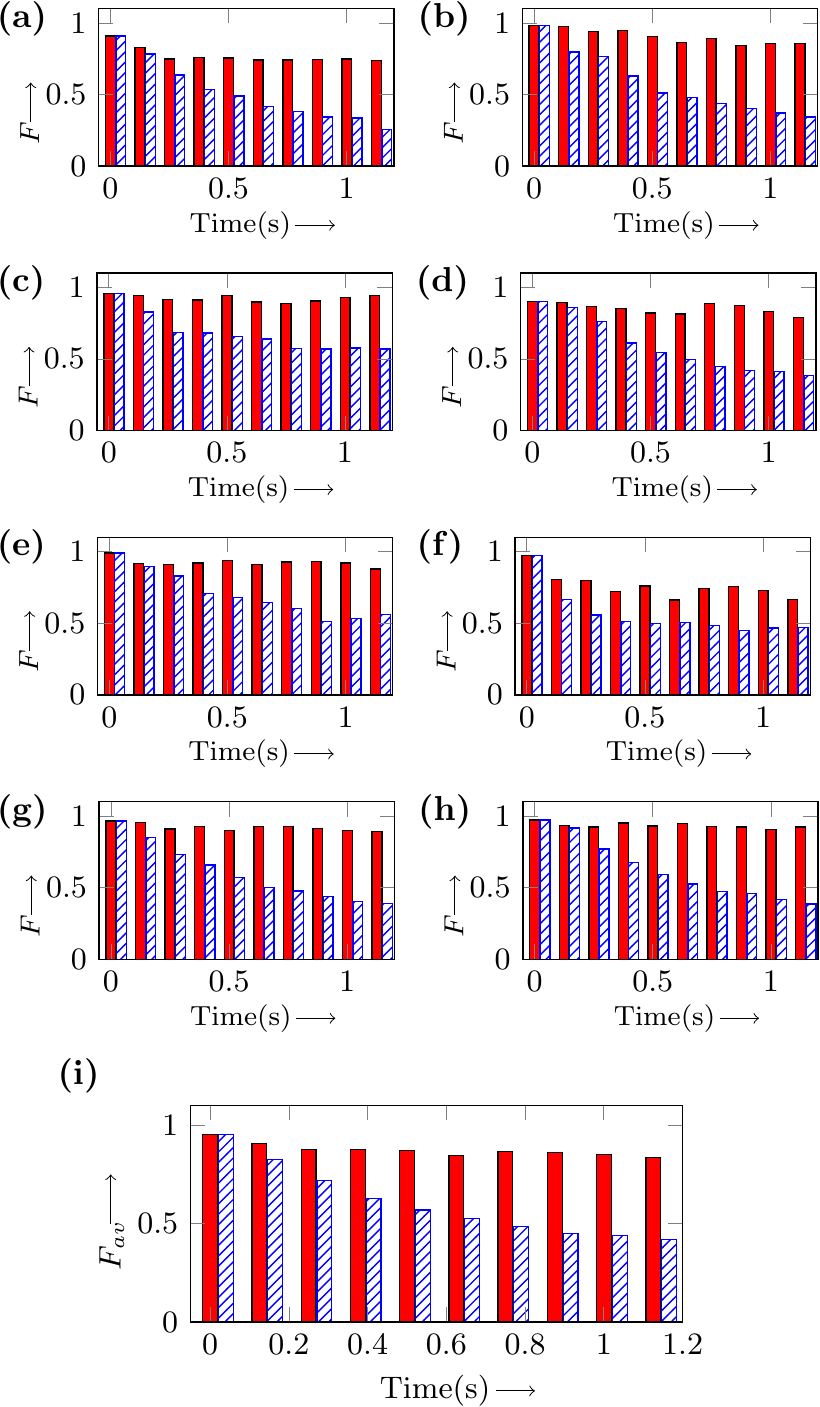}
\caption{ Bar plots of fidelity versus time of 
eight randomly generated states (labeled RS-$i$, $i=1..8$),
without any protection (cross-hatched bars) 
and after applying NUDD protection (red solid bars):
(a) RS-1, (b) RS-2, (c) RS-3, (d) RS-4, (e) RS-5, 
(f) RS-6, (g) RS-7 and (h) RS-8.
(i) Bar plot showing average fidelity of all eight randomly
generated states, at each time point.
}
\label{random_fid_nudd}
\end{figure}
A general state in the two-qubit subspace ${\cal
P} =\{\vert 01\rangle,\vert 10\rangle\}$ can be
written in the form \begin{equation} \vert \psi
\rangle= \cos{\frac{\theta}{2}} \vert
01\rangle+e^{-\iota \phi} \sin{\frac{\theta}{2}}
\vert 10\rangle\ 
\label{blocheqn_nudd}
\end{equation} 
These states were
experimentally created by using the values of
$\theta$ and $\phi$  
(Eqn.~(\ref{blocheqn_nudd})) as the flip angles of the
rf pulses in the NMR pulse sequence
(see Fig.~\ref{randomckt_nudd} for visualization).  The eight
randomly generated representative two-qubit states are shown in
Fig.~\ref{blochfig_nudd}.
The entire
three-layered NUDD sequence was applied 10 times
on each of the eight random states.  The time $t$
for the sequence was kept at $t=0.05$ s and one run
of the NUDD sequence took $0.12756$ s.  The plots
of fidelity versus time are shown as bar graphs in
Fig.~\ref{random_fid_nudd}, with the cross-hatched bars
representing state fidelity without any protection
and the solid bars representing state fidelity after
NUDD protection. The signal attenuation has been
ignored and the density matrix has been renormalized at every
time point in Fig.~\ref{random_fid_nudd}.
The final bar plot in Fig.~\ref{random_fid_nudd}(i) shows the average
fidelity of all the randomly generated states at
each time point.  The results of protecting these
random states via three-layered NUDD are tabulated
in Table~\ref{table1_nudd}. Each state has been tagged
by a label RS-$i$ (RS denoting ``Random State''
and $i=1,..8$), with its $\theta,\phi$ values
displayed in the next column.  The fourth column
displays the values of the natural decoherence
time (in seconds) of each state without
NUDD protection, estimated by
computing the time upto which state fidelity
does not fall below 0.8. The last column in the table
displays the time for which the state remains
protected after applying NUDD, estimated by
computing the time upto which state fidelity
does not fall below 0.8.  While the NUDD scheme is
able to protect specific states in the subspace
with varying degrees of success (as evidenced from
the entries in the last column of in
Table~\ref{table1_nudd}), on an average as seen from
the bar plot of the average fidelity in
Fig.~\ref{random_fid_nudd}(i), the scheme performs
quite well.
\section{Conclusions}
\label{concl_nudd}
In this chapter, a three-layer nested UDD sequence was experimentally implemented 
on an NMR quantum information processor and
explored its efficiency in protecting arbitrary 
states in a two-dimensional subspace of two qubits.
The nested UDD layers were applied in a particular sequence,
and the full NUDD scheme was able to achieve second
order decoupling of the system and bath.  The scheme is sufficiently
general as it does not assume prior information
about the explicit form of the system-bath
coupling.  The experiments were highly demanding,
with the control operations being complicated and
involving manipulations of both qubits simultaneously.  
However, our results demonstrate
that such systematic NUDD schemes 
can be experimentally implemented, and are able to
protect multiqubit states in systems that are
arbitrarily coupled to quantum baths. 

In addition, to demonstrate that
the NUDD scheme does not depend on the actual form of an
initial superposition state. This scheme was applied to
arbitrary states with randomly sampled
coefficients and the experimental results shows that advantage of 
the NUDD schemes lies in the fact that one is 
sure that some amount of state protection will 
always be achieved. Furthermore, one need not know
anything about the state to be protected or the
nature of the quantum channel
responsible for its decoherence. All one needs to
know is the subspace to which the state belongs.
In summary, if the QIP experimentalist has full
knowledge of the state to be protected, it is
better to use UDD schemes that are not nested.
However, if there is only partial knowledge of the state,
 the QIP experimentalist would do better to use
 these ``generic'' NUDD schemes. If we do not know
 the subspace to which the state belongs, we need to consider the full space,
 which increases the number of nesting layers,
 and experimentally implementation becomes a difficult task. 
 The study in this chapter points the way to the realistic
 protection of fragile quantum states up to high orders
 and against arbitrary noise.
\newpage

\chapter{Experimentally preserving time-invariant discord  
using dynamical decoupling}\label{discordchapter}
\section{Introduction}
\label{intro_discord}
Interaction of quantum systems with their environment 
causes the destruction
of intrinsic quantum properties such as quantum superposition,
quantum entanglement and more general quantum correlations~\cite{breuer-book}.
It has been observed that quantum entanglement may disappear completely but 
there still may exist quantum correlations in a quantum system. 
Quantum correlations are thus more fundamental 
than entanglement~\cite{lanyon-prl-08}. 
The quantification of quantum correlations,
distinction from their classical counterparts,
and their behavior under decoherence, is of paramount
importance to quantum information processing~\cite{nielsen-book-02}.
Several measures of nonclassical correlations have been
developed~\cite{lanyon-prl-08} and their signatures
experimentally measured on an NMR
setup~\cite{soarespinto-pra-10,silva-prl-13,silva-prl-16}.  Quantum
discord is a measure of nonclassical correlations that are
not accounted for by quantum
entanglement~\cite{olliver-prl-01}.  While the intimate
connection of quantum entanglement with quantum nonlocality
is well understood and entanglement has long been considered
a source of quantum computational
speedup~\cite{horodecki-rmp-09}, the importance of quantum
discord, its intrinsic quantumness and its potential use in
quantum information processing protocols, is being explored
in a number of contemporary studies~\cite{modi-rmp-12}.

A surprising recent finding suggests that for certain quantum states
up to some time $\bar{t}$, quantum correlations are not
destroyed by decoherence whereas classical correlations
decay. After time $\bar{t}$ the situation is
reversed and the  quantum correlations begin to decay~\cite{maziero-pra-09,mazzola-prl-10}.
This inherent immunity of such quantum correlations to environmental noise
throws up new possibilities for the characterization of
quantum behavior and its exploitation for quantum
information processing.  The peculiar ``frozen'' behavior of quantum
discord in the presence of noise was experimentally
investigated using photonic qubits~\cite{xu-nc-10} and  NMR
qubits~\cite{auccaise-prl-11,paula-prl-13}.  The class of
initial quantum states that exhibit this sudden transition in
their decay rates was theoretically studied under the action
of standard noise channels, and it was inferred that the type of
states that display such behavior depends on the
nature of the decohering channel being
considered~\cite{fanchini-pra-10}.  Dynamical decoupling (DD)
methods have been proposed to protect quantum discord from
environment-induced errors~[\cite{fanchini-pra-12}] and  a recent
work showed that interestingly, DD schemes
can also influence the timescale over which time-invariant
quantum discord remains oblivious to
decoherence~\cite{addis-pra-15}.

In this chapter we demonstrate the remarkable preservation of
time-invariant quantum discord upon applying
time-symmetric DD schemes
of the bang-bang variety, on a two-qubit NMR
quantum information processor. We begin by looking at a measure
for quantum and classical correlations for
a special class of two-qubit quantum states,
namely, Bell-diagonal (BD) states. Then the noise
affecting a spin system is characterized by experimentally
measuring  the relaxation of the spins. 
Considering the spin system used which is a heteronuclear
and the experimental results of the relaxation parameters,
a model is proposed that the decoherence channel acting 
on the two qubits is mainly a phase damping channel acting
independent on each qubit. According to this model zero 
quantum coherences and double quantum
coherences decay with same rates which is equal to the sum of single quantum coherences
decay rates. Experimental results validate this model. 
The BD state experimentally prepared, its decay is observed
and found the quantum correlations freeze for a time interval.
The results obtained from the experiments and theoretical 
noise model are compared.
In the final section of this chapter,
several robust DD sequences 
are applied to the system in the BD state,
resulting in a significantly prolonged freezing time of quantum correlations.

\section{Measure for correlations of two qubits}
Consider a bipartite system made of an system A and B
into a Hilbert space ${\cal H}_A \otimes {\cal H}_B$ with ${\cal H}_A $
and ${\cal H}_B $ is a Hilbert state of system A and B respectively.
 The total correlations of a state in a 
bipartite quantum system are measured by the
quantum mutual information ${\cal I}(\rho_{AB})$ defined as
\begin{equation}
\label{It}
 {\cal I}(\rho_{AB})=S(\rho_{A})+S(\rho_{B})-S(\rho_{AB})
\end{equation}

where  $S(\rho)=-\tr\{ \ \rho \log_2 \rho \ \}$
is the von Neumann entropy.

Then the quantum discord is defined as 
\begin{equation}
\label{discord}
 {\cal D}(\rho_{AB})= {\cal I}(\rho_{AB})-{\cal C}(\rho_{AB})
\end{equation}
 where ${\cal C}(\rho_{AB})$ is a classical correlations
 of the state~\cite{mazzola-prl-10,luo-pra-08}.
Next we specify the quantity used for measuring the classical correlations.
Such a quantity is based on the generalization of the concept of
conditional entropy. We know that performing measurements
on system B affects our knowledge of system A.
How much system A is modified by a measurement of B
depends on the type of measurement performed on B. Here
the measurement is considered of von Neumann type. It
is described by a complete set of orthonormal projectors
$\{\Pi_k\}$ on subsystem B corresponding to the outcome k. The
classical correlations ${\cal{C}}{(\rho_{AB})}$ 
are then defined as
\begin{equation}
\label{clasico}
 {\cal{C}}(\rho_{AB})=\max_{\{\varPi_{k} \}}[S(\rho_{A})-S(\rho_{\rho_A}|\{\varPi_{k} \})]
\end{equation}

where the maximum is taken over the set of the projective measurements
$\{\varPi_{k} \}$ and $S(\rho_{\rho_A}|\{\varPi_{k} \})=\sum_{k}p_{k}S(\rho_{k})$ is
the conditional entropy of A, given the knowledge of the state of B, 
with  $p_{k}=\tr_{AB}(\rho_{AB}\varPi_k)$ and $\rho_{k}=\tr_{B}(\varPi_k \rho_{AB} \varPi_k)/p_{k}$. 

Consider a two-qubit quantum system with its state space 
spanned by the states $\{\vert00\rangle,\vert01\rangle,$ 
$\vert10\rangle,\vert11\rangle\}$ and the eigenstates of the Pauli operators
$\sigma_z^1\otimes\sigma_z^2$.
Any state for such a system is locally equivalent to 
\begin{equation}
\rho=\frac{1}{4}\left(I+\vec{a}.\vec{\sigma}\otimes {I}+I\otimes\vec{b}.\vec{\sigma}+\sum^{3}_{j=1}c_j\sigma_j\otimes\sigma_j\right)
\end{equation}

where ${I}$ is the identity operator, $ \vec{\sigma}=(\sigma_x,\sigma_y,\sigma_z)$ with 
Pauli spin observables in the $x$,$y$,$z$ direction, $\vec{u}=(u_x, u_y, u_z)$ $\in$ $R^3$, $\vec{u} \vec{\sigma}=u_x \sigma_x+u_y \sigma_y+u_z \sigma_z$ ($u=a,b,c$) . Here we are only interested 
in the family of BD states which is described by a density matrix

\begin{equation}
 \rho_{AB}= \frac{1}{4}\left({I}+\sum_{j=1}^3c_j \sigma_j \otimes \sigma_j \right)
 \label{abeq}
\end{equation}
where $c_j$ are real constants such that $\rho_{AB}$ is a valid density operator.
The eigenvalues of the density matrix $\rho_{AB}$ are given by 

\begin{eqnarray}
 \lambda_0&=&\frac{1}{4}(1-c_1-c_2-c_3), \nonumber \\ 
 \lambda_1&=&\frac{1}{4}(1-c_1+c_2+c_3), \nonumber \\ 
 \lambda_2&=&\frac{1}{4}(1+c_1-c_2+c_3), \nonumber \\
 \lambda_3&=&\frac{1}{4}(1+c_1+c_2-c_3). \nonumber \\
\end{eqnarray}
The reduced density matrices of $\rho_{AB}$ as given in Eq.~\ref{abeq}
are $\rho_A=\frac{{I}}{2}$ and $\rho_B=\frac{{I}}{2}$.
The total correlations $\cal{I}(\rho)$ are given by:

\begin{equation}
 {\cal{I}(\rho)}=2+\sum_{l=0}^3 \lambda_l \log_2\lambda_l
 \label{totalco}
\end{equation}
The classical correlations for the BD states given as

\begin{equation}
 {\cal C}[\rho_{AB}]=\sum_{j=1}^{2}
\frac{1+(-1)^{j}\chi}{2}\log_{2}[1+(-1)^{j}\chi], 
\label{claco}
\end{equation}

where $\chi=\max \{\vert c_1 \vert ,\vert c_2 \vert ,\vert c_3\vert \}$.
The maximization procedure with
respect to the projective measurements, present in the
definition of the classical correlations of Eq.{(\ref{clasico})},
can be performed explicitly for the system here considered
noticing that the complete set of the orthogonal projectors is 
given by $\varPi_j=\vert\theta_j\rangle\langle\theta_j\vert$, with j=1,2, $\vert\theta_1\rangle=
{\rm cos} \theta \vert0\rangle+e^{-i\phi} {\rm sin} \theta \vert1\rangle$,
$\vert\theta_2\rangle=
{\rm sin} \theta \vert0\rangle+e^{-i\phi} {\rm cos} \theta \vert1\rangle$, and the state 
of the system always remains of the form given by Eq.{(\ref{abeq})} during time evolution~\cite{mazzola-prl-10}.
\section{Characterization of noise channels}
In high field NMR, the Zeeman interaction causes
a splitting of the energy levels according to 
the field direction and the difference between 
magnetic quantum numbers
\begin{equation}
 \Delta m_{rs}= m_r-m_s
\end{equation}
defines the order of the coherence.
If $\Delta m_{rs}=0$ the coherence is a zero quantum (ZQ) coherence,
if $\Delta m_{rs}=\pm1$ the coherence is a single quantum (SQ) coherence,
and if $\Delta m_{rs}=\pm2$ the coherence is a double quantum (DQ) coherence.
In general, a density matrix element $\rho_{rs}$ 
represents $p$-quantum coherence $(p = m_r-m_s )$~\cite{levitt-book-2008,ernst-book-87}.
For our two-qubit system we use for the experiments,
the experimentally measured longitudinal NMR spin relaxation times are ${T}_1^{\rm H} \approx 7.9$ s
and ${T}_1^{\rm C}\approx$ 16.6 s, which are much longer than the measured 
effective NMR transverse spin relaxation rates ${T}_2^{\rm H}$ = $0.513 \pm 0.01$ s 
and ${T}_2^{\rm C}= 0.193 \pm 0.005$ s.
Since we study the decoherence of the state for an evolution time of approximately 0.5 s,
we neglect the effects of the amplitude-damping 
channel whose effects are associated with ${T}_1$ and
assume that the main noise channel for our system is the
phase-damping channel. The effective transverse spin 
relaxation rates were measured by applying a $90^{\circ}$ excitation
pulse followed by a decay interval (with no refocusing pulse) 
and by fitting the resulting decay of the magnetization.
The decay rate of SQ coherence of spin(i) is given by $1/T_2^i$. 
Our experimental system is heteronuclear, with
two different nuclear species (proton and carbon), 
having very different Larmor resonance frequencies and hence
large chemical shift differences. We hence hypothesize that each
spin decoheres in an independent phase damping channel
which is not correlated with that of the other spin. 
Therefore, we
model the phase damping channel for this system
as a homogeneous dephasing channel acting independently
on each qubit~\cite{childs-pra-01}.
We use the operator-sum 
representation formalism and the associated phase-damping Kraus
 operators~\cite{nielsen-book-02}.The Kraus operators for a two-qubit
 system under phase damping channel are: 
\begin{eqnarray}
\label{pd2_chap5}
E_1(t)&=& \frac{1}{2}(1+e^{-\gamma_1 t})^{\frac{1}{2}} (1+e^{-\gamma_2 t})^{\frac{1}{2}} I\otimes I,
\nonumber \\
E_2(t)&=& \frac{1}{2}(1+e^{-\gamma_1 t})^{\frac{1}{2}} (1-e^{-\gamma_2 t})^{\frac{1}{2}} I\otimes \sigma_z,
\nonumber \\
E_3(t)&=& \frac{1}{2}(1-e^{-\gamma_1 t})^{\frac{1}{2}} (1+e^{-\gamma_2 t})^{\frac{1}{2}} \sigma_z\otimes I,
\nonumber \\
E_4(t)&=& \frac{1}{2}(1-e^{-\gamma_1 t})^{\frac{1}{2}} (1-e^{-\gamma_2 t})^{\frac{1}{2}} \sigma_z\otimes \sigma_z,
\end{eqnarray}
where $\gamma_{1}=1/T^i_2$ is the decay constant of the $i^{th}$ spin,
$I$ is the identity matrix and $\sigma_z$ is a Pauli matrix.
\begin{equation}
{\sum_jE_j(t)E_j(t)^{\dagger}}=1,
\end{equation}

\begin{equation}
\label{evobd1}
 \rho_{BD}(t)=\sum_jE_j(t)\rho_{AB}(0)E_j(t)^{\dagger}.
\end{equation}

Once we fix the model, we can apply it to any state.
One of the qualitative predictions of the model is that the
DQ and ZQ decay rates are equal.
We measured these rates in two independent experiments
and compared them with those predicted by the above
model (whose parameters are completely fixed). The 
experimentally measured DQ and ZQ decay rates were $\gamma_{DQ} = 6.395 \pm 0.23$ s$^{-1}$
and $\gamma_{ZQ} = 6.138 \pm 0.275$  s$^{-1}$ respectively, and
the theoretically predicted common decay was $\gamma_1+\gamma_2=\frac{1}{T^H_2}+\frac{1}{T^C_2}=7.21 \pm 0.173$ s$^{-1}$,
showing the model works.

\subsection{Bell-diagonal states under a dephasing channel}
For a two-qubit system where each qubit
is affected by an independent local dephasing channel,
to see the evolution
under the phase-damping channel acting independently on each qubit,
we use the operator-sum representation formalism and the associated phase-damping Kraus
operators~{\cite{nielsen-book-02}}. The evolution in a noisy environment is governed by 
the phase-damping channel in Eq.~(\ref{pd2_chap5}),
and it turns out for Bell-diagonal states $\rho_{AB}$
\begin{eqnarray}
 c_1(t)&=& c_1(0)exp[-(\gamma_1+\gamma_2 ) t], \nonumber \\
 c_2(t)&=&c_2(0)exp[-(\gamma_1+\gamma_2 ) t], \nonumber \\
 c_3(t)&=&c_3(0).
 \label{cs}
\end{eqnarray}

\section{Time invariant quantum correlations}
We now consider a system in the special class of Bell-diagonal states
for which $c_1(0)=\pm1$ and $c_2(0)=\mp c_3(0)$, 
with $\vert c_3 \vert < 1 $ under the dephasing channel as define above in Eq. ~(\ref{pd2_chap5}). 
At time $t$ the total correlations  are given by
\begin{eqnarray}
\label{totalbd}
 {{\cal I}[\rho_{AB}(t)]}&=&\sum_{j=1}^2 \frac{1+(-1)^jc_3(t)}{2} \log_2[1+(-1)^jc_3(t)]+ \nonumber \\
  &&\sum_{j=1}^2 \frac{1+(-1)^jc_1(t)}{2} \log_2[1+(-1)^jc_1(t)],
\end{eqnarray}
the classical correlations are given by 
 \begin{equation}
 \label{classicalbd}
 {{\cal C}[\rho_{AB}(t)]}= \sum_{j=1}^2 \frac{1+(-1)^j\chi(t)}{2} \log_2[1+(-1)^j\chi(t)],
\end{equation}
 where $\chi(t)=\max \{\vert c_1(t) \vert ,\vert c_2(t) \vert ,\vert c_3(t)\vert \}$,
and the quantum correlations are given by
\begin{equation}
\label{discord1}
 {\cal D}[\rho_{AB}(t)]= {\cal I}[\rho_{AB}(t)]-{\cal C}[\rho_{AB}(t)].
\end{equation}

We can see that since our system is mainly affected by 
the phase damping channel,
for
\begin{equation}
 t<\bar{t}=-\frac{ln(|c_3|)}{2\gamma}
\end{equation}
quantum correlations remain constant
 \begin{equation}
 \label{invdiscord1}
 {{\cal D}[\rho_{AB}(t<\bar{t})]}= \sum_{j=1}^2 \frac{1+(-1)^jc_3(0)}{2} \log_2[1+(-1)^jc_3(0)],
\end{equation}
and after $t>\bar{t}$ the quantum correlations start decreasing 
towards zero and the classical correlations of system remain constant which is given by:
\begin{equation}
 \label{invdiscord2}
 {{\cal C}[\rho_{AB}(t>\bar{t})]}= \sum_{j=1}^2 \frac{1+(-1)^jc_3(0)}{2} \log_2[1+(-1)^jc_3(0)].
\end{equation}
\section{Experimental realization of time-invariant discord} 
\subsection{NMR System}
We create and preserve time-invariant discord in a two-qubit
NMR system  of chloroform-$^{13}$C,
with the ${}^{1}$H and ${}^{13}$C nuclear spins encoding the
two qubits (Fig.~\ref{molecule_nudd}).  The ensemble of nuclear
spins is placed in a longitudinal strong static magnetic
field ($B_0 \approx 14.1$T) oriented along the $z$
direction.  The ${}^{1}$H and ${}^{13}$C nuclear spins
precess around $B_0$ at Larmor frequencies of $\approx 600$
MHz and $\approx 150$ MHz, respectively. The evolution of
spin magnetization is controlled by applying rf-field pulses
in the $x$ and $y$ directions.
The internal Hamiltonian of the system  in the rotating
frame is given by Eq.(\ref{hamiltonian_nudd})

The two-qubit system
was initialized into the pseudopure state $\vert 00 \rangle$
by the spatial averaging technique~\cite{cory-physicad}.
Density matrices were reconstructed from
experimental data by using a reduced set of
quantum state tomography (QST)
operations combined with
the maximum likelihood method~\cite{singh-pla-16} as described in 
Chapter~\ref{chapter_mle} to avoid any negative eigen values.
The fidelity $F$ of all the experimental
density matrices reconstructed
in this work was computed
using the Eq.(\ref{mle_fidelity_2}).
The experimentally created
pseudopure state $\vert 00 \rangle$ was tomographed with a
fidelity of $0.99$,  and the NMR signal of this state was
used as a reference for computation of state fidelity in all
subsequent time-invariant discord experiments.
\subsection{Observing time-invariant discord}

\begin{figure}[h]
\centering
\includegraphics[angle=0,scale=1.1]{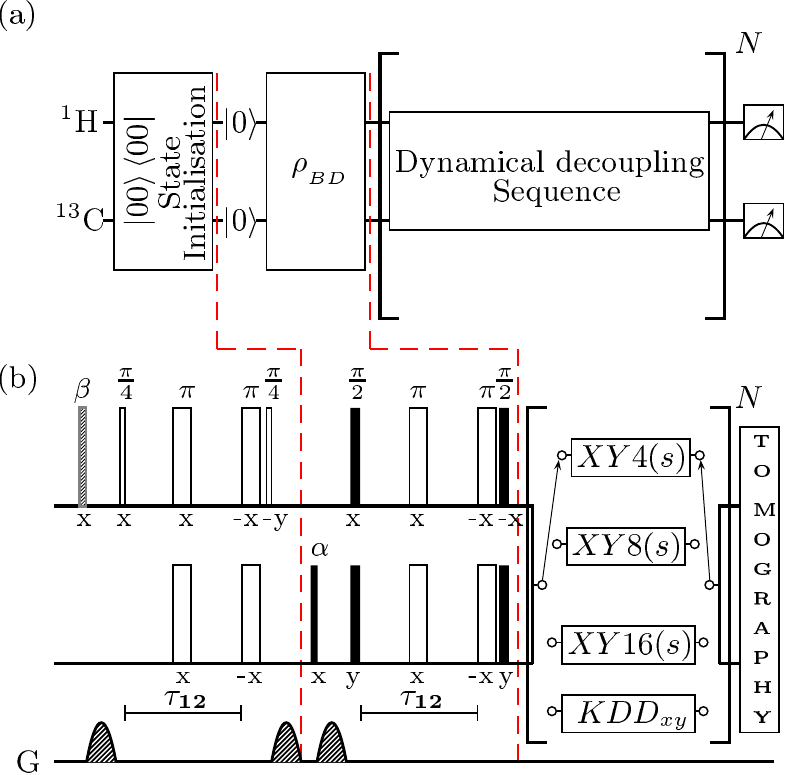}
\caption{
(a) Quantum circuit for the initial pseudopure
state preparation, followed by the block for 
BD state preparation. The next block depicts
the DD scheme used to
preserve quantum discord. The
entire DD sequence is repeated $N$ times before
measurement.
(b) NMR pulse sequence corresponding to the
quantum circuit. The rf pulse flip angles are set
to $\alpha=46^{\circ}$ and $\beta= 59.81^{\circ}$,
while all other pulses are labeled with their
respective angles and phases.
}
\label{ckt}
\end{figure}

A class of Bell-diagonal (BD) states with maximally mixed marginals
defined in terms of Pauli operators $\sigma_i$ are
\begin{equation}
\label{eqbd}
\rho_{\rm BD}=\frac{1}{4}\left(\,{I} \right. +
{\sum_{i=1}^3} \left. c_i\ \sigma_i
\otimes\sigma_i \right)
\end{equation}
where the coefficients $c_i$ with  $0 \le \vert c_i \vert \le 1$ determine the
state completely and can be computed 
as $c_i = \langle \sigma_i \otimes \sigma_i
\rangle$.

We aim to prepare an initial BD state
with the  parameters $c_1(0) = 1$, $c_2(0) = 0.7$,
$c_3(0) = -0.7$.  The NMR pulse
sequence for the preparation of this state 
from the  $\vert 00 \rangle$ pseudopure state
is given in Fig.~\ref{ckt}(b).  
Preparing the BD state involves 
manipulation of NMR multiple-quantum coherences by applying
rotations in the zero-quantum and double-quantum 
spin magnetization subspaces.
Since the molecule
is a heteronuclear spin system,
high-power, short-duration rf pulses were used for
gate implementation (with rf pulses of flip
angles $\alpha=45.57^{\circ}$ and $\beta=
59.81^{\circ}$ in Fig.~\ref{ckt} having pulse
lengths of $6.85 \mu$s and $5.02 \mu$s,
respectively).  The experimentally achieved
$\rho^{\rm E}_{\rm BD}$ (reconstructed using the
maximum likelihood method~\cite{singh-pla-16} had
parameters $c_1(0) = 1.0$, $c_2(0) = 0.680$ and
$c_3(0) = -0.680$ and a computed fidelity of
$0.99$.  The experimentally reconstructed density
matrix (using quantum state tomography and maximum
likelihood) was found to be:
\\ 

 $\mathbf{\rho^{\rm E}_{\rm BD}} ==\left(\begin{array}{cccc}
 0.080 & 0.003+0.000i & 0.003+0.000i & 0.080+0.000i \\
 0.003-0.000i & 0.420 & 0.420+0.001i & 0.003+0.000i \\
0.003-0.000i & 0.420-0.001i &0.420 & 0.003+0.000i \\
0.080-0.000i & 0.003-0.000i & 0.003-0.000i &0.080 \\
\end{array}\right)$
\\

Correlation functions for the BD states
can be computed readily to give 
the classical correlations (${\cal C}[\rho(t)]$),
the quantum discord (${\cal D}[\rho(t)]$), and
total correlations (${\cal
I}[\rho(t)]$~\cite{mazzola-prl-10}: 
\begin{eqnarray}
{\cal C}[\rho(t)]&=&\sum_{j=1}^{2}
\frac{1+(-1)^{j}\chi(t)}{2}\log_{2}[1+(-1)^{j}\chi(t)] 
\nonumber \\
{\cal I}[\rho(t)]&=&
\sum_{j=1}^{2}\frac{1+(-1)^{j}
c_1(t)}{2}\log_{2}[1+(-1)^{j}c_1(t)] \nonumber \\
&&+
\sum_{j=1}^{2}\frac{1+(-1)^{j}
c_3}{2}\log_{2}[1+(-1)^{j}c_3]
\nonumber \\
{\cal D}(\rho)&\equiv&{\cal I}(\rho)-{\cal C}(\rho) 
\label{relations}
\end{eqnarray}
where $\chi(t)=\max\{|c_{1}(t)|,|c_{2}(t)|,|c_{3}(t)|\}$.

\begin{figure} 
\centering
\includegraphics[angle=0,scale=1.5]{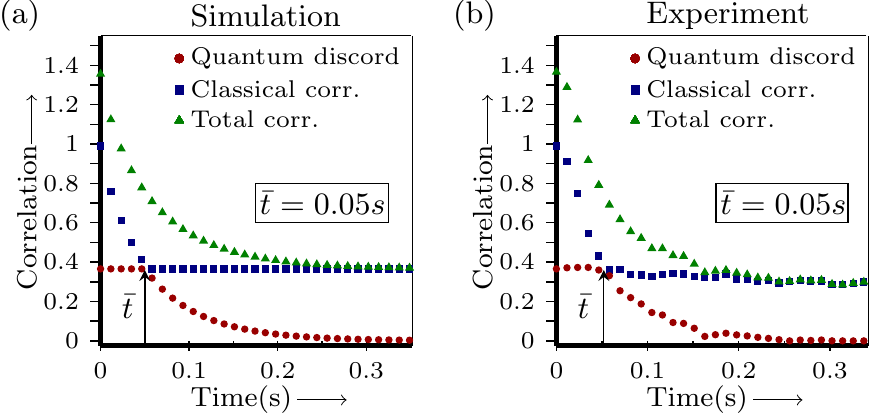}
\caption{ Time evolution of 
total correlations (triangles), 
classical correlations (squares) 
and quantum discord (circles) of 
the BD state:  (a) Simulation, (b)
Experimental plot without applying any preservation.
}
\label{tidplot_ab}
\end{figure}

For the class of BD states with coefficients
$c_1=\pm 1, c_2=\mp c_3, \vert c_3 \vert < 1$,
the evolution under a noisy environment is governed
by the noise operators given in Eq.{\ref{pd2_chap5}}, and
it turns out that $c_1(t)=c_1(0) exp{[-(\gamma_1+\gamma_2) t]},
c_2(t) = c_2(0) exp{[-(\gamma_1+\gamma_2)t]}$,
 $c_3(0) \equiv c_3$, and   
the quantum discord does not decay up to some
finite time $\bar{t}$~\cite{mazzola-prl-10}.  
Initially ($t=0$),  the
computed correlations in the $\rho^{\rm E}_{\rm
BD}$ state (taken as an average of five density
matrices initially prepared in the same state)
turned out to be ${\cal C}[\rho(0)]= 0.999 \pm
0.169 $, ${\cal D}[\rho(0)]= 0.366 \pm 0.024$ and
${\cal I}[\rho(0)]= 1.366 \pm 0.017$.  The
simulated and experimental plots of the dynamics
of the quantum discord, classical correlations and
total correlations are shown in~Fig.\ref{tidplot_ab}~(a) and (b)
respectively, and shows a distinct transition from the classical to
the quantum decoherence regimes at $t = \bar{t}$. 
The transition time up to which quantum
discord remains constant is
$\bar{t}=\frac{1}{2\gamma} \ln
\left|\frac{c_{1}(0)}{c_{3}(0)}\right|$.

\begin{figure}
\centering
\includegraphics[angle=0,scale=1.5]{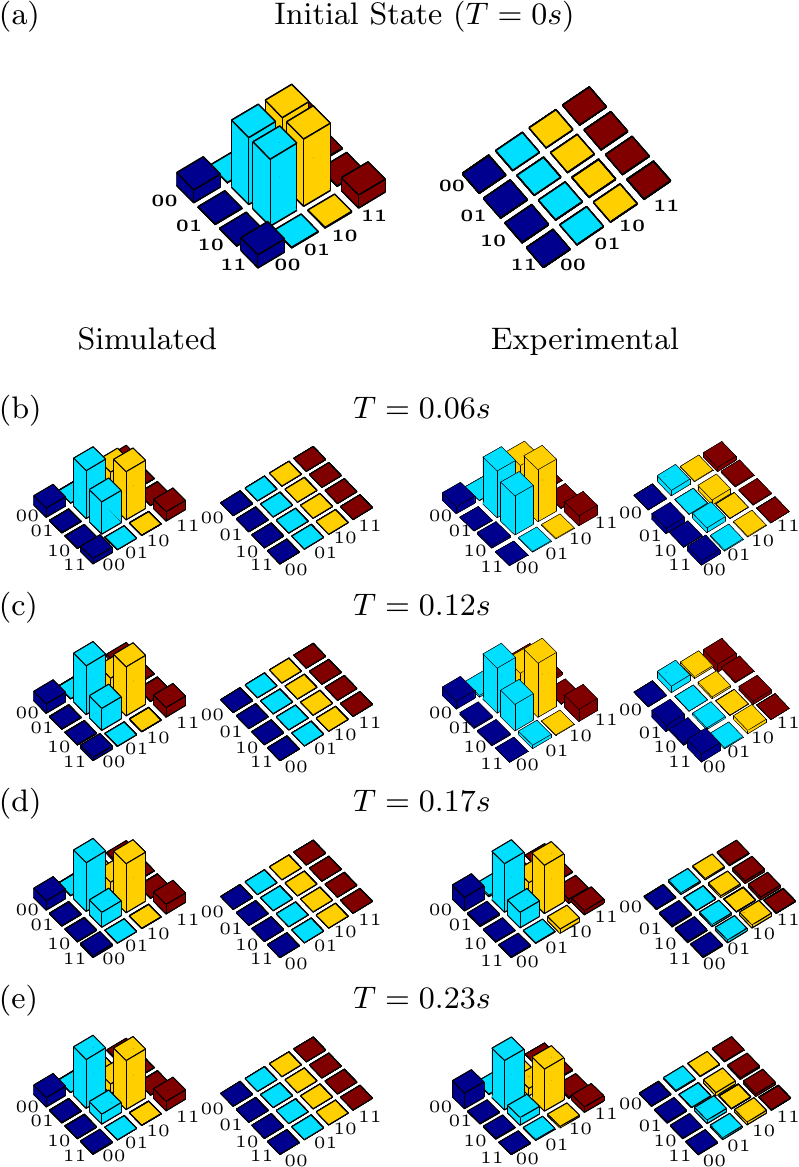}
\caption{
Real (left) and imaginary (right) parts of the experimental
tomographs of the (a) Bell Diagonal (BD) state, with a computed
fidelity of 0.99.  (b)-(e) depict the state at $T = 0.06,
0.12, 0.17, 0.23 $s, with the tomographs on the left and the
right representing the simulated and experimental state, respectively. 
The rows and columns are labeled in the computational basis ordered from
$\vert 00 \rangle$ to $\vert 11 \rangle$.
}
\label{tidtomo1}
\end{figure}

We allowed the experimentally prepared state 
$\rho^{\rm E}_{\rm BD}$, 
to evolve freely in time and determined the
parameters $c_i$ at each time point.  We used these
experimentally determined coefficients
to compute the classical correlations (${\cal
C}[\rho(t)]$), the quantum discord (${\cal
D}[\rho(t)]$), and total correlations (${\cal
I}[\rho(t)]$) at each time point.
The transition time up to which quantum discord
remains constant was experimentally determined to 
be $\bar{t} = 0.052$s. The experimental results correlate well with
the theoretical noise model which allows to calculate $\bar{t}$ and
the value comes out to be 0.054 s. The tomographed density matrix,
reconstructed at different time points, shows
that state evolution remains confined to the subspace of
BD states, as is evident from the experimentally
reconstructed density matrices of the state at different time
points displayed in Fig.\ref{tidtomo1}.

\section{Protection of time-invariant discord using dynamical decoupling schemes}

DD schemes, consisting of
repeated sets of $\pi$ pulses with tailored
inter-pulse delays and phases, have played
an important role in dealing with the debilitating effects of
decoherence~\cite{uhrig-prl-09}. Several NMR QIP
experiments have successfully used DD-type schemes to
preserve quantum states~\cite{roy-pra-11,singh-pra-14}.
These schemes are expected to provide an 
advantage over traditional Carr-Purcell-Meiboom-Gill (CPMG) refocusing schemes
and here we compared the two experimentally.

\begin{figure} 
\centering
\includegraphics[angle=0,scale=1.5]{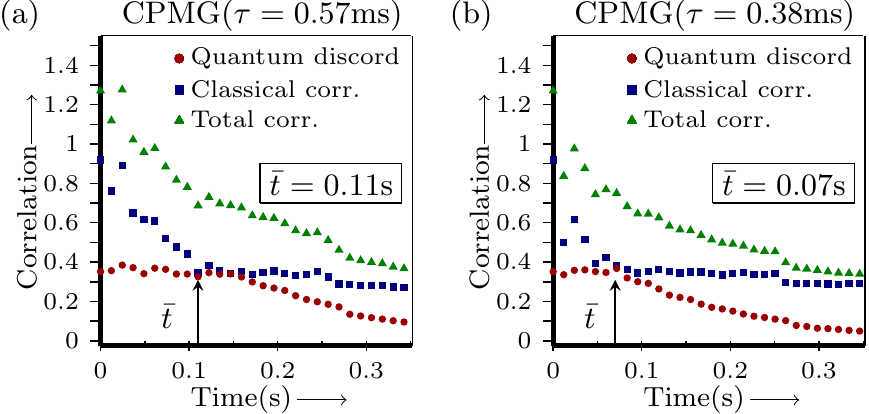}
\caption{ Time evolution of 
total correlations (triangles), 
classical correlations (squares) 
and quantum discord (circles) of 
the BD state: Experimental plots using CPMG preserving sequences
(a) CPMG with $\tau=0.57$ ms and (b) CPMG with $\tau=0.38$ ms.
}
\label{tidplot_cpmg}
\end{figure} 

We first protect these states and their related quantum
correlations using standard CPMG schemes of the type
($\tau-\pi-\tau-\pi$). The results are shown in Fig.{\ref{tidplot_cpmg}}(a) and
(b) for two different values of $\tau$ (the time interval between
two consecutive $\pi$ pulses). It turns out that the CPMG
schemes which are based on the traditional Hahn spin echo are able
to provide protection to some extent and the life time of
the time invariant quantum discord grows from 0.05 s to
0.11 s as we increase the number of $\pi$ pulses. However,
as we try to increase the number of $\pi$ pulses further the
results are counter-productive as shown in Fig.{\ref{tidplot_cpmg}}(b).
Therefore, beyond a point we have to give up the standard
Hahn spin echo strategy.

\begin{figure}

\centering
\includegraphics[angle=0,scale=1.3]{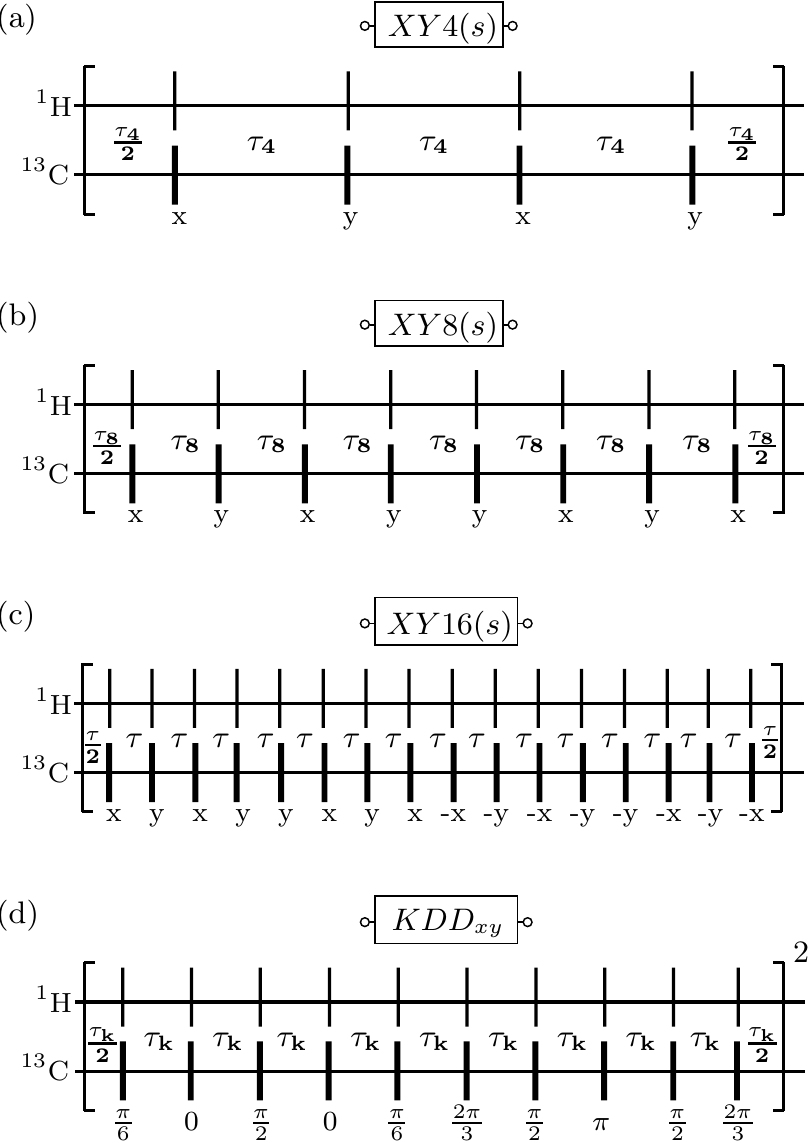}
\caption{
NMR pulse sequence corresponding to DD schemes
(a) XY4(s), (b) XY8(s), (c) XY16(s), and (d) KDD$_{xy}$ , with
time delays between pulses denoted by $\tau_4$ , $\tau_8$, $\tau$, $\tau_k$ , respectively.
All the pulses are of flip angle $\pi$ and are labeled with their
respective phases. The pulses are applied simultaneously on
both qubits. The superscript `2' in the KDD xy sequence denotes
that one unit cycle of this sequence contains two blocks of
the ten-pulse block represented schematically, i.e., a total of
twenty pulses. The shorter duration proton pulses and the
longer duration carbon pulses are centered on each other and
the various time delays ($\tau_4$,  $\tau_8$ , $\tau$, $\tau_k$) in all the DD schemes are
tailored to the gap between two consecutive carbon pulses.}
\label{ddscheme}
\end{figure}


Next we implemented more sophisticated DD sequences to
explore if we can further protect the BD state, and hence
the related quantum correlations typified via the survival
time of the quantum discord $\bar{t}$. If the $\pi$ pulses in a DD
sequence are non-ideal (either due to finite pulse lengths
or flip angle and off-resonant driving errors), it leads to
imperfect system-bath decoupling. Several schemes have
been designed to make DD sequences robust against pulse
imperfections by achieving ideal pulse rotations. Standard
CPMG-based DD sequences used $\pi$ pulses applied along
the same rotation axis, which then preserve coherence
along only one spin component. The XY family of DD
sequences applies pulses along two orthogonal ($x$, $y$) axes,
which preserves coherences about both spin rotation axes.
The basic XY4(s) DD sequence is a four-pulse sequence
with phases $x - y - x - y$, with pulses applied at the center
of each time period such that the whole sequence is 
time-symmetric with respect to its center~\cite{souza-prl-11}.
The XY8(s) DD sequence uses the XY4(s) sequence as a building block,
by combining the XY4(s) sequence with its time-reversed
version, so that the whole eight-pulse sequence is
explicitly time-symmetric. Each cycle of the symmetrized DD
sequences is applied several times to achieve higher-order
decoupling~\cite{ahmed-pra-13}. The Knill dynamical decoupling (KDD)
sequence combines the rotation pattern of the XY4(s) 
sequence with composite pulses, replacing each $\pi$ pulse in
the DD sequence with a composite sequence of five pulses
with different phases [\cite{ahmed-pra-13}]:

\begin{equation}
\label{kddphi}
 {\rm KDD}_\phi =
(\pi)_{\pi/6+\phi}-(\pi)_{\phi}-
(\pi)_{\pi/2+\phi}-(\pi)_{\phi}-(\pi)_{\pi/6+\phi}.
\end{equation}

The additional phases in the KDD$_{xy}$ ($xy$ in the subscript
denoting pulses applied along both axes) lead to better
compensation for pulse errors, combining two of
the basic five-pulse blocks given in Eq.{(\ref{kddphi})} that are shifted in
phase by $ \frac{\pi}{2}$ i.e [KDD$_0$-KDD$_{\frac{\pi}{2}}$]. The four symmetrized
DD schemes used in our experiments to preserve time-
invariant discord, namely XY4(s), XY8(s), XY16(s) and
KDD$_{xy}$ , are schematically represented in Fig.{\ref{ddscheme}}. The `2'
in the superscript of the KDD$_{xy}$ scheme denotes that one
unit cycle of this scheme contains two of ten-pulse
blocks, for a total of twenty pulses. We applied $\pi$ pulses
simultaneously on both spins, with pulse lengths of $15.1$ $\mu$s
and $26.8$ $\mu$s for the proton and carbon spins, respectively.
The proton and carbon pulses are centered on each other
and the time delay between pulses was set at the gap 
between two consecutive carbon pulses. Each DD sequence
was applied a repeated number of times (N being as large
as experimentally possible), for good coherence preservation.

\begin{figure} 
\centering
\includegraphics[angle=0,scale=1.5]{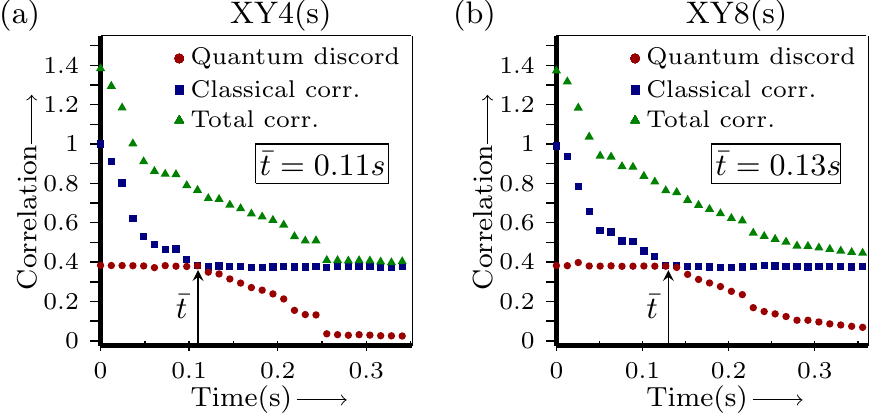}
\caption{ Time evolution of 
total correlations (triangles), 
classical correlations (squares) 
and quantum discord (circles) of 
the BD state: Experimental plots using (a) XY4(s) with $\tau_4= 0.58$ ms and (b) XY8(s)
with $\tau_8 = 0.29$ ms.
}
\label{tidplot_xy48}
\end{figure} 


\begin{figure}
\centering
\includegraphics[angle=0,scale=1.65]{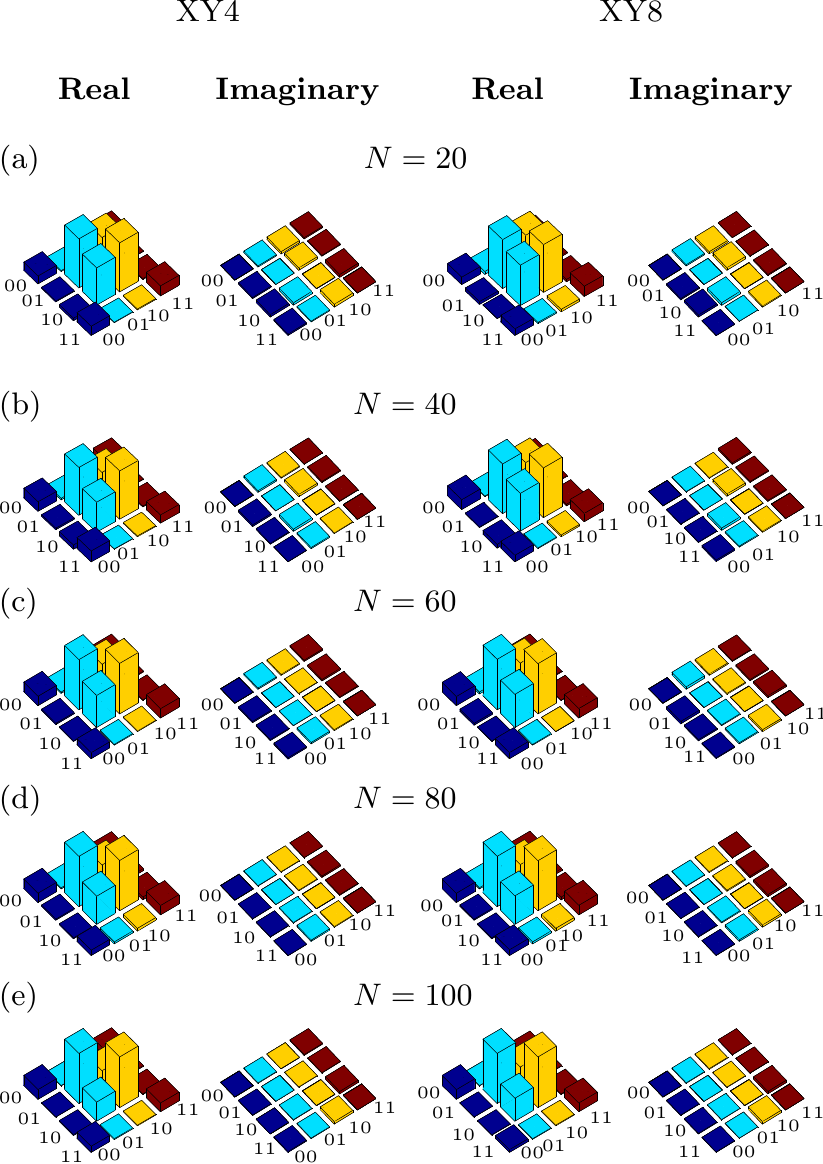}
\caption{
Real (left) and imaginary (right) parts of the experimental
tomographs of the (a)-(e) depict the BD state at $N = 20,
40, 60, 80,100 $, with the tomographs on the left and the
right representing the BD state after applying the XY4 and 
XY8 scheme, respectively.  The rows and
columns are labeled in the computational basis ordered from
$\vert 00 \rangle$ to $\vert 11 \rangle$.
}
\label{tid_tomo2}
\end{figure}

\begin{figure} 
\centering
\includegraphics[angle=0,scale=1.5]{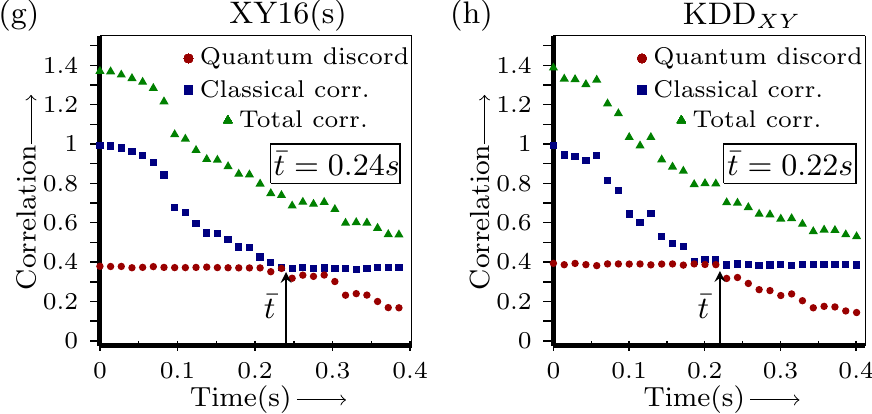}
\caption{ Time evolution of 
total correlations (triangles), 
classical correlations (squares) 
and quantum discord (circles) of 
the BD state: Experimental plots using (a) XY16(s) with $\tau= 0.145$ ms and (b) KDD$_{xy}$
with $\tau_k = 0.116$ ms.
}
\label{tidplot_xy16kd}
\end{figure}

\begin{figure}
\centering
\includegraphics[angle=0,scale=1.5]{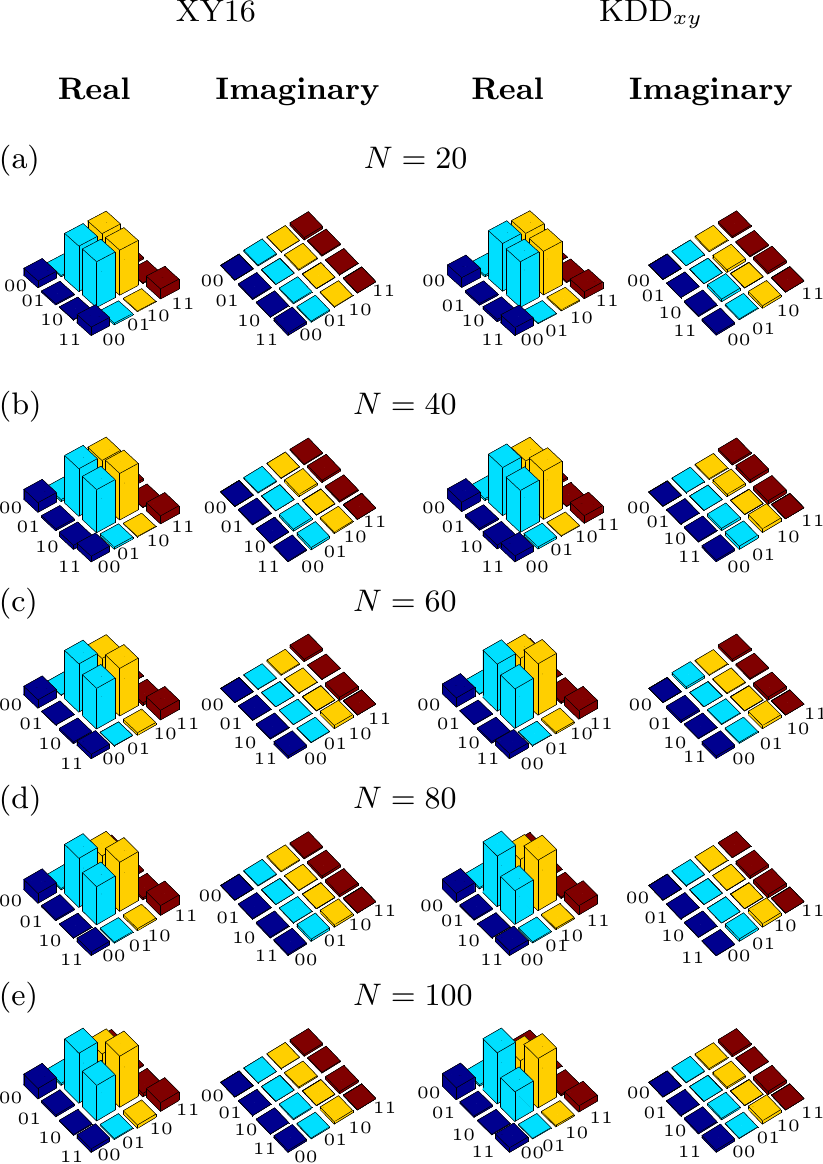}
\caption{Real (left) and imaginary (right) parts of the experimental
tomographs  in (a)-(e) depict the Bell Diagonal (BD) state at $N = 20,
40, 60, 80,100 $, with the tomographs on the left and the
right representing the BD state after applying the XY16 and 
KDD$_{xy}$ preserving DD schemes, respectively.  The rows and
columns are labeled in the computational basis ordered from
$\vert 00 \rangle$ to $\vert 11 \rangle$.
}
\label{tid_tomo3}
\end{figure}

We implemented the symmetrized XY4(s) DD scheme
for $\tau_4= 0.58$ ms and an experimental time for one run
of $2.43$ ms. The time for which quantum discord 
persists using the XY4(s) scheme is $\bar{t}= 0.11$ s, 
which is double the time as compared to `no-preservation' (Fig.~\ref{tidplot_xy48}(a)).
We implemented the XY8(s) DD scheme for $\tau_8 = 0.29$ ms
and an experimental time for one run of 2.52 ms. The
time for which quantum discord persists using the XY8(s)
scheme is $\bar{t}= 0.13 s$, nearly the same as the XY4(s) scheme
(Fig.~\ref{tidplot_xy48}(b)). The XY16(s) DD scheme provides even better
compensation than the XY8(s) sequence. We implemented
the XY16(s) scheme for $\tau = 0.145$ ms and an 
experimental time for one run of $2.75$ ms. The time for
which quantum discord persists for the XY16(s) scheme is
$\bar{t}= 0.24$ s, which is four times the persistence time of the
discord when no preservation is applied~(Fig.\ref{tidplot_xy16kd}(a)). We
implemented the KDD$xy$ sequence with $\tau= 0.116$ ms and
an experimental time for one run of 2.86 ms. In this case
too, the time for which quantum discord persists  $\bar{t}= 0.22$ s 
is quadrupled, as compared to no-preservation~(Fig.\ref{tidplot_xy16kd}(b)).
The $\tau$ delays are chosen in such a way that for a given sequence
the protection is maximum. If we decrease $\tau$ delay intervals further, 
errors due to $\pi$ pulses dominates. The DD sequence was looped for five times between each
time point in~Figs.\ref{tidplot_cpmg}, \ref{tidplot_xy48} and \ref{tidplot_xy16kd},
which typically means repeating a DD sequence 175-200 times during an experiment
covering all time points. The experimental tomographs
of the BD state at different instances of time after using
XY4(s), XY8(s), XY16(s) and KDD$_{xy}$ DD preservation
schemes to protect time-invariant discord are given in 
~Fig.\ref{tid_tomo2} and ~Fig.\ref{tid_tomo3}. All the 
experimental tomographs display excellent preservation of
the state, with very little leakage.

\begin{figure} 
\centering
\includegraphics[angle=0,scale=1]{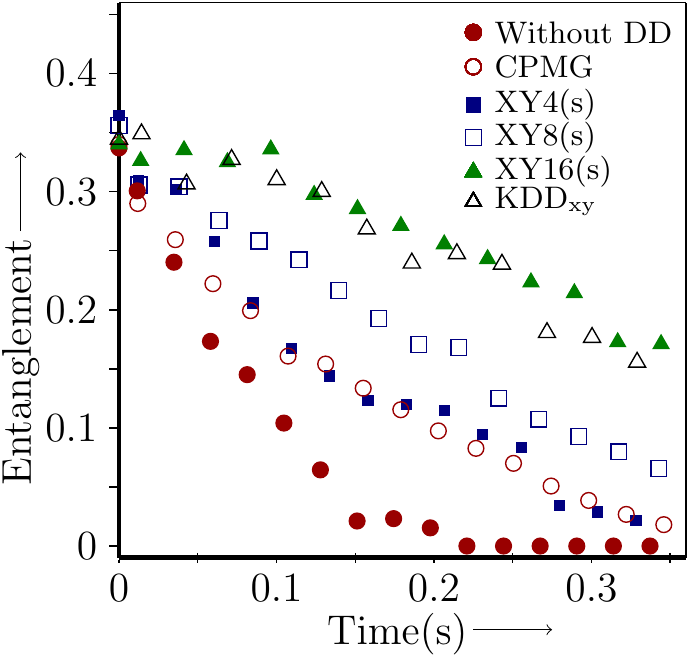}
\caption{Plot of time evolution of entanglement without applying any preservation (filled circle),
CPMG with $\tau=0.57$ ms (empty circle), XY4(s) (filled rectangle), XY8(s) (empty rectangle),
XY16(s) (filled triangle) and KDD$_{xy}$ (empty triangle),respectively.}
\label{ent_plot}
\end{figure}

In ~Fig.\ref{ent_plot}, we plotted the protection 
of quantum entanglement 
in the two-qubit BD state via CPMG and DD schemes, 
using negativity as an entanglement measure. It
turns out that the results of entanglement protection are
similar to that of discord protection namely, that CPMG
and XY4(s) schemes protect entanglement to some extent
while the more involved DD schemes are able to protect
entanglement for relatively long times up to 0.3 s (as compared 
to its natural decay time of 0.12 s without any DD
protection). However, the phenomenon of freezing of 
entanglement is not observed, unlike the case for quantum
discord.

\subsection{Conclusions}
The two-qubit system we used in our experiments
is mainly affected by an independent phase damping
channel on each qubit for the experimental time regime
under consideration. One of the main sources of relaxation in NMR 
is dipolar relaxation. Such noise can be ideally
suppressed by the CPMG sequence. The suppression of noise
is expected to increase by reducing the delay between the 
$\pi$ pulses, but due to imperfect $\pi$ pulses after a certain
time point instead of suppressing noise, starts contributing 
to the noise. Also, the CPMG sequence is designed to protect 
magnetization along one particular axis, not for a general axis. 
To mitigate all these problems we used XY4(s), XY8(s), XY16(s)
and KDD$_{xy}$ DD sequences, which are robust against these
imperfections and protect magnetization along the general axis. 
Our results show that time-invariant quantum discord, which
remains unaffected under certain decoherence regimes, can be
preserved for very long times using DD schemes. 
Our experiments have important implications in
situations where persistent quantum correlations have to be
maintained to carry out quantum information
processing tasks.

\chapter{ Dynamics of tripartite entanglement
under decoherence and protection using dynamical
decoupling}\label{3qubitchapter}
\section{Introduction}
\label{intro_3q}
In Chapter~{\ref{discordchapter}}, we considered the situation where we have
knowledge of the state of the system as well as its interaction with the
environment. In such situations, since the noise model is known, decoupling
strategies can be designed to cancel this noise. Using these decoupling
strategies, we experimentally extended the lifetime of time invariant discord
of two-qubit Bell-diagonal states.
In this chapter we extend the same idea for 
experimental preservation of three-qubit entangled states.
Quantum entanglement is considered to lie at the crux of
QIP~\cite{nielsen-book-02} and while two-qubit entanglement can be completely
characterized, multipartite entanglement is more difficult to quantify and is
the subject of much recent research~\cite{horodecki-rmp-09}.  Entanglement can
be rather fragile under decoherence and various multiparty entangled states
behave very differently under the same decohering
channel~\cite{dur-prl-04}.  It is hence of paramount importance to
understand and control the dynamics of multipartite entangled states in 
multivarious noisy
environments~\cite{mintert-pr-05,aolita-prl-08,aolita-rpp-15}.  
A three-qubit system is a good model system to study the diverse response of
multipartite entangled states to decoherence and  the
entanglement dynamics of three-qubit GHZ and W states have been theoretically
well studied~\cite{borras-pra-09,weinstein-pra-10}.
Under an arbitrary (Markovian) decohering environment, it was shown that W
states are more robust than GHZ states for certain kinds of channels while the
reverse is true for other kinds of
channels~\cite{carvalho-prl-04,siomau-eurphysd-10,siomau-pra-10,ali-jpb-14}.

On the experimental front, tripartite entanglement was generated using 
photonic qubits and the robustness of W state entanglement was
studied in optical 
systems~\cite{lanyon-njp-09,zang-scirep-15,he-qip-15,zang-optics-16}.  The 
dynamics of multi-qubit entanglement under
the influence of decoherence was experimentally characterized using a string of
trapped ions~\cite{barreiro-nature} and in
superconducting qubits~\cite{wu-qip-16}. 
In the context of NMR quantum information
processing, three-qubit entangled states were experimentally
prepared~\cite{suter-3qubit,dogra-pra-15,manu-pra-14,nelson-pra-00}, and their decay rates
compared with bipartite entangled states~\cite{kawamura-ijqc-06}.

With a view to protecting entanglement,
dynamical decoupling (DD) schemes have been successfully
applied to decouple a multiqubit system from both transverse
dephasing and longitudinal relaxation
baths~\cite{viola-review,uhrig-njp-08,kuo-jmp-12,zhen-pra-16}.  UDD
schemes have been used in the context of entanglement
preservation~\cite{song-ijqi-13,franco-prb-14}, and it was
shown theoretically that Uhrig DD schemes are able to
preserve the entanglement of two-qubit Bell states and
three-qubit GHZ states for quite long
times~\cite{agarwal-scripta}.

In this chapter, first the Lindblad master equation is solved analytically 
 for the robustness of  three different tripartite entangled states, namely,
the GHZ, W and $\rm W{\bar W}$ states with different noise channels.
Then the robustness against decoherence of these three different tripartite entangled
states are experimentally explored.  The ${\rm W{\bar W}}$ state is a novel
tripartite entangled state which belongs to the GHZ entanglement class in the
sense that it is SLOCC equivalent to the GHZ state, however
stores its entanglement in ways very similar to that of the W 
 state~\cite{Devi2012,das-pra-15}. 
Next, the experimental data are best modeled by considering
the main noise channel to be an uncorrelated phase damping
channel acting independently on each qubit, along with a generalized
amplitude damping channel. Next, the entanglement of these states is protected using
two different DD sequences: the symmetrized XY-16(s) and the Knill dynamical
 decoupling (KDD) sequences, and evaluated their efficacy of protection.

\section{Dynamics of tripartite entanglement}
\label{entang-decoh_3q}
\subsection{Tripartite entanglement under different noise channels}
We considered four different noise channels:
phase damping (Pauli $\sigma_z$),
amplitude damping (Pauli $\sigma_x$), bit-phase
flip (Pauli $\sigma_y$) and a uniform depolarizing
channel  along the lines suggested
in Reference~\cite{jung-pra-08}.
The master equation is given by~\cite{lindblad}:
\begin{equation}
\frac{\partial \rho}{\partial t} 
= -i[H_s,\rho] + \sum_{i,\alpha}
\left[
L_{i,\alpha} \rho L_{i,\alpha}^{\dagger}
- \frac{1}{2} \{ L^{\dagger}_{i,\alpha} 
L_{i,\alpha},\rho
\}
\right]
\label{mastereqn_3qchap} 
\end{equation}
where $H_s$ is the system Hamiltonian, 
$L_{i,\alpha} \equiv \sqrt{\kappa_{i,\alpha}} 
\sigma^{(i)}_{\alpha}$ is the Lindblad operator
acting on the $i$th qubit and $\sigma^{(i)}_{\alpha}$
is the Pauli operator on the $i$th qubit,
$\alpha=x,y,z$; the constant
$\kappa_{i,\alpha}$ turns out to be the inverse of the
decoherence time.
This master equation approach has been shown to be
equivalent to the standard operator sum representation
method for
open quantum systems, where
density operator evolution is given in terms of
the standard Kraus operators for various noise
channels~\cite{nielsen-book-02}.  For two qubits, all entangled states are
negative under partial transpose (NPT) and for such NPT states, the minimum
eigenvalues of the partially transposed density operator is a measure of
entanglement~\cite{peres-prl-96}.  This idea has been extended to three qubits,
and entanglement can be quantified for a three-qubit system using the
well-known tripartite negativity ${\cal N}^{(3)}_{123}$
measure~\cite{vidal-pra-02,weinstein-pra-10}:
\begin{equation}
{\cal{N}}^{(3)}_{123}= [{\cal N}_{1}{\cal
N}_{2}{\cal N}_{3}]^{1/3} 
\end{equation}
where the negativity of a qubit ${\cal N}_i$ 
refers to the most negative eigenvalue of
the partial transpose of the density matrix with respect to the qubit $i$.
\begin{figure}[H]
\centering
\includegraphics[angle=0,scale=1.26]{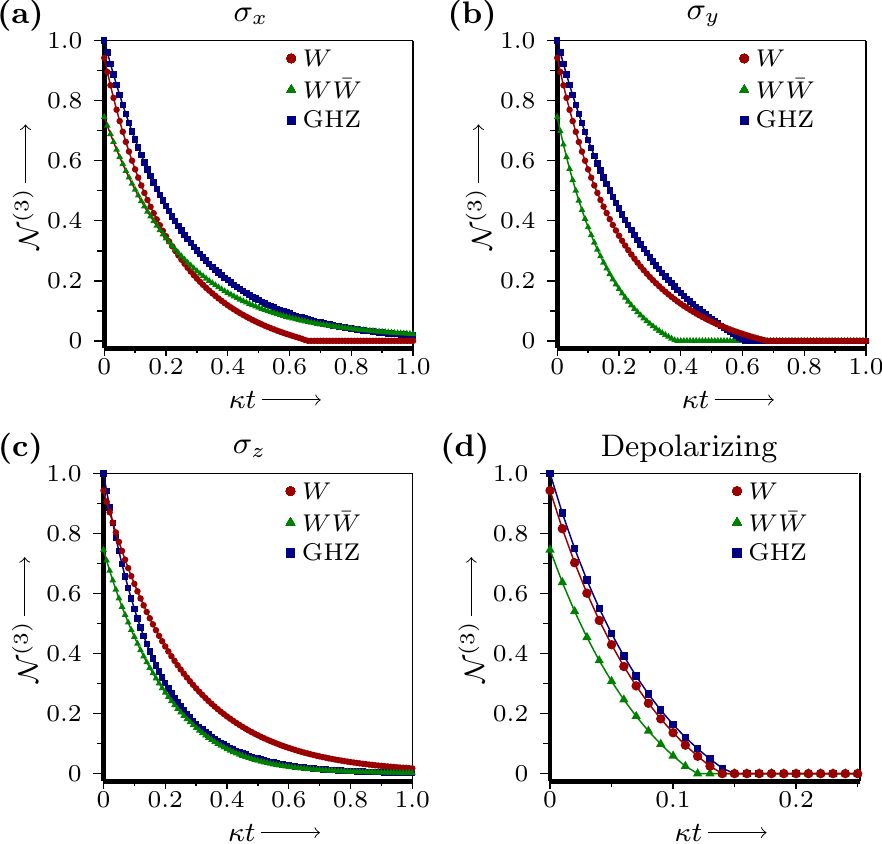}
\caption{Simulation of decay of tripartite entanglement parameter
negativity ${\cal N}^{(3)}$ of the GHZ state (blue squares),
the W state (red
circles) and the $W\bar{W}$ state (green triangles) under the 
action of (a) amplitude damping (Pauli
$\sigma_x$) channel, 
(b) bit-phase flip (Pauli $\sigma_y$)
channel (c) phase damping (Pauli $\sigma_z$)
channel and
(d) isotropic
noise (depolarizing) channel. The $\kappa$ parameter
denotes inverse of the decoherence time.}
\label{noise-simulation_3qchap}
\end{figure}

The analytical results of the Lindblad equation under the action of the 
different noise
channels on
tripartite entanglement are displayed in
Fig.~\ref{noise-simulation_3qchap}, where
Figs.~\ref{noise-simulation_3qchap}(a)-(d) show 
the decohering
effects of the amplitude
damping (Pauli $\sigma_x$) channel, the bit-phase flip 
(Pauli $\sigma_y$) channel,
the phase damping (Pauli $\sigma_z$) channel and the
depolarizing (isotropic noise) channel respectively,
on the GHZ, W and ${\rm W \bar{W}}$ states.
The simulation indicates that the
W state is more robust against the phase-damping 
channel as compared to the GHZ state, while the
reverse is true for the amplitude-damping channel. 
Both states decohere to nearly the same extent
under the action of the bit-phase flip channel and the
depolarizing channel.
The ${\rm W \bar{W}}$
state decoheres more rapidly under the actions of the
bit-phase flip channel and the depolarizing channel,
as compared to the other GHZ and the W states.
For short timescales its decoherence behavior
mimics the W state under the action of the 
amplitude damping channel while for longer timescales
it decoheres similar to the GHZ state. 
On the other hand, under the action of the
phase-damping channel the ${\rm W \bar{W}}$ state
initially decoheres faster than the GHZ state
and later closely follows the GHZ decay behavior.

\subsection{NMR system}

\begin{figure}[h]
\centering
\includegraphics[angle=0,scale=1]{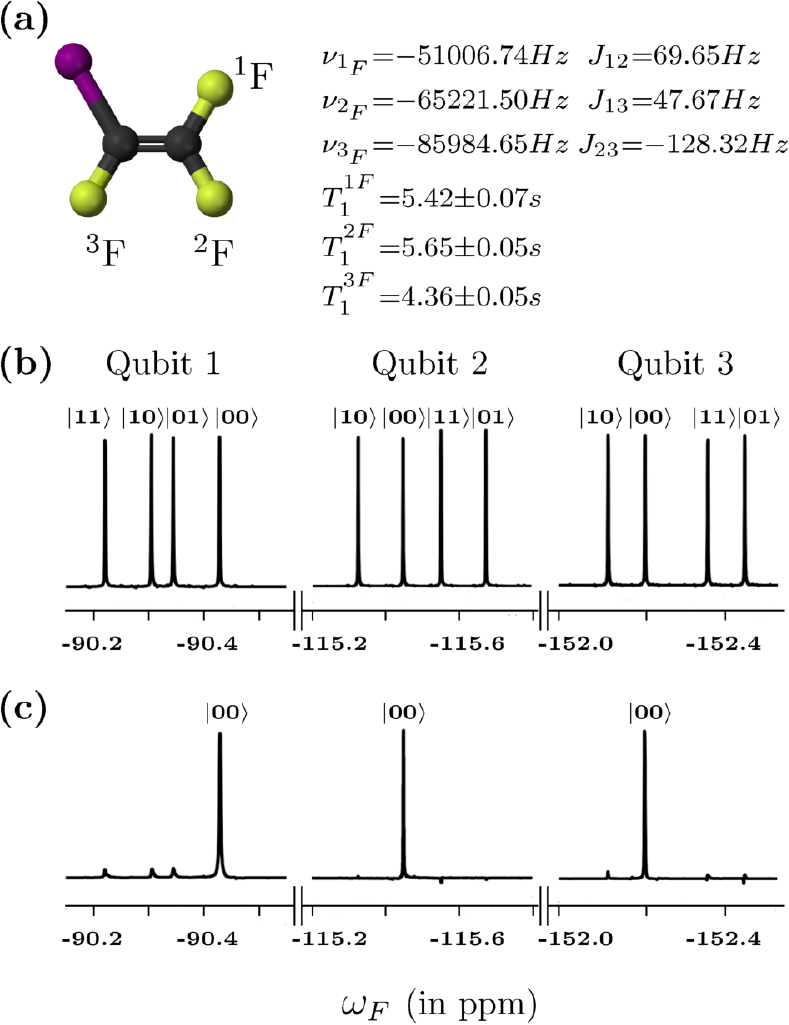}
\caption{(a) Molecular structure of trifluoroiodoethylene molecule and
tabulated system parameters  with chemical shifts $\nu_i$
and scalar couplings J$_{ij}$ (in Hz), and spin-lattice
relaxation times $T_{1}$ and spin-spin relaxation times
T$_{2}$ (in seconds).  (b) NMR spectrum obtained after a
$\pi/2$ readout pulse on the thermal equilibrium state.  and
(c) NMR spectrum of the pseudopure $\vert 000 \rangle$
state.  The resonance lines of each qubit are labeled by the
corresponding logical states of the other qubit.
}
\label{systemmol_3qchap}
\end{figure}
We use the
three ${}^{19}$F nuclear spins of the 
trifluoroiodoethylene (C$_2$F$_3$I) molecule
to encode the three qubits. On an NMR
spectrometer operating at 600 MHz, the fluorine
spin resonates at a Larmor frequency of 
$\approx 564$ MHz.
The molecular structure of the three-qubit system with
tabulated system parameters and the NMR spectra of the
qubits at thermal equilibrium and prepared in the
pseudopure state $\vert 000 \rangle$ are shown in
Figs.~\ref{systemmol_3qchap}(a), (b), and (c), respectively.
The Hamiltonian of a 
weakly-coupled three-spin system in a frame 
rotating at $\omega_{{\rm rf}}$ (the frequency of the
electromagnetic field $B_1(t)$ applied to manipulate spins
in a static magnetic field $B_0$)
is given by~\cite{ernst-book-87}:
\begin{equation}
{\cal H} = -\sum_{i=1}^3 (\omega_i -
\omega_{{\rm rf}}) I_{iz} 
+ \sum_{i<j,j=1}^3 2 \pi J_{ij} I_{iz} I_{jz}
\end{equation}
where $I_{iz}$ is the spin angular momentum
operator in the $z$ direction for ${}^{19}$F; the
first term in the Hamiltonian denotes the 
Zeeman interaction between the fluorine spins and the
static magnetic field $B_0$ with $\omega_i = 2 \pi \nu_i$ being
the Larmor frequencies; the second term represents
the spin-spin interaction  with $J_{ij}$ being the 
scalar coupling constants.
The three-qubit equilibrium density matrix (in the high
temperature and high field approximations) is in a highly
mixed state given by:
\begin{eqnarray}
\rho_{eq}&=&\tfrac{1} {8}(I+\epsilon \ \Delta \rho_{eq})
\nonumber \\
\Delta\rho_{{\rm eq}} &\propto& \sum_{i=1}^{3} I_{iz}
\end{eqnarray}
with a thermal polarization $\epsilon \sim 10^{-5}$, $I$
being
the $8 \times 8$ identity operator and 
$\Delta \rho_{{\rm eq}}$ being the deviation part of
the density matrix.
The system was first initialized into the $\vert 000\rangle$
pseudopure state using the spatial averaging
technique~\cite{cory-physicad},
with the density operator given by
\begin{equation}
\rho_{000}=\frac{1-\epsilon}{8}I 
+ \epsilon \vert 000\rangle\langle000 \vert
\end{equation}

\begin{figure}[h]
\centering
\includegraphics[angle=0,scale=1.54]{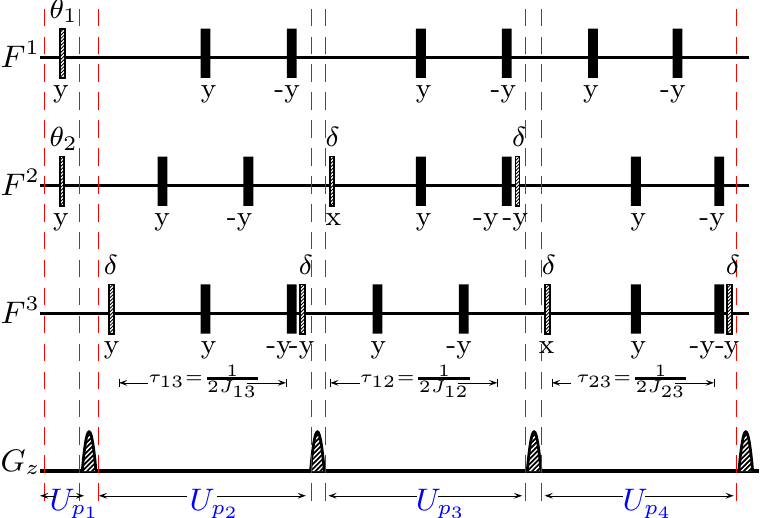}
\caption{NMR pulse sequence used to prepare
pseudopure state $\rho_{000}$ starting from 
thermal equilibrium.The
pulses represented by black filled rectangles are of angle
$\pi$. The other rf flip angles are set to $\theta_1=\frac{5\pi}{12}$, 
$\theta_2=\frac{\pi}{6}$ and $\delta=\frac{\pi}{4}$. The phase of
each rf pulse is written below each pulse bar. The evolution
interval $\tau_{ij}$ is set to a multiple of the scalar
coupling strength ($J_{ij}$).
}
\label{ppure-fig_3qchap}
\end{figure}

The specific sequence of rf pulses, $z$ gradient pulses and
time evolution periods we used to 
prepare the pseudopure state
$\rho_{000}$ starting from thermal equilibrium is shown in
Fig.~\ref{ppure-fig_3qchap}.
All the rf pulses used
in the pseudopure state preparation scheme were constructed
using the Gradient Ascent Pulse Engineering (GRAPE)
technique~\cite{tosner-jmr-09} and were designed to be
robust against rf inhomogeneity, with an average fidelity of
$ \ge 0.99$. Wherever possible, two independent
spin-selective rf pulses were combined using a specially
crafted single GRAPE pulse;  for instance the first two rf
pulses to be applied before the first field gradient pulse,
were combined into a single pulse specially crafted pulse
($U_{p_{1}}$ in Fig.~\ref{ppure-fig_3qchap}), of duration
$600 \mu$s. The combined pulses $U_{p_{2}}$, $U_{p_{3}}$
and $U_{p_{4}}$ applied later in the
sequence were of a total duration $\approx 20$ ms.

All experimental density matrices were reconstructed using a
reduced tomographic protocol and by using maximum likelihood
estimation~\cite{leskowitz-pra-04,singh-pla-16}
as described in Chapter~\ref{chapter_mle} with the set
of operations $\{ III, IIY, IYY, YII, XYX, XXY,$
$ XXX\}$; $I$
is the identity (do-nothing operation) and $X (Y)$ denotes a
single spin operator implemented by a spin-selective $\pi/2$
pulse.  We constructed these spin-selective pulses for
tomography using GRAPE, with the length of each pulse
 $\approx 600 \mu$s.  
The experimentally created
pseudopure state $\vert 000\rangle$ was tomographed with 
a fidelity of 0.99 and the total time taken to prepare the
state was $\approx 60$ ms. The fidelity of an experimental
density matrix was computed using Eq.~(\ref{mle_fidelity_2}).
\begin{figure}[H]
\centering
\includegraphics[angle=0,scale=1.2]{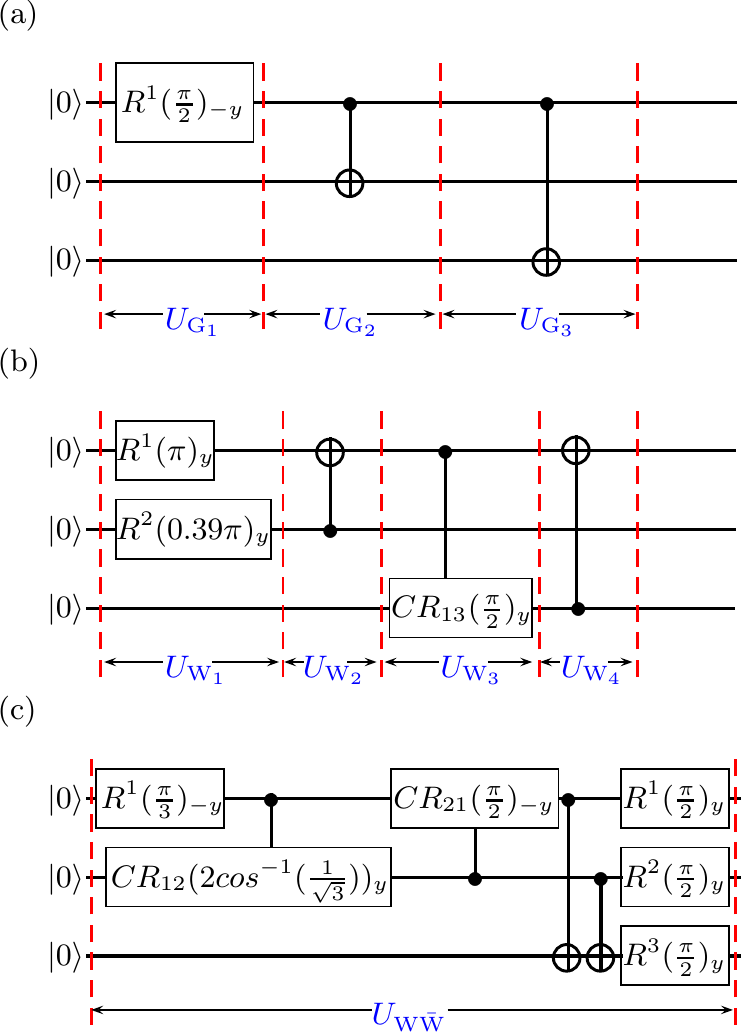}
\caption{(Quantum circuit showing 
the sequence of implementation of the
single-qubit local rotation gates (labeled by 
$R$), two-qubit controlled-rotation gates (labeled
by $CR$) and controlled-NOT
gates required
to construct the (a) GHZ state (b) W state and (c) 
${\rm W\bar{W}}$ state.}
\label{ckt_3qchap}
\end{figure}
\subsection{Construction of tripartite entangled states}
\label{construct_3qchap}
Tripartite entanglement has been well characterized and it
is known that the two different classes of tripartite
entanglement, namely GHZ-class and W-class, are
inequivalent. While both classes are maximally entangled,
there are differences in the their
type of entanglement: the W-class entanglement is more
robust against particle loss than the GHZ-class (which
becomes separable if one particle is lost) and it is
also known that the W state has the maximum possible
bipartite entanglement in its reduced two-qubit 
states~\cite{guhne-review}. The entanglement in the
${\rm W \bar{W}}$ state (which belongs to the GHZ-class
of entanglement) shows a surprising result, that it is
reconstructible from its reduced two-qubit states (similar
to the W-class of states). 
We now turn to the construction of tripartite entangled
states on the three-qubit NMR system. The quantum circuits
to prepare the three qubits in a GHZ-type state, a W state
and a ${\rm W \bar{W}}$ state are shown in Figs.~\ref{ckt_3qchap}
(a), (b) and (c), respectively. Several of the quantum gates
in these circuits were optimized using the GRAPE algorithm
and we were able to achieve a high gate fidelity and
smaller pulse lengths.
\begin{figure}[H]
\centering
\includegraphics[angle=0,scale=1.1]{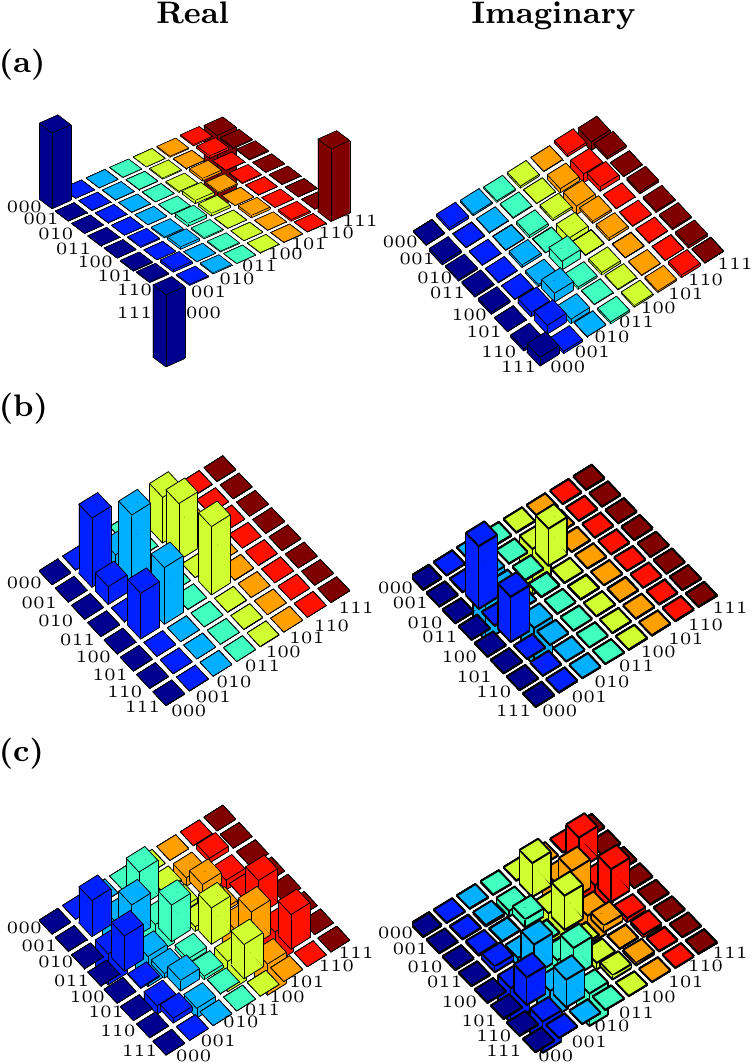}
\caption{The real (left) and imaginary (right) parts of the
experimentally tomographed (a) GHZ-type state, with
a fidelity of $0.97$. (b)  
W state, with a fidelity of $0.96$ and (c)  
$W\bar{W}$ state with a fidelity of $0.94$.
The rows and columns encode the
computational basis in binary order from
$\vert 000 \rangle$ to $\vert 111 \rangle$.}
\label{nodd_3q}
\end{figure}

The GHZ-type
$\frac{1}{\sqrt{2}}(\vert000\rangle-\vert111\rangle)$ state
was prepared from the $\vert000\rangle$ pseudopure state by
a sequence of three quantum gates (labeled as $U_{G_1},
U_{G_2}, U_{G_3}$ in Fig.~\ref{ckt_3qchap}(a)): first a selective
rotation of $\left [\frac{\pi}{2}\right]_{-y}$ on the first
qubit, followed by a CNOT$_{12}$ gate, and finally a
CNOT$_{13}$ gate.  The step-by-step sequential gate
operation leads to: 
\begin{eqnarray}
\vert 0 0 0 \rangle &\stackrel{{R^1{\left
(\frac{\pi}{2}\right)_{-y}}}}{\longrightarrow}&
\frac{1}{\sqrt{2}}\left(\vert 0 0 0 \rangle -
\vert 1 0 0 \rangle \right) \nonumber \\
&\stackrel{\rm CNOT_{12}}{\longrightarrow}&
\frac{1}{\sqrt{2}}\left(\vert 0 0 0 \rangle -
\vert 1 1 0 \rangle\right) \nonumber \\
&\stackrel{\rm CNOT_{13}}{\longrightarrow}&
\frac{1}{\sqrt{2}}\left(\vert 0 0 0 \rangle -
\vert 1 1 1 \rangle \right)
\end{eqnarray}
All the pulses for the three gates used for GHZ state
construction were designed using the GRAPE algorithm and had
a fidelity $\ge$ 0.995. The GRAPE pulse duration
corresponding to the gate $U_{G_1}$ is $600$ $\mu$s, while
the $U_{G_2}$ and $U_{G_3}$ gates had pulse durations of
$24$ ms. The GHZ-type state was prepared with a fidelity of
0.97. The W state was prepared from the initial $\vert 000 \rangle$ by a
sequence of four unitary operations (labeled as $U_{W_1}, U_{W_2}, 
U_{W_3}, U_{W_4}$ in Fig.~\ref{ckt_3qchap}(b) and the sequential
gate operation leads to:
\begin{eqnarray}
\vert 000 \rangle &
\stackrel{R^{1}\left({\pi}\right)_y}{\longrightarrow} &
\vert 100 \rangle \nonumber \\
 & \stackrel{\rm R^{2}\left({0.39\pi}\right)_y}
{\longrightarrow} &
\sqrt{\frac{2}{3}}
\vert 100 \rangle + \frac{1}{\sqrt{3}} \vert 110 \rangle
\nonumber \\
& \stackrel{\rm CNOT_{21}} {\longrightarrow} &
\sqrt{\frac{2}{3}} \vert 100
\rangle +
\frac{1}{\sqrt{3}}  \vert 010 \rangle
\nonumber \\
& \stackrel{\rm
CR_{13}{\left(\frac{\pi}{2}\right)_y}}{\longrightarrow} &
\frac{1}{\sqrt{3}} [\vert 100 \rangle + 
\vert 101 \rangle + \vert 010 \rangle]
\nonumber \\
& \stackrel{\rm CNOT_{31}} {\longrightarrow} &
\frac{1}{\sqrt{3}}[
\vert 100 \rangle +  \vert
001 \rangle +
\vert 010 \rangle] 
\label{weqn_3q}
\end{eqnarray}
The different unitaries were individually optimized using
GRAPE and the  pulse duration for $U_{W_1}$, $U_{W_2}$, $U_{W_3}$, and
$U_{W_4}$ turned out to be $600 \mu $s, $24 $ms, $16 $ms, and $20 $ms,
respectively and the fidelity of the final state was
estimated to be 0.94.

The ${\rm W\bar{W}}$ state was constructed by
applying the following sequence of gate operations on  
the $\vert000\rangle$ state:
\begin{eqnarray}
\vert 000 \rangle &
\stackrel{R^{1}\left(\frac{\pi}{3}\right)_{-y}}{\longrightarrow}
&
\frac{\sqrt{3}}{2}\vert 000 \rangle-\frac{1}{2}\vert 100
\rangle \nonumber \\
 & \stackrel{\scriptstyle {\rm
CR}_{12}\left(0.61\pi\right)_y}
{\longrightarrow} &
\frac{\sqrt{3}}{2}\vert 0 0 0 \rangle -
\frac{1}{2\sqrt{3}} \vert 1 0 0 \rangle -
\sqrt{\frac{1}{6}} \vert 1 1 0 \rangle \nonumber \\
& \stackrel{\scriptstyle {\rm
CR}_{21}\left(\frac{\pi}{2}\right)_{-y}}
{\longrightarrow} &
\frac{1}{2}(\sqrt{3}\vert 0 0 0 \rangle -
\frac{1}{\sqrt{3}} (\vert 1 0 0 \rangle +
\vert 1 1 0 \rangle +  \nonumber \\
& & \vert 0 1 0 \rangle))    \nonumber \\
& \stackrel{\rm CNOT_{13}}{\longrightarrow} &  \frac{1}{2}(
\sqrt{3}\vert 0 0 0 \rangle -
\frac{1}{\sqrt{3}} (\vert 1 0 1 \rangle
+ \vert 1 1 1 \rangle + \nonumber \\
& &\vert 0 1 0 \rangle)) \nonumber \\
& \stackrel{\rm CNOT_{23}}{\longrightarrow} &  \frac{1}{2}(
\sqrt{3}\vert 0 0 0 \rangle -
\frac{1}{\sqrt{3}} (\vert 1 0 1 \rangle
+ \vert 1 1 0 \rangle + \nonumber \\
& & \vert 0 1 1 \rangle))\nonumber \\
& \stackrel{\rm
R^{123}\left(\frac{\pi}{2}\right)_y}{\longrightarrow} &
\frac{1}{\sqrt{6}}(
\vert 0 0 1 \rangle + \vert 0 1 0 \rangle +
\vert 0 1 1 \rangle + \nonumber \\
& &  \vert 1 0 0 \rangle + \vert 1 0 1 \rangle + \vert 1 1 0
\rangle)
\label{wwbareqn_3q}
\end{eqnarray}
The unitary operator for the entire preparation
sequence (labeled $U_{W\bar{W}}$ in Fig.~\ref{ckt_3qchap}(c))
comprising a spin-selective rotation operator:~two
controlled-rotation gates, two controlled-NOT gates
and one non-selective rotation by $\frac{\pi}{2}$ on all
the three qubits,
was created by a specially crafted single
GRAPE pulse (of pulse length $48$ms) and applied to the initial state
$\vert000\rangle$. The final state had a computed fidelity of
0.95.
\subsection{Decay of tripartite entanglement}
\label{decay_3q}
\begin{figure}[H]
\centering
\includegraphics[angle=0,scale=1.35]{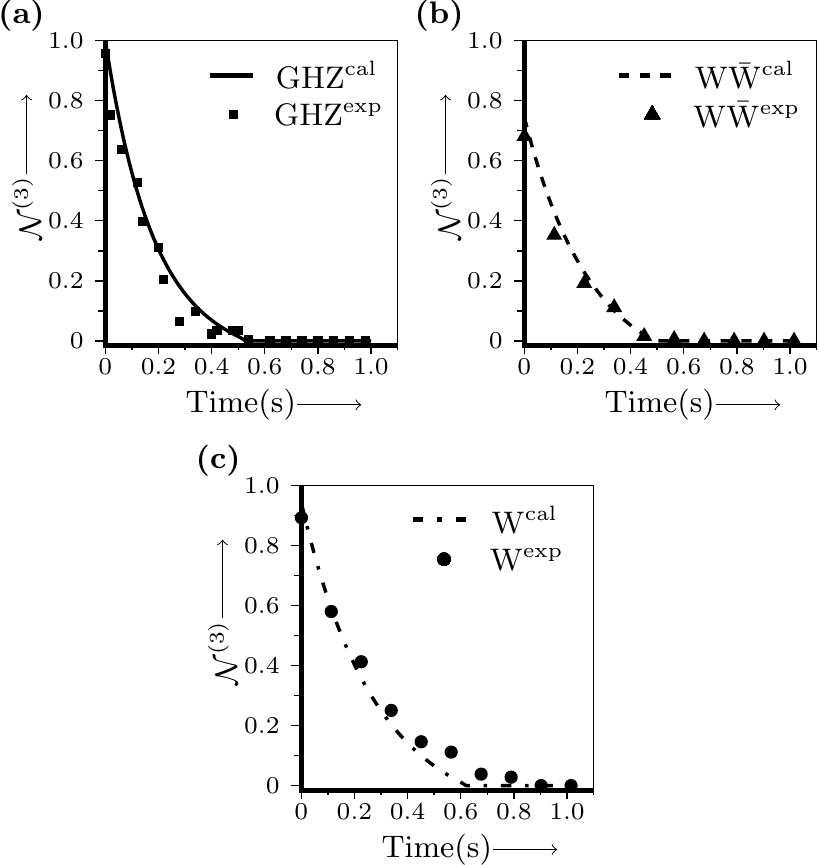}
\caption{Time dependence of the tripartite negativity
${\cal N}^{(3)}$ for the three-qubit system initially
experimentally 
prepared in the (a) GHZ state (squares) (b) W state (circles) 
and (c) ${\rm W {\bar W}}$
state (triangles) (the superscript ${\rm exp}$ denotes
``experimental data''). 
The fits are the calculated decay of negativity ${\cal N}^{(3)}$ of
the GHZ state (solid line), the $W\bar{W}$ state (dashed line) and 
the W state (dotted-dashed line), 
under the action of the modeled NMR noise 
channel (the superscript ${\rm cal}$ denotes ``calculated fit''). 
The W state is most robust against the NMR noise
channel, whereas
the GHZ state is most fragile.}
\label{3qdecay_3qchap}
\end{figure}

\begin{figure}[H]
\centering
\includegraphics[angle=0,scale=1.35]{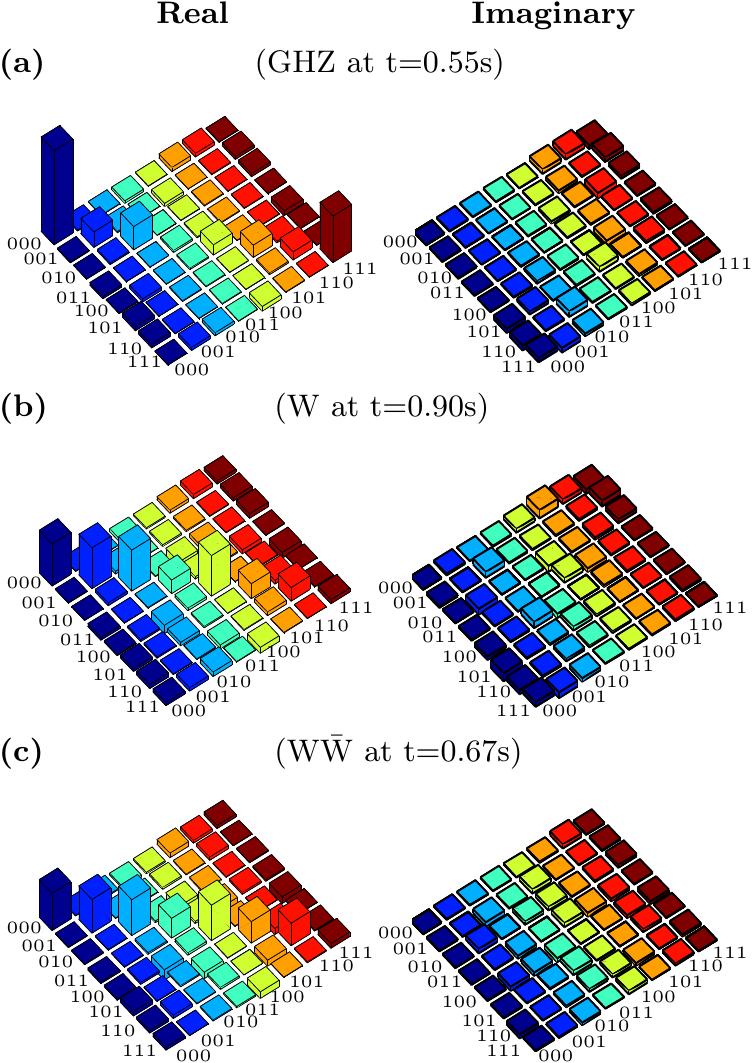}
\caption{The real (left) and imaginary (right) parts of the
experimentally tomographed 
density matrix of the state at the
time instances when 
the tripartite negativity ${\cal N}^{(3)}_{123}$ approaches
zero for the (a) GHZ state at $t=0.55$ s (b) 
W state at $t=0.90$ s and (c) 
${\rm W\bar{W}}$ state at $t=0.67$ s.
The rows and columns encode the
computational basis in binary order, from
$\vert 000 \rangle$ to $\vert 111 \rangle$.}
\label{tomodecay_3qchap}
\end{figure}

We next turn to the dynamics of tripartite entanglement under decoherence
channels acting on the system. We studied the time evolution of the 
tripartite negativity ${\cal N}^{(3)}_{123}$ for the tripartite entangled 
states, as computed from the experimentally
reconstructed density matrices at each time instant.  The experimental results
are depicted in Fig~\ref{3qdecay_3qchap} (a), (b) and (c) for the GHZ state, the ${\rm
W \bar{W}}$ state, and the W state, respectively.  Of the three entangled
states considered in this study, the GHZ and W states are maximally entangled
and hence contain the most amount of tripartite negativity, while the ${\rm W
\bar{W}}$ state is not maximally entangled and hence has a lower tripartite
negativity value.  The experimentally prepared GHZ state initially has a ${\cal
N}^{(3)}_{123}$ of 0.96 (quite close to its theoretically expected value of
1.0).  The GHZ state decays rapidly, with its negativity approaching zero in
0.55 s.  The experimentally prepared ${\rm W \bar{W}}$ state initially has a
${\cal N}^{(3)}_{123}$ of 0.68 (close to its theoretically expected value of
0.74), with its negativity approaching zero at 0.67 s.  The experimentally
prepared W state initially has a ${\cal N}^{(3)}_{123}$ of 0.90 (quite close to
its theoretically expected value of 0.94).  The W state is quite long-lived,
with its entanglement persisting up to 0.9 s.  The tomographs of the
experimentally reconstructed density matrices of the GHZ, W and ${\rm W
\bar{W}}$ states at the time instances when the tripartite negativity parameter
${\cal N}^{(3)}_{123}$ approaches zero for each state, are displayed in
Fig.~\ref{tomodecay_3qchap}.
We explored
the noise channels acting on our three-qubit
NMR entangled states which best fit our experimental data,
by analytically solving a master equation in
the Lindblad form, along the lines suggested
in Reference~\cite{jung-pra-08}.
The master equation is given by Eq.~(\ref{mastereqn_3qchap}).
We consider a decoherence model wherein a nuclear spin is acted on by two noise
channels namely a phase damping channel (described by the T$_2$ relaxation in
NMR) and a generalized amplitude damping  channel (described by the T$_1$
relaxation in NMR)~\cite{childs-pra-03}.  As the fluorine spins  in our
three-qubit system have widely differing chemical shifts, we assume that each
qubit interacts independently with its own environment.  The experimentally
determined T$_1$ NMR relaxation rates are T$_1^{1F}=5.42\pm 0.07$ s,
T$_1^{2F}=5.65\pm 0.05$ s and T$_1^{3F}=4.36\pm 0.05$ s, respectively.  The
T$_2$ relaxation rates were experimentally measured by first rotating the spin
magnetization into the transverse plane by a $90^{\circ}$ rf pulse followed by
a delay and fitting the resulting magnetization decay.  The experimentally
determined T$_2$ NMR relaxation rates are T$_2^{1F}=0.53\pm 0.02$s,
T$_2^{2F}=0.55\pm 0.02$ s,  and T$_2^{3F}=0.52\pm 0.02$ s, respectively.  We
solved the master equation (Eq.~(\ref{mastereqn_3qchap})) for the GHZ, W and ${\rm
W\bar{W}}$ states with the Lindblad operators $L_{i,x} \equiv
\sqrt{\frac{\kappa_{i,x}}{2}}\sigma^{(i)}_{x}$ and $L_{i,z} \equiv
\sqrt{\frac{\kappa_{i,z}}{2}} \sigma^{(i)}_{z}$, where
$\kappa_{i,x}=\frac{1}{T_1^i}$ and $\kappa_{i,z}=\frac{1}{T_2^i}$.

\noindent{ Under the simultaneous action of all the NMR noise
channels, the GHZ state decoheres as:}
\begin{equation} 
\rho_{GHZ}=\left( \begin{array}{cccccccc} \alpha_1  & 0 & 0
& 0 & 0 & 0 & 0 & \beta_1 \\ 0 & \alpha_2  & 0 & 0 & 0 & 0 & \beta_2 & 0 \\ 0 &
0 & \alpha_3  & 0 & 0 & \beta_3 & 0 & 0 \\ 0 & 0 & 0 & \alpha_4  & \beta_4 & 0
& 0 & 0 \\ 0 & 0 & 0 & \beta_4 & \alpha_4  & 0 & 0 & 0 \\ 0 & 0 & \beta_3 & 0 &
0 & \alpha_3  & 0 & 0 \\ 0 & \beta_2 & 0 & 0 & 0 & 0 & \alpha_2  & 0 \\ \beta_1
& 0 & 0 & 0 & 0 & 0 & 0 & \alpha_1 \\ \end{array} \right) 
\end{equation}
where 
\begin{eqnarray} 
\alpha_1 &=&
\frac{1}{8}(1+e^{-(\kappa_{x,1}+\kappa_{x,2})t}+e^{-(\kappa_{x,1}+\kappa_{x,3})t}
+e^{-(\kappa_{x,2}+\kappa_{x,3})t}) 
\nonumber \\ \alpha_2 &=&
\frac{1}{8}(1+e^{-(\kappa_{x,1}+\kappa_{x,2})t}-e^{-(\kappa_{x,1}+\kappa_{x,3})t}
-e^{-(\kappa_{x,2}+\kappa_{x,3})t}) \nonumber \\ \alpha_3 &=&
\frac{1}{8}(1-e^{-(\kappa_{x,1}+\kappa_{x,2})t}+e^{-(\kappa_{x,1}+\kappa_{x,3})t}
-e^{-(\kappa_{x,2}+\kappa_{x,3})t}) \nonumber \\ 
\alpha_4 &=&
\frac{1}{8}(1-e^{-(\kappa_{x,1}+\kappa_{x,2})t}-e^{-(\kappa_{x,1}+\kappa_{x,3})t}
+e^{-(\kappa_{x,2}+\kappa_{x,3})t}) \nonumber \\ 
\beta_1 &=&
\frac{1}{8}(e^{-(\kappa_{1,x}+\kappa_{2,x}+\kappa_{3,x}+\kappa_{1,z}+\kappa_{2,z}+\kappa_{3,z})t}
\nonumber \\ &&
\quad(e^{\kappa_{1,x}t}+ e^{\kappa_{2,x}t}+e^{\kappa_{3,x}t}
+e^{(\kappa_{1,x}+\kappa_{2,x}+\kappa_{3,x})t}) \nonumber \\
\beta_2 &=&
\frac{1}{8}(e^{-(\kappa_{1,x}+\kappa_{2,x}+\kappa_{3,x}+\kappa_{1,z}+\kappa_{2,z}+\kappa_{3,z})t}
\nonumber \\ &&
\quad (-e^{\kappa_{1,x}t}-e^{\kappa_{2,x}t}+e^{\kappa_{3,x}t}
+e^{(\kappa_{1,x}+\kappa_{2,x}+\kappa_{3,x})t}) \nonumber \\
\end{eqnarray}
\begin{eqnarray} 
\beta_3 &=&
\frac{1}{8}(e^{-(\kappa_{1,x}+\kappa_{2,x}+\kappa_{3,x}+\kappa_{1,z}+\kappa_{2,z}+\kappa_{3,z})t}
\nonumber \\ 
\quad &&(-e^{\kappa_{1,x}t}+ e^{\kappa_{2,x}t}-e^{\kappa_{3,x}t}
+e^{(\kappa_{1,x}+\kappa_{2,x}+\kappa_{3,x})t}) \nonumber \\
\beta_4 &=&
\frac{1}{8}(e^{-(\kappa_{1,x}+\kappa_{2,x}+\kappa_{3,x}+\kappa_{1,z}+\kappa_{2,z}+\kappa_{3,z})t}
\nonumber \\ &&
\quad (e^{\kappa_{1,x}t}- e^{\kappa_{2,x}t}-e^{\kappa_{3,x}t}
+e^{(\kappa_{1,x}+\kappa_{2,x}+\kappa_{3,x})t}) \nonumber \\
\end{eqnarray}

Under the simultaneous action of all the NMR noise channels, the W
state decoheres as:
\begin{equation}
\rho_{W} = \left(
\begin{array}{cccccccc}
 \alpha_{1}  & 0 & 0 & \beta_{1} & 0 & \beta_{5} & \beta_{1} &0 \\
 0 & \alpha_{2}  & \beta_{2} & 0 & \beta_{6} & 0 & 0& \beta_{10} \\
 0 & \beta_{2} & \alpha_{3}  & 0 & \beta_{11}& 0 & 0 & \beta_{7} \\
 \beta_{1} & 0 & 0 & \alpha_{4}  & 0 & \beta_{12} & \beta_{8} & 0 \\
 0 & \beta_{6} & \beta_{11} & 0 & \alpha_{5}  & 0 & 0 & \beta_{3} \\
 \beta_{5} & 0 & 0 & \beta_{12} & 0 & \alpha_{6}  & \beta_{4} & 0 \\
 \beta_{1} & 0 & 0 &\beta_{8} & 0 & \beta_{4} & \alpha_{7}  & 0 \\
0 & \beta_{10}&\beta_{7} & 0 & \beta_{3} & 0 & 0 & \alpha_{8} \\
\end{array}
\right) \nonumber \\
\end{equation}
Where
\begin{eqnarray}
\alpha_{1}&=&\frac{1}{8}- \frac{1}{24} e^{-(\kappa_{x,1}+\kappa_{x,2}+\kappa_{x,3})t} (3+e^{\kappa_{x,1}t} 
 +e^{\kappa_{x,2}t}- 
\nonumber \\ &&
e^{(\kappa_{x,1}+\kappa_{x,2})t}+e^{\kappa_{x,3}t} 
 - e^{(\kappa_{x,1}+\kappa_{x,3})t} - e^{(\kappa_{x,2}+\kappa_{x,3})t}) \nonumber \\
\alpha_{2}&=&\frac{1}{8}+ \frac{1}{24} e^{-(\kappa_{x,1}+\kappa_{x,2}+\kappa_{x,3})t} (3+e^{\kappa_{x,1}t} 
+e^{\kappa_{x,2}t}-
\nonumber \\ && 
e^{(\kappa_{x,1}+\kappa_{x,2})t}-e^{\kappa_{x,3}t}
+ e^{(\kappa_{x,1}+\kappa_{x,3})t} +
e^{(\kappa_{x,2}+\kappa_{x,3})t}) \nonumber \\ 
\alpha_{3}&=&\frac{1}{8}+ \frac{1}{24}
e^{-(\kappa_{x,1}+\kappa_{x,2}+\kappa_{x,3})t}
(3+e^{\kappa_{x,1}t} 
-e^{\kappa_{x,2}t}
\nonumber \\ && 
+ e^{(\kappa_{x,1}+\kappa_{x,2})t}+e^{\kappa_{x,3}t} 
- e^{(\kappa_{x,1}+\kappa_{x,3})t} + e^{(\kappa_{x,2}+\kappa_{x,3})t}) \nonumber \\
\alpha_{4}&=&\frac{1}{8}- \frac{1}{24}
e^{-(\kappa_{x,1}+\kappa_{x,2}+\kappa_{x,3})t}
(3+e^{\kappa_{x,1}t} 
 -e^{\kappa_{x,2}t}
\nonumber \\ && 
+ e^{(\kappa_{x,1}+\kappa_{x,2})t}-e^{\kappa_{x,3}t}
+ e^{(\kappa_{x,1}+\kappa_{x,3})t} - e^{(\kappa_{x,2}+\kappa_{x,3})t}) 
\nonumber \\
\alpha_{5}&=&\frac{1}{8}+ \frac{1}{24} e^{-(\kappa_{x,1}+\kappa_{x,2}+\kappa_{x,3})t} (3-e^{\kappa_{x,1}t} 
+e^{\kappa_{x,2}t}
\nonumber \\ && 
+ e^{(\kappa_{x,1}+\kappa_{x,2})t}+e^{\kappa_{x,3}t} 
+ e^{(\kappa_{x,1}+\kappa_{x,3})t} - e^{(\kappa_{x,2}+\kappa_{x,3})t}) \nonumber \\
\end{eqnarray}
\begin{eqnarray}
\alpha_{6}&=&\frac{1}{8}+ \frac{1}{24} e^{-(\kappa_{x,1}+\kappa_{x,2}+\kappa_{x,3})t} (-3+e^{\kappa_{x,1}t} 
-e^{\kappa_{x,2}t}
\nonumber \\ && 
- e^{(\kappa_{x,1}+\kappa_{x,2})t}+e^{\kappa_{x,3}t} 
+ e^{(\kappa_{x,1}+\kappa_{x,3})t} - e^{(\kappa_{x,2}+\kappa_{x,3})t}) \nonumber \\ 
\alpha_{7}&=&\frac{1}{8}+ \frac{1}{24} e^{-(\kappa_{x,1}+\kappa_{x,2}+\kappa_{x,3})t} (-3+e^{\kappa_{x,1}t}+ 
e^{\kappa_{x,2}t}
\nonumber \\ &&
+ e^{(\kappa_{x,1}+\kappa_{x,2})t}-e^{\kappa_{x,3}t} 
- e^{(\kappa_{x,1}+\kappa_{x,3})t} - e^{(\kappa_{x,2}+\kappa_{x,3})t}) \nonumber \\
\alpha_{8}&=&\frac{1}{8}- \frac{1}{24} e^{-(\kappa_{x,1}+\kappa_{x,2}+\kappa_{x,3})t} (-3+e^{\kappa_{x,1}t}+ 
e^{\kappa_{x,2}t}
\nonumber \\ &&
+ e^{(\kappa_{x,1}+\kappa_{x,2})t}+e^{\kappa_{x,3}t} 
+ e^{(\kappa_{x,1}+\kappa_{x,3})t} + e^{(\kappa_{x,2}+\kappa_{x,3})t}) \nonumber 
\end{eqnarray}
\begin{eqnarray}
\beta_{1}&=&\frac{1}{12} (e^{-(\kappa_{x,1}+\kappa_{x,2}+\kappa_{x,3}+\kappa_{z,2}+\kappa_{z,3})t}\nonumber \\
&&(1+e^{(\kappa_{x,1})t})(-1+e^{(\kappa_{x,2}+\kappa_{x,3})t}))\nonumber \\
\beta_{2}&=&\frac{1}{12} (e^{-(\kappa_{x,1}+\kappa_{x,2}+\kappa_{x,3}+\kappa_{z,2}+\kappa_{z,3})t}\nonumber \\
&&(1+e^{(\kappa_{x,1})t})(1+e^{(\kappa_{x,2}+\kappa_{x,3})t}))\nonumber \\
\beta_{3}&=&\frac{1}{12} (e^{-(\kappa_{x,1}+\kappa_{x,2}+\kappa_{x,3}+\kappa_{z,2}+\kappa_{z,3})t}\nonumber \\
&&(-1+e^{(\kappa_{x,1})t})(-1+e^{(\kappa_{x,2}+\kappa_{x,3})t}))\nonumber \\
\beta_{4}&=&\frac{1}{12} (e^{-(\kappa_{x,1}+\kappa_{x,2}+\kappa_{x,3}+\kappa_{z,2}+\kappa_{z,3})t}\nonumber \\
&&(-1+e^{(\kappa_{x,1})t})(1+e^{(\kappa_{x,2}+\kappa_{x,3})t}))\nonumber 
\end{eqnarray}
\begin{eqnarray}
\beta_{5}&=&\frac{1}{12} (e^{-(\kappa_{x,1}+\kappa_{x,2}+\kappa_{x,3}+\kappa_{z,1}+\kappa_{z,3})t}\nonumber \\
&&(1+e^{(\kappa_{x,2})t})(-1+e^{(\kappa_{x,1}+\kappa_{x,3})t}))\nonumber \\
\beta_{6}&=&\frac{1}{12} (e^{-(\kappa_{x,1}+\kappa_{x,2}+\kappa_{x,3}+\kappa_{z,1}+\kappa_{z,3})t}\nonumber \\
&&(1+e^{(\kappa_{x,2})t})(1+e^{(\kappa_{x,1}+\kappa_{x,3})t}))\nonumber \\
\beta_{7}&=&\frac{1}{12} (e^{-(\kappa_{x,1}+\kappa_{x,2}+\kappa_{x,3}+\kappa_{z,1}+\kappa_{z,3})t}\nonumber \\
&&(-1+e^{(\kappa_{x,2})t})(-1+e^{(\kappa_{x,1}+\kappa_{x,3})t}))\nonumber \\
\beta_{8}&=&\frac{1}{12} (e^{-(\kappa_{x,1}+\kappa_{x,2}+\kappa_{x,3}+\kappa_{z,1}+\kappa_{z,3})t}\nonumber \\
&&(-1+e^{(\kappa_{x,2})t})(1+e^{(\kappa_{x,1}+\kappa_{x,3})t}))\nonumber \\
\beta_{9}&=&\frac{1}{12} (e^{-(\kappa_{x,1}+\kappa_{x,2}+\kappa_{x,3}+\kappa_{z,1}+\kappa_{z,2})t}\nonumber \\
&&(-1+e^{(\kappa_{x,1}+\kappa_{x,2})t})(1+e^{(\kappa_{x,3})t}))\nonumber \\
\end{eqnarray}
\begin{eqnarray}
\beta_{10}&=&\frac{1}{12} (e^{-(\kappa_{x,1}+\kappa_{x,2}+\kappa_{x,3}+\kappa_{z,1}+\kappa_{z,2})t}\nonumber \\
&&(-1+e^{(\kappa_{x,1}+\kappa_{x,2})t})(-1+e^{(\kappa_{x,3})t}))\nonumber \\
\beta_{11}&=&\frac{1}{12} (e^{-(\kappa_{x,1}+\kappa_{x,2}+\kappa_{x,3}+\kappa_{z,1}+\kappa_{z,2})t}\nonumber \\
&&(1+e^{(\kappa_{x,1}+\kappa_{x,2})t})(1+e^{(\kappa_{x,3})t}))\nonumber \\
\beta_{12}&=&\frac{1}{12} (e^{-(\kappa_{x,1}+\kappa_{x,2}+\kappa_{x,3}+\kappa_{z,1}+\kappa_{z,2})t}\nonumber \\
&&(1+e^{(\kappa_{x,1}+\kappa_{x,2})t})(-1+e^{(\kappa_{x,3})t}))
\end{eqnarray}

Under the simultaneous action of all the 
NMR noise channels, the ${\rm W \bar{W}}$
state decoheres as:
\begin{equation}
\rho_{{\rm W \bar{W}}} =\left(
\begin{array}{cccccccc}
 \alpha_{1}  & \beta_{1} & \beta_{2} & \beta_{3} &\beta_{4} & \beta_{5} & \beta_{6} &\beta_{7} \\
  \beta_{1}  & \alpha_{2}  & \beta_{8} &\beta_{9} & \beta_{10} & \beta_{11} & \beta_{12} & \beta_{13} \\
  \beta_{2}  & \beta_{8} & \alpha_{3}  &  \beta_{14}  & \beta_{15}&  \beta_{16}  &  \beta_{11} & \beta_{5} \\
 \beta_{3} &  \beta_{9}  & \beta_{14}  & \alpha_{4}  &  \beta_{17}  & \beta_{15} & \beta_{10} &  \beta_{4} \\
  \beta_{4}  & \beta_{10} & \beta_{15} &  \beta_{17}  & \alpha_{4}  & \beta_{15}  &  \beta_{9}  & \beta_{18} \\
 \beta_{5} &  \beta_{11}  &  \beta_{16}  & \beta_{15} & \beta_{15}  & \alpha_{3}  & \beta_{8} &  \beta_{2}  \\
 \beta_{6} &  \beta_{12}  &  \beta_{11} &\beta_{10} &  \beta_{9} & \beta_{8} & \alpha_{2}  &  \beta_{1}  \\
 \beta_{7}  & \beta_{13}&\beta_{5} &  \beta_{4}  & \beta_{18} &  \beta_{2} &  \beta_{1} & \alpha_{1} \\
\end{array}
\right) 
\end{equation}
where
\begin{eqnarray}
\alpha_{1}&=&\frac{1}{24}(3-e^{-(\kappa_{x,1}+\kappa_{x,2})t}-e^{-(\kappa_{x,1}+\kappa_{x,3})t} \nonumber \\
&&- e^{-(\kappa_{x,2}+\kappa_{x,3})t})  \nonumber \\
\alpha_{2}&=&\frac{1}{24}(3-e^{-(\kappa_{x,1}+\kappa_{x,2})t}+e^{-(\kappa_{x,1}+\kappa_{x,3})t} \nonumber \\
&& + e^{-(\kappa_{x,2}+\kappa_{x,3})t})  \nonumber \\
\alpha_{3}&=&\frac{1}{24}(3+e^{-(\kappa_{x,1}+\kappa_{x,2})t}-e^{-(\kappa_{x,1}+\kappa_{x,3})t} \nonumber \\
&& + e^{-(\kappa_{x,2}+\kappa_{x,3})t})  \nonumber \\
\alpha_{4}&=&\frac{1}{24}(3+e^{-(\kappa_{x,1}+\kappa_{x,2})t}+e^{-(\kappa_{x,1}+\kappa_{x,3})t} \nonumber \\
\end{eqnarray}
\begin{eqnarray}
&& - e^{-(\kappa_{x,2}+\kappa_{x,3})t})  \nonumber \\
\beta_{1}&=&\frac{1}{12} e^{-(\kappa_{x,1}+\kappa_{x,2}+2\kappa_{z,3})t} \nonumber \\
&&(e^{(\kappa_{x,1}+\kappa_{x,2}+\kappa_{z,3})t}-e^{ \kappa_{z,3} t}) \nonumber \\
\beta_{2}&=&\frac{1}{12} e^{-(\kappa_{x,1}+\kappa_{x,3}+2\kappa_{z,2})t} \nonumber \\
&&(e^{(\kappa_{x,1}+\kappa_{x,3}+\kappa_{z,2})t}-e^{ \kappa_{z,2} t}) \nonumber \\
\beta_{3}&=&\frac{1}{12} e^{-(\kappa_{x,2}+\kappa_{x,3}+2(\kappa_{z,2}+\kappa_{z,3}))t} \nonumber \\
&&(e^{(\kappa_{x,2}+\kappa_{x,3}+\kappa_{z,2}+\kappa_{z,3})t}-e^{ (\kappa_{z,2}+\kappa_{z,3}) t}) \nonumber \\
\beta_{4}&=&\frac{1}{12} e^{-(\kappa_{x,2}+\kappa_{x,3}+2\kappa_{z,1})t} \nonumber \\
&&(-e^{ \kappa_{z,1} t}+e^{(\kappa_{x,2}+\kappa_{x,3}+\kappa_{z,1})t}) \nonumber \\
\beta_{5}&=&\frac{1}{12} e^{-(\kappa_{x,1}+\kappa_{x,3}+2(\kappa_{z,1}+\kappa_{z,3}))t} \nonumber \\
&&(-e^{ (\kappa_{z,1}+\kappa_{z,3}) t}+e^{(\kappa_{x,1}+\kappa_{x,3}+\kappa_{z,1}+\kappa_{z,3})t}) \nonumber \\
\end{eqnarray}

\begin{eqnarray}
\beta_{6}&=&\frac{1}{12} e^{-(\kappa_{x,1}+\kappa_{x,2}+2(\kappa_{z,1}+\kappa_{z,2}))t} \nonumber \\
&&(-e^{ (\kappa_{z,1}+\kappa_{z,2}) t}+e^{(\kappa_{x,1}+\kappa_{x,2}+\kappa_{z,1}+\kappa_{z,2})t}) \nonumber \\
\beta_{7}&=&-\frac{1}{24} e^{-(\kappa_{x,1}+\kappa_{x,2}+\kappa_{x,3}+\kappa_{z,1}+\kappa_{z,2}+\kappa_{z,3})t} \nonumber \\
&& (e^{\kappa_{x,1} t}+e^{\kappa_{x,2} t}+e^{\kappa_{x,3} t}-3e^{(\kappa_{x,1}+\kappa_{x,2}+\kappa_{x,3})t}) \nonumber \\
\beta_{8}&=&\frac{1}{12} e^{-(\kappa_{x,2}+\kappa_{x,3}+2(\kappa_{z,2}+\kappa_{z,3}))t} \nonumber \\
&&(e^{ (\kappa_{z,2}+\kappa_{z,3}) t}+e^{(\kappa_{x,2}+\kappa_{x,3}+\kappa_{z,2}+\kappa_{z,3})t}) \nonumber \\
\beta_{9}&=&\frac{1}{12} e^{-(\kappa_{x,1}+\kappa_{x,3}+2\kappa_{z,2})t} \nonumber \\
&&(e^{ \kappa_{z,2} t}+e^{(\kappa_{x,1}+\kappa_{x,3}+\kappa_{z,2})t}) \nonumber \\
\beta_{10}&=&\frac{1}{12} e^{-(\kappa_{x,1}+\kappa_{x,3}+2(\kappa_{z,1}+\kappa_{z,3}))t} \nonumber \\
&&(e^{ (\kappa_{z,1}+\kappa_{z,3}) t}+e^{(\kappa_{x,1}+\kappa_{x,3}+\kappa_{z,1}+\kappa_{z,3})t}) \nonumber \\
\beta_{11}&=&\frac{1}{12} e^{-(\kappa_{x,2}+\kappa_{x,3}+2\kappa_{z,1})t}(e^{ \kappa_{z,1} t}+e^{(\kappa_{x,2}+\kappa_{x,3}+\kappa_{z,1})t}) \nonumber \\
\beta_{12}&=&\frac{1}{24} e^{-(\kappa_{x,1}+\kappa_{x,2}+\kappa_{x,3}+\kappa_{z,1}+\kappa_{z,2}+\kappa_{z,3})t} \nonumber \\
\end{eqnarray}
\begin{eqnarray}
&& (e^{\kappa_{x,1} t}+e^{\kappa_{x,2} t}-e^{\kappa_{x,3} t}+3e^{(\kappa_{x,1}+\kappa_{x,2}+\kappa_{x,3})t}) \nonumber \\
\beta_{13}&=&\frac{1}{12} e^{-(\kappa_{x,1}+\kappa_{x,2}+2(\kappa_{z,1}+\kappa_{z,2}))t} \nonumber \\
&&(-e^{ (\kappa_{z,1}+\kappa_{z,2}) t}+e^{(\kappa_{x,1}+\kappa_{x,2}+\kappa_{z,1}+\kappa_{z,2})t}) \nonumber \\
\beta_{14}&=&\frac{1}{12} e^{-(\kappa_{x,1}+\kappa_{x,2}+2\kappa_{z,3})t} \nonumber \\
&&(e^{ \kappa_{z,3} t}+e^{(\kappa_{x,1}+\kappa_{x,2}+\kappa_{z,3})t}) \nonumber \\
\beta_{15}&=&\frac{1}{12} e^{-(\kappa_{x,1}+\kappa_{x,2}+2(\kappa_{z,1}+\kappa_{z,2}))t} \nonumber \\
&&(e^{ (\kappa_{z,1}+\kappa_{z,2}) t}+e^{(\kappa_{x,1}+\kappa_{x,2}+\kappa_{z,1}+\kappa_{z,2})t}) \nonumber \\
\beta_{16}&=&\frac{1}{24} e^{-(\kappa_{x,1}+\kappa_{x,2}+\kappa_{x,3}+\kappa_{z,1}+\kappa_{z,2}+\kappa_{z,3})t} \nonumber \\
&& (e^{\kappa_{x,1} t}-e^{\kappa_{x,2} t}+e^{\kappa_{x,3} t}+3e^{(\kappa_{x,1}+\kappa_{x,2}+\kappa_{x,3})t}) \nonumber \\
\beta_{17}&=&\frac{1}{24} e^{-(\kappa_{x,1}+\kappa_{x,2}+\kappa_{x,3}+\kappa_{z,1}+\kappa_{z,2}+\kappa_{z,3})t} \nonumber \\
&& (-e^{\kappa_{x,1} t}+e^{\kappa_{x,2} t}+e^{\kappa_{x,3} t}+3e^{(\kappa_{x,1}+\kappa_{x,2}+\kappa_{x,3})t}) \nonumber \\
\beta_{18}&=&\frac{1}{12} e^{-(\kappa_{x,2}+\kappa_{x,3}+2(\kappa_{z,2}+\kappa_{z,3}))t} \nonumber \\
&&(e^{ (\kappa_{z,2}+\kappa_{z,3}) t}+e^{(\kappa_{x,2}+\kappa_{x,3}+\kappa_{z,2}+\kappa_{z,3})t}) 
\end{eqnarray}
With this
model, the GHZ  state decays at the rate $\gamma^{al}_{GHZ}=6.33\pm 0.06
s^{-1}$, and its entanglement approaches zero in 0.53 s.  The $W\bar{W}$ state
decays at the rate  $\gamma^{al}_{W\bar{W}}=5.90 \pm 0.10  s^{-1}$, and its
entanglement approaches zero in 0.50 s.  The $W$ state decays at the rate
$\gamma^{al}_{W}=4.84\pm 0.07 s^{-1}$, and its entanglement approaches zero in
0.62 s. Solving the master equation (Eq.~(\ref{mastereqn_3qchap})) ensures that the
off-diagonal elements of the corresponding $\rho$ matrices satisfy
a set of coupled equations, from which the explicit values of 
$\alpha$s and $\beta$s can be computed. The equations are solved in
the high-temperature limit. For an ensemble of NMR spins at room temperature 
this implies that the energy $E << k_{B} T$ where $k_B$ is the Boltzmann
constant and $T$ refers to the temperature, ensuring a Boltzmann
distribution of spin populations at thermal equilibrium. The results of
the analytical calculation and the experimental data match well, as shown in
Fig.~\ref{3qdecay_3qchap}.

\section{Protecting three-qubit entanglement via dynamical
decoupling}
\label{ddprotect_3q}
As the tripartite entangled states under investigation 
are robust against noise to varying extents, we wanted to
discover if either the amount of entanglement in these
states could be protected or their entanglement could be
preserved for longer times, using dynamical decoupling (DD)
protection schemes. 
While DD sequences are effective
in decoupling system-environment interactions, often errors
in their implementation arise either due to errors in the
pulses or errors due to off-resonant driving~\cite{suter-review}.
\begin{figure}[H]
\centering
\includegraphics[angle=0,scale=1.4]{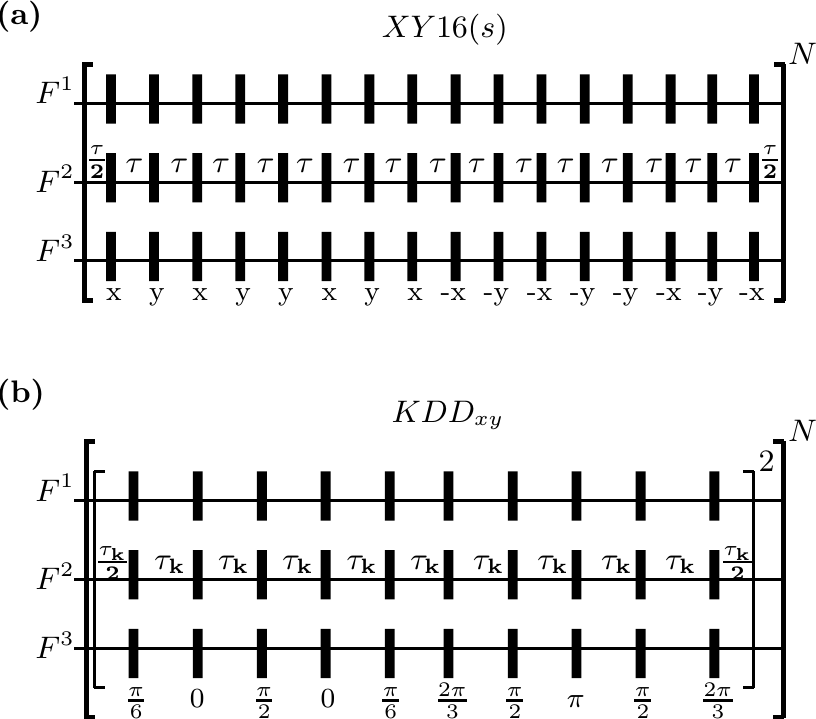}
\caption{NMR pulse sequence corresponding to (a) 
XY-16(s)
and (b) KDD$_{xy}$ DD schemes 
(the superscript 2 implies that  
the set of pulses inside the bracket is applied twice, to
form one cycle of the DD scheme).  The
pulses represented by black filled rectangles (in both
schemes) are of angle
$\pi$, and are applied simultaneously on all 
three qubits (denoted by $F^{i}, i=1,2,3$).
The angle below each pulse denotes the phase with which it
is applied. Each DD cycle is repeated $N$ times, with
$N$ large to achieve good system-bath decoupling.}
\label{dd-fig_3qchap}
\end{figure}
Two approaches have been used to design robust DD sequences
which are impervious to pulse imperfections:
the first approach replaces the $\pi$ rotation
pulses with composite pulses inside the DD sequence, while
the second approach focuses on 
optimizing phases of the pulses in the DD sequence. In
this work, we use DD sequences 
that use pulses with phases applied along
different rotation axes: the XY-16(s) and the 
Knill Dynamical Decoupling (KDD)
schemes~\cite{souza-pra-12}.
In conventional DD schemes the $\pi$ pulses are applied
along one axis (typically $x$) and as a consequence, only
the coherence along that axis is well protected. The XY
family of DD schemes applies pulses along two perpendicular
($x,y$) axes, which protects coherence equally along both
these axes~\cite{souza-phil}.
The XY-16(s) sequence is 
constructed by combining an XY-8(s) cycle with
its phase-shifted copy,
where the (s) denotes the ``symmetric'' version i.e. 
the cycle is time-symmetric with respect to its center. The
XY-8 cycle is itself created by combining a 
basic XY-4
cycle with its time-reversed copy. One full unit cycle of
the XY-16(s) sequence comprises sixteen $\pi$ pulses
interspersed with free evolution time periods, and
each cycle
is repeated $N$ times for better decoupling.
The KDD sequence 
has  additional phases which further
symmetrize pulses in the $x-y$ plane and compensate
for pulse errors; each $\pi$ pulse
in a basic XY DD sequence is replaced by five $\pi$ pulses,
each of a different 
phase~\cite{ryan-prl-10,souza-prl-11}:
\begin{equation}
{\rm KDD}_{\phi} \equiv
(\pi)_{\frac{\pi}{6}+\phi}-
(\pi)_{\phi} -(\pi)_{\frac{\pi}{2}+\phi}-(\pi)_{\phi}
-(\pi)_{\frac{\pi}{6}+\phi}
\label{basickdd_3qchap}
\end{equation}
where $\phi$ denotes the phase of the pulse; we set
$\phi=0$ in our experiments. The KDD$_{\phi}$ sequence of
five pulses given in Eq.~(\ref{basickdd_3qchap}) 
protects coherence along only one axis. To
protect coherences along both the $(x,y)$ axes, 
we use the KDD$_{xy}$ sequence, which combines two
basic five-pulse blocks shifted in phase by $\pi/2$ i.e
$[{\rm KDD}_{\phi} - {\rm KDD}_{\phi+\pi/2}]$.
One unit cycle of the KDD$_{xy}$ sequence contains two
of these pulse-blocks shifted in phase, for a total
of twenty $\pi$ pulses.
The XY-16(s) and KDD$_{xy}$ DD sequences  are given
in Figs.~\ref{dd-fig_3qchap}(a) and (b) respectively, where the
black filled rectangles represent $\pi$ pulses on all three
qubits and $\tau$ ($\tau_k$) indicates a free evolution
time period.  We note here that the 
chemical shifts of the
three fluorine qubits 
in our particular
molecule cover a very large frequency
bandwidth, 
making it difficult to implement an accurate non-selective
pulse simultaneously on all  the qubits.  To circumvent this
problem, we crafted a special excitation pulse of duration $\approx
400 \mu$s  consisting of a set of three Gaussian shaped pulses
that are applied at different spin frequency offsets and are
frequency modulated to achieve simultaneous
excitation~\cite{das-pra-15}.

\begin{figure}[H]
\centering
\includegraphics[angle=0,scale=1.4]{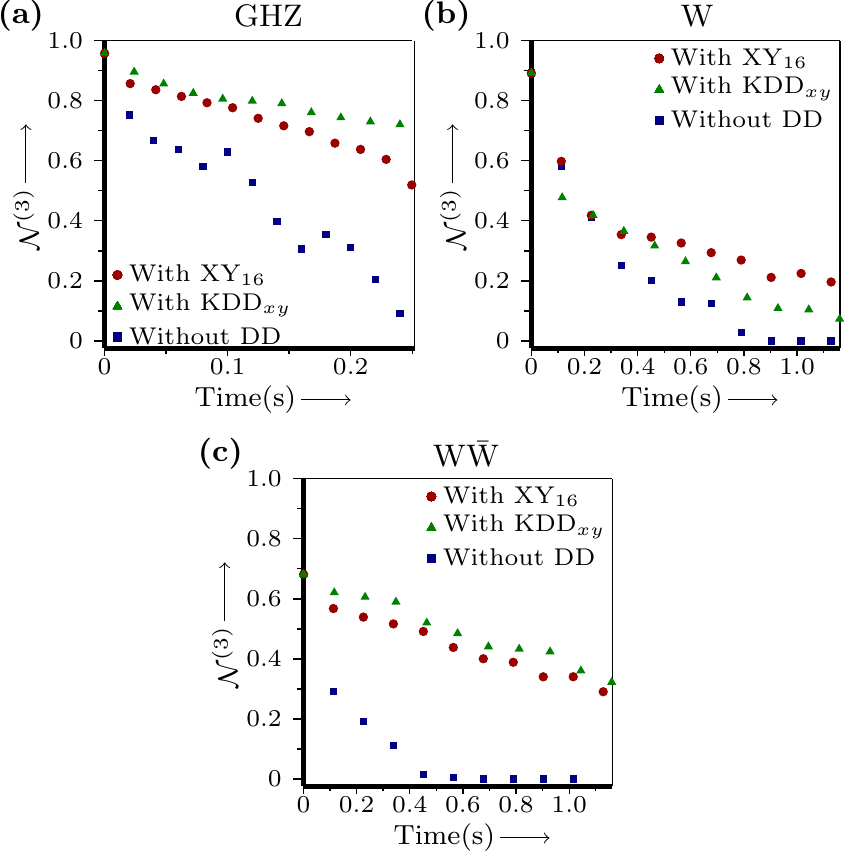}
\caption{Plot of the tripartite  negativity (${\cal N}^{(3)}_{123}$) with
time, computed for the (a) GHZ-type state, (b) W state
and (c) $W\bar{W}$ state. The negativity was computed for
each state without applying any protection and after
applying the XY-16(s) and KDD$_{xy}$ dynamical decoupling
sequences.Note that the time scale for part (a) is
different from (b) and (c)}
\label{entang-dd_3q}
\end{figure}

Figs.\ref{entang-dd_3q}(a),(b) and (c) show the results 
of protecting the GHZ, W and ${\rm W\bar{W}}$ states
respectively, using the XY-16(s) and the KDD$_{xy}$
DD sequences.  

\noindent{\bf GHZ state protection:}
The XY-16(s) protection scheme was implemented on the GHZ state with
an inter-pulse delay of $\tau = 0.25$ ms and one run of the
sequence took $10.40$ ms (including the length of
the sixteen $\pi$ pulses). The value of the negativity
${\cal N}^3_{123}$ 
remained close to 
0.80 and 0.52 for up to $80$ms and $240$ ms respectively
when XY-16 protection was applied, while 
for the unprotected state 
the state fidelity is
quite low and 
${\cal N}^3_{123}$ decayed  to 
a low value of 0.58 and 0.09  at $80$ms and $240$ ms, respectively
(Fig.~\ref{entang-dd_3q}(a)).  The KDD$_{xy}$ protection scheme on this
state was implemented with an inter-pulse delay $\tau_k
=0.20$ ms and one run of the sequence took $12$ ms
(including the length of the twenty $\pi$ pulses).  The value of
the negativity ${\cal N}^3_{123}$ remained close to 0.80 and 0.72 for
up to $140$ms and $240$ ms when KDD$_{xy}$ protection was applied
(Fig.~\ref{entang-dd_3q}(a)).

\noindent{\bf W state protection:}
The XY-16(s) protection scheme was implemented on the
W state with an
inter-pulse delay $\tau = 3.12$ ms 
and one run of the
sequence took $56.40$ ms (including the length of the
sixteen $\pi$ pulses). 
The value of the negativity ${\cal
N}^3_{123}$ remained close to 0.30 for up to $0.68$ s when XY-16
protection was applied, whereas ${\cal N}^3_{123}$ reduced 
to 0.1
at $0.68$ s 
 when no state protection is applied
(Fig.~\ref{entang-dd_3q}(b)).
The KDD$_{xy}$ protection scheme was implemented
on the W state with an inter-pulse delay 
$\tau_k = 2.5$ ms and one run of the
sequence took $58$ ms (including the length of the
twenty $\pi$ pulses). 
The value of the negativity ${\cal
N}^3_{123}$ remained close to 0.21 for upto $0.70$ s when KDD$_{xy}$
protection was applied 
(Fig.~\ref{entang-dd_3q}(b)).

\noindent{\bf ${\rm W \bar{W}}$ state protection:}
The XY-16(s) protection sequence was
implemented on the ${\rm W \bar{W}}$ state
with an inter-pulse delay of $\tau = 3.12$ ms and one run of the
sequence took $56.40$ ms (including the length of the
sixteen $\pi$ pulses). 
The value of the negativity ${\cal
N}^3_{123}$ remained close to  0.5 for upto $0.45$ s when
XY-16(s) protection was applied, whereas ${\cal N}^3_{123}$ 
reduced almost to zero ($\approx 0.02$)
at $0.45$ s 
when no protection was applied
(Fig.~\ref{entang-dd_3q}(c)).
The KDD$_{xy}$ protection sequence was
applied with an inter-pulse delay of 
$\tau_k = 2.5$ ms and one run of the
sequence took $58$ ms (including the length of the
twenty $\pi$ pulses). 
The value of the negativity ${\cal
N}^3_{123}$ remained close to 0.52 for upto $0.46$ s when KDD$_{xy}$
protection was applied
(Fig.~\ref{entang-dd_3q}(c)).

The results of UDD-type of protection summarized
above demonstrate that state protection worked to varying
degrees and protected the entanglement of the tripartite
entangled states to different extents, depending on the
type of state to be protected. The GHZ state showed maximum
protection and the ${\rm W\bar{W}}$ state also showed
a significant amount of protection, while the W state
showed a marginal improvement under
protection.
We note here that the lifetime of
the GHZ state is not significantly enhanced by using
DD state protection; what is noteworthy is that state
fidelity remains high (close to 0.8) under DD
protection, whereas the
state quickly gets disentangled (fidelity drops to 0.4) when
no protection is applied. This implies that under DD
protection, there is no leakage from the state to other
states in the Hilbert space of the three qubits. 
\section{Conclusions}
\label{concl_3q}
An experimental study of the 
dynamics of tripartite entangled states in a three-qubit
NMR system was undertaken. The results are relevant in the
context of other studies which
showed that different entangled states exhibit varying
degrees of robustness against diverse noise channels. The W state
was found to be the most robust against
the decoherence channel acting on the three NMR qubits, the
GHZ state was the most fragile and decayed very quickly,
while the ${\rm W \bar{W}}$ state was more robust than the
GHZ state but less robust than the W state. For entanglement
protection dynamical decoupling sequences were implemented on these states.
Both DD schemes were able to achieve a good degree of entanglement protection.
The GHZ state was dramatically protected, with its entanglement persisting for nearly
 double the time. The W state showed a marginal improvement,
 which was to be expected since these DD schemes are designed to protect mainly
 against dephasing noise, and the 
experimental results indicated that the W state 
 is already robust against this type of decohering channel.
 Interestingly, although the $\rm W {\bar W}$ state belongs to the GHZ
 entanglement-class, our experiments revealed that its entanglement
 persists for a longer time than the GHZ state, while
 the DD schemes are able to preserve its entanglement to a reasonable
 extent.  The decoherence characteristics of the ${\rm W \bar{W}}$ state hence
 suggest a way of protecting fragile GHZ-type 
 states against noise by transforming the type of
 entanglement (since a GHZ-class state can be transformed via local operations
 to a ${\rm W\bar{W}}$ state). These aspects of the entanglement
 dynamics of the ${\rm W {\bar W}}$ state require more
 detailed studies for a better understanding.

This work has provided new insights into the way entanglement behaves
under decoherence and can be protected from the same. This kind of
protection strategies will provide a benchmark for extension to higher
qubit systems and to the study of different classes of entangled
states. The ${\rm W {\bar W}}$ state is in the GHZ class, however its
dynamics under decoherence resembles the W state, thereby providing it
a certain amount of immunity under dephasing. This aspect, coupled with
the protection schemes, can lead to new ways of entanglement protection.
\newpage
\chapter{Summary and future outlook}
\label{conclusion}

The first project in this thesis focused on the reconstruction of a density
matrix.  The density matrix which was reconstructed using the standard quantum
state tomography (QST) method had a few negative eigenvalues, which makes it an
invalid density matrix.  The maximum likelihood (ML) estimation method
estimates the quantum state density matrix by determining the parameters that
are most likely to match the experimentally generated data. The evaluation
starts with initial parameters based on \emph{a priori} knowledge of
experimentally reconstructed density matrix.  By putting a constraint at every
stage of the estimation process, the density matrix was obtained to be positive
and normalized.  The standard QST method led to negative eigenvalues in the
reconstructed density matrix and hence an overestimation of the entanglement
parameter quantifying the residual entanglement in the state.  The ML
estimation method on the other hand, by virtue of its leading to a physical
density matrix reconstruction every time, gives us a true measure of residual
entanglement, and hence can be used to quantitatively study the decoherence of
multiqubit entanglement. In this thesis ML estimation method is 
applied on two- and three-qubit systems. The ML estimation
can extended further for higher qubit, but this kind of characterization
of quantum states becomes too costly as the number of qubits increase. 
To solve this problem one can look for the use of graphics processing 
units (GPUs) by incorporating them in the optimization process, as it is recently
used for speedup in quantum optimal control~\cite{leung-pra-17}. 

The later part of the thesis focused on mitigating the unwanted effects of
system-environment interaction. The efficacy of different system-environment
decoupling strategies were explored experimentally.  These strategies were
built depending on our knowledge about the system-environment interactions and
the state to be preserved.

\par
The first situation considered was one where the state of the system was known,
and the system-environment interaction was unknown.
The state was protected against evolution using the super-Zeno scheme.
The next situation considered was one where only the subspace to which the state 
belongs was known and an arbitrary state belonging to known subspace
was protected using the nested Uhrig dynamical decoupling (NUDD) scheme.
The advantage of the NUDD schemes lies in the fact that one is 
sure that some amount of state protection will always be achieved.

\par

Next, a system was considered whose state as well as
interaction with the environment is known. A Bell-diagonal (BD) state was
experimentally prepared on a two-qubit system. The noise model considered
was that of each qubit was mainly affected by an independent phase damping channel, 
and quantum discord of the system was observed. Quantum discord remained 
constant for some time $\bar{t}$, after which it began to decay.
The Carr-Purcell-Meiboom-Gill (CPMG) preserving sequence was applied on the BD
state and it was found that reducing the delay between the $\pi$
pulses increased the lifetime of the time-invariant discord, up to a certain point.
After that, the errors due to pulse imperfections dominated due to which
the lifetime of the time-invariant discord started decreasing. 
To overcome this problem XY16 and KDD$_{xy}$ dynamical decoupling sequences 
were used which are robust against such errors. These sequences substantially extended
the lifetime of time-invariant discord. These experiments have important implications in situations where
persistent quantum correlations have to be maintained in order to 
carry out quantum information processing tasks. Such situations usually arises when the number of gates
are large or quantum gates are compare to coherence time.  

\par 
Finally, the decay of tripartite entangled states was studied by 
experimentally preparing GHZ, W and $\rm W\bar{W}$ states on a three-qubit  
NMR quantum information processor. The natural decoherence present 
in the three-spin spin system was best modeled by
considering the main noise channel to be an uncorrelated phase damping channel 
acting independently on each qubit, with a small contribution from the generalized
amplitude damping channel. The W state was found to be the most robust against this type
of noise, whereas the GHZ state was the most fragile; the $\rm W\bar{W}$ state was more robust
than the GHZ state but less robust than the W state. 
The dynamical decoupling sequences XY16 and KDD$_{xy}$ 
were applied on these states and a significant protection of
entanglement for the GHZ and $\rm W\bar{W}$ states
was observed. These quantum state protection strategies can be explored 
for higher qubit systems and can be used to optimize gates
such that the effect of the decoherence is minimized.


\begin{thebibliography}{100}

\bibitem{nielsen-book-02}
M.~A. Nielsen and I.~L. Chuang,
\newblock {\em Quantum computation and quantum information},
\newblock Cambridge University Press, Cambridge UK, 2000.

\bibitem{suter-book-04}
J.~Stolze and D.~Suter,
\newblock {\em Quantum computing: A short course from theory to experiment},
\newblock Wiley-VCH, 1 edition, 2004.

\bibitem{feynmann-ijqi-1982}
R.~P. Feynman,
\newblock Simulating physics with computers,
\newblock Int.J.Theor.Phys. {\bf 21}(6), 467--488 (1982).

\bibitem{Deutsch97}
D.~Deutsch,
\newblock Quantum theory, the church-turing principle and the universal quantum
  computer,
\newblock Proc. Royal Soc. A {\bf 400}(1818), 97--117 (1985).

\bibitem{shor}
P.~W. Shor,
\newblock Polynomial-time algorithms for prime Factorization and discrete
  logarithms on a quantum computer,
\newblock SIAM J. Comput. {\bf 26}(5), 1484--1509 (1997).

\bibitem{grover-prl-97}
L.~K. Grover,
\newblock Quantum mechanics helps in searching for a needle in a haystack,
\newblock Phys. Rev. Lett. {\bf 79}, 325--328 (1997).

\bibitem{divincenzo}
D.~P. DiVincenzo,
\newblock The physical implementation of quantum computation,
\newblock FORTSCHR PHYS {\bf 48}(9-11), 771--783 (2000).

\bibitem{barz-jpb-15}
S.~Barz,
\newblock Quantum computing with photons: introduction to the circuit model,
  the one-way quantum computer, and the fundamental principles of photonic
  experiments,
\newblock J. Phys. B: At. Mol. Opt. Phys {\bf 48}(8), 083001 (2015).

\bibitem{grangier-fp-00}
P.~Grangier, G.~Reymond, and N.~Schlosser,
\newblock Implementations of quantum computing using cavity quantum
  electrodynamics schemes,
\newblock FORTSCHR PHYS {\bf 48}(9-11), 859--874 (2000).

\bibitem{hanffer-pr-08}
H.~Haffner, C.~Roos, and R.~Blatt,
\newblock Quantum computing with trapped ions,
\newblock Phys. Rep. {\bf 469}(4), 155 -- 203 (2008).

\bibitem{devoret-sc-13}
M.~H. Devoret and R.~J. Schoelkopf,
\newblock Superconducting Circuits for Quantum Information: An Outlook,
\newblock Science {\bf 339}(6124), 1169--1174 (2013).

\bibitem{weber-pnas-10}
J.~R. Weber, W.~F. Koehl, J.~B. Varley, A.~Janotti, B.~B. Buckley, C.~G. Van~de
  Walle, and D.~D. Awschalom,
\newblock Quantum computing with defects,
\newblock Proc. Natl. Acad. Sci. USA {\bf 107}(19), 8513--8518 (2010).

\bibitem{serra-proc-12}
R.~M. Serra and I.~S. Oliveira,
\newblock Nuclear magnetic resonance quantum information processing,
\newblock Phil. T. Roy. Soc. A {\bf 370}(1976), 4615--4619 (2012).

\bibitem{walther-rpp-06}
H.~Walther, B.~T.~H. Varcoe, B.-G. Englert, and T.~Becker,
\newblock Cavity quantum electrodynamics,
\newblock Rep. Prog. Phys. {\bf 69}(5), 1325 (2006).

\bibitem{monz-prl-11}
T.~Monz, P.~Schindler, J.~T. Barreiro, M.~Chwalla, D.~Nigg, W.~A. Coish,
  M.~Harlander, W.~H\"ansel, M.~Hennrich, and R.~Blatt,
\newblock 14-Qubit Entanglement: Creation and Coherence,
\newblock Phys. Rev. Lett. {\bf 106}, 130506 (2011).

\bibitem{averin-prl-92}
D.~V. Averin and Y.~V. Nazarov,
\newblock Single-electron charging of a superconducting island,
\newblock Phys. Rev. Lett. {\bf 69}, 1993--1996 (Sep 1992).

\bibitem{Barends-prl-13}
R.~Barends, J.~Kelly, A.~Megrant, D.~Sank, E.~Jeffrey, Y.~Chen, Y.~Yin,
  B.~Chiaro, J.~Mutus, C.~Neill, P.~O'Malley, P.~Roushan, J.~Wenner, T.~C.
  White, A.~N. Cleland, and J.~M. Martinis,
\newblock Coherent Josephson Qubit Suitable for Scalable Quantum Integrated
  Circuits,
\newblock Phys. Rev. Lett. {\bf 111}, 080502 (Aug 2013).

\bibitem{Schirhagl-2014}
R.~Schirhagl, K.~Chang, M.~Loretz, and C.~L. Degen,
\newblock Nitrogen-Vacancy Centers in Diamond: Nanoscale Sensors for Physics
  and Biology,
\newblock Annual Review of Physical Chemistry {\bf 65}(1), 83--105 (2014),
\newblock PMID: 24274702.

\bibitem{Hui2016}
Y.~Y. Hui, C.-A. Cheng, O.~Y. Chen, and H.-C. Chang,
\newblock {\em Bioimaging and Quantum Sensing Using NV Centers in Diamond
  Nanoparticles}, pages 109--137,
\newblock Springer International Publishing, Cham, 2016.

\bibitem{suter-pnmrs-17}
D.~Suter and F.~Jelezko,
\newblock Single-spin magnetic resonance in the nitrogen-vacancy center of
  diamond,
\newblock Progress in Nuclear Magnetic Resonance Spectroscopy {\bf 98-99}, 50
  -- 62 (2017).

\bibitem{ibmq}
I.~B.~M. quantum~computing platform,
\newblock IBM Corporation, Armonk, New York, United States, 2017.

\bibitem{oliveira-book-07}
I.~S. Oliveira, T.~J. Bonagamba, R.~S. Sarthour, J.~C. Freitas, and E.~R.
  deAzevedo,
\newblock {\em NMR quantum information processing},
\newblock Elsevier, Amsterdam, The Netherlands, 2007.

\bibitem{levitt-book-2008}
M.~H. Levitt,
\newblock {\em Spin dynamics: Basics of nuclear magnetic resonance},
\newblock John Wiley and Sons, Chichester England, 2008.

\bibitem{cory-physicad}
D.~Cory, M.~Price, and T.~Havel,
\newblock Nuclear magnetic resonance spectroscopy: An experimentally accessible
  paradigm for quantum computing,
\newblock Physica D {\bf 120}, 82 (1998).

\bibitem{tosner-jmr-09}
Z.~Tosner, T.~Vosegaard, C.~Kehlet, N.~Khaneja, S.~J. Glaser, and N.~C.
  Nielsen,
\newblock Optimal control in NMR spectroscopy: Numerical implementation in
  SIMPSON,
\newblock J. Magn. Reson. {\bf 197}, 120 (2009).

\bibitem{glaser-book-07}
N.~C. Nielsen, C.~Kehlet, S.~J. Glaser, and N.~Khaneja,
\newblock {\em Optimal control methods in NMR spectroscopy},
\newblock eMagRes., John Wiley $\&$ Sons, Ltd, 2007.

\bibitem{manu-pra-12}
V.~S. Manu and A.~Kumar,
\newblock Singlet-state creation and universal quantum computation in NMR using
  a genetic algorithm,
\newblock Phys. Rev. A {\bf 86}, 022324 (2012).

\bibitem{manu-pra-14}
V.~S. Manu and A.~Kumar,
\newblock Quantum simulation using fidelity-profile optimization,
\newblock Phys. Rev. A {\bf 89}, 052331 (2014).

\bibitem{vandersypen-review}
L.~M.~K. Vandersypen and I.~L. Chuang,
\newblock NMR techniques for quantum control and computation,
\newblock Rev. Mod. Phys. {\bf 76}, 1037--1069 (2005).

\bibitem{jones-nature-98}
J.~A. Jones, M.~Mosca, R.~H. Hansen, C.~S. Yannoni, M.~H. Sherwood, and I.~L.
  Chuang,
\newblock Implementation of a quantum search algorithm on a quantum computer,
\newblock Nature {\bf 393}(883), 344--346 (1998).

\bibitem{dorai-pra-2000}
K.~Dorai, Arvind, and A.~Kumar,
\newblock Implementing quantum-logic operations, pseudopure states, and the
  Deutsch-Jozsa algorithm using noncommuting selective pulses in NMR,
\newblock Phys. Rev. A {\bf 61}, 042306 (2000).

\bibitem{dorai-pra-2001}
K.~Dorai, Arvind, and A.~Kumar,
\newblock Implementation of a Deutsch-like quantum algorithm utilizing
  entanglement at the two-qubit level on an NMR quantum-information processor,
\newblock Phys. Rev. A {\bf 63}, 034101 (2001).

\bibitem{stadelhofer-pra-2005}
R.~Stadelhofer, D.~Suter, and W.~Banzhaf,
\newblock Quantum and classical parallelism in parity algorithms for ensemble
  quantum computers,
\newblock Phys. Rev. A {\bf 71}, 032345 (2005).

\bibitem{xio-pra-05}
L.~Xiao and J.~A. Jones,
\newblock Error tolerance in an NMR implementation of Grover's fixed-point
  quantum search algorithm,
\newblock Phys. Rev. A {\bf 72}, 032326 (2005).

\bibitem{li-prx-17}
J.~Li, R.~Fan, H.~Wang, B.~Ye, B.~Zeng, H.~Zhai, X.~Peng, and J.~Du,
\newblock Measuring Out-of-Time-Order Correlators on a Nuclear Magnetic
  Resonance Quantum Simulator,
\newblock Phys. Rev. X {\bf 7}, 031011 (Jul 2017).

\bibitem{du-prl-10}
J.~Du, N.~Xu, X.~Peng, P.~Wang, S.~Wu, and D.~Lu,
\newblock NMR Implementation of a Molecular Hydrogen Quantum Simulation with
  Adiabatic State Preparation,
\newblock Phys. Rev. Lett. {\bf 104}, 030502 (Jan 2010).

\bibitem{shankar-pla-14}
R.~Shankar, S.~S. Hegde, and T.~S. Mahesh,
\newblock Quantum simulations of a particle in one-dimensional potentials using
  NMR,
\newblock Phys. Lett. A {\bf 378}(1), 10 -- 15 (2014).

\bibitem{anjusha-pla-16}
A.~V.S., S.~S. Hegde, and T.~Mahesh,
\newblock NMR investigation of the quantum pigeonhole effect,
\newblock Phys. Lett. A {\bf 380}(4), 577 -- 580 (2016).

\bibitem{peng-prl-09}
X.~Peng, J.~Zhang, J.~Du, and D.~Suter,
\newblock Quantum simulation of a system with competing two- and three-body
  interactions,
\newblock Phys. Rev. Lett. {\bf 103}, 140501 (Sep 2009).

\bibitem{zhang-prl-11}
J.~Zhang, T.-C. Wei, and R.~Laflamme,
\newblock Experimental quantum simulation of entanglement in many-body systems,
\newblock Phys. Rev. Lett. {\bf 107}, 010501 (Jun 2011).

\bibitem{arindam-jmr-05}
A.~Ghosh and A.~Kumar,
\newblock Relaxation of pseudo pure states: The role of cross-correlations,
\newblock J. Magn. Reson. {\bf 173}(1), 125--133 (2005).

\bibitem{hegde-pra-14}
S.~S. Hegde and T.~S. Mahesh,
\newblock Engineered decoherence: Characterization and suppression,
\newblock Phys. Rev. A {\bf 89}, 062317 (2014).

\bibitem{kawamura-ijqc-06}
M.~Kawamura, T.~Morimoto, Y.~Mori, R.~Sawae, K.~Takarabe, and Y.~Manmoto,
\newblock Decoherence of a Greenberger-Horne-Zeilinger state in a five-qubit
  NMR quantum computer,
\newblock Intl. J. Quant. Chem. {\bf 106}(15), 3108--3112 (2006).

\bibitem{rao-pra-13}
K.~R.~K. Rao, H.~Katiyar, T.~S. Mahesh, A.~Sen~(De), U.~Sen, and A.~Kumar,
\newblock Multipartite quantum correlations reveal frustration in a quantum
  Ising spin system,
\newblock Phys. Rev. A {\bf 88}, 022312 (Aug 2013).

\bibitem{dogra-pla-16}
S.~Dogra, K.~Dorai, and Arvind,
\newblock Experimental demonstration of quantum contextuality on an NMR qutrit,
\newblock Phys. Lett. A {\bf 380}(22), 1941 -- 1946 (2016).

\bibitem{katiyar-epl-16}
H.~Katiyar, C.~S.~S. Kumar, and T.~S. Mahesh,
\newblock NMR investigation of contextuality in a quantum harmonic oscillator
  via pseudospin mapping,
\newblock EPL {\bf 113}(2), 20003 (2016).

\bibitem{chen-pra-11}
H.~Chen, X.~Kong, B.~Chong, G.~Qin, X.~Zhou, X.~Peng, and J.~Du,
\newblock Experimental demonstration of a quantum annealing algorithm for the
  traveling salesman problem in a nuclear-magnetic-resonance quantum simulator,
\newblock Phys. Rev. A {\bf 83}, 032314 (2011).

\bibitem{criger4620}
B.~Criger, G.~Passante, D.~Park, and R.~Laflamme,
\newblock Recent advances in nuclear magnetic resonance quantum information
  processing,
\newblock Phil. Trans. R. Soc. A {\bf 370}(1976), 4620--4635 (2012).

\bibitem{Jones-review}
J.~A. Jones,
\newblock Quantum computing with NMR,
\newblock Prog. Nucl. Magn. Reson. Spectrosc. {\bf 59}(2), 91 -- 120 (2011).

\bibitem{dorai-review}
K.~Dorai, T.S.Mahesh, Arvind, and A.~Kumar,
\newblock Quantum computation using NMR,
\newblock Curr. Sci. {\bf 79}, 1447--1458 (2000).

\bibitem{pan-pra-14}
J.~Pan, Y.~Cao, X.~Yao, Z.~Li, C.~Ju, H.~Chen, X.~Peng, S.~Kais, and J.~Du,
\newblock Experimental realization of quantum algorithm for solving linear
  systems of equations,
\newblock Phys. Rev. A {\bf 89}, 022313 (Feb 2014).

\bibitem{li-prl-15}
Z.~Li, X.~Liu, N.~Xu, and J.~Du,
\newblock Experimental realization of a quantum support vector machine,
\newblock Phys. Rev. Lett. {\bf 114}, 140504 (Apr 2015).

\bibitem{ramanathan2004}
C.~Ramanathan, N.~Boulant, Z.~Chen, D.~G. Cory, I.~Chuang, and M.~Steffen,
\newblock NMR quantum information processing,
\newblock Quant. Inf. Proc. {\bf 3}(1), 15--44 (2004).

\bibitem{suter-review}
D.~Suter and G.~A. \'Alvarez,
\newblock Colloquium: Protecting quantum information against environmental
  noise,
\newblock Rev. Mod. Phys. {\bf 88}, 041001 (2016).

\bibitem{banaszek-pra-99}
K.~Banaszek, G.~M. D'Ariano, M.~G.~A. Paris, and M.~F. Sacchi,
\newblock Maximum-likelihood estimation of the density matrix,
\newblock Phys. Rev. A {\bf 61}, 010304 (1999).

\bibitem{hradil-pra-97}
Z.~Hradil,
\newblock Quantum state estimation,
\newblock Phys. Rev. A {\bf 55}, R1561 (1997).

\bibitem{kohout-prl-10}
R.~Blume-Kohout,
\newblock Hedged maximum likelihood quantum state estimation,
\newblock Phys. Rev. Lett. {\bf 105}, 200504 (2010).

\bibitem{plenio-mle}
T.~Baumgratz, A.~Nusseler, M.~Cramer, and M.~B. Plenio,
\newblock A scalable maximum likelihood method for quantum state tomography,
\newblock New. J. Phys. {\bf 15}, 125004 (2013).

\bibitem{singh-pla-16}
H.~Singh, Arvind, and K.~Dorai,
\newblock Constructing valid density matrices on an NMR quantum information
  processor via maximum likelihood estimation,
\newblock Phys. Lett. A {\bf 380}(38), 3051 -- 3056 (2016).

\bibitem{dhar-prl-06}
D.~Dhar, L.~K. Grover, and S.~M. Roy,
\newblock Preserving quantum states using inverting pulses: A super-Zeno
  effect,
\newblock Phys. Rev. Lett. {\bf 96}, 100405 (2006).

\bibitem{mukhtar-pra-10-1}
M.~Mukhtar, W.~T. Soh, T.~B. Saw, and J.~Gong,
\newblock Universal dynamical decoupling: Two-qubit states and beyond,
\newblock Phys. Rev. A {\bf 81}, 012331 (2010).

\bibitem{ting-chinese-09}
R.~Ting-Ting, L.~Jun, S.~Xian-Ping, and Z.~Ming-Sheng,
\newblock Preservation of quantum states via a super-Zeno effect on ensemble
  quantum computers,
\newblock Chin. Phys. B {\bf 18}(11), 4711 (2009).

\bibitem{singh-pra-14}
H.~Singh, Arvind, and K.~Dorai,
\newblock Experimental protection against evolution of states in a subspace via
  a super-Zeno scheme on an NMR quantum information processor,
\newblock Phys. Rev. A {\bf 90}, 052329 (2014).

\bibitem{mukhtar-pra-10-2}
M.~Mukhtar, W.~T. Soh, T.~B. Saw, and J.~Gong,
\newblock Protecting unknown two-qubit entangled states by nesting Uhrig's
  dynamical decoupling sequences,
\newblock Phys. Rev. A {\bf 82}, 052338 (2010).

\bibitem{zhen-pra-16}
X.-L. Zhen, F.-H. Zhang, G.~Feng, H.~Li, and G.-L. Long,
\newblock Optimal experimental dynamical decoupling of both longitudinal and
  transverse relaxations,
\newblock Phys. Rev. A {\bf 93}, 022304 (2016).

\bibitem{singh-pra-17}
H.~Singh, Arvind, and K.~Dorai,
\newblock Experimental protection of arbitrary states in a two-qubit subspace
  by nested Uhrig dynamical decoupling,
\newblock Phys. Rev. A {\bf 95}, 052337 (2017).

\bibitem{childs-pra-03}
A.~M. Childs, I.~L.Chuang, and D.~W. Leung,
\newblock Realization of quantum process tomography in NMR,
\newblock Phys. Rev. A {\bf 64}, 012314 (2001).

\bibitem{souza-prl-11}
A.~M. Souza, G.~A. \'Alvarez, and D.~Suter,
\newblock Robust dynamical decoupling for quantum computing and quantum Memory,
\newblock Phys. Rev. Lett. {\bf 106}, 240501 (2011).

\bibitem{souza-pra-12}
A.~M. Souza, G.~A. \'Alvarez, and D.~Suter,
\newblock Effects of time-reversal symmetry in dynamical decoupling,
\newblock Phys. Rev. A {\bf 85}, 032306 (2012).

\bibitem{mazzola-prl-10}
L.~Mazzola, J.~Piilo, and S.~Maniscalco,
\newblock Sudden transition between classical and quantum decoherence,
\newblock Phys. Rev. Lett. {\bf 104}, 200401 (2010).

\bibitem{singh-epl-17}
H.~Singh, Arvind, and K.~Dorai,
\newblock Experimentally freezing quantum discord in a dephasing environment
  using dynamical decoupling,
\newblock EPL {\bf 118}(5), 50001 (2017).

\bibitem{wang-prl-16}
X.-L. Wang, L.-K. Chen, W.~Li, H.-L. Huang, C.~Liu, C.~Chen, Y.-H. Luo, Z.-E.
  Su, D.~Wu, Z.-D. Li, H.~Lu, Y.~Hu, X.~Jiang, C.-Z. Peng, L.~Li, N.-L. Liu,
  Y.-A. Chen, C.-Y. Lu, and J.-W. Pan,
\newblock Experimental Ten-Photon Entanglement,
\newblock Phys. Rev. Lett. {\bf 117}, 210502 (2016).

\bibitem{Negrevergne-prl-06}
C.~Negrevergne, T.~S. Mahesh, C.~A. Ryan, M.~Ditty, F.~Cyr-Racine, W.~Power,
  N.~Boulant, T.~Havel, D.~G. Cory, and R.~Laflamme,
\newblock Benchmarking Quantum Control Methods on a 12-Qubit System,
\newblock Phys. Rev. Lett. {\bf 96}, 170501 (2006).

\bibitem{macmohan-book-08}
D.~McMahon,
\newblock {\em Quantum computing explained},
\newblock Wiley-IEEE Computer Society Press, 2008.

\bibitem{barenco-pra-95}
A.~Barenco, C.~H. Bennett, R.~Cleve, D.~P. DiVincenzo, N.~Margolus, P.~Shor,
  T.~Sleator, J.~A. Smolin, and H.~Weinfurter,
\newblock Elementary gates for quantum computation,
\newblock Phys. Rev. A {\bf 52}, 3457--3467 (1995).

\bibitem{keeler-book}
J.~Keeler,
\newblock {\em Understanding NMR spectroscopy},
\newblock Wiley, Cambridge, 2002.

\bibitem{bloch-pr-1940}
F.~Bloch and A.~Siegert,
\newblock Magnetic resonance for nonrotating fields,
\newblock Phys. Rev. {\bf 57}, 522--527 (1940).

\bibitem{cory-pnas-97}
D.~G. Cory, A.~F. Fahmy, and T.~F. Havel,
\newblock Ensemble quantum computing by NMR spectroscopy,
\newblock Proc. Natl. Acad. Sci. {\bf 94}(5), 1634--1639 (1997).

\bibitem{gershenfeld-sc-97}
N.~A. Gershenfeld and I.~L. Chuang,
\newblock Bulk spin-resonance quantum computation,
\newblock Science {\bf 275}(5298), 350--356 (1997).

\bibitem{vandersypen-nature-01}
L.~M.~K. Vandersypen, M.~Steffen, G.~Breyta, C.~S. Yannoni, M.~H. Sherwood, and
  I.~L. Chuang,
\newblock Experimental realization of Shor's quantum factoring algorithm using
  nuclear magnetic resonance,
\newblock Nature {\bf 414}(883), 883--887 (2001).

\bibitem{braunstein-prl-99}
S.~L. Braunstein, C.~M. Caves, R.~Jozsa, N.~Linden, S.~Popescu, and R.~Schack,
\newblock Separability of very noisy mixed States and implications for NMR
  quantum computing,
\newblock Phys. Rev. Lett. {\bf 83}, 1054--1057 (1999).

\bibitem{yu-pra-05}
T.~M. Yu, K.~R. Brown, and I.~L. Chuang,
\newblock Bounds on the entanglement attainable from unitary transformed
  thermal states in liquid-state nuclear magnetic resonance,
\newblock Phys. Rev. A {\bf 71}, 032341 (2005).

\bibitem{soares-pinto-proc-12}
D.~O. Soares-Pinto, R.~Auccaise, J.~Maziero, A.~Gavini-Viana, R.~M. Serra, and
  L.~C. C{\'e}leri,
\newblock On the quantumness of correlations in nuclear magnetic resonance,
\newblock Phil. T. Roy. Soc. A {\bf 370}(1976), 4821--4836 (2012).

\bibitem{fortunato-jcp-02}
E.~M. Fortunato, M.~A. Pravia, N.~Boulant, G.~Teklemariam, T.~F. Havel, and
  D.~G. Cory,
\newblock Design of strongly modulating pulses to implement precise effective
  Hamiltonians for quantum information processing,
\newblock J. Chem. Phys. {\bf 116}, 7599 (2002).

\bibitem{Nelder}
J.~A. Nelder and R.~Mead,
\newblock A simplex method for function minimization,
\newblock Comput. J. {\bf 7}(4), 308--313 (1965).

\bibitem{holland-book}
J.~H. Holland,
\newblock {\em Adaptation in natural and artificial systems: An introductory
  analysis with applications to biology, control, and artificial intelligence},
\newblock MIT press, 1992.

\bibitem{suter-ia-08}
R.~Stadelhofer, W.~Banzhaf, and D.~Suter,
\newblock Evolving blackbox quantum algorithms using genetic programming,
\newblock Artif. Intell. Eng. Des. Anal. Manuf. {\bf 22}(3), 285--297 (2008).

\bibitem{hardy-2010}
Y.~Hardy and W.-H. Steeb,
\newblock Genetic algorithms and optimization problems in quantum computing,
\newblock Int. J. Mod. Phy. C {\bf 21}(11), 1359--1375 (2010).

\bibitem{bang-korean}
J.~Bang and S.~Yoo,
\newblock A genetic-algorithm-based method to find unitary transformations for
  any desired quantum computation and application to a one-bit oracle decision
  problem,
\newblock J. Korean Phys. Soc. {\bf 65}(12), 2001--2008 (2014).

\bibitem{navarro-pra-06}
J.~C. N.-M. noz, H.~C. Rosu, and R.~L\'opez-Sandoval,
\newblock Genetic algorithm optimization of entanglement,
\newblock Phys. Rev. A {\bf 74}, 052308 (2006).

\bibitem{lidar-pra}
G.~Quiroz and D.~A. Lidar,
\newblock Optimized dynamical decoupling via genetic algorithms,
\newblock Phys. Rev. A {\bf 88}, 052306 (2013).

\bibitem{khaneja-jmr-05}
N.~Khaneja, T.~Reiss, C.~Kehlet, T.~S.-Herbrüggen, and S.~J. Glaser,
\newblock Optimal control of coupled spin dynamics: Design of NMR pulse
  sequences by gradient ascent algorithms,
\newblock J. Magn. Reson. {\bf 172}(2), 296 -- 305 (2005).

\bibitem{Khitrin2011}
A.~K. Khitrin, M.~Michalski, and J.-S. Lee,
\newblock Reversible projective measurement in quantum ensembles,
\newblock Quant. Inf. Proc. {\bf 10}(4), 557--566 (2011).

\bibitem{lee-apl-06}
J.-S. Lee and A.~K. Khitrin,
\newblock Projective measurement in nuclear magnetic resonance,
\newblock App. Phys. Lett. {\bf 89}(7), 074105 (2006).

\bibitem{leskowitz-pra-04}
G.~M. Leskowitz and L.~J. Mueller,
\newblock State interrogation in nuclear magnetic resonance quantum-information
  processing,
\newblock Phys. Rev. A {\bf 69}, 052302 (2004).

\bibitem{long-joptb}
G.~Long, H.~Yan, and Y.~Sun,
\newblock Analysis of density matrix reconstruction in NMR quantum computing,
\newblock J. Opt. B {\bf 3}, 376 (2001).

\bibitem{ernst-book-87}
R.~R. Ernst, G.~Bodenhausen, and A.~Wokaun,
\newblock {\em Principles of nuclear magnetic resonance in one and two
  dimensions},
\newblock Oxford University Press, New York, 1987.

\bibitem{palmer-95}
J.~Cavanaugh, W.~J. Fairbrother, A.~G.~P. III, and N.~J. Skelton,
\newblock {\em Protein NMR spectroscopy: Principles and practice},
\newblock Academic Press Inc., San Diego USA, 1995.

\bibitem{kumar-pnmrs-00}
A.~Kumar, R.~C.~R. Grace, and P.~Madhu,
\newblock Cross-correlations in NMR,
\newblock Prog. Nucl. Magn. Reson. Spectrosc. {\bf 37}(3), 191 -- 319 (2000).

\bibitem{streltsov-rmp-17}
A.~Streltsov, G.~Adesso, and M.~B. Plenio,
\newblock Colloquium,
\newblock Rev. Mod. Phys. {\bf 89}, 041003 (Oct 2017).

\bibitem{katiyar-pra-12}
H.~Katiyar, S.~S. Roy, T.~S. Mahesh, and A.~Patel,
\newblock Evolution of quantum discord and its stability in two-qubit NMR
  systems,
\newblock Phys. Rev. A {\bf 86}, 012309 (2012).

\bibitem{preskill385}
J.~Preskill,
\newblock Reliable quantum computers,
\newblock Proc. Royal Soc. A {\bf 454}(1969), 385--410 (1998).

\bibitem{duan-prl-97}
L.-M. Duan and G.-C. Guo,
\newblock Preserving coherence in quantum computation by pairing quantum bits,
\newblock Phys. Rev. Lett. {\bf 79}, 1953--1956 (1997).

\bibitem{geoffrey-book-03}
R.~F. G.~L.~Sewell, F.~Benatti,
\newblock {\em Irreversible quantum dynamics},
\newblock Lecture Notes in Physics 622, Springer-Verlag Berlin Heidelberg, 1
  edition, 2003.

\bibitem{viola-review}
L.~Viola,
\newblock Advances in decoherence control,
\newblock J. Mod. Opt. {\bf 51}(16-18), 2357--2367 (2004).

\bibitem{viola-prl-99-2}
L.~Viola, E.~Knill, and S.~Lloyd,
\newblock Dynamical decoupling of open quantum systems,
\newblock Phys. Rev. Lett. {\bf 82}, 2417--2421 (1999).

\bibitem{hahn-pr-50}
E.~L. Hahn,
\newblock Spin echoes,
\newblock Phys. Rev. {\bf 80}, 580--594 (1950).

\bibitem{meiboom-rsi-58}
S.~Meiboom and D.~Gill,
\newblock Modified spin-echo method for measuring nuclear relaxation times,
\newblock Rev. Sci. Instrum. {\bf 29}(8), 688--691 (1958).

\bibitem{uhrig-prl-07}
G.~S. Uhrig,
\newblock Keeping a quantum bit alive by optimized $\ensuremath{\pi}$-pulse
  sequences,
\newblock Phys. Rev. Lett. {\bf 98}, 100504 (2007).

\bibitem{carr-pra-54}
H.~Y. Carr and E.~M. Purcell,
\newblock Effects of Diffusion on Free Precession in Nuclear Magnetic Resonance
  Experiments,
\newblock Phys. Rev. {\bf 94}, 630--638 (1954).

\bibitem{yang-prl-08}
W.~Yang and R.-B. Liu,
\newblock Universality of Uhrig dynamical decoupling for suppressing qubit pure
  dephasing and relaxation,
\newblock Phys. Rev. Lett. {\bf 101}, 180403 (2008).

\bibitem{massar-prl-95}
S.~Massar and S.~Popescu,
\newblock Optimal extraction of information from finite quantum ensembles,
\newblock Phys. Rev. Lett. {\bf 74}, 1259 (1995).

\bibitem{derka-prl-98}
R.~Derka, V.~Buzek, and A.~K. Ekert,
\newblock Universal algorithm for optimal estimation of quantum states from
  finite ensembles via realizable generalized measurement,
\newblock Phys. Rev. Lett. {\bf 80}, 1571 (1998).

\bibitem{rehacek-book}
M.~Paris and J.~\ifmmode \check{R}\else \v{R}\fi{}eh\'a\ifmmode~\check{c}\else
  \v{c}\fi{}ek,
\newblock {\em Lecture notes in physics - Quantum state estimation},
\newblock Springer, Berlin Heidelberg, 2004.

\bibitem{plenio-naturecomm}
M.~Cramer, M.~B. Plenio, S.~T. Flammia, R.~Somma, D.~Gross, S.~D. Bartlett,
  O.~Landon-Cardinal, D.~Poulin, and Y.-K. Liu,
\newblock Efficient quantum state tomography,
\newblock Nat. Commun. {\bf 149}, 1147 (2010).

\bibitem{rehacek-pra-15}
J.~\ifmmode \check{R}\else \v{R}\fi{}eh\'a\ifmmode~\check{c}\else \v{c}\fi{}ek,
  Y.~S. Teo, and Z.~Hradil,
\newblock Determining which quantum measurement performs better for state
  estimation,
\newblock Phys. Rev. A {\bf 92}, 012108 (2015).

\bibitem{chuang-proc-98}
I.~L. Chuang, N.~Gershenfeld, M.~Kubinec, and D.~W. Leung,
\newblock Bulk quantum computation with nuclear magnetic resonance: Theory and
  experiment,
\newblock Phil. T. Roy. Soc. A {\bf 454}, 447 (1998).

\bibitem{lee-pla-02}
J.-S. Lee,
\newblock The quantum state tomography on an NMR system,
\newblock Phys. Lett. A {\bf 305}(6), 349 -- 353 (2002).

\bibitem{james-pra-01}
D.~F.~V. James, P.~G. Kwiat, W.~J. Munro, and A.~G. White,
\newblock Measurement of qubits,
\newblock Phys. Rev. A {\bf 64}, 052312 (2001).

\bibitem{tan-joptb-99}
S.~M. Tan,
\newblock An inverse problem approach to optical homodyne tomography,
\newblock J. Opt. B {\bf 44}(11-12), 2233--2259 (1997).

\bibitem{hradil-pra-00}
Z.~Hradil, J.~Summhammer, G.~Badurek, and H.~Rauch,
\newblock Reconstruction of the spin state,
\newblock Phys. Rev. A {\bf 62}, 014101 (2000).

\bibitem{rehacek-pra-01}
J.~\ifmmode \check{R}\else \v{R}\fi{}eh\'a\ifmmode~\check{c}\else \v{c}\fi{}ek,
  Z.~Hradil, and M.~Je\ifmmode~\check{z}\else \v{z}\fi{}ek,
\newblock Iterative algorithm for reconstruction of entangled states,
\newblock Phys. Rev. A {\bf 63}, 040303 (2001).

\bibitem{miranowicz-pra-14}
A.~Miranowicz, K.~Bartkiewicz, J.~Perina, M.~Koashi, N.~Imoto, and F.~Nori,
\newblock Optimal two-qubit tomography based on local and global measurements:
  Maximal robustness against errors as described by condition numbers,
\newblock Phys. Rev. A {\bf 90}, 062123 (2014).

\bibitem{rehacek-pra-07}
J.~\ifmmode \check{R}\else \v{R}\fi{}eh\'a\ifmmode~\check{c}\else \v{c}\fi{}ek,
  Z.~Hradil, E.~Knill, and A.~I. Lvovsky,
\newblock Diluted maximum-likelihood algorithm for quantum tomography,
\newblock Phys. Rev. A {\bf 75}, 042108 (2007).

\bibitem{huszar-pra-12}
F.~Husz\'ar and N.~M.~T. Houlsby,
\newblock Adaptive Bayesian quantum tomography,
\newblock Phys. Rev. A {\bf 85}, 052120 (2012).

\bibitem{opatrny-pra-97}
T.~Opatrny, D.~G. Welsch, and W.~Vogel,
\newblock Least-squares inversion for density- matrix reconstruction,
\newblock Phys. Rev. A {\bf 56}, 1788 (1997).

\bibitem{kaznady-pra-09}
M.~S. Kaznady and D.~F.~V. James,
\newblock Numerical strategies for quantum tomography: Alternatives to full
  optimization,
\newblock Phys. Rev. A {\bf 79}, 022109 (2009).

\bibitem{teo-prl-11}
Y.~S. Teo, Z.~Huangjun, B.-G. Englert, J.~\ifmmode \check{R}\else
  \v{R}\fi{}eh\'a\ifmmode~\check{c}\else \v{c}\fi{}ek, and Z.~Hradil,
\newblock Quantum-state reconstruction by maximizing likelihood and entropy,
\newblock Phys. Rev. Lett. {\bf 107}, 020404 (2011).

\bibitem{teo-pra-12}
Y.~S. Teo, B.~Stoklasa, B.-G. Englert, J.~\ifmmode \check{R}\else
  \v{R}\fi{}eh\'a\ifmmode~\check{c}\else \v{c}\fi{}ek, and Z.~Hradil,
\newblock Incomplete quantum state estimation: A comprehensive study,
\newblock Phys. Rev. A {\bf 85}, 042317 (2012).

\bibitem{chen-pra-13}
J.~Chen, H.~Dawkins, Z.~Ji, N.~Johnston, D.~Kribs, F.~Shultz, and B.~Zeng,
\newblock Uniqueness of quantum states compatible with given measurement
  results,
\newblock Phys. Rev. A {\bf 88}, 012109 (2013).

\bibitem{smithey-prl-93}
D.~T. Smithey, M.~Beck, M.~G. Raymer, and A.~Faridami,
\newblock Measurement of the wigner distribution and the density matrix of a
  light mode using optical homodyne tomography: Application to squeezed states
  and the vacuum,
\newblock Phys. Rev. Lett. {\bf 70}, 1244 (1993).

\bibitem{thew-pra-02}
R.~T. Thew, K.~Nemoto, A.~G. White, and W.~J. Munro,
\newblock Qudit quantum-state tomography,
\newblock Phys. Rev. A {\bf 66}, 012303 (2002).

\bibitem{garge-jmr-03}
H.~Grage and M.~Akke,
\newblock A statistical analysis of NMR spectrometer noise,
\newblock J. Magn. Reson. {\bf 162}(1), 176 -- 188 (2003).

\bibitem{matlab}
MATLAB,
\newblock {\em Version 8.5.0 (R2015a)},
\newblock MathWorks Inc., Natick, Massachusetts, 2015.

\bibitem{weinstein-prl-01}
Y.~S. Weinstein, M.~A. Pravia, E.~M. Fortunato, S.~Lloyd, and D.~G. Cory,
\newblock Implementation of the quantum Fourier transform,
\newblock Phys. Rev. Lett. {\bf 86}, 1889 (2001).

\bibitem{uhlmann-fidelity}
A.~Uhlmann,
\newblock The transition probability in the state space of a *-algebra,
\newblock Rep. Math. Phys. {\bf 9}, 273 (1976).

\bibitem{jozsa-fidelity}
R.~Jozsa,
\newblock Fidelity for mixed quantum states,
\newblock J. Mod. Opt. {\bf 41}, 2315 (1994).

\bibitem{dogra-pra-15}
S.~Dogra, K.~Dorai, and Arvind,
\newblock Experimental construction of generic three-qubit states and their
  reconstruction from two-party reduced states on an NMR quantum information
  processor,
\newblock Phys. Rev. A {\bf 91}, 022312 (2015).

\bibitem{das-pra-15}
D.~Das, S.~Dogra, K.~Dorai, and Arvind,
\newblock Experimental construction of a W superposition state and its
  equivalence to the Greenberger-Horne-Zeilinger state under local filtration,
\newblock Phys. Rev. A {\bf 92}, 022307 (2015).

\bibitem{viola-pra-98}
L.~Viola and S.~Lloyd,
\newblock Dynamical suppression of decoherence in two-state quantum systems,
\newblock Phys. Rev. A {\bf 58}, 2733--2744 (1998).

\bibitem{viola-prl-99}
L.~Viola, S.~Lloyd, and E.~Knill,
\newblock Universal control of decoupled quantum systems,
\newblock Phys. Rev. Lett. {\bf 83}, 4888--4891 (1999).

\bibitem{sudarshan-jomp-77}
B.~Misra and E.~C.~G. Sudarshan,
\newblock The Zeno's paradox in quantum theory,
\newblock J. Math. Phys. {\bf 18}(4), 756--763 (1977).

\bibitem{sudarshan}
C.~B. Chiu, E.~C.~G. Sudarshan, and B.~Misra,
\newblock Time evolution of unstable quantum states and a resolution of Zeno's
  paradox,
\newblock Phys. Rev. D {\bf 16}, 520 (1977).

\bibitem{facchi-jmp-10}
P.~Facchi and M.~Ligabo,
\newblock Quantum Zeno effect and dynamics,
\newblock J. Math. Phys. {\bf 51}, 022103 (2010).

\bibitem{facchi-jphyconf-09}
P.~Facchi, G.~Marmo, and S.~Pascazio,
\newblock Quantum Zeno dynamics and quantum Zeno subspaces,
\newblock J. Phys.: Conf. Ser. {\bf 196}, 012017 (2009).

\bibitem{facchi-prl-02}
P.~Facchi and S.~Pascazio,
\newblock Quantum Zeno subspaces,
\newblock Phys. Rev. Lett. {\bf 89}, 080401 (2002).

\bibitem{busch-jphyconf-10}
J.~Busch and A.~Beige,
\newblock Protecting subspaces by acting on the outside,
\newblock J. Phys.: Conf. Ser. {\bf 254}, 012009 (2010).

\bibitem{erez-pra-04}
N.~Erez, Y.~Aharonov, B.~Reznik, and L.~Vaidman,
\newblock Correcting quantum errors with the Zeno effect,
\newblock Phys. Rev. A {\bf 69}, 062315 (2004).

\bibitem{sabrina-prl-08}
S.~Maniscalco, F.~Francica, R.~L. Zaffino, N.~L. Gullo, and F.~Plastina,
\newblock Protecting entanglement via the quantum Zeno effect,
\newblock Phys. Rev. Lett. {\bf 100}, 090503 (2008).

\bibitem{oliveira-pra-08}
J.~G.~O. Jr., R.~R. Jr., and M.~C. Nemes,
\newblock Protecting, enhancing, and reviving entanglement,
\newblock Phys. Rev. A {\bf 78}, 044301 (2008).

\bibitem{silva-prl-12}
G.~A. Paz-Silva, A.~T. Rezakhani, J.~M. Dominy, and D.~A. Lidar,
\newblock Zeno effect for quantum computation and control,
\newblock Phys. Rev. Lett. {\bf 108}, 080501 (2012).

\bibitem{uhrig-njp-08}
G.~S. Uhrig,
\newblock Exact results on dynamical decoupling by $\pi$ pulses in quantum
  information processes,
\newblock New. J. Phys. {\bf 10}(8), 083024 (2008).

\bibitem{hou-annal-12}
Y.-C. Hou, G.-F. Zhang, Y.~Chen, and H.~Fan,
\newblock Preservation of entanglement in a two-qubit-spin coupled system,
\newblock Ann. Phys. {\bf 327}, 292 (2012).

\bibitem{itano-pra-90}
W.~M. Itano, D.~J. Heinzen, J.~J. Bollinger, and D.~J. Wineland,
\newblock Quantum Zeno effect,
\newblock Phys. Rev. A {\bf 41}, 2295 (1990).

\bibitem{bernu-prl-08}
J.~Bernu, S.~Deleglise, C.~Sayrin, S.~Kuhr, I.~Dotsenko, M.~Brune, J.~M.
  Raimond, and S.~Haroche,
\newblock Freezing coherent field growth in a cavity by the quantum Zeno
  effect,
\newblock Phys. Rev. Lett. {\bf 101}, 180402 (2008).

\bibitem{franson-pra-04}
J.~D. Franson, B.~C. Jacobs, and T.~B. Pittman,
\newblock Quantum computing using single photons and the Zeno effect,
\newblock Phys. Rev. A {\bf 70}, 062302 (2004).

\bibitem{tong-pra-14}
Q.-J. Tong, J.-H. An, L.~C. Kwek, H.-G. Luo, and C.~H. Oh,
\newblock Simulating Zeno physics by a quantum quench with superconducting
  circuits,
\newblock Phys. Rev. A {\bf 89}, 060101 (2014).

\bibitem{xiao-pla-06}
L.~Xiao and J.~A. Jones,
\newblock NMR analogues of the quantum Zeno effect,
\newblock Phys. Lett. A {\bf 359}, 424 (2006).

\bibitem{kondo-qph-14}
Y.~Kondo, Y.~Matsuzaki, K.~Matsushima, and J.~G. Filgueiras,
\newblock Using the quantum Zeno effect for suppression of decoherence,
\newblock New. J. Phys. {\bf 18}(1), 013033 (2016).

\bibitem{zheng-pra-13}
W.~Zheng, D.~Z. Xu, X.~Peng, X.~Zhou, J.~Du, and C.~P. Sun,
\newblock Experimental demonstration of the quantum Zeno effect in NMR with
  entanglement-based measurements,
\newblock Phys. Rev. A {\bf 87}, 032112 (2013).

\bibitem{brazil-norm}
A.~Gavini-Viana, A.~M. Souza, D.~O. Soares-Pinto, J.~Teles, R.~S. Sarthour,
  E.~R. deAzevedo, T.~J. Bonagamba, and I.~S. Oliveira,
\newblock Normalization procedure for relaxation studies in NMR quantum
  information processing,
\newblock Quant. Inf. Proc. {\bf 9}, 575 (2009).

\bibitem{hodgson-pra-10}
T.~E. Hodgson, L.~Viola, and I.~D'Amico,
\newblock Towards optimized suppression of dephasing in systems subject to
  pulse timing constraints,
\newblock Phys. Rev. A {\bf 81}, 062321 (2010).

\bibitem{schroeder-pra-11}
C.~A. Schroeder and G.~S. Agarwal,
\newblock Optimized pulse sequences for suppressing unwanted transitions in
  quantum systems,
\newblock Phys. Rev. A {\bf 83}, 012324 (2011).

\bibitem{yang-fpc-11}
W.~Yang, Z.-Y. Wang, and R.-B. Liu,
\newblock Preserving qubit coherence by dynamical decoupling,
\newblock Front. Phys. China {\bf 6}(1), 2--14 (2011).

\bibitem{liu-nc-13}
G.-Q. Liu, H.~C. Po, J.~Du, R.-B. Liu, and X.-Y. Pan,
\newblock Noise-resilient quantum evolution steered by dynamical decoupling,
\newblock Nat. Commun. {\bf 4}, 2254 (2013).

\bibitem{uhrig-prl-09}
G.~S. Uhrig,
\newblock Concatenated control sequences based on optimized dynamic decoupling,
\newblock Phys. Rev. Lett. {\bf 102}, 120502 (2009).

\bibitem{khodjasteh-pra-11}
K.~Khodjasteh, T.~Erd\'elyi, and L.~Viola,
\newblock Limits on preserving quantum coherence using multipulse control,
\newblock Phys. Rev. A {\bf 83}, 020305 (2011).

\bibitem{pan-jpb-11}
Y.~Pan, Z.-R. Xi, and J.~Gong,
\newblock Optimized dynamical decoupling sequences in protecting two-qubit
  states,
\newblock J. Phys. B: At. Mol. Opt. Phys {\bf 44}(17), 175501 (2011).

\bibitem{cong-ijqi-11}
S.~Cong, L.~Chan, and J.~Liu,
\newblock An optimized dynamical decoupling strategy to suppress decoherence,
\newblock Int. J. Quantum Inf. {\bf 09}, 1599--1615 (2011).

\bibitem{alvarez-pra-12}
G.~A. \'Alvarez, A.~M. Souza, and D.~Suter,
\newblock Iterative rotation scheme for robust dynamical decoupling,
\newblock Phys. Rev. A {\bf 85}, 052324 (2012).

\bibitem{west-njp-12}
J.~R. West and B.~H. Fong,
\newblock Exchange-only dynamical decoupling in the three-qubit decoherence
  free subsystem,
\newblock New. J. Phys. {\bf 14}(8), 083002 (2012).

\bibitem{ahmed-pra-13}
M.~A.~A. Ahmed, G.~A. \'Alvarez, and D.~Suter,
\newblock Robustness of dynamical decoupling sequences,
\newblock Phys. Rev. A {\bf 87}, 042309 (2013).

\bibitem{kuo-jmp-12}
W.-Jung, G.~Q., G.~A. Paz-Silva, and D.~A. Lidar,
\newblock Universality proof and analysis of generalized nested Uhrig dynamical
  decoupling,
\newblock J. Math. Phys. {\bf 53}(12) (2012).

\bibitem{wang-pra-11}
Z.-Y. Wang and R.-B. Liu,
\newblock Protection of quantum systems by nested dynamical decoupling,
\newblock Phys. Rev. A {\bf 83}, 022306 (2011).

\bibitem{jing2015}
J.~Jing and L.-A. Wu,
\newblock Overview of quantum memory protection and adiabaticity induction by
  fast signal control,
\newblock Science Bulletin {\bf 60}(3), 328 -- 335 (2015).

\bibitem{jiang-pra-11}
L.~Jiang and A.~Imambekov,
\newblock Universal dynamical decoupling of multiqubit states from environment,
\newblock Phys. Rev. A {\bf 84}, 060302 (2011).

\bibitem{sharf-pra-00}
Y.~Sharf, T.~F. Havel, and D.~G. Cory,
\newblock Spatially encoded pseudopure states for NMR quantum-information
  processing,
\newblock Phys. Rev. A {\bf 62}, 052314 (2000).

\bibitem{breuer-book}
F.~P. Heinz-Peter~Breuer,
\newblock {\em The theory of open quantum systems},
\newblock Oxford University Press, 2002.

\bibitem{lanyon-prl-08}
B.~P. Lanyon, M.~Barbieri, M.~P. Almeida, and A.~G. White,
\newblock Experimental quantum computing without entanglement,
\newblock Phys. Rev. Lett. {\bf 101}, 200501 (2008).

\bibitem{soarespinto-pra-10}
D.~O. Soares-Pinto, L.~C. C\'eleri, R.~Auccaise, F.~F. Fanchini, E.~R.
  deAzevedo, J.~Maziero, T.~J. Bonagamba, and R.~M. Serra,
\newblock Nonclassical correlation in NMR quadrupolar systems,
\newblock Phys. Rev. A {\bf 81}, 062118 (2010).

\bibitem{silva-prl-13}
I.~A. Silva, D.~Girolami, R.~Auccaise, R.~S. Sarthour, I.~S. Oliveira, T.~J.
  Bonagamba, E.~R. deAzevedo, D.~O. Soares-Pinto, and G.~Adesso,
\newblock Measuring bipartite quantum correlations of an unknown state,
\newblock Phys. Rev. Lett. {\bf 110}, 140501 (2013).

\bibitem{silva-prl-16}
I.~A. Silva, A.~M. Souza, T.~R. Bromley, M.~Cianciaruso, R.~Marx, R.~S.
  Sarthour, I.~S. Oliveira, R.~Lo~Franco, S.~J. Glaser, E.~R. deAzevedo, D.~O.
  Soares-Pinto, and G.~Adesso,
\newblock Observation of Time-Invariant Coherence in a Nuclear Magnetic
  Resonance Quantum Simulator,
\newblock Phys. Rev. Lett. {\bf 117}, 160402 (2016).

\bibitem{olliver-prl-01}
H.~Ollivier and W.~H. Zurek,
\newblock Quantum Discord: A measure of the quantumness of correlations,
\newblock Phys. Rev. Lett. {\bf 88}, 017901 (Dec 2001).

\bibitem{horodecki-rmp-09}
R.~Horodecki, P.~Horodecki, M.~Horodecki, and K.~Horodecki,
\newblock Quantum entanglement,
\newblock Rev. Mod. Phys. {\bf 81}, 865--942 (2009).

\bibitem{modi-rmp-12}
K.~Modi, A.~Brodutch, H.~Cable, T.~Paterek, and V.~Vedral,
\newblock The classical-quantum boundary for correlations: Discord and related
  measures,
\newblock Rev. Mod. Phys. {\bf 84}, 1655--1707 (2012).

\bibitem{maziero-pra-09}
J.~Maziero, L.~C. C\'eleri, R.~M. Serra, and V.~Vedral,
\newblock Classical and quantum correlations under decoherence,
\newblock Phys. Rev. A {\bf 80}, 044102 (2009).

\bibitem{xu-nc-10}
J.~Xu, X.~Xu, C.~Li, C.~Zhang, X.~Zou, and G.~Guo,
\newblock Experimental investigation of classical and quantum correlations
  under decoherence,
\newblock Nat. Commun. {\bf 1} (2010).

\bibitem{auccaise-prl-11}
R.~Auccaise, L.~C. C\'eleri, D.~O. Soares-Pinto, E.~R. deAzevedo, J.~Maziero,
  A.~M. Souza, T.~J. Bonagamba, R.~S. Sarthour, I.~S. Oliveira, and R.~M.
  Serra,
\newblock Environment-induced sudden transition in quantum discord dynamics,
\newblock Phys. Rev. Lett. {\bf 107}, 140403 (2011).

\bibitem{paula-prl-13}
F.~M. Paula, I.~A. Silva, J.~D. Montealegre, A.~M. Souza, E.~R. deAzevedo,
  R.~S. Sarthour, A.~Saguia, I.~S. Oliveira, D.~O. Soares-Pinto, G.~Adesso, and
  M.~S. Sarandy,
\newblock Observation of environment-induced double sudden transitions in
  geometric quantum correlations,
\newblock Phys. Rev. Lett. {\bf 111}, 250401 (2013).

\bibitem{fanchini-pra-10}
F.~F. Fanchini, T.~Werlang, C.~A. Brasil, L.~G.~E. Arruda, and A.~O. Caldeira,
\newblock Non-Markovian dynamics of quantum discord,
\newblock Phys. Rev. A {\bf 81}, 052107 (2010).

\bibitem{fanchini-pra-12}
F.~F. Fanchini, E.~F. de~Lima, and L.~K. Castelano,
\newblock Shielding quantum discord through continuous dynamical decoupling,
\newblock Phys. Rev. A {\bf 86}, 052310 (2012).

\bibitem{addis-pra-15}
C.~Addis, G.~Karpat, and S.~Maniscalco,
\newblock Time-invariant discord in dynamically decoupled systems,
\newblock Phys. Rev. A {\bf 92}, 062109 (2015).

\bibitem{luo-pra-08}
S.~Luo,
\newblock Quantum discord for two-qubit systems,
\newblock Phys. Rev. A {\bf 77}, 042303 (2008).

\bibitem{childs-pra-01}
A.~M. Childs, I.~L. Chuang, and D.~W. Leung,
\newblock Realization of quantum process tomography in NMR,
\newblock Phys. Rev. A {\bf 64}, 012314 (2001).

\bibitem{roy-pra-11}
S.~S. Roy, T.~S. Mahesh, and G.~S. Agarwal,
\newblock Storing entanglement of nuclear spins via Uhrig dynamical decoupling,
\newblock Phys. Rev. A {\bf 83}, 062326 (2011).

\bibitem{dur-prl-04}
W.~Dur and H.-J. Briegel,
\newblock Stability of macroscopic entanglement under decoherence,
\newblock Phys. Rev. Lett. {\bf 92}, 180403 (2004).

\bibitem{mintert-pr-05}
F.~Mintert, A.~Carvalho, M.~Ku\'{\i}{s}, and A.~Buchleitner,
\newblock Measures and dynamics of entangled states,
\newblock Phys. Rep. {\bf 415}(4), 207 -- 259 (2005).

\bibitem{aolita-prl-08}
L.~Aolita, R.~Chaves, D.~Cavalcanti, A.~Ac\'{\i}n, and L.~Davidovich,
\newblock Scaling laws for the decay of multiqubit entanglement,
\newblock Phys. Rev. Lett. {\bf 100}, 080501 (2008).

\bibitem{aolita-rpp-15}
L.~Aolita, F.~d.~Melo, and L.~Davidovich,
\newblock Open-system dynamics of entanglement: A key issues review,
\newblock Rep. Prog. Phys. {\bf 78}(4), 042001 (2015).

\bibitem{borras-pra-09}
A.~Borras, A.~P. Majtey, A.~R. Plastino, M.~Casas, and A.~Plastino,
\newblock Robustness of highly entangled multiqubit states under decoherence,
\newblock Phys. Rev. A {\bf 79}, 022108 (2009).

\bibitem{weinstein-pra-10}
Y.~S. Weinstein,
\newblock Entanglement dynamics in three-qubit $X$ states,
\newblock Phys. Rev. A {\bf 82}, 032326 (2010).

\bibitem{carvalho-prl-04}
A.~R.~R. Carvalho, F.~Mintert, and A.~Buchleitner,
\newblock Decoherence and multipartite entanglement,
\newblock Phys. Rev. Lett. {\bf 93}, 230501 (2004).

\bibitem{siomau-eurphysd-10}
M.~Siomau and S.~Fritzsche,
\newblock Entanglement dynamics of three-qubit states in noisy channels,
\newblock Eur. Phys. J. D. {\bf 60}(2), 397--403 (2010).

\bibitem{siomau-pra-10}
M.~Siomau and S.~Fritzsche,
\newblock Evolution equation for entanglement of multiqubit systems,
\newblock Phys. Rev. A {\bf 82}, 062327 (2010).

\bibitem{ali-jpb-14}
M.~Ali and O.~Guhne,
\newblock Robustness of multiparticle entanglement: Specific entanglement
  classes and random states,
\newblock J. Phys. B: At. Mol. Opt. Phys {\bf 47}(5), 055503 (2014).

\bibitem{lanyon-njp-09}
B.~P. Lanyon and N.~K. Langford,
\newblock Experimentally generating and tuning robust entanglement between
  photonic qubits,
\newblock New. J. Phys. {\bf 11}(1), 013008 (2009).

\bibitem{zang-scirep-15}
X.-P. Zang, M.~Yang, F.~Ozaydin, W.~Song, and Z.-L. Cao,
\newblock Generating multi-atom entangled W states via light-matter interface
  based fusion mechanism,
\newblock Sci. Rep. {\bf 5}, 16245 (2015).

\bibitem{he-qip-15}
X.-L. He and C.-P. Yang,
\newblock Deterministic transfer of multiqubit GHZ entangled states and quantum
  secret sharing between different cavities,
\newblock Quant. Inf. Proc. {\bf 14}(12), 4461--4474 (2015).

\bibitem{zang-optics-16}
X.-P. Zang, M.~Y., F.~Ozaydin, W.~Song, and Z.-L. Cao,
\newblock Deterministic generation of large scale atomic W states,
\newblock Opt. Express {\bf 24}(11), 12293--12300 (2016).

\bibitem{barreiro-nature}
J.~T. Barreiro, P.~Schindler, O.~Guhne, T.~Monz, M.~C., C.~F. Roos,
  M.~Hennrich, and R.~Blatt,
\newblock Experimental multiparticle entanglement dynamics induced by
  decoherence,
\newblock Nature {\bf 6}, 943--946 (2010).

\bibitem{wu-qip-16}
J.-L. Wu, C.~Song, J.~Xu, L.~Yu, X.~Ji, and S.~Zhang,
\newblock Adiabatic passage for one-step generation of $n$-qubit
  Greenberger--Horne--Zeilinger states of superconducting qubits via quantum
  Zeno dynamics,
\newblock Quant. Inf. Proc. {\bf 15}(9), 3663--3675 (2016).

\bibitem{suter-3qubit}
X.~Peng, J.~Zhang, J.~Du, and D.~Suter,
\newblock Ground-state entanglement in a system with many-body interactions,
\newblock Phys. Rev. A {\bf 81}, 042327 (2010).

\bibitem{nelson-pra-00}
R.~J. Nelson, D.~G. Cory, and S.~Lloyd,
\newblock Experimental demonstration of Greenberger--Horne--Zeilinger
  correlations using nuclear magnetic resonance,
\newblock Phys. Rev. A {\bf 61}, 022106 (2000).

\bibitem{song-ijqi-13}
H.~Song, Y.~Pan, and X.~Zairong,
\newblock Dynamical control of quantum correlations in a common environment,
\newblock Int. J. Quantum Inf. {\bf 11}(01), 1350012 (2013).

\bibitem{franco-prb-14}
R.~L. Franco, A.~D'Arrigo, G.~Falci, G.~Compagno, and E.~Paladino,
\newblock Preserving entanglement and nonlocality in solid-state qubits by
  dynamical decoupling,
\newblock Phys. Rev. B {\bf 90}, 054304 (2014).

\bibitem{agarwal-scripta}
G.~S. Agarwal,
\newblock Saving entanglement via a nonuniform sequence of $\pi$ pulses,
\newblock Phys. Scr. {\bf 82}(3), 038103 (2010).

\bibitem{Devi2012}
A.~R.~U. Devi, Sudha, and A.~K. Rajagopal,
\newblock Majorana representation of symmetric multiqubit states,
\newblock Quant. Inf. Proc. {\bf 11}(3), 685--710 (2012).

\bibitem{jung-pra-08}
E.~Jung, M.-R. Hwang, Y.~H. Ju, M.-S. Kim, S.-K. Yoo, H.~Kim, D.~Park, J.-W.
  Son, S.~Tamaryan, and S.-K. Cha,
\newblock Greenberger-Horne-Zeilinger versus $W$ states: Quantum teleportation
  through noisy channels,
\newblock Phys. Rev. A {\bf 78}, 012312 (2008).

\bibitem{lindblad}
G.~Lindblad,
\newblock On the generators of quantum dynamical semigroups,
\newblock Commun. Math. Phys. {\bf 48}, 119 (1976).

\bibitem{peres-prl-96}
A.~Peres,
\newblock Separability criterion for density matrices,
\newblock Phys. Rev. Lett. {\bf 77}, 1413--1415 (1996).

\bibitem{vidal-pra-02}
G.~Vidal and R.~F. Werner,
\newblock Computable measure of entanglement,
\newblock Phys. Rev. A {\bf 65}, 032314 (2002).

\bibitem{guhne-review}
O.~G\"{u}hne and G.~T\'oth,
\newblock Entanglement detection,
\newblock Phys. Rep. {\bf 474}(1-6), 1--75 (2009).

\bibitem{souza-phil}
A.~M. Souza, G.~A. \'Alvarez, and D.~Suter,
\newblock Robust dynamical decoupling,
\newblock Phil. Trans. R. Soc. A {\bf 370}, 4748--4769 (2012).

\bibitem{ryan-prl-10}
C.~A. Ryan, J.~S. Hodges, and D.~G. Cory,
\newblock Robust decoupling techniques to extend quantum coherence in diamond,
\newblock Phys. Rev. Lett. {\bf 105}, 200402 (2010).

\bibitem{leung-pra-17}
N.~Leung, M.~Abdelhafez, J.~Koch, and D.~Schuster,
\newblock Speedup for quantum optimal control from automatic differentiation
  based on graphics processing units,
\newblock Phys. Rev. A {\bf 95}, 042318 (2017).

\end{thebibliography}
\end{document}